\documentclass[chicago,apjl]{emulateapj}
\usepackage{apjfonts}
\usepackage{amssymb}
\usepackage{natbib}
\shorttitle{KBSS-MOSFIRE}
\shortauthors{Steidel et al.}
\tighten 
\newcommand{\msun}{\ensuremath{\rm M_\odot}}

\newcommand{\Ha}{\ensuremath{\rm H\alpha}}
\newcommand{\Hb}{\ensuremath{\rm H\beta}}

\newcommand{\kms}{\rm km~s\ensuremath{^{-1}\,}}

\newcommand{\minpoint}{\mbox{$'\mskip-4.7mu.\mskip0.8mu$}}

\newcommand{\secpoint}{\mbox{$''\mskip-7.6mu.\,$}}

\def\ltsima{$\; \buildrel < \over \sim \;$}
\def\simlt{\lower.5ex\hbox{\ltsima}}
\def\gtsima{$\; \buildrel > \over \sim \;$}
\def\simgt{\lower.5ex\hbox{\gtsima}}

\begin{document}

\title{Strong Nebular Line Ratios in the Spectra of $z \sim 2-3$ Star-Forming Galaxies: \\ First Results from KBSS-MOSFIRE\altaffilmark{1}}

\slugcomment{DRAFT: \today}
\author{\sc Charles C. Steidel\altaffilmark{2}, Gwen C. Rudie\altaffilmark{2,3,4}, Allison L. Strom\altaffilmark{2}, Max Pettini\altaffilmark{5}, \\ Naveen A. Reddy\altaffilmark{6,11}, Alice E. Shapley\altaffilmark{7}, Ryan F. Trainor\altaffilmark{2}, Dawn K. Erb\altaffilmark{8}, Monica L. Turner\altaffilmark{9}, \\ Nicholas P. Konidaris\altaffilmark{2}, Kristin R. Kulas\altaffilmark{7,10}, Gregory Mace\altaffilmark{7},
Keith Matthews\altaffilmark{2},  Ian S. McLean\altaffilmark{7}
} 

\altaffiltext{1}{Based on data obtained at the
W.M. Keck Observatory, which
is operated as a scientific partnership among the California Institute of
Technology, the
University of California, and NASA, and was made possible by the generous
financial
support of the W.M. Keck Foundation.
}
\altaffiltext{2}{Cahill Center for Astronomy and Astrophysics, California Institute of Technology, 1216 E. California Blvd., MS 249-17, Pasadena, CA 91125, USA}
\altaffiltext{3}{Carnegie Observatories, 813 Santa Barbara Street, Pasadena, CA 91101, USA}
\altaffiltext{4}{Carnegie-Princeton Fellow}
\altaffiltext{5}{Institute of Astronomy, Madingley Road, Cambridge CB3 0HA, UK}
\altaffiltext{6}{Department of Physics and Astronomy, University of California, Riverside, 900 University Avenue, Riverside, CA 92521, USA}
\altaffiltext{7}{University of California, Los Angeles, Department of Physics and Astronomy, 430 Portola Plaza, Los Angeles, CA 90095, USA}
\altaffiltext{8}{Center for Gravitation, Cosmology, and Astrophysics, Department of Physics, University of Wisconsin-Milwaukee, 1900 E. Kenwood Blvd., Milwaukee, WI 53211, USA}
\altaffiltext{9}{Leiden Observatory, Leiden University, PO Box 9513, 2300 RA Leiden, The Netherlands}
\altaffiltext{10}{NASA Postdoctoral Program Fellow, Ames Research Center, MS 211-1, Moffett Field, CA 94035}
\altaffiltext{11}{Alfred P. Sloan Foundation Research Fellow}

\begin{abstract}
We present initial results of a deep near-IR spectroscopic survey covering the 15 fields of the Keck Baryonic 
Structure Survey (KBSS) using MOSFIRE on the Keck 1 telescope, focusing on a sample of 251 galaxies with redshifts $2.0< z < 2.6$, star-formation rates 
$2\simlt {\rm SFR} \simlt 200$ M$_{\odot}$ yr$^{-1}$, and stellar masses ${\rm 8.6 < log(M_{\ast}/M_{\odot})< 11.4}$,
with high-quality spectra in both H- and K-band atmospheric windows. We show unambiguously that the locus of $z\sim 2.3$ 
galaxies in the ``BPT'' nebular diagnostic diagram exhibits a disjoint, yet similarly tight, 
relationship between the ratios [NII]$\lambda 6585$/\Ha\ and [OIII]/\Hb\ as compared to local galaxies. 
Using photoionization models, we argue that the offset of the $z \sim 2.3$ locus relative to $z\sim 0$ 
is explained by a combination of harder ionizing radiation field, higher ionization parameter, and higher N/O
at a given O/H than applies to most local galaxies, and that the position of a galaxy along the $z\sim2.3$ star-forming BPT 
locus is surprisingly {\it insensitive} to gas-phase oxygen abundance.  The observed nebular emission line ratios 
are most easily reproduced by models in which the net ionizing radiation field resembles 
a blackbody with effective temperature $T_{\rm eff}=50000-60000$ K 
and N/O close to the solar value at all O/H. We critically assess 
the applicability of commonly-used strong line indices for estimating gas-phase metallicities, and consider
the implications of the small intrinsic scatter in the empirical relationship between excitation-sensitive line indices
and $M_{\ast}$ (i.e., the ``mass-metallicity'' relation), at $z \simeq 2.3$. 
\end{abstract}

\keywords{cosmology: observations --- galaxies: evolution --- galaxies: high-redshift --- 
galaxies: abundances -- galaxies: starburst --- galaxies: fundamental parameters }

\section{Introduction}

\label{sec:intro}

In principle, deep near-IR spectroscopy of
high-\textit{z} galaxies offers the possibility of applying the wealth of locally-calibrated and tested rest-frame
optical nebular emission line
diagnostics to directly probe \ion{H}{2} region physics in galaxies as they were forming. 
In practice, however, this potentially-powerful method--  building on well-established
techniques developed over the course of several decades for nearby galaxies-- has been relatively slow to develop. In spite
of substantial observational effort (e.g., \citealt{pettini98,pettini01,erb04,sep+04,erb+06b,kriek08,maiolino08,forst09,mannucci10,henry13,cullen14,troncoso14,wuyts14}),
samples of high redshift galaxies for which a reasonably complete set of strong lines has been measured remain
very small. Moreover, except for gravitationally-lensed examples (e.g., \citealt{teplitz01,hainline09,finkelstein09,jones10,richard11,rigby11,wuyts12,christensen12,jones13,amorin14,james14}), 
the low S/N of the near-IR spectra has limited both the dynamic range and the significance of
observed line ratios for individual objects. The advent of efficient multi-object near-IR spectrographs on 8-10m class telescopes
has long promised to revolutionize nebular spectroscopy of high-redshift galaxies by vastly enlarging the sample
sizes and making very deep spectroscopy observationally practical. 

The suite of nebular emission lines available in the rest-frame optical (i.e., $0.3 \simlt \lambda \simlt 1$ $\mu$m) 
includes probes of density ([OII]$\lambda\lambda3727,3729$ and [SII] $\lambda\lambda6718$,6732),
electron temperature ([O III]$\lambda\lambda$4960,5008/[O III]$\lambda$4364) and ionization state 
(e.g., [O III]$\lambda\lambda$4960,5008/[O II]$\lambda\lambda$3727,3729) as well as the so-called ``strong-line'' metallicity indicators, e.g. those based on 
${\rm \left([OIII]+[OII]\right)/H\beta}$ (``R23''; \citealt{pagel79,kewley02}), ${\rm [N II]\lambda6585/H\alpha}$ (``N2'') and ${\rm ([O III]\lambda5008/\Hb) /([N II]\lambda6585/\Ha)}$ (``O3N2''; \citet{pettini04} [PP04]).
In addition, the \citet{baldwin81} (see also \citealt{vo87}) diagnostic line ratios (``BPT'': [N II]/H$\alpha$ and 
[O III]/\Hb) are commonly used to establish the dominant excitation mechanism of nebular emission in galaxies, providing a 
relatively ``clean'' separation of galaxies whose spectra are dominated by AGN-ionized gas from those ionized primarily by the UV radiation field of young stars 
(e.g., \citealt{kewley01,kauffmann03, brinchmann08}).
Using large samples, primarily drawn from the SDSS spectroscopic database, it has been shown that star-forming galaxies occupy
a relatively tight locus in the BPT plane.  As the earliest samples of high-redshift galaxies with
the relevant measurements became available,
however, there were already indications that distant star-forming galaxies occupy a region of the BPT plane
distinct from that of the vast majority of star-forming galaxies in the local universe 
(\citealt{shapley05z1,erb+06a,liu+shapley08,brinchmann08}). If the initial observations were to hold up
when confronted with much larger samples, it would suggest that using nebular line ratios to measure metallicity and other physical
properties of the high-$z$ \ion{H}{2} regions may be more complex than might have been hoped. 

It is well-known that various nebular diagnostics using strong optical emission lines in galaxy spectra 
can differ substantially-- the most obvious example is systematic differences of up to $\sim 0.77$ dex 
in oxygen abundance
for ostensibly the same set of low-redshift galaxies 
(see e.g. \citealt{kewley+ellison08,maiolino08}.)  
The very different abundance scales depend to a large extent on whether the calibration has been done
using theoretical models (which tend to infer higher O abundances) or empirically, using sensitive observations of weak 
electron temperature sensitive emission lines-- the so-called ``direct'', or ``$T_{\rm e}$'' method. The direct
method is generally considered to provide more reliable results when available, but has the practical disadvantage that it 
requires the detection of very weak emission lines, already challenging for nearby galaxies, becoming rapidly 
more difficult with increasing redshift as the lines become apparently fainter and are redshifted into spectral regions
plagued by much higher terrestrial background. It has also been argued that $T_{\rm e}$-based metallicities may be biased
low due to temperature gradients and/or by the details of the electron energy distribution (e.g., \citealt{stasinska05,dopita13}).  

To place the situation for the determination of nebular oxygen abundances in context, at the highest stellar
masses,  
the asymptotic (i.e., maximum) gas-phase metallicity of star-forming galaxies 
ranges from below solar to nearly 3 times solar (see \citealt{kewley+ellison08}).  
Given the problematic differences in metallicity scale among the many locally-calibrated and/or theoretically derived
``strong-line'' indicators, attempts have been made to implement new calibrations for which all of the various
strong-line methods yield consistent metallicities when applied to large samples of local galaxies (\citealt{kewley+ellison08,
maiolino08}).  
The results have been successful, in the sense that it is possible to force the calibrations
to give the same results (to within $\simeq 0.03$ dex) for the same sample of galaxies (\citealt{kewley+ellison08}),
thus providing confidence that one can at least measure {\it relative} oxygen abundances at $z \simeq 0$. However,
even putting aside our ignorance of the ``correct'' \ion{H}{2} region abundance scales at $z \simeq 0$, it is a separate
issue as to whether the ``re-normalization'' of the strong-line techniques can (or should) be applied to samples of high redshift
galaxies-- clearly this is a desirable possibility, but it has not yet been demonstrated. 
The root of the problem, which is the main topic of this paper, is that measuring line ratios 
and then applying regression formulae established at $z\simeq 0$ will work only if the physics of high-$z$ \ion{H}{2} regions 
resembles that of local star-forming galaxies. If there are substantive physical differences, blind application
of local calibrations will introduce systematics in inferred metallicity; the origins of any systematics are likely
to be fundamental to understanding what drives star formation in rapidly-evolving galaxies at high redshifts. 

At most redshifts $z > 1$, only a subset of   
optical emission lines used by the so-called ``strong-line'' techniques in the local
universe are accessible to ground-based spectroscopy due to significant gaps in the near-IR atmospheric transmission as well
as the increasingly prohibitive thermal background at observed wavelengths of $\simgt 2.3-2.4$ $\mu$m.  
%Unless the physics of HII regions is sufficiently alike those of the low-redshift calibration samples, mapping 
%line intensity ratios in the spectra of high-redshift galaxies to oxygen abundance using zero-redshift 
%calibrations (which themselves have required substantial adjustment to achieve internal consistency) is  
%arguably quite dangerous.  
Potentially most-problematic is comparison of metallicities inferred from 
one set of strong-line indicators for a sample in a particular redshift range, with those based on a different set of
lines at a second redshift.  In such a case, it would be impossible to distinguish between evolution of gas-phase
metallicities and changes (for other physical reasons) in the dependence of the measured line intensity ratios on metallicity. 

\begin{figure}[htbp]
\centerline{\includegraphics[width=9.0cm]{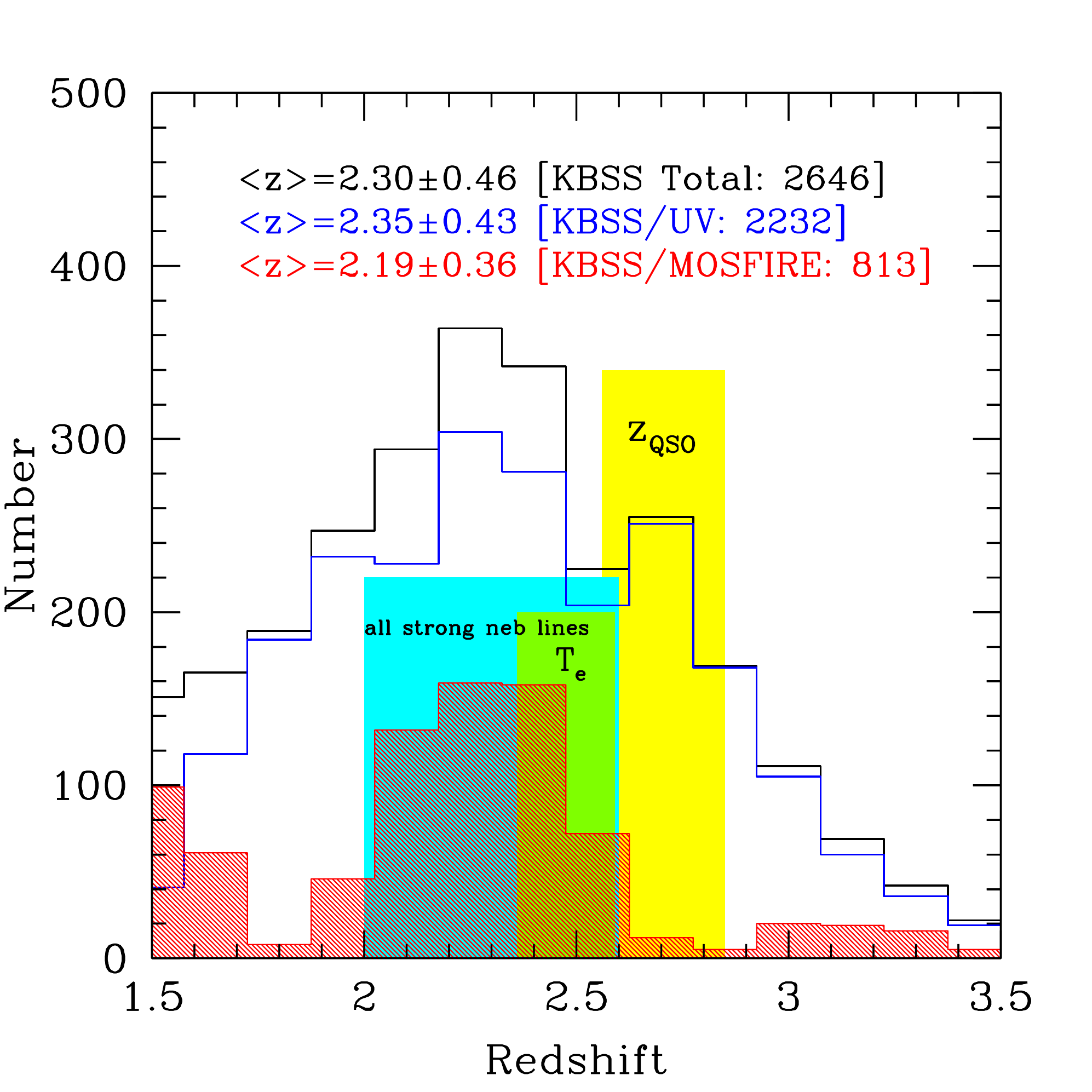}}
\caption{Redshift histogram in the KBSS survey regions as of 2014 June. The unshaded black
histogram shows the redshift distribution of 2646 spectroscopically confirmed (using MOSFIRE and/or LRIS-B)
galaxies in the 15 KBSS survey fields; the 
blue histogram shows the redshift distribution of the subset with rest-UV spectra from
Keck/LRIS-B, 
the (red) shaded histogram shows the
distribution of nebular redshifts obtained with MOSFIRE, and the black unshaded histogram is the KBSS
total with spectroscopic redshifts. The cyan-shaded region schematically illustrates 
the redshift range over which the targeted suite of strong nebular emission lines falls within
the ground-based near-IR atmospheric windows ($2 \simlt z \simlt 2.6$); the green shading (labeled ``$T_{\rm e}$'')
corresponds to the
subset of the cyan region over which the electron temperature sensitive
[OIII]$\lambda 4364$ line is accessible in H band ($2.36 \le z \le 2.57$; see section~\ref{sec:obs}). The yellow shading shows
the redshift range of the very bright background QSOs in the KBSS fields; while these are not directly
relevant to the topic of this paper, their lines of sight provide extremely sensitive measurements
of \ion{H}{1} and metals in the circum-galactic (CGM) and intergalactic medium (IGM) surrounding KBSS survey
galaxies (see \citealt{rudie12a,rakic12,trainor12,turner14}). 
\label{fig:zhist_kbss}
}
\end{figure}

A better statistical lever-arm, initially independent of the low-redshift calibrations, can be constructed using observations
of high redshift galaxies selected in special redshift intervals for which a relatively complete set of  
the rest-optical nebular lines falls fortuitously within the near-IR atmospheric 
windows for ground-based spectroscopy. Perhaps the best such interval is 
$2.0 \simlt z \simlt 2.6$ (e.g., \citealt{erb+06b}), where \Ha, [NII], and [SII] fall  
in the K band, [OIII] and \Hb\ fall in H, and [OII] and [NeIII] in J. In large part for this reason, 
the Keck Baryonic Structure Survey (KBSS; see \citealt{rudie12a,trainor12,rakic12}) has focused on galaxies in this redshift range  
over the past several years (e.g., \citealt{adelberger04,steidel04,erb+06a,erb+06b,erb+06c,shapley05,reddy08a}). 
Figure~\ref{fig:zhist_kbss} shows the current KBSS spectroscopic redshift distribution and schematically illustrates 
the high priority redshift ranges targeted by the current work.  

KBSS provides
a wealth of multi-wavelength ancillary data as well as a large
sample of spectroscopically-identified galaxies (primarily using Keck/LRIS-B) with $1.5 \simlt z \simlt 3.5$.    
In this paper,  we
present initial results based on new multiplexed near-IR (rest-frame optical) spectroscopy obtained in the KBSS survey regions, focusing on
$2.0 \le z \le 2.6$, for the reasons outlined above and summarized in Figure~\ref{fig:zhist_kbss}. 

The paper is organized as follows: section~\ref{sec:obs} describes the new observations and the properties of the initial KBSS-MOSFIRE sample; 
section 3 compares the locus
of relative emission line intensities of $z \sim 2.3$ galaxies with samples of galaxies in the local universe, showing 
very distinct differences between the two. Section~\ref{sec:bpt_interpretation} attempts to explain the principal cause of the change in the diagnostics, 
with the aid of photoionization models. Section~\ref{sec:agn} briefly examines the extent to which the observed strong emission line ratios (the ``BPT'' diagram)
can be used at high redshift to discriminate between hot young stars and AGN as ionizing sources;
Section~\ref{sec:analogs} identifies likely local analogs of the high-redshift sources and compares them to the most
extreme galaxies in the high redshift sample, as a means of forecasting what more sensitive
observations might yield. Section~\ref{sec:mass_met} revisits the 
relationship between stellar mass and inferred metallicity (the ``Mass-Metallicity'' relation, or ``MZR'') at $z \sim 2.3$, and briefly
addresses the extent to which the new KBSS-MOSFIRE sample supports the concept of a ``fundamental metallicity relation'' similar
to that observed in the local universe. Finally, section 8 summarizes the main results, and discusses their implications for
metal enrichment and star formation in galaxies near the peak of the galaxy formation epoch.  

Throughout the paper, we assume a $\Lambda-$CDM cosmology with $H_0=70$ \kms Mpc$^{-1}$, $\Omega_{\Lambda}=0.7$,
and $\Omega_{\rm m} = 0.3$, a \citet{chabrier03} stellar initial mass function (IMF), and the solar metallicity scale
of \citet{asplund09}, for which ${\rm 12+log(O/H) = 8.69}$

\begin{figure*}[htbp]
\centering
\includegraphics[width=17cm]{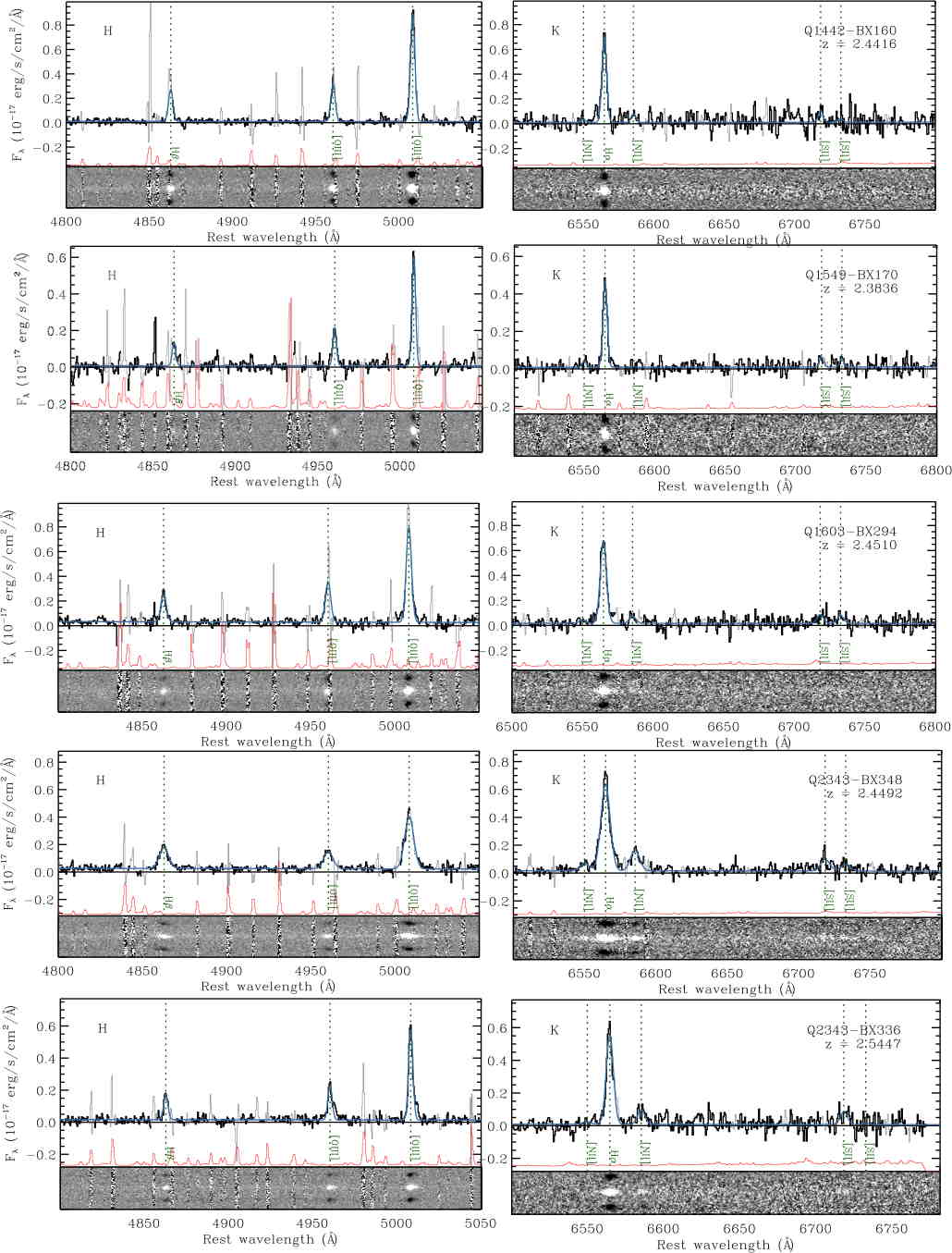}
\caption{Portions of MOSFIRE H-band (left) and K-band (right) spectra for 10 of the KBSS galaxies listed in Tables~\ref{tab:n2ha_and_o3n2} 
and \ref{tab:n2ha_lim_o3hb_det}. The flux-calibrated spectra are presented {\it unsmoothed}, with their original pixel sampling, with
the wavelength scale shifted to each galaxy's rest frame. The best-fit 
line profiles are superposed (blue), while the $1\sigma$ error spectrum (red) is offset, but on the same flux scale, as its corresponding
galaxy spectrum. 
The stacked two-dimensional spectra from which the 1-d spectra were extracted are shown in grayscale, over the same range of rest-wavelength. 
Each reduced 2-D spectrogram exhibits a positive (central) image and 2 flanking negative images due to the differencing of spatially dithered
exposures (see section~\ref{sec:observations}) that is part of the background subtraction procedure.} 
\end{figure*}
\begin{figure*}
\setcounter{figure}{1} 
\includegraphics[width=17cm]{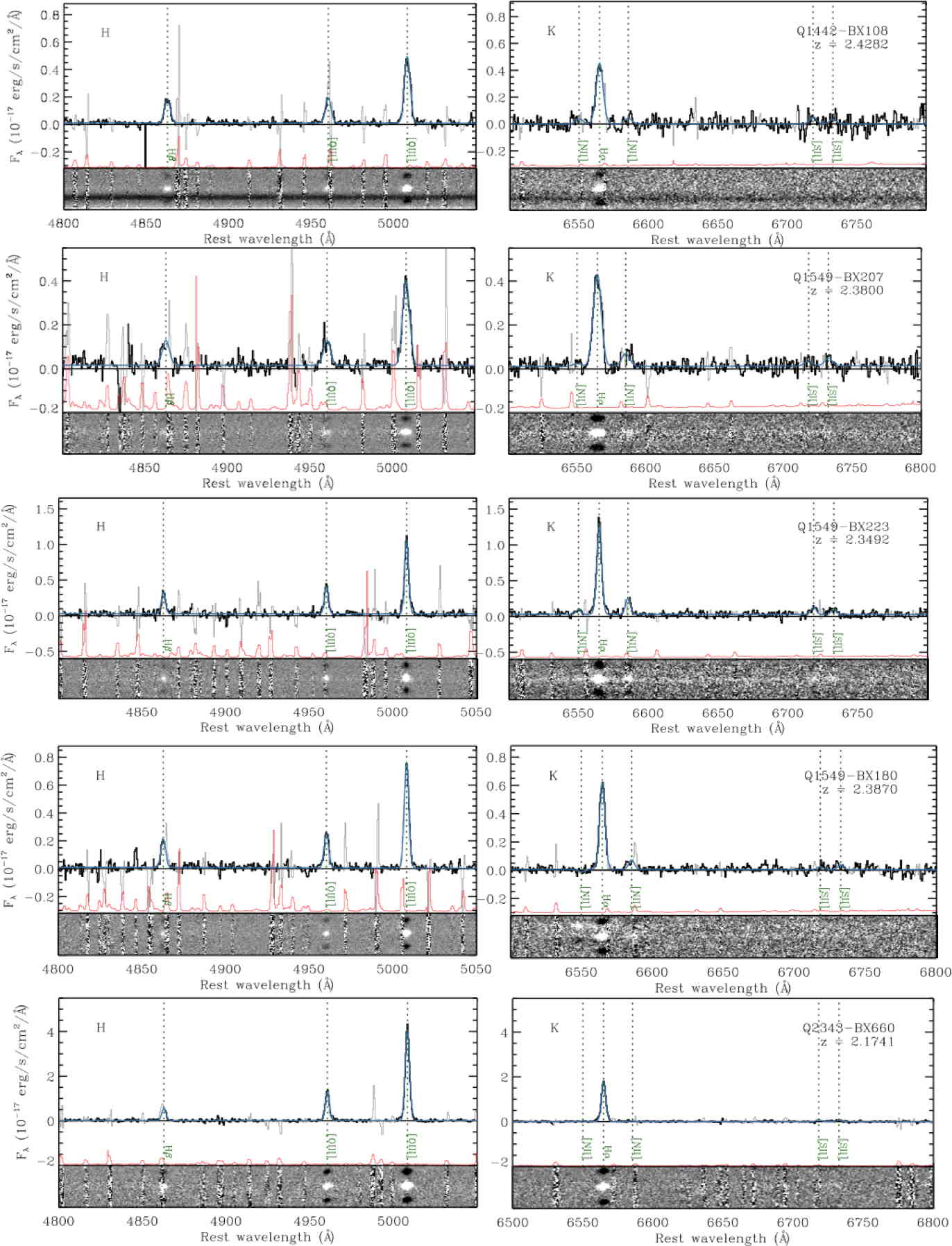}
\caption{(Continued)}
\label{fig:spectra}
\end{figure*}

\section{Observations and Data}

\label{sec:obs}

All near-IR spectroscopic observations described in this paper were obtained using 
the Multi-Object Spectrometer for InfraRed Exploration (MOSFIRE; \citealt{mclean10,mclean12}), the recently
commissioned near-IR imaging spectrometer on the Keck 1 10m telescope at the W.M. Keck Observatory on Mauna Kea.  
Some of the data were obtained during MOSFIRE commissioning science verification in 2012 May and June, 
with the remainder obtained during early science observations in 2012 September and October and 2013 March, May, June,
and November. 

\subsection{Target Selection and Survey Strategy}

Over the course of MOSFIRE commissioning and early science observations, we developed
an observing strategy that takes advantage of the unique capabilities of the instrument 
in order to achieve multiple scientific goals.  
The combination of the compact KBSS field geometry (typically 7\minpoint5 by 5\minpoint5)
with the flexibility of MOSFIRE's electronically 
re-configurable cryogenic focal plane mask (the ``Configurable Slit Unit'', or CSU) 
lends itself to a ``tiered'' approach to the near-IR survey, combining routine and difficult
observations on the same masks. Because most of the KBSS fields are only slightly larger than the $6\minpoint1\times6\minpoint1$
MOSFIRE field of view,
there is significant spatial overlap of every mask within a given field. By repeatedly observing masks with
similar footprint but distinct sets of objects, we ensure that all high priority targets are observed
and that the geometrical constraints imposed by slitmasks do not limit the sampling of targets on small
angular scales. 
At the same time, if very deep spectra are required to detect weak emission lines
(e.g., auroral [OIII] $\lambda 4364$, or [NII]$\lambda 6585$ in galaxies with very low metallicity), 
the same target is repeated on multiple masks, thereby accumulating much longer total 
total integration times (more than 10 hours in some cases).

The objects in the parent catalog for each KBSS field were assigned numerical priorities 
based on multiple criteria: the highest priorities were given to galaxies known from previous 
spectroscopic observations to lie in narrow redshift range  
$2.36 \le z \le 2.57$ -- the range over which the set of strong emission lines (including [\ion{S}{2}]$\lambda\lambda6718$, 6732) 
as well as [OIII]$\lambda 4364$
are all accessible within the near-IR atmospheric windows (Figure~\ref{fig:zhist_kbss}). 
These would be the initial candidates to appear 
on multiple masks, since (for example) the flux of the $T_{\rm e}$-sensitive ${\rm [OIII]\lambda 4364}$ line is expected
to be $\simgt 50$ times smaller than that of [OIII]$\lambda 5008$.  
The design of a series of masks in a given field proceeded by keeping the highest priority targets on each, 
and assigning the rest of the available slit ``real estate'' to different targets according to their relative numerical priorities.  
A typical mask included 10-15 such fixed targets, out of a total $\simeq 30-35$ slits.  

For the other 20-25 targets on each mask, initial priorities were assigned  
based on the following criteria, from 
highest to lowest: (1) those with existing high quality UV spectra and known redshifts $2 \simlt z \simlt 2.6$, 
weighted according to their angular separation from the central QSO sightline (2) those
flagged as probable high-stellar-mass targets
in the redshift range $1.5 \le z \le 2.5$, selected using joint optical/near-IR photometric
criteria (3) ${\cal R} \le 25.5$ UV color selected galaxies expected to have redshifts
within the optimal $2 \simlt z \simlt 2.6$ range (in practice, these are the ``BX'' and ``MD'' objects 
defined by \cite{adelberger04,steidel04} and \cite{steidel03}) (4) rest-UV color-selected galaxies 
judged likely to have redshifts $z < 2$ or $z > 2.6$ from their rest-UV colors,  
but not yet confirmed spectroscopically, and (5) UV color-selected galaxies with
${\cal R} > 25.5$ (when the depth of the optical photometry allows). Note that category (2) includes galaxies 
that satisfy the UV color selection 
criteria and have red optical/IR colors $({\cal R}-K_s)_{\rm AB} > 2$, 
as well as those with UV colors redder than those of BX/MD galaxies
and ${\rm ({\cal R}-K_s)_{AB} > 2}$. Empirically, we have found that the latter criteria, indicated with the prefix ``RK'', 
identify more heavily-reddened galaxies, with relatively large M$_{\ast}$ and $1.4 \simlt z \simlt 2.5$,  
that would otherwise not be included in our spectroscopic samples. 
The ``RK'' sample and its statistical properties are discussed in more detail by Strom et al (in prep.)    
The main purpose of including category (2) targets is to improve the sample statistics for 
galaxies with ${\rm log(M_{\ast}/M_{\odot}) > 10.5}$.  

Since MOSFIRE mask configurations can be updated easily (and electronically) during an observing
run, and the MOSFIRE-DRP (developed by us)   
produces pipeline-processed 2-D ``stacks'' in nearly-real-time,
the overall efficiency and scientific return of the survey is optimized through 
quantitative assessment of the data immediately after an observing
sequence has completed (generally 20x180s for K band, 30x120s for J or H-band). The results 
are then used to modify target lists for subsequent masks, 
performing a running ``triage'', in which we remove objects with $z < 2$ or $z > 3$
after they have been successfully identified (replacing them with a new set of targets according to the 
aforementioned priorities), and evaluate the need for additional integration time for targets 
in the optimal redshift range. 

A significant fraction of targets (both within and outside of the optimal redshift range $2.0 \simlt z \simlt 2.6$)
requires only 1 hour total integration in a given band to produce spectra
of sufficient quality to yield precise nebular redshifts, line widths, and strong-line ratios.
However, many targets prove more difficult;  
our unique iterative procedure is used to ensure that the highest priority or most difficult targets
receive the longest total integration times (up to $\simgt 10$ hours),  that galaxies having useful 
diagnostics in multiple atmospheric bands
are observed in multiple bands, and that minimal time is spent observing objects that do not further the
scientific goals. 
Thus, the total integration times for observations presented here span a wide range: 
${\rm 3578s < t_{\rm exp} < 29100s}$ in H-band and ${\rm 3578s < t_{exp} < 43700s}$ in K-band. 
The median (average) total integration times for galaxies appearing
in Tables 1 and 2 were 8350s (10780s) and 8950s (11520s) in H and K bands, respectively; objects listed in Table 3 
have median (average) K-band integration times of 5368s (6810s).  

MOSFIRE observations have been acquired in all 15 KBSS survey regions, though at the time the current sample
was finalized a few of the fields had been observed to the desired depth in only 1 band, usually H\footnote{Experience has shown that H-band observations, in addition
to having the best sensitivity per unit integration time, are most likely
to yield spectroscopic redshifts for galaxies without previous spectroscopic identifications, since \Ha\ 
falls in the band for $1.2 \simlt z \simlt 1.74$, [\ion{O}{3}]$\lambda 5008$ for $1.85 \simlt z \simlt 2.59$, and
[OII]$\lambda\lambda 3727,3729$ for $2.92 \simlt z \simlt 3.83$. }.

\subsection{MOSFIRE Instrumental Details}

\subsubsection{Overview}
\label{sec:observations}

MOSFIRE obtains spectra of up to 46 objects simultaneously within a 6\minpoint1 x 6\minpoint1 field of view at
the f/15 Cassegrain focus of the Keck 1 10m telescope. 
For H and K band spectroscopic observations, a custom made gold-coated reflection grating with 110.5 lines mm$^{-1}$ 
is used in orders 4 and 3, respectively, providing wavelength coverage of 
$1.465-1.799\mu$m (H) and $1.953-2.398\mu$m (K) for slits in the center of the field of view.  
Although the MOSFIRE CSU can configure up to 46 slits anywhere across the 6\minpoint1 field, in practice
masks observed for our program included $28-34$ targets distributed within the central 6\minpoint1 by 3\minpoint0
field of view of the instrument, in order to ensure a large swath of common wavelength coverage for each slit. 
All masks were designed with slit widths of 0\secpoint7, with lengths in the
range 7\secpoint0$-$23\secpoint0\footnote{MOSFIRE slit lengths are quantized, with lengths $(8.0\times N-1.0)$ arcsec, 
where $N$ is the integer number of masking bars comprising
the slit.}. With 0\secpoint7 slits, MOSFIRE achieves spectral resolution of $R = 3690$ (3620) in the $K$ (H) atmospheric
bands, sampled with 2.172 (1.629) \AA\ pix$^{-1}$ in the dispersion direction, and 0\secpoint18 pix$^{-1}$ spatially.   
The anamorphic magnification of the spectrometer layout is such that a spectral resolution element is sampled
with $\sim 2.7$ pixels at the MOSFIRE detector.

MOSFIRE observations were acquired using a 2-position nod sequence separated by 3\secpoint0 along slits; individual
integrations were 180s and 120s for K and H band, respectively, usually obtained in sequences of $\sim$ one hour total integration time. 
MOSFIRE's Hawaii-2RG detector was read out using Fowler sampling with 16 read pairs, resulting in effective read 
noise of $\simeq 5.3$ electrons (rms).  The decision to use 2-3 minute individual integration times between nods was based
on significant
experimentation with temporal sampling, readout modes, and dither strategies optimized 
for faint-object spectroscopy; we have recommended the same strategy to other MOSFIRE users via the 
MOSFIRE web documentation\footnote{
\href{http://www2.keck.hawaii.edu/inst/mosfire/exposure\_recipes.html}{http://www2.keck.hawaii.edu/inst/mosfire/exposure\_recipes.html}}.

By design, 
the integration times used for individual MOSFIRE exposures 
are sufficient to yield background-limited
performance in spectral regions free of strong OH night sky lines, but are short enough to mitigate the effects 
of the strong and highly-variable OH emission lines on accurate background subtraction. 
The dark current of the MOSFIRE detector is negligible
($< 0.008$ electrons s$^{-1}$ pixel$^{-1}$) relative to the inter-OH background ($\simeq 0.2-0.3$ electrons s$^{-1}$ pixel$^{-1}$.)   
Using the ABAB dither sequence of short individual exposures and combining frames taken in positions A and B separately,
quite good background subtraction is obtained by simple subtraction (i.e., A-B or B-A) because they have been obtained 
quasi-simultaneously; residuals are generally seen only in the OH emission lines, which vary significantly
on timescales shorter than the 120s (or 180s) nod time. The differencing has the advantage of removing many systematics
that would otherwise cause problems for background subtraction, and requires no fitting or re-sampling of the data. 
In spite of its Cassegrain location, MOSFIRE is very stable thanks to active flexure compensation that maintains 
the spectral format fixed with respect to the detector at the level of better than 0.05 pixels (rms) over the course of 
a typical 1-2 hour exposure sequence. 

\subsubsection{Pipeline Data Reduction}

\label{sec:twodspec}

The MOSFIRE data reduction pipeline (DRP; described in more detail below) performs the background subtraction in two stages, of
which the first is the 
simple pairwise subtraction of the interleaved, dis-registered stacks just described. This is followed by fitting a 2-D b-spline
model to the background residuals only, using a method similar to that described by \cite{kelson03}.  
We have found that the combination yields background residual errors consistent with counting statistics 
even in the vicinity of strong OH emission lines.

\begin{figure}[htpb]
\centerline{\includegraphics[width=9cm]{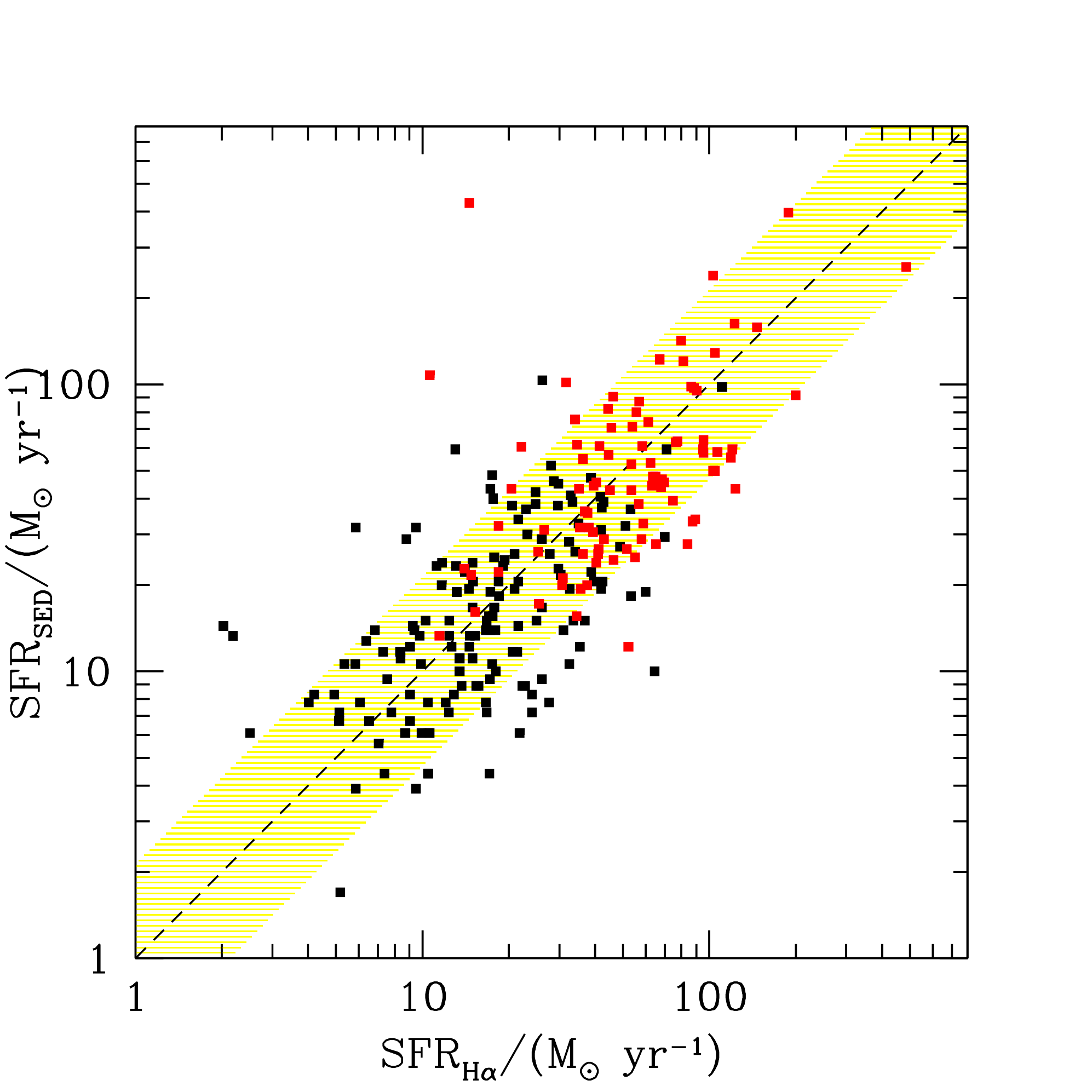}}
\caption{Comparison of star formation rates estimated from SED fitting (SFR$_{\rm SED}$) 
with those based on the \Ha\ luminosity (SFR$_{\Ha}$). The dashed line 
indicates equality between SFR$_{\Ha}$ and SFR$_{\rm SED}$. Both estimates assume a
\cite{chabrier03} IMF, with \cite{calzetti00} attenuation for the stellar continuum and nebular 
extinction as given in equations~\ref{eqn:part1} and \ref{eqn:part2}. Galaxies with ${\rm E(B-V)_{cont} > 0.2}$ are
shown with red points.  
The median
SFR$_{\rm sed}$ and SFR$_{\rm \Ha}$ agree to within $\sim 5$\%, with a median absolute deviation of $\simeq 0.20$ dex
(indicated with the shaded region). 
\label{fig:sfr_v_sfr}
}
\end{figure}

MOSFIRE data were reduced using the publicly-available data reduction pipeline (DRP) developed by the instrument team\footnote{See
\href{http://code.google.com/p/mosfire/}{http://code.google.com/p/mosfire/}}. The
MOSFIRE DRP produces flat-fielded, wavelength calibrated, rectified, and stacked 2-D spectrograms for each slit on
a given mask. The 2-D wavelength solutions
in H band were obtained from the night sky OH emission lines for each slit, while in K band a combination of night sky and Ne arc lamp spectra
was used. The typical wavelength solution residuals were $0.08$ \AA\  (K) and
$0.06$ \AA\ (H), or $\simeq 1.1$ \kms (rms). All spectra were reduced to vacuum wavelengths and corrected for the heliocentric
velocity at the start of each exposure sequence prior to being combined using inverse-variance weighting to form the final 2-D spectra.  
One-dimensional (1-D) spectra, together with their associated 1$\sigma$ error vectors, were extracted from the final background-subtracted,
rectified spectrograms using ``MOSPEC'', an IDL-based 1-D spectral analysis tool developed specifically for analysis of
MOSFIRE spectra of faint galaxies (see Strom et al., in prep., for a full description.) 
Figure~\ref{fig:spectra} shows relevant portions of reduced 1-D and 2-D spectra for 10 of the galaxies in the KBSS-MOSFIRE sample 
discussed below, chosen to span the range of line flux, excitation, and total integration time among the full sample discussed below. 

\subsubsection{Extraction and 1-D Spectral Analysis}
\label{sec:onedspec}

In brief, MOSPEC extracts the 1-D spectrum and its associated $1\sigma$ error spectrum, applying flux calibration 
and telluric correction based on wide-slit and narrow-slit observations, respectively, of A0V stars (i.e., Vega analogs). 
Continuum levels were estimated using the best-fit stellar population synthesis model spectrum after re-normalizing it to match
the median observed continuum level; in cases where continuum was not significantly detected, a low-order polynomial fit
excluding the positions of known emission lines was used instead. The advantage of the first method is that the contribution
of stellar Balmer absorption features coincident with Balmer emission lines is accounted for consistently, since the emission line
intensities are measured relative to the continuum level including the suppression by stellar absorption features. 
The typical effect of including the Balmer absorption on the measurement of the \Hb\ emission line
strength is to increase it by $< 10\%$, and is generally negligible for objects having strong emission lines but very weak continua. 
Once the continuum level is established, MOSPEC 
performs a simultaneous fit to a 
user-specified set of emission lines; outputs include redshift, line flux, line width, and the associated 
uncertainties.  The relative intensities of the [NII]$\lambda \lambda 6549$, 6585 and [OIII]$\lambda\lambda 4960$, 5008 were held fixed
at 1:3.  The fitted line profiles within a given observed band were constrained
to have the same redshift and velocity width; Gaussian profiles were 
found to provide good fits to the data except in cases with very high S/N ($\simgt 50$), 
where small departures of line shapes from Gaussian may yield statistically
significant residuals relative to the models. In such cases, the line intensities and significance were also estimated from direct integration
of the line profiles and error vectors within $\pm 3.0 \sigma$ of the line center derived from the Gaussian fit. 
In most such cases, the best-fit Gaussian line profile and direct integration yield line intensities that agree to better than a few percent.
For well-detected objects, the statistical uncertainty on measured redshifts was $\sigma_z \simeq 1-10$ \kms\ (depending on line
width and S/N), and the agreement in redshift between the independently-fitted H and K-band spectra was typically $\Delta z \simlt 0.0001$,
i.e., $\Delta {\rm v} \simlt 10$ \kms\ (rms). Similarly, the independently-fitted line widths in the H and K bands 
agree with one another to within $5-10$ \kms\ for typical line widths $\sigma_v \simeq 100$ \kms.

\subsubsection{Sensitivity}
\label{sec:sensitivity}

The sensitivity of the MOSFIRE spectra for detection of relatively narrow nebular emission lines is 
of course strongly wavelength-dependent even when sky subtraction systematics have been eliminated; the detection
sensitivity for a given spectral feature can also be time-dependent, as the intensity of OH emission lines 
can vary by up to an order of magnitude over the course of an observing night, and a line's velocity 
with respect to OH emission lines changes with variations in the heliocentric velocity at the time of the observations. 
MOSFIRE's relatively high spectral resolution, dark inter-line background (0.2-0.3 e$^{-}$ pix$^{-1}$), high system
throughput ($\simeq 40$\% on the grating blaze in H and K bands), and fast optics (so that background-limited observations
are achieved in short integrations times) have all been optimized by design for spectroscopy
of faint objects. Thus, we find that, in spectral regions free of strong OH emission and under typical  
observing conditions, the limiting (5$\sigma$, 1 hour) emission line flux (assuming the median line width of FWHM$ \simeq 240$ \kms\ and 
a typical spatial extraction aperture of 7-9 pixels [$\simeq 1\secpoint25-1\secpoint62$]) is $3.5\times10^{-18}$ ergs s$^{-1}$ cm$^{-2}$ ($4.5-14 \times 10^{-18}$ ergs s$^{-1}$ cm$^{-2}$) in H-band (K-band\footnote{In K-band, the sky continuum level rises monotonically for $\lambda \simgt 2.2\mu$m; the 
brighter limiting flux is appropriate to the red edge of the band, near $2.4\mu$m.}).  
The corresponding limiting fluxes for the median total integration times discussed above ($\sim 2.3$ and  $\sim 2.5$ hours in H, K, respectively) 
are $\simeq 2.6\times10^{-18}$ (H-band) and $\simeq 2.7-8.5 \times 10^{-18}$ ergs s$^{-1}$ cm$^{-2}$ (K-band). These sensitivities are within
$\simeq 10$\% of those predicted by the MOSFIRE exposure time calculator XTcalc\footnote{\href{http://www2.keck.hawaii.edu/inst/mosfire/etc.html}{http://www2.keck.hawaii.edu/inst/mosfire/etc.html}.} which we developed during commissioning and subsequently made publicly available.

\subsubsection{Spectroscopic Sample Definition}
\label{sec:sample_definition}

In this paper, we focus on the subset of KBSS-MOSFIRE galaxies with nebular redshifts 
$1.95 \simlt z \simlt 2.65$ and sufficiently deep 
K- and H-band spectra to allow significant detections or useful limits for a minimum set of emission lines\footnote{All wavelengths are in
vacuum.}: \Ha~$\lambda 6564.61$, [NII]~$\lambda
6585.27$ (in the K band), and \Hb~$\lambda 4862.72$, [OIII]~$\lambda\lambda 4960.30$, 5008.24 (in the H-band). 
A measurement was considered a ``detection'' when the statistical significance of both [OIII]$\lambda5008$ and \Ha\ was
$> 5\sigma$, and that of both \Hb\ and [NII]$\lambda 6585$ $> 2 \sigma$. Undetected \Hb\ and/or [NII] lines were flagged as limits
and assigned a flux upper limit of $+2\sigma$.   
The 168 galaxies in the targeted redshift range for which all features satisfy the criteria for detection  
are listed in Table~\ref{tab:n2ha_and_o3n2}.
%, and are represented in Figure~\ref{fig:bpt} as solid (dark) circles with error bars.  

In practice, the most difficult of the BPT lines to detect is the [NII]$\lambda 6585$ feature, whose intensity is 
typically only 10\% that of \Ha, and can be substantially weaker 
in the most metal-poor galaxies (e.g., \citealt{erb+06a}; see 
Figure~\ref{fig:bpt}). Because of this, the KBSS-MOSFIRE sample contains a significant number of galaxies for which 
[OIII], \Ha, and \Hb\ are well-detected according to the above criteria, 
but only upper limits have been measured for [NII]$\lambda 6585$. 
These 51 galaxies are listed in Table~\ref{tab:n2ha_lim_o3hb_det}.
%, and are indicated in Figure~\ref{fig:bpt} with lighter filled triangles and left-pointing 
%arrows indicating $2\sigma$ upper limits on log([NII]/\Ha).  
%The current KBSS-MOSFIRE 
%sample with both H and K band measurements includes only one case with a measurement of [\ion{N}{2}]/\Ha\ and [\ion{O}{3}] but only a
%lower limit on [\ion{O}{3}]/\Hb: Q1603-BX191, which is identified as an AGN (see discussion in section~\ref{sec:agn}.) 

For some purposes in what follows below, we have made use of additional KBSS-MOSFIRE galaxies 
with $1.95 \simlt z \simlt 2.65$ and measurements of \Ha\ and [\ion{N}{2}] from MOSFIRE K-band observations, for which comparable H-band
observations have not yet been obtained. These 32 galaxies are listed separately in Table~\ref{tab:n2ha}.

\subsection{Stellar Masses and Star Formation Rates}

\label{sec:mstar_and_sfr}

We assigned stellar masses (M$_{\ast}$) to the KBSS-MOSFIRE galaxies using
population synthesis SED fits based on photometry in the
optical ($U_nG{\cal R}$), near-IR ($K_s$, J, and, for a subset, WFC3-IR F160W), and {\it Spitzer}/IRAC (3.6$\mu$m and/or 4.5$\mu$m, for
all but one of the KBSS fields) bands. 
Prior to performing the SED fits, the near-IR photometry was corrected for the contribution of \Ha\ and [OIII] emission lines to the broadband fluxes using 
the spectroscopically measured values.
The population synthesis method used is described in detail by (e.g.) \cite{shapley05,erb+06b,reddy12}; for the current sample we 
adopted the best-fit stellar masses using the \cite{bc03} models assuming constant SFR and 
extinction according to \cite{calzetti00}.  As discussed in detail by \citet{shapley05} and \citet{erb+06b}, typical uncertainties in
${\rm log(M_{\ast}/M_{\odot})}$ are estimated to be $\pm 0.1-0.2$ dex. 
Inferred stellar masses and SFRs throughout  
this paper assume a \citet{chabrier03} IMF for ease of comparison with the majority of other galaxy samples considered. For a \citet{salpeter55} IMF, 
both values would be larger by a factor of 1.8.
SFRs were derived  
from the observed \Ha\ line fluxes after correcting for slit losses and nebular extinction, as described below. 

\subsubsection{Slit Loss Corrections}
\label{sec:slit_loss}

The typical galaxy in our spectroscopic sample has an intrinsic half-light radius $r_{\rm e} \simeq 1.6$ kpc, or  
$\simeq 0\secpoint2$ at $z \simeq 2.3$ (\citealt{law2012}), so that light losses at the 0\secpoint7 MOSFIRE
entrance slits are modulated primarily by the seeing during an observation, which 
was generally in the range $0\secpoint35-0\secpoint7$, with a median value of $\simeq 0\secpoint6$ 
(FWHM). For a point source centered in a 0\secpoint7 slit, the fraction of light falling
outside the slit is $\simeq 20$\% for Gaussian seeing with FWHM$=0\secpoint6$, assuming that the extraction aperture
in the slit direction is sized to include the whole spatial profile. In practice, 
most of the $z \simeq 2.3$ galaxies are only marginally
resolved in $\simeq 0\secpoint6$ seeing, with spatial profiles that may be both non-Gaussian and asymmetric, so
that slit losses for galaxies are expected to be larger than for true point sources. 
Wherever possible, two estimates of the slit loss correction (SC) were made for each object;
the first, which we call the ``2-D profile method'', used a Gaussian fit to the observed
spatial profile of \Ha\ emission along the slit to calculate the fraction of the total contained within the
aperture defined by the slit width of 0\secpoint7 and the extraction aperture, which was adjusted interactively to include the full
spatial profile along the slit, with a median value of 8 pixels
($\simeq 1\secpoint44$).  
The 2-D profile method has the advantage
that it accounts for the actual size of the galaxy image at the slit, averaged over the full duration of an observation, but has the disadvantage that the true 2-D spatial
profile of \Ha\ emission is generally unknown and may not be symmetric as assumed. 
A second estimate of the slit loss was made for objects having significant continuum flux measured in the spectra ($\sim 70$\% of the
sample). In this case, the slit loss correction ${\rm SC_{sed}}$ was 
was obtained using the  
scale factor between the observed spectroscopic K band continuum level and the median flux density 
of the best-fit stellar population synthesis model
over the same spectral range, measured in the continuum fitting procedure described above.  
This method of measuring slit losses
(essentially, by comparing to external photometry) accounts for both slit losses and (if relevant) any differences in observing
conditions between the science observations and the spectrophotometric calibration star, whereas the 2-D profile method 
alone provides only a relative, ``geometric'' correction to the observed flux.  
However, ${\rm SC_{sed}}$ explicitly assumes that the spatial distribution of line emission (the quantity one is interested in
correcting) is the same as that of the near-IR continuum starlight (to which one is fitting the SED models), which need not be the case.  
In addition, for continuum-faint galaxies, the determination of ${\rm SC_{sed}}$ can be quite noisy in the face of
systematics in the background subtraction on a given slit.    

For objects yielding measurements of both ``geometric'' and ``absolute'' slit correction estimates, they agree
reasonably well, with median values ${\rm \langle SC_{2D}\rangle = 1.54 \pm 0.24}$ and  ${\rm \langle SC_{sed} \rangle 
= 2.11\pm0.56}$, and ${\rm \langle SC_{sed}/SC_{2D} \rangle =  1.33\pm 0.26} $ (errors are the inter-quartile range). 
Not surprisingly, the slit loss correction factor
depends on near-IR luminosity (i.e., stellar mass, to zeroth order), with brighter galaxies requiring larger slit 
loss corrections due to their generally larger $r_e$. For example, galaxies with continuum detections and ${\rm log(M_{\ast}/M_{\odot}) < 9.5}$ have
${\rm \langle SC_{sed} \rangle = 1.71 \pm 0.74}$\footnote{Note that only 20 of 43 galaxies at low mass have believable spectroscopic continuum
detections, compared to 38 of 47 in the high mass subsample.}, whereas    
those with ${\rm log(M_{\ast}/M_{\odot}) > 10.5}$ have ${\rm \langle SC_{sed} \rangle = 2.25\pm 0.39}$; here the error in the
low-mass sub-sample is dominated by noise associated with the spectroscopic continuum measurements. The values of ${\rm SC_{2D}}$ are
generally much less noisy than ${\rm SC_{sed}}$ for continuum-faint objects, since they rely only on the detection of the \Ha\ emission line.  
Clearly, slit loss corrections remain a significant source of uncertainty in measuring SFR, 
probably at the $\pm 25$\% level for individual galaxies. However, we argue below that they are probably
small compared to the uncertainties associated with extinction estimates.  

For the purposes of this paper, we applied correction factors to the observed \Ha\ fluxes as follows:  
\begin{eqnarray}
{\rm SC(\Ha)} &=& {\rm 1.6~; ~ log(M_{\ast}/M_{\odot}) < 10.0}\label{eqn:sc_part1} \\
{\rm SC(\Ha)} &=& {\rm 2.0~; ~ log(M_{\ast}/M_{\odot}) \ge 10.0} \label{eqn:sc_part2}    
\end{eqnarray}

A relatively bright star ($K_s \simlt 19$) has been included on all KBSS-MOSFIRE 
masks since mid-2013 (and on many masks observed prior to that time); these stars were assigned a normal 0\secpoint7 slit and
were reduced in the same way as the galaxies on the mask. Their measured fluxes (i.e., prior to slit loss correction)   
are typically a factor of $\simeq 1.2-1.4$ smaller than those expected based on the broad-band
photometry of the same stars, and thus consistent with the adopted galaxy slit loss corrections.
We also compared the 
observed \Ha+[NII]$\lambda 6585$ fluxes for 18 of the KBSS-MOSFIRE 
targets (all in the Q1700 field) with measurements made from deep, continuum-subtracted narrow-band \Ha\ observations, 
discussed previously by \citet{reddy10} and \citet{erb+06c}, finding 
that $f_{\rm NB}/f_{\rm mos} = 2.06 \pm 0.54$ (median and inter-quartile range) where $f_{\rm NB}$ is the photometric
line flux from the narrow-band observations and  $f_{\rm mos}$ is the observed line
flux measured from the MOSFIRE spectra.  

\subsubsection{Extinction Corrections}
\label{sec:extinction}

Extinction corrections were applied to the \Ha\ fluxes using the value of  
${\rm E(B-V)_{\rm cont}}$ from the
SED fits, which assumed the \citet{calzetti00} starburst attenuation relation (see e.g. \citealt{erb+06c,reddy10}); 
the present KBSS-MOSFIRE sample has ${\rm 0 \le E(B-V)_{cont} \le 0.8}$ with a median ${\rm E(B-V)_{\rm cont}\simeq 0.2}$. 
It is conventional to assume that nebular emission lines are affected differently by dust compared to the UV stellar
continuum, and therefore subject to a different attenuation relation. \citet{calzetti00} found a relationship between the reddening of the stars and that of the ionized gas in nearby starburst galaxies,
\begin{equation}
{\rm E(B-V)_{neb} = 2.27~E(B-V)_{cont} }
\label{eqn:calzetti}
\end{equation}
where the color excess for the stellar continuum ${\rm E(B-V)_{cont}}$ can be interpreted with the 
\citet{calzetti00} attenuation relation, but ${\rm E(B-V)_{neb}}$ is derived from a line-of-sight
attenuation relation (e.g., the diffuse Galactic ISM extinction curve of \citealt{cardelli89})
\footnote{The relation in equation~\ref{eqn:calzetti} is often misunderstood to mean that 
the attenuation of \Ha\ emission in magnitudes 
is higher by a factor of 2.27 than for a continuum photon at the same wavelength; however, it is important to note
that equation~\ref{eqn:calzetti} {\it assumes} that the two values of the color excess are 
applied in combination with {\it different reddening curves}.
Although often done, it is incorrect (or at least inconsistent with the original derivation 
and intended use of equation~\ref{eqn:calzetti}) to 
use the continuum reddening curve to estimate the attenuation of emission lines. 
Under the common assumptions that 
${\rm E(B-V)_{neb} = C~E(B-V)_{cont}}$  
(where C is a constant) and that ${\rm E(B-V)_{neb}}$ can be multiplied with the 
selective extinction coefficient at 6564\AA\ in the 
\citet{calzetti00} attenuation relation, one obtains the same \Ha\ attenuation as given 
by proper interpretation of equation~\ref{eqn:calzetti} 
with ${\rm 1.36 \simlt C \simlt 1.72}$ (see equations~\ref{eqn:ha_extinction} and \ref{eqn:ha_extinction_smc}.)}. 
%A number of authors have applied extinction corrections to \Ha\ using ${\rm E(B-V)_{cont}}$ based on SED fitting, then 
%applying the correction using the same \citet{calzetti00} continuum extinction curve. Strictly speaking, this
%is incorrect, given the way that equation~\ref{eqn:calzetti} was derived. However, we note that 
%applying the ``same'' E(B-V) value and the same \citet{calzetti00} extinction curve to estimate the attenuation of 
%\Ha\ is equivalent to   
%to assuming that is equivalent to assumin; however, it is important to note that
In the original calibration of equation~\ref{eqn:calzetti}, 
a standard Galactic ISM reddening curve was used with measurements of H recombination line 
ratios to derive ${\rm E(B-V)_{neb}}$;  
for the \citet{cardelli89} Galactic extinction curve, 
${\rm A(0.656~\mu m)_{GAL}/E(B-V) = 2.52}$ 
[the average ``SMC bar'' extinction curve of \citet{gordon03} has ${\rm A(0.656~\mu m)_{SMC}/E(B-V) = 2.00}$] 
whereas the \citet{calzetti00} continuum reddening
curve has ${\rm A(0.656~\mu m)/E(B-V) = 3.33}$.  
Equation~\ref{eqn:calzetti} implies that, to use a measurement of 
the color excess ${\rm E(B-V)_{cont}}$ to estimate the attenuation 
of the \Ha\ emission line in magnitudes,  
\begin{equation} 
{\rm A(\Ha) = 2.52\times 2.27~E(B-V)_{cont}=5.72 \times E(B-V)_{cont} }
\label{eqn:ha_extinction}
\end{equation}
assuming Galactic extinction, or  
\begin{equation}
{\rm A(\Ha) =2.00\times2.27~E(B-V)_{cont} = 4.54\times E(B-V)_{cont} }
\label{eqn:ha_extinction_smc}
\end{equation}
for SMC extinction (\citealt{gordon03}) applied to the nebular emission.

However, the relationship 
between ${\rm E(B-V)_{neb}}$ and ${\rm E(B-V)_{cont}}$, and the appropriate extinction curve to be used with
either, remains uncertain for high redshift star-forming galaxies. It has been shown that  
the assumption that ${\rm E(B-V)_{neb} = E(B-V)_{cont}}$ together with the \citet{calzetti00} continuum attenuation
relation (i.e., that ${\rm A(\Ha) = 3.33~E(B-V)_{cont}}$) yields SFRs consistent with those measured
from stacks of X-ray, mid-IR, and far-IR observations of similarly-selected $z\sim 2$ galaxies \citep{reddy04,erb+06c,reddy10,reddy12b}.
Other analyses, however, suggest higher nebular 
extinction (see, e.g., \citealt{forst09,price13}), particularly for more metal-rich 
and/or higher mass galaxies, even after accounting for the extinction curve/color excess interpretation issues mentioned above.  

For definiteness, we have assumed the following:
\begin{eqnarray}
{\rm A(\Ha)} &=& {\rm 4.54~E(B-V)_{cont}~; ~ E(B-V)_{cont} \le 0.20}\label{eqn:part1} \\
{\rm A(\Ha)} &=& {\rm 5.72~E(B-V)_{cont}~; ~ E(B-V)_{cont} > 0.20} \label{eqn:part2}    
\end{eqnarray}
equivalent to using an SMC-like extinction curve for galaxies with continuum reddening equal to or
below the
median value, and Galactic diffuse ISM extinction curve\footnote{The LMC average extinction curve is nearly identical to
that of the Galaxy over the relevant wavelength range.} for those above. 
Using these corrections,  
we find that the median log(SFR$_{\Ha}$) and median log(SFR$_{\rm SED}$) for the KBSS-MOSFIRE 
sample agree to within 0.02 dex; for individual galaxies, the two SFR estimates have a median absolute deviation
$\simeq 0.20$ dex (see Figure~\ref{fig:sfr_v_sfr}.).
Of course, we do not know that agreement between SFR$_{\rm SED}$ and SFR$_{\Ha}$ means that either is ``correct'', 
but at the very least we can say
that they are surprisingly consistent both as an ensemble and  
on an object-by-object basis. 

Observations of Balmer line ratios (e.g., \Ha/\Hb) can be used to measure ${\rm E(B-V)_{neb}}$  
directly, and the current KBSS-MOSFIRE sample with $2 \simlt z \simlt 2.6$ includes more than 200
galaxies for which both \Ha\ and \Hb\ line fluxes have S/N$>5$; however, 
greater attention to relative calibration between K-band and H-band spectra is required before the line ratios can be confidently
used for extinction measurements, and thus we defer quantitative discussion to future work. Nevertheless, we find 
a median $I(\Ha)/I(\Hb) = 3.89\pm0.65$, or E(B$-$V)$_{\rm neb} = 0.29\pm0.16$ when evaluated assuming the \citet{cardelli89}
Galactic extinction curve; the corresponding value of the median continuum color excess for the same set of galaxies
is E(B-V)$_{\rm cont}=0.17$. 
%, or 
%%${\rm E(B-V)_{neb} = 1.70 E(B-V)_{cont}}$. 

After the corrections for slit losses and extinction were applied, \Ha\ fluxes were converted to luminosities assuming  
a $\Lambda$-CDM cosmology with $\Omega_m=0.3$, $\Omega_{\Lambda} = 0.7$, and $h=0.7$ 
and the \cite{kennicutt98} conversion between \Ha\ luminosity
and SFR (with adjustment to the Chabrier IMF). 
In what follows, we use the values of SFR$_{\Ha}$ listed in Tables 1-3,  
since they are less strongly covariant with inferred $M_{\ast}$ (see, e.g., \citealt{reddy12}) compared to SFR$_{\rm SED}$; however, none of the results presented in this paper 
would alter significantly if SFR$_{\rm SED}$ were used instead. 

Similarly, it is important to emphasize that, with the exception of the inferred SFR (and thus also sSFR), most
of the results presented in this paper 
do not rely on the absolute calibration of emission line fluxes or their extinction corrections; rather, they depend primarily on the intensity ratios of
lines observed simultaneously (i.e., in the same atmospheric band) and are sufficiently close to one another in wavelength that differential 
slit losses or extinction should be negligible. 

\begin{figure}[htpb]
\centerline{\includegraphics[width=9cm]{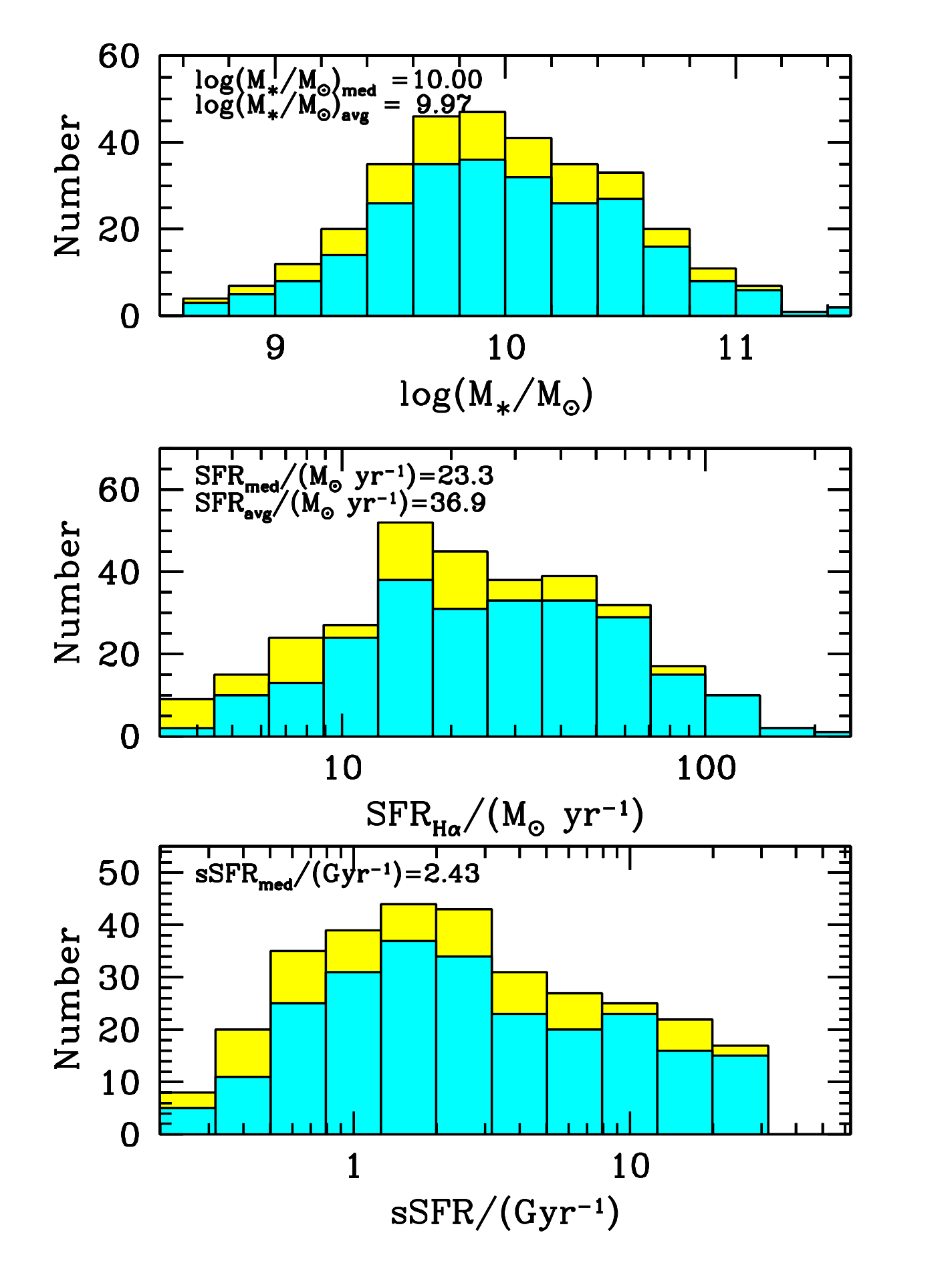}}
\caption{
Histograms of stellar mass, star formation rate, and specific star formation rate for the KBSS-MOSFIRE sample 
with $1.95 \le z \le 2.60$.  The yellow (light) histogram includes a total of 321 galaxies with \Ha\ 
and stellar mass measurements, independent of whether or not [NII]$\lambda 6585$, [OIII], or \Hb\ are detected. The cyan (darker) histogram
includes the 242 galaxies appearing in Tables 1-3, excluding 9 objects flagged as AGN [see section \ref{sec:agn}]. 
The mean and median values are given on each panel for the parent sample; the subset with N2 and/or [OIII]/\Hb\  measurements is statistically 
indistinguishable.  
\label{fig:sfr_v_mstar}
}
\end{figure}

\subsection{Current Sample Statistics} 

\label{sec:sample_statistics}

As summarized in Figure~\ref{fig:sfr_v_mstar}, the $z\sim 2.3$ KBSS-MOSFIRE sample includes galaxies with
${\rm 8.6 \simlt log(M_{\ast}/M)_{\odot} \simlt 11.4}$ and ${\rm 2 \simlt SFR_{\Ha} \simlt 500}$ \msun\ yr$^{-1}$.  
Specific star formation rates (sSFR${\rm \equiv SFR_{\Ha}/M_{\ast}}$) range over more than two orders of magnitude,
with a median value of $2.4$ Gyr$^{-1}$, in good agreement with median values estimated when SFR is measured
using mid- and far-IR luminosities in addition to the UV \citep{reddy12,reddy12b}. 
Note that Figure~\ref{fig:sfr_v_mstar} compares histograms of ${\rm M_{\ast}}$, SFR$_{\Ha}$, and sSFR for the sample 
of 242 galaxies appearing in Tables 1-3 (excluding 9 flagged as AGN; see section~\ref{sec:agn}) with a ``parent'' 
KBSS-MOSFIRE sample of 321 galaxies in the same redshift range with 
$\ge 5\sigma$ 
\Ha\ detections, without regard to whether or not additional emission lines have been observed or detected. Thus, galaxies in the \Ha\ sample
but not appearing in Tables 1-3 are single-line detections, observed only in K band  
and of lower overall S/N, usually because their spectra are based on relatively short total integration times obtained
for redshift identification.  

Of the 251 objects included in Tables 1-3, 189 (75.3\%) had prior redshift identifications 
from optical (rest-frame far-UV) spectra obtained with Keck 1/LRIS-B, 30 (12.0\%) had been observed previously with LRIS-B 
without yielding a redshift identification,  and 32 (12.7\%) had never before been observed spectroscopically.  
Among all of the KBSS-MOSFIRE observations so far, when the redshift was known from optical spectroscopy to be in the
targeted range $2 \le z \le 2.6$, more than 90\% yielded successful detections of rest-frame optical nebular lines; when
the redshift was not known {\it a priori}, a similar fraction yielded new spectroscopic redshifts from the MOSFIRE H-band 
and/or or K-band spectra.  

As mentioned above, the selection criteria used for KBSS-MOSFIRE are broader than those of purely rest-UV-color selected 
samples over the same range of redshifts discussed by (e.g.) \citet{steidel04,erb+06a,erb+06b,reddy10,reddy12}; specifically,
targets were included whose observed rest-UV and UV/optical colors indicate more heavily reddened galaxies compared to those selected by 
the ``BX'' and ``MD'' criteria. We have also found that the spectroscopic success rate for optically-faint (${\cal R} \simgt 25$) 
galaxies within the UV-color selected samples is higher using near-IR spectroscopy, where most galaxies have strong nebular emission
lines, than for optical spectroscopy, where most galaxies have no strong emission lines, and thus identification depends on
much weaker absorption lines observed against a faint stellar continuum.  
Compared to the optical spectroscopic sample of $2 \le z \le 2.6$ UV color-selected
galaxies in the same 15 KBSS fields (1202 galaxies at the time of this writing), the KBSS-MOSFIRE sample includes a slightly larger
fraction of galaxies with ${\rm log(M_{\ast}/M_{\ast}) > 10.5}$ (19.5\% vs.  17.4\%) and with masses 
${\rm log(M_{\ast}/M_{\ast}) < 9.5}$ (19.5\% vs. 18.5\%). 
The median SFR$_{\rm SED}$ is 23.3 ${\rm M_{\odot}}$ yr$^{-1}$ in the KBSS-MOSFIRE sample, to be compared with 19.5 ${\rm M_{\odot}}$ yr$^{-1}$ 
in the full rest-UV spectroscopic sample. In summary, the sensitivity of MOSFIRE for near-IR spectroscopy has produced a spectroscopic
sample that is essentially unbiased with respect to the parent photometric sample, at least in terms of SFR and $M_{\ast}$; 
this was not the case for the earlier NIRSPEC sample 
at similar redshifts (\citealt{erb+06a,erb+06b}).  

Realistically, any spectroscopic sample at high redshift, whether based on near-IR or optical spectra, 
suffers from incompleteness with respect to SFR, which will in turn affect the sample's distribution of
$M_{\ast}$. At the low mass end, for example, even with zero extinction our photometric selection criteria limit
the galaxies to $G \simlt 26$, which corresponds to ${\rm SFR \simgt 1.3}$ M$_{\odot}$ yr$^{-1}$ using
the standard conversion of rest-frame 1500 \AA\ luminosity to SFR (e.g., \citealt{mpd98}) at $z \sim 2.3$; our
detection limit for \Ha\ corresponds to approximately the same SFR for zero extinction.  The same practical 
limits would apply to even the deepest near-IR-selected samples.  At the high stellar mass end,
greater extinction (nebular and/or UV) may more than compensate for larger overall SFRs, so that the 
resulting selection function with respect to M$_{\ast}$ or SFR becomes potentially complex. 
The KBSS-MOSFIRE sample
is undoubtedly missing high M$_{\ast}$ galaxies with very low SFR, which constitute a substantial fraction ($\simeq 40$\%) of 
galaxies with ${\rm log (M_{\ast}/M_{\odot}) > 11.0}$ and $z\sim 2.3$ according to, e.g. \citet{kriek08}. 
At the low-mass end, it would be missing most galaxies with uncorrected \Ha\ line fluxes $\simlt 5\times 10^{-18}$ ergs s$^{-1}$
cm$^{-2}$, corresponding to SFR$\simlt 4$M$_{\odot}$ yr$^{-1}$ at $z \sim 2.3$ after typical correction for slit losses and
extinction for galaxies with ${\rm log (M_{\ast}/M_{\odot}) \sim 9}$. 

\subsection{Targets with Previous Near-IR Spectroscopic Observations}

\label{sec:previous}

Of the 251 targets listed in Tables 1-3, 25 galaxies ($\simeq 10$\%)  were also included
in the NIRSPEC sample of \citet{erb+06a,erb+06b,erb+06c}, though only 2 of the 25 had been spectroscopically observed 
in more than one near-IR atmospheric band.  In general, the MOSFIRE spectra of the same targets 
are of much higher S/N and have $\simeq 3$ times higher spectral resolution; however, the nebular redshifts of
objects in both samples measurements agree well, with $\langle c(z_{\rm MOS}-z_{\rm NS})/(1+z_{\rm MOS}) \rangle = -15\pm41$ \kms (rms).    
Nine of the current KBSS-MOSFIRE sample were observed by \citet{law09}, and one by \citet{law2012}, using the OSIRIS 
integral-field spectrometer and the Keck 2 Laser Guide Star
Adaptive Optics facility; two galaxies in the current sample were observed as part of the SINS survey using SINFONI on the VLT (\citealt{forst09}). 
NIRSPEC-based nebular redshifts of 34 objects in the current KBSS-MOSFIRE sample were used to measure galaxy systemic redshifts
by \citet{steidel2010}  
in their analysis of the kinematics of galaxy-scale outflows at $z \sim 2.3$. 

Stellar masses and SFRs (based on SED fitting, including some of the earliest Spitzer/IRAC photometry) 
have been presented by \citealt{shapley05} for 17 of the current
KBSS-MOSFIRE targets in the Q1700 survey field; many of the current Q1700 galaxies were included in more recent work by
\citet{kulas13}, based on independent measurements of a subset of the current MOSFIRE data in that field.  

The last column in each of Tables 1-3 includes the references to earlier work, where relevant.  

\begin{figure*}[hpbt]
\centering
\includegraphics[width=16cm]{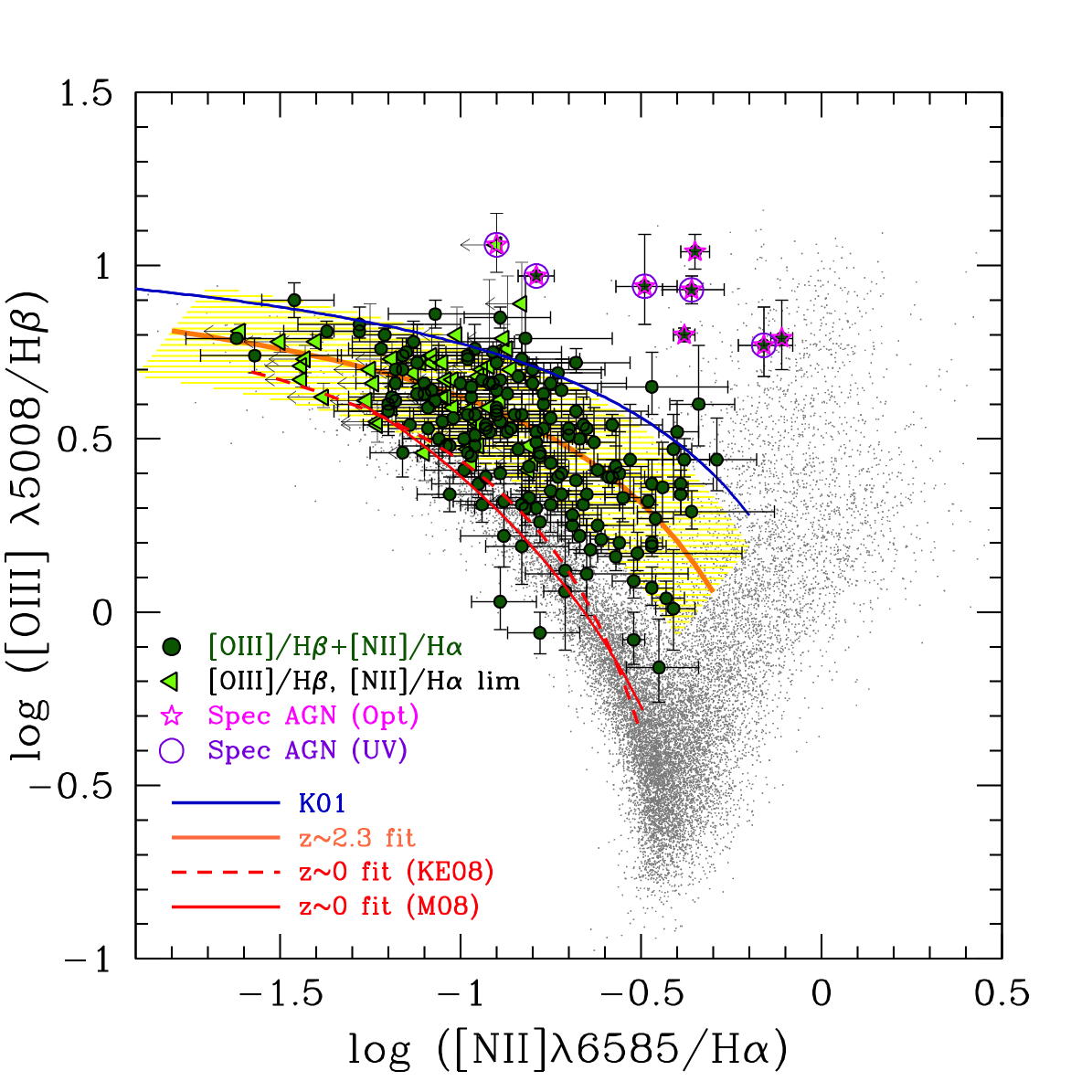}
\caption{``BPT'' diagram for 219 objects with $\langle z \rangle = 2.34\pm0.16$ in the KBSS-MOSFIRE survey (large points with error bars) 
in comparison with the SDSS ($z \simeq 0$) sample (e.g., \citealt{tremonti04}; locus of gray points).  
The 168 objects with measurements in both [NII]/\Ha\ and [OIII]\Hb\ are indicated with dark green points, while
an additional 51 galaxies with [OIII]/\Hb\ detections and upper limits (2$\sigma$) for [NII]/\Ha\ are light triangles 
with left-pointing arrows. 
The red curves trace the ``metallicity sequence'' of SDSS star-forming galaxies, showing the expected location
of galaxies in the BPT plane for oxygen abundances of 0.2-1.0 solar -- the solid curve is based on the calibration of  
\cite{maiolino08},  while the dashed curve represents the same
metallicity sequence implied by 
the strong-line calibration of \cite{kewley+ellison08}. Both curves have been adjusted to the N2 metallicity scale of
PP04 for consistency.   
The blue solid curve is 
the ``maximum starburst'' model of \cite{kewley01}. The orange curve is the best-fit BPT sequence for the KBSS-MOSFIRE sample (equation~\ref{eqn:bpt_fit})
,
with the yellow shaded region tracing the inferred intrinsic dispersion of $\pm 0.1$ dex. 
Eight objects among the 219 have been identified as 
AGN based on their rest-UV and/or rest-optical spectra (see discussion in section~\ref{sec:agn}); these
are indicated with magenta ``stars''. AGN identified by both rest-UV and rest-optical spectra are indicated
by circles surrounding the stars.  
\label{fig:bpt}
}
\end{figure*}

\begin{figure}[htpb]
\centering
\includegraphics[width=8.5cm]{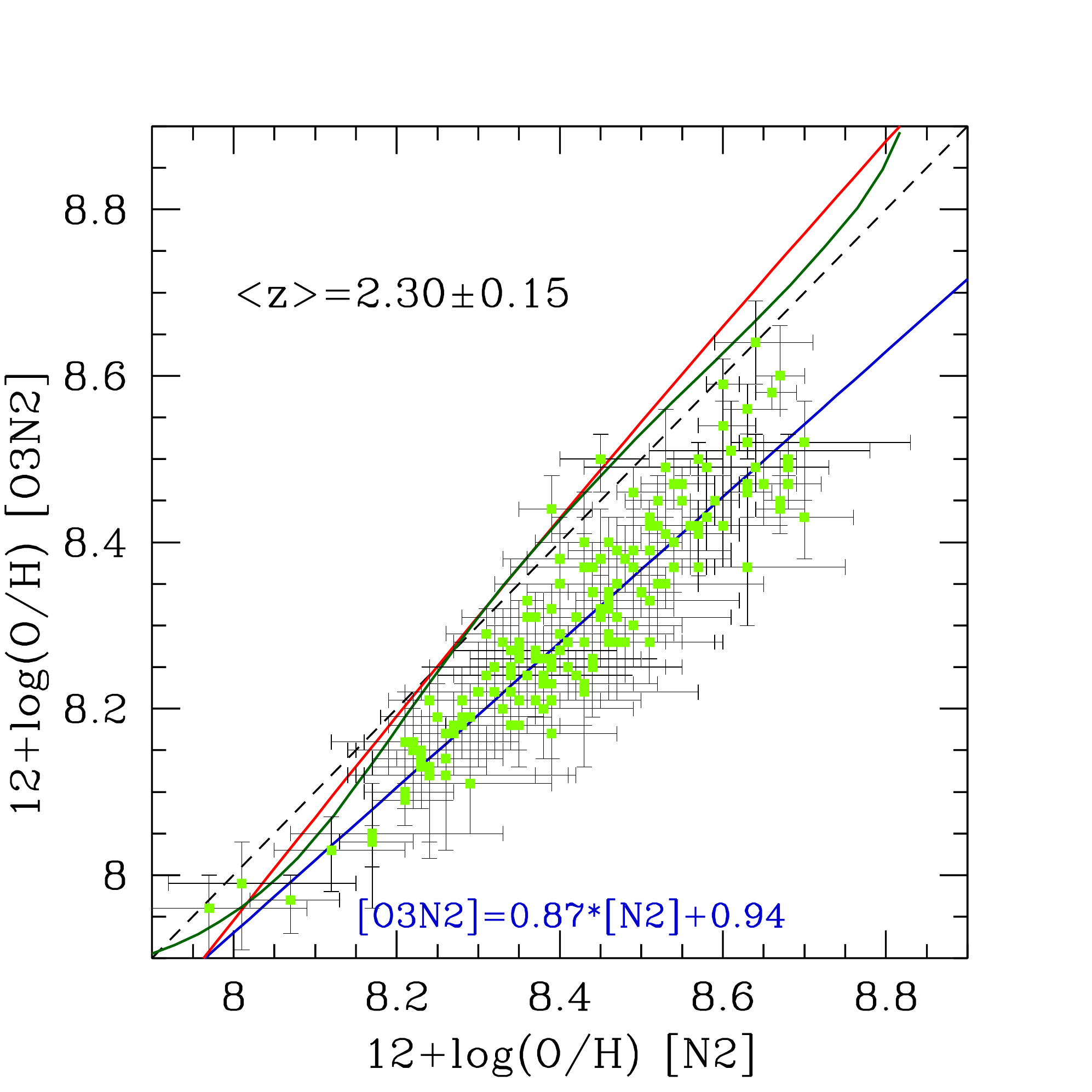}
\caption{Comparison of the inferred oxygen abundance of the $\langle z \rangle = 2.3$ sample based
on the PP04 calibrations of the N2 and O3N2 strong-line indices (points with error bars, where the error bars account for measurement
errors only and not for uncertainties in the calibrations). The dashed line indicates the expected relation if the two methods
were to give the same value of 12+log(O/H). 
The red and green curves represent the best-fit
regression formulae for recent re-calibrations of strong line indicators based on the low redshift sample (\citealt{kewley+ellison08}
and \citealt{maiolino08}, respectively)-- the observed scatter in the low-redshift training sets is $\sim 0.03-0.04$ dex. 
The systematically lower values of O3N2-based oxygen abundances as compared to those of N2 
are evident, consistent
with an offset of ${\rm \Delta(12+log(O/H) = 0.13\pm0.01}$ dex. 
The blue line is the best linear fit to the relation
between the two inferred values at $z=2.3$. The scatter about this relation, after accounting for measurement errors, is $\simeq 0.04$ dex.   
\label{fig:met_compare}
}
\end{figure}

\section{The ``BPT'' Diagram at $\langle z \rangle = 2.3$}
\label{sec:bpt}

\subsection{The Locus of Star-forming Galaxies in the BPT Plane}

Perhaps the most remarkable aspect of the BPT diagram for local star-forming galaxies is the narrow locus along which
most star forming galaxies are found, sometimes referred to as the ``HII region abundance sequence'' (\citealt{dopita2000})
because the left-hand branch can be interpreted as a sequence in overall ionized-gas-phase metallicity. 
The tightness of the sequence is controlled by the range within a galaxy sample of some combination of the hardness and intensity of
the ionizing stellar radiation field and the properties of the ambient ISM being ionized. At $z \simeq 0$, more than 90\% of star-forming 
galaxies fall within $\pm0.1$ dex of the ridge-line of the sequence (\citealt{kewley13}); for the SDSS data set used in Figure~\ref{fig:bpt}, 
the scatter in log([OIII]/\Hb) at fixed log([NII]/\Ha) is $\simeq 0.11$ dex after accounting for measurement errors. 

Figure~\ref{fig:bpt}
shows definitively what had already been suggested by  
the relatively small number of earlier measurements for galaxies at $z\sim 1-2.5$ \citep{shapley05z1,erb+06a,liu+shapley08}: the nebular spectra
of high redshift star-forming galaxies
occupy an almost entirely distinct region of the BPT diagram as compared
to local galaxies. 
It has been shown (e.g., \citealt{kewley01,kauffmann03}) that, for local galaxies, the locus of points along the star-forming branch of the BPT diagram 
can be fit well by a function
\begin{equation}
{\rm  
log([OIII]/\Hb) = \frac{0.61}{log~([NII]/\Ha)+0.08} + 1.10 }
\end{equation} 
(e.g.,\citealt{kewley13}). 
Fitting the same functional form to the KBSS-MOSFIRE sample in Table~\ref{tab:n2ha_and_o3n2} yields
\begin{equation}
{\rm log([OIII]/\Hb) = \frac{0.67}{log~([NII]/\Ha)-0.33}~ + ~ 1.13 } .
\label{eqn:bpt_fit}
\end{equation}
Formally, $\chi^2/\nu = 13.6$ for the best-fit model with respect to the data, or a weighted error of $\simeq 0.15$ dex. For comparison
to the BPT locus of local star-forming galaxies, it is of interest to estimate the intrinsic scatter (in the absence
of measurement errors) of the locus about the best-fit model.  
%We note that the $z \sim 2.3$ data are equally well-fit by a linear relationship between the
%two quantities,
%\begin{equation}
%{\rm log([OIII]/\Hb) = 0.22  log~([NII]/\Ha) -0.39,}
%\end{equation}
%over the range ${\rm -1.4 < log([NII]/\Ha) < -0.4}$, with $\sigma_{\rm obs} \simeq 0.16$, and reduced $\chi^2$=8.6. 
To accomplish this, we assumed that the error bars on each point $\sigma_{{\rm m},i}$ are the true measurement errors 
but that the total variance for each point $\sigma^2_{{\rm tot},i} = \sigma^2_{{\rm m},i} + \sigma^2_{\rm sc}$, where    
$\sigma_{\rm sc}$ represents the intrinsic scatter, and is assumed to be a constant (i.e., independent of the measurement errors). 
The value adopted for $\sigma_{\rm sc}$ is that which yields $\chi^2/\nu \approx 1$; we find that $\sigma_{\rm sc} \approx 0.12$ dex--
remarkably similar to the scatter observed in the SDSS galaxy sample relative to the best-fit locus 
(which generally has negligible measurement errors by comparison). 
%Alternatively, a relatively model-independent estimate is summarized in Table~\ref{tab:bpt_scatter}, which compares the distribution
%of log([OIII]/\Hb) in bins of log([NII]/\Ha) for the SDSS and KBSS-MOSFIRE $z=2.3$ samples. 
%Clearly the observed scatter at $z \sim 2.3$ is affected by larger measurement errors, but after approximate correction for this component of the
%variance,  the {\it intrinsic} scatter of ${\rm log([OIII]/\Hb)}$ at
%fixed log([\ion{N}{2}]/\Ha) is again comparable to that of the SDSS sample. 
Figure~\ref{fig:bpt} (light shaded region)
shows that the vast majority of data points (as well as the points with upper limits on ${\rm log~([NII]/\Ha)}$) 
are consistent with a swath in which both
${\rm log~([NII]/\Ha)}$ and $\rm log~([OIII]/\Hb)$ vary by $\pm 0.12$ dex with respect to the best-fit model in equation~\ref{eqn:bpt_fit}.  

Formally, it is difficult to distinguish whether the shift in the locus is primarily due to changes in [OIII]/\Hb, 
[NII]/\Ha, or both.
The shift 
has implications, independent of its physical origin, for the use of strong-line nebular
diagnostics beyond the local universe. 
As shown in Figure~\ref{fig:bpt}, the calibrations (or re-calibrations)
of the strong line indices imply a one-dimensional curve in the BPT plane, since galaxies of a given 
value of 12+log(O/H) map uniquely to values of [NII]/\Ha\ and
[OIII]/\Hb, with metallicity increasing toward the ``right'' and ``down'' along the sequence. 
The red curves superposed on the $z \simeq 0$ locus in the BPT plane trace the metallicity sequence predicted 
by recently re-calibrated strong-line indicators that make use of the same line ratios that appear in the BPT diagram, for galaxies 
with oxygen abundances from 0.2-1.0 times solar (${\rm 8.0 \le 12+log(O/H) \le 8.7}$;
the solid curve is the best fit
regression formula advocated by \cite{maiolino08} 
while the dashed curve is the same locus predicted by the conversion 
formulae of \cite{kewley+ellison08}\footnote{Both curves were corrected to reflect oxygen abundances consistent with the N2 abundance scale
with the PP04 calibration, for consistency.}. Not surprisingly, both curves follow the ridge line in BPT space traced by the SDSS sample rather
accurately-- reflecting the fact that essentially the same set of galaxies was used to establish the best-fit joint calibrations
for the relevant strong-line indices.  

The important point is that according to the local calibrations, overall changes in [O/H] would simply move objects
along these curves; it then follows that any galaxy whose BPT parameters do {\it not} fall along the calibration
sequence cannot yield consistent values of 12+log(O/H).  Stated simply, Figure~\ref{fig:bpt} shows
that there is a problem applying a calibration based on local galaxies to a high-redshift sample, even for those that have been re-normalized to consistent
metallicity scales for $z \simeq 0$ (e.g., \citealt{maiolino08,kewley+ellison08}).  
In practice this means that the ``measured'' 12+log(O/H) from strong line ratios will
depend systematically on which emission lines are measured. For example, many measurements at $z < 2.6$ rely on the
N2 metallicity calibration, since applying it requires observations in only one atmospheric band and the ratio is insensitive to
nebular extinction; at $z \simgt 3$, on the other hand, estimates are more likely to be based on R23 and other permutations of [OII], [OIII], and
\Hb, since [NII] and \Ha\ cannot be observed from the ground. In the latter case, additional issues come into play, e.g., nebular extinction, accurate relative
flux calibrations, and the well-known non-monotonic
behavior of the line indices.

Figure~\ref{fig:met_compare} illustrates the problem in the context of the $z \sim 2.3$ sample: using locally-established
metallicity calibrations leads to systematically different metallicities even for the closely-related N2 and
O3N2 methods (both calibrations from PP04), which were calibrated primarily using the ``direct'' or ``$T_{\rm e}$'' method and the same set of local \ion{H}{2} regions.  Interestingly, the {\it scatter} in the locus
of inferred metallicities for the $z \sim 2.3$ sample remains small ($\simlt 0.04$ dex after accounting for the contribution
of measurement errors to the observed scatter), 
suggesting that a re-calibration at high redshift 
of the strong-line indicators may produce an equally good, albeit different, mapping of metallicity to strong
line intensity ratios\footnote{In section~\ref{sec:pp04_calibration}, we revisit the calibrations of the PP04 N2 and O3N2 metallicity relations
and their implications for the high-redshift sample.}.
The linear regression in Figure~\ref{fig:met_compare} serves as an initial estimate of how the conversion might work at $z \sim 2.3$; 
it will be used in section~\ref{sec:mass_met} below. 

The question remains whether either of the estimates of $12+{\rm log~(O/H)}$ is reliable when applied to galaxies at $z \simeq 2.3$; 
the answer depends strongly on what factor is primarily responsible for the shift in the BPT sequence
at $z \sim 2.3$, and whether it is reasonable to interpret the locus as an abundance sequence as at low redshift.   
We address this question in section~\ref{sec:bpt_interpretation} below. 

\begin{figure}[htpb]
\centerline{\includegraphics[width=8cm]{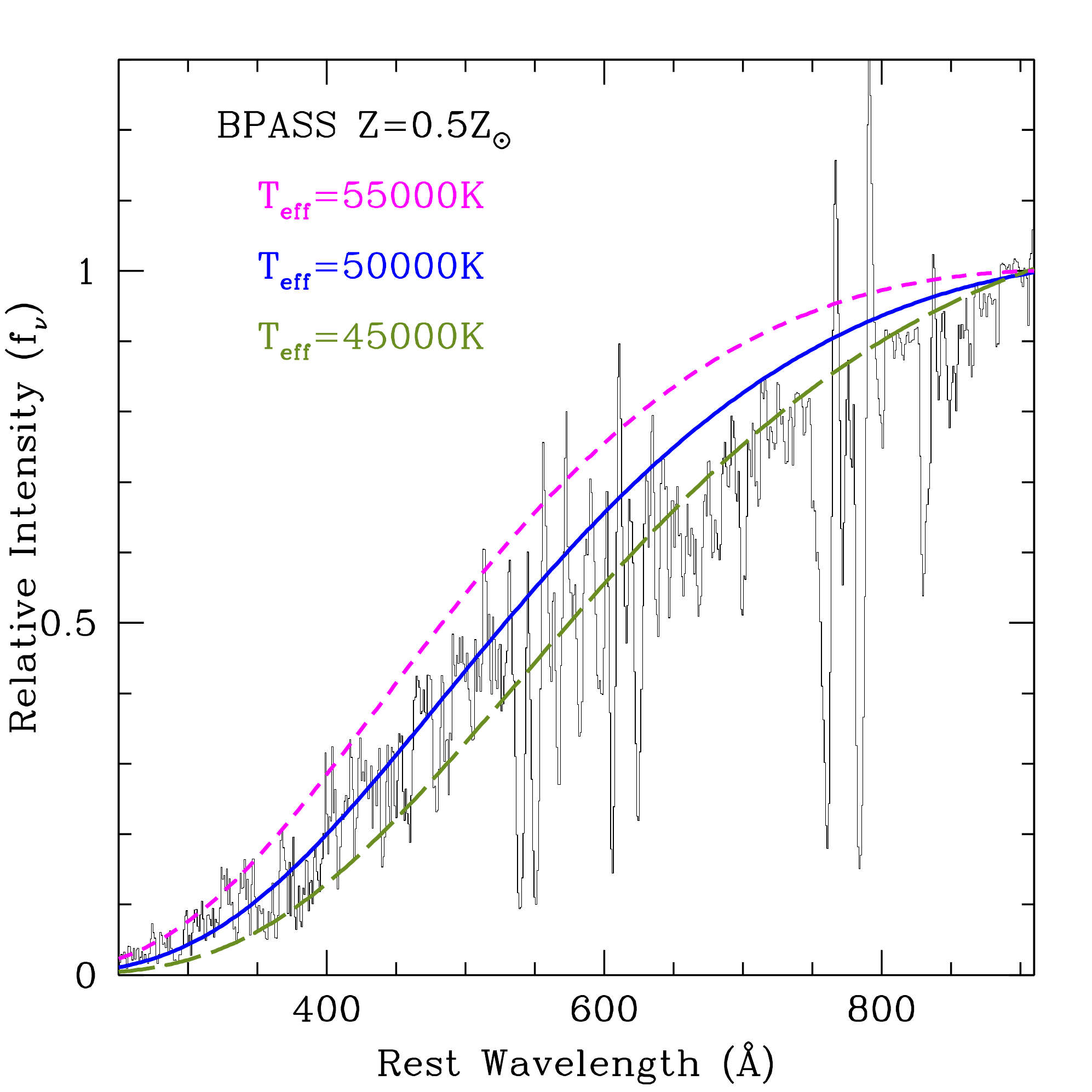} } 
\caption{Comparison of blackbody spectra with $T_{\rm eff} = 45000-55000$ K and a ``Binary Population and Spectral Synthesis'' (BPASS) 
population synthesis model with continuous
star formation, $Z=0.5~Z_{\odot}$, an age of $10^8$ years, and including the effects of binaries (\citealt{eldridge09}). 
The spectra have been normalized to match
at rest wavelength of $912$ \AA\ (1 Ryd). The $T_{\rm eff} = 50000$ K blackbody (blue solid curve) is a good match to the theoretical
spectrum, whose metallicity 12+log(O/H)=8.4 is typical of those inferred for the KBSS-MOSFIRE sample. }
\label{fig:bpass_compare}
\end{figure}

\begin{figure}[htbp]
\centerline{\includegraphics[width=8.5cm]{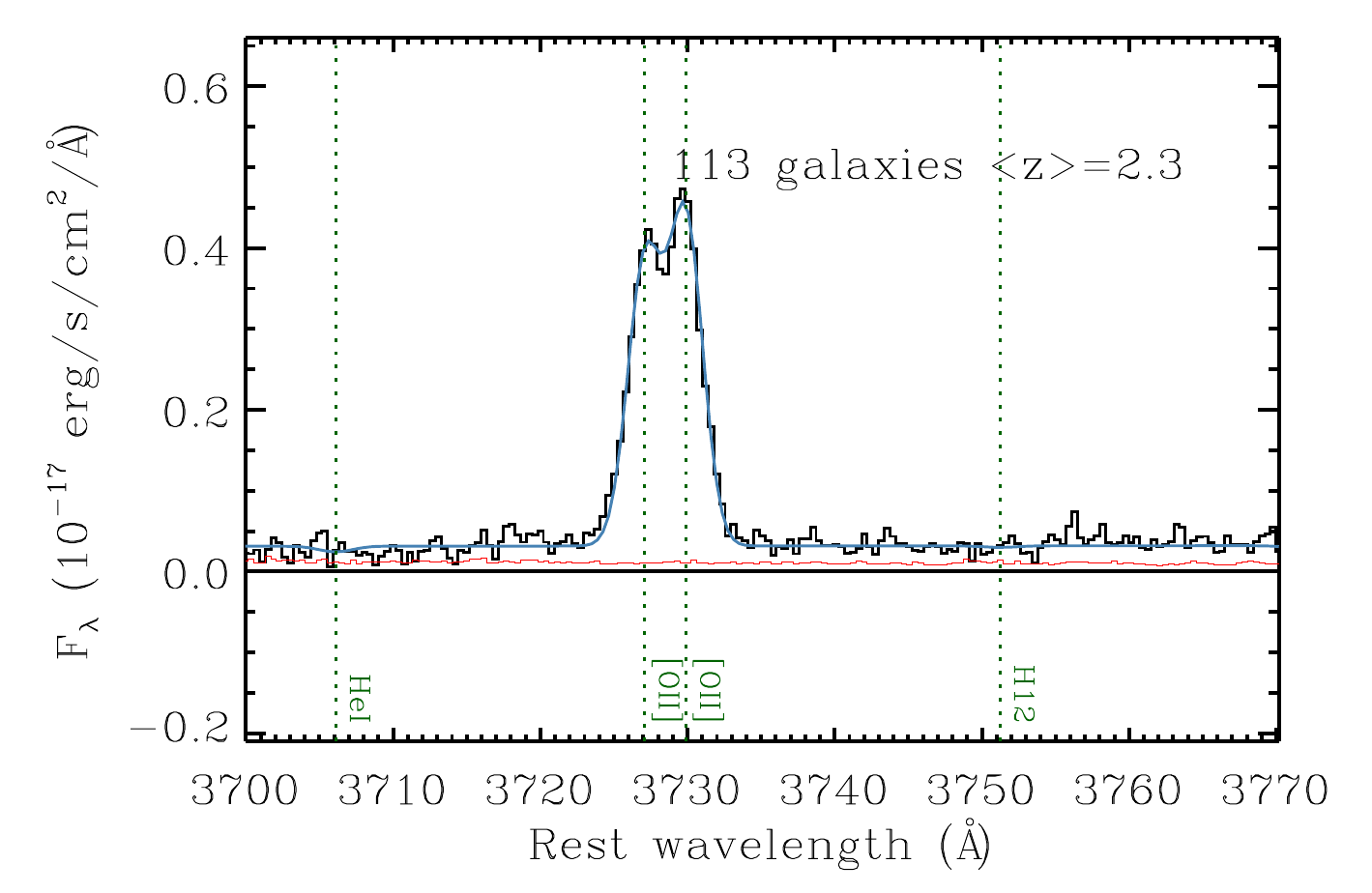}}
\caption{Stacked J-band spectrum of 113 KBSS-MOSFIRE $z \sim 2.3$ galaxies showing the resolved [OII] doublet, with $I(3727)/I(3729)=0.86$.  
The median value of the ratio for individual galaxies is identical, corresponding to 
a median electron density $n_{\rm e} \simeq 220$ cm$^{-3}$. 
}
\label{fig:o2ratio}
\end{figure}

\begin{figure*}[thbp]
\centerline{\includegraphics[width=9.0cm]{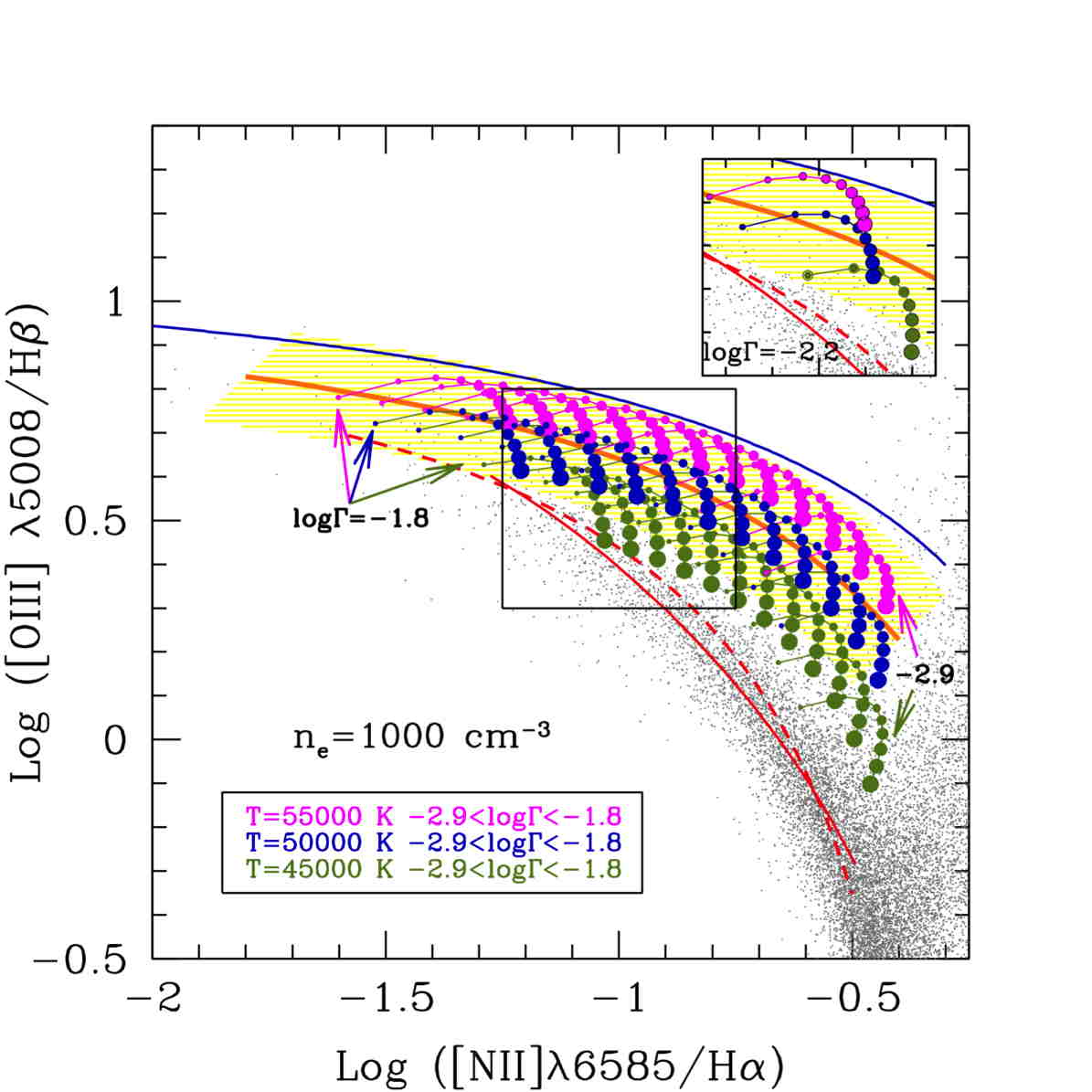}\includegraphics[width=9.0cm]{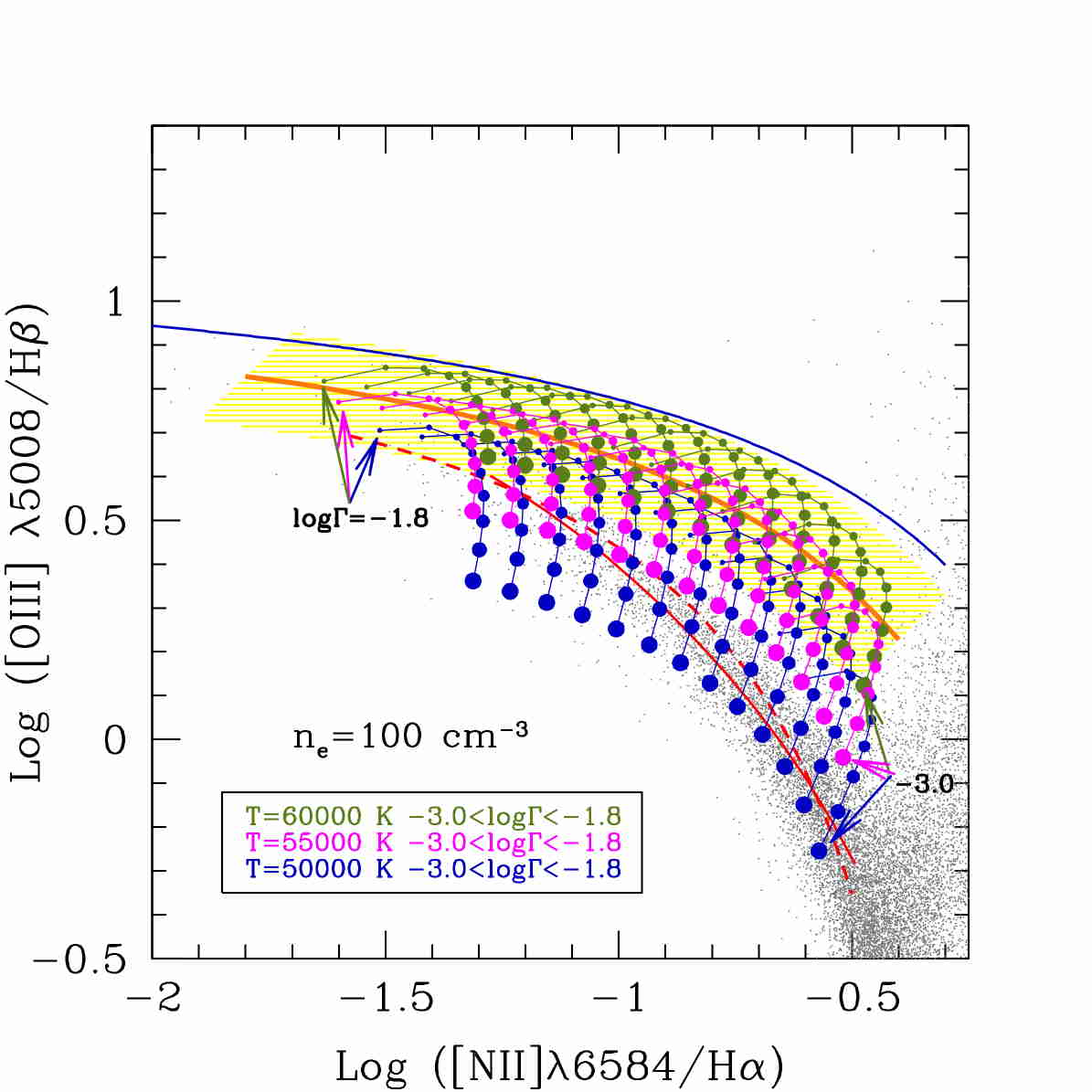}}
\caption{(Left) Predicted locus in the BPT diagram for CLOUDY models described in the text; the colors
are coded according to the assumed shape (parametrized by $T_{\rm eff}$) of the ionizing radiation field, for ionization parameters in
the indicated range and assuming $n_{\rm e} \sim 1000$ cm$^{-3}$. The shaded region and the solid and dashed
curves are the same as in Figure~\ref{fig:bpt}.  The curves between points, with the same color coding as
the points themselves, connect model runs with the same value of log$\Gamma$, at intervals of ${\rm \Delta log \Gamma = 0.1}$ dex in
between each.  
For each value of log$\Gamma$, 
the connected points range in metallicity ${\rm Z/Z_{\odot}=0.2-1.0}$ in steps of ${\rm \Delta (Z/Z_{\odot}) = 0.1}$, where
the point size scales with ${\rm Z/Z_{\odot}}$. The inset panel re-plots the region within the black box, but for a single
value of the ionization parameter. This view shows the modest dependence on gas-phase metallicity at fixed $\Gamma$ and $T_{\rm eff}$. 
(Right) As in the left panel, but for models with $n_e=100$
cm$^{-3}$}
\label{fig:bpt_mod}
\end{figure*}
\begin{figure*}[htb]
\centerline{\includegraphics[width=9cm]{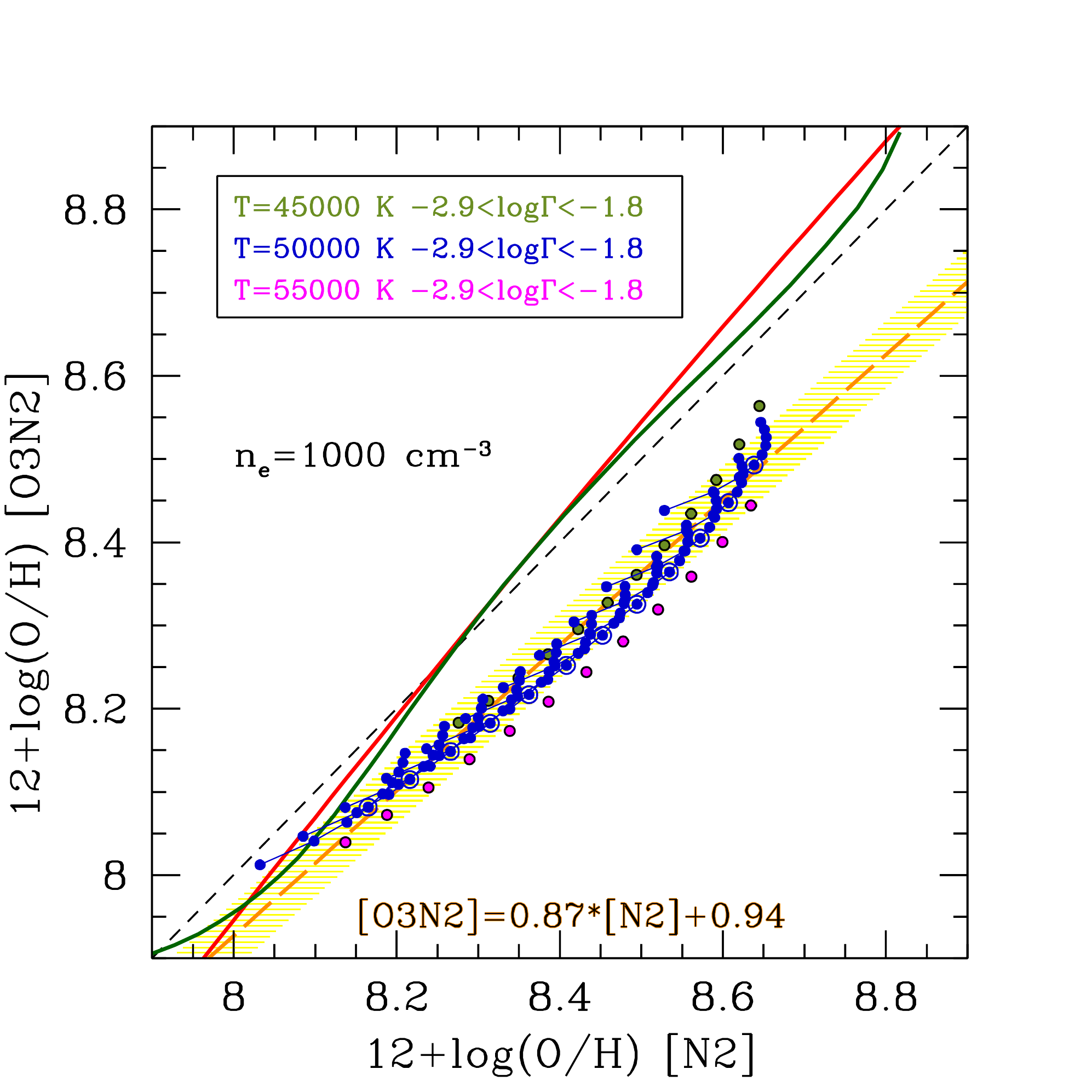}\includegraphics[width=9cm]{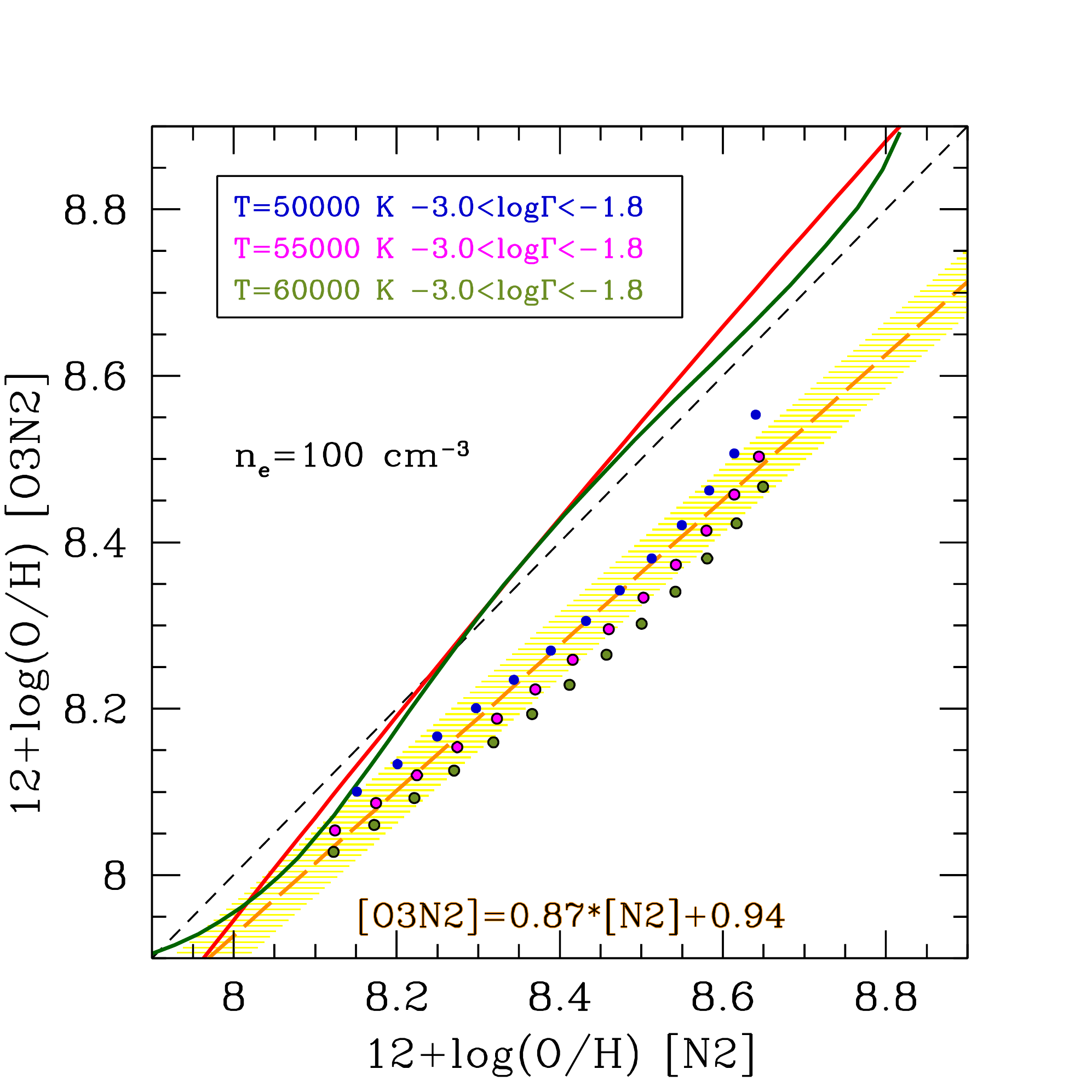}}
\caption{ (Left) The predicted run of the N2 and O3N2 metallicity estimates for the same set of model parameters as in 
Figure~\ref{fig:bpt_mod}, which should be compared with Figure~\ref{fig:met_compare}. The
yellow shaded region represents the $\simeq 0.04$ dex scatter inferred from the $z\sim 2.3$ observations relative to
the best-fit linear relation (orange dashed line). Note that the run of values for these metallicity
estimators is actually almost entirely due to a variation in ionization parameter, rather than gas-phase metallicity,
and the scatter about the linear relation is dominated by the differences in $T_{\rm eff}$ considered. 
As in Figure~\ref{fig:bpt_mod}a, models with $T_{\rm eff} = 50,000$ K are the best overall fit to the observations. 
(Right) Same as the left-hand panel, for models with $n_e=100$ cm$^{-3}$. Here only the points for ${\rm Z/Z_{\odot}=0.5}$
are plotted, for clarity, to better illustrate the $\Gamma$ and $T_{\rm eff}$ dependence (at fixed gas-phase metallicity).}
\label{fig:met_comp_mod}
\end{figure*}

\section{Physical Interpretation of the $z \sim 2.3$ BPT Diagram}

\label{sec:bpt_interpretation}

The physical cause of the offset of high-redshift galaxies in the BPT plane 
has recently been explored by a number of authors 
through examination of the relatively small number of nearby galaxies occupying similar
positions in the BPT diagram (e.g., \citealt{liu+shapley08,brinchmann08}), using theoretical models (e.g., \citealt{kewley01,erb2010,kewley13}),
or a combination of the two (e.g., \citealt{shirazi13,kewley13b}). \citet{kewley13} in particular have explored in some detail
how altering various physical parameters (metallicity, hardness of ionizing radiation field, electron density, prevalence of 
shocks or AGN versus stellar photoionization) in high redshift HII regions would affect galaxies' position in the BPT
diagram.  
A common conclusion of most of the recent work is that the main driver of the offset is a higher effective ionization parameter, or
the dimensionless ratio of the number density of H-ionizing photons to that of H atoms in the \ion{H}{2} gas,
\begin{equation}
\Gamma \equiv \frac{n_{\gamma}}{n_H} \approx \frac{n_{\gamma}}{n_e} 
\label{eqn:gamma}
\end{equation}
where $n_H$ is the number density of hydrogen atoms and $n_{\gamma}$ is the equivalent density of photons capable of ionizing hydrogen 
impinging on the face of the gas layer. $\Gamma$ is analogous to the commonly-used parameter $U$ (or $q=cU$, where $c$ is the speed of light), 
except that the latter are generally 
used in the context of a spherical geometry as in the case of an idealized \ion{H}{2} region Stromgren sphere surrounding a point-like ionizing source
such as a single O-star. $\Gamma$ is intended to be more general, and to avoid the connotation of a particular geometrical configuration.  
The effective ionization parameter $\Gamma$ obviously depends on both the shape and intensity of 
the radiation field and the physical density in the ionized gas; the former will in turn depend on 
the physical density of star formation and the ionizing-luminosity-weighted effective temperature mix of the stars producing the ionizing photons.  
It ($\Gamma$) will also depend on the relative three-dimensional distribution of massive stars, ionized gas, and neutral ISM within a galaxy; in
some locations, a packet of gas may be ionized by multiple sources impinging from different distances and directions, each of which has been 
subject to different modulations of intensity and shape by intervening material.  

\subsection{Photoionization Models}
\label{sec:bpt_models}

To gain some intuition, we ran a large grid of model \ion{H}{2} regions using CLOUDY\footnote{Calculations were performed
with version 13.02 of CLOUDY, last described by \citet{ferland13}.} 
in which gas-phase metallicity, ionization parameter, physical density, and the effective temperature ($T_{\rm eff}$) 
of the stellar ionizing sources
were allowed to vary. We initially assumed solar abundance ratios for all elements, but allowed the overall metallicity to range
from 0.2 to 1.0 times solar.  For simplicity, we began by modeling the UV radiation field shape with a blackbody of temperature
$T_{\rm eff} =45,000~K$, motivated by the shape of the rest-frame far-UV spectra of $z \sim 3$ LBGs in a very deep spectroscopic survey 
(Steidel et al 2015, in prep.). We then extracted the predicted intensity ratios of nebular emission lines, as well
as the corresponding 
values of the N2 and O3N2 estimates of ${\rm 12+~log(O/H)}$, for comparison with the model metallicity. We sought 
the range of model parameters that could reproduce the high redshift BPT data, including the observed trend 
in the metallicity indicators (Figure~\ref{fig:met_compare}).  We found that higher effective temperatures
for the ionizing radiation field were needed to reproduce the [OIII]/\Hb\ ratios of the bulk of the $z \sim 2.3$
galaxies, so the grids were expanded to include blackbody energy distributions with ${\rm 40,000~K \le T_{\rm eff} \le 60,000~K}$.

Note that we have deliberately chosen not to use theoretical stellar models in the CLOUDY runs 
because of the large uncertainties in the ionizing spectra of O stars
and the very high density and complex morphology of star formation within the high-redshift galaxies. Instead, we 
emphasize that we are interested in constraining the effective shape of the ionizing radiation field and the average ratio of ionizing
photons to ISM density within the ionized regions {\it required to reproduce the observations}.  
In spite of the relative simplicity of our models, we argue that assuming blackbody ionizing spectra is
reasonable. For example, Figure~\ref{fig:bpass_compare} shows that blackbody ionizing spectra 
represent a reasonable approximation to the shape of the 1-4 Ryd stellar continuum of modern stellar population synthesis 
models (\citealt{eldridge09}.) Similarly, we show in section~\ref{sec:implications} below that 
the low-redshift BPT sequence can be adequately reproduced 
assuming a blackbody ionizing spectrum with $T_{\rm eff} \simeq	 42000$ K. 

\begin{figure*}[htbp]
\centerline{\includegraphics[width=8.5cm]{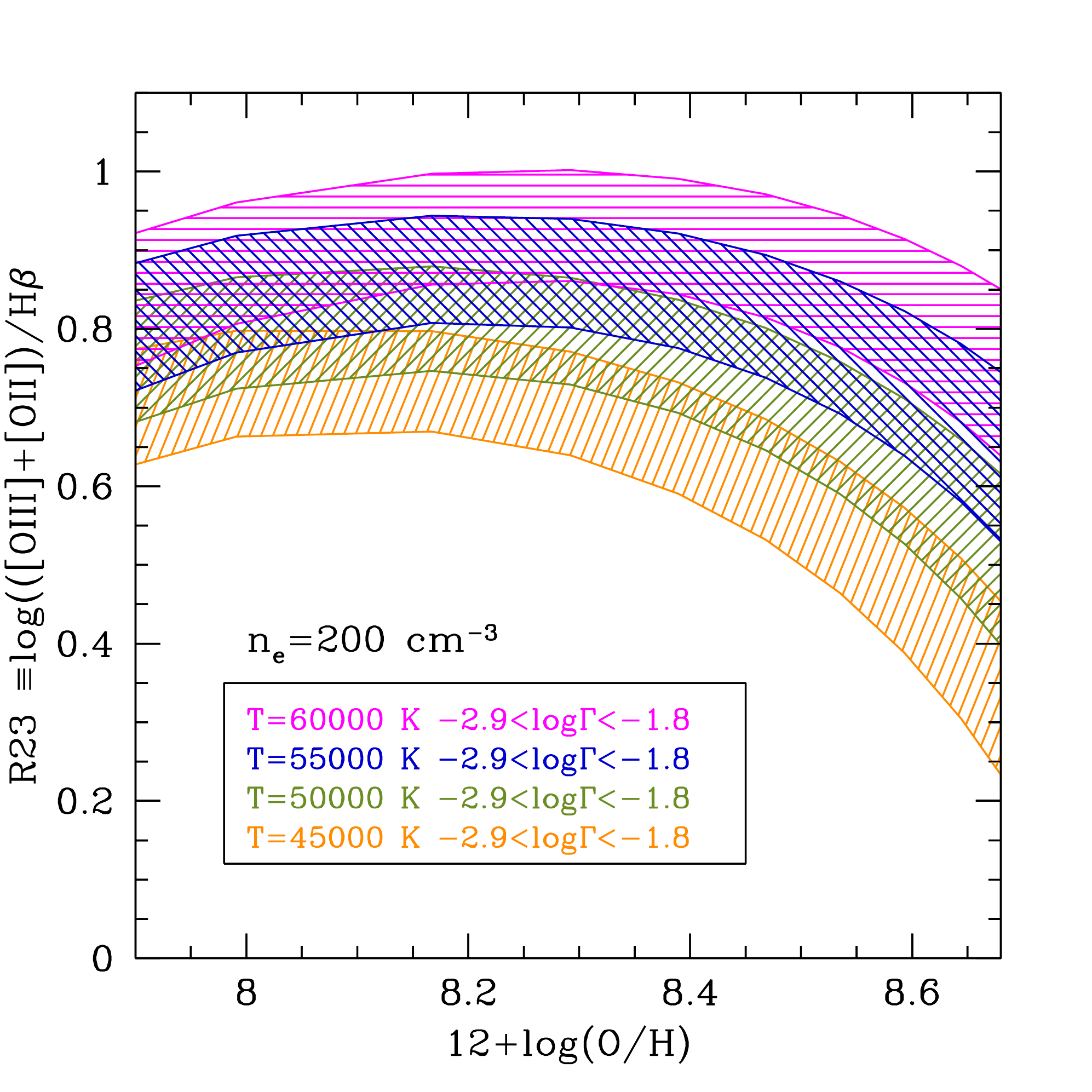}\includegraphics[width=8.5cm]{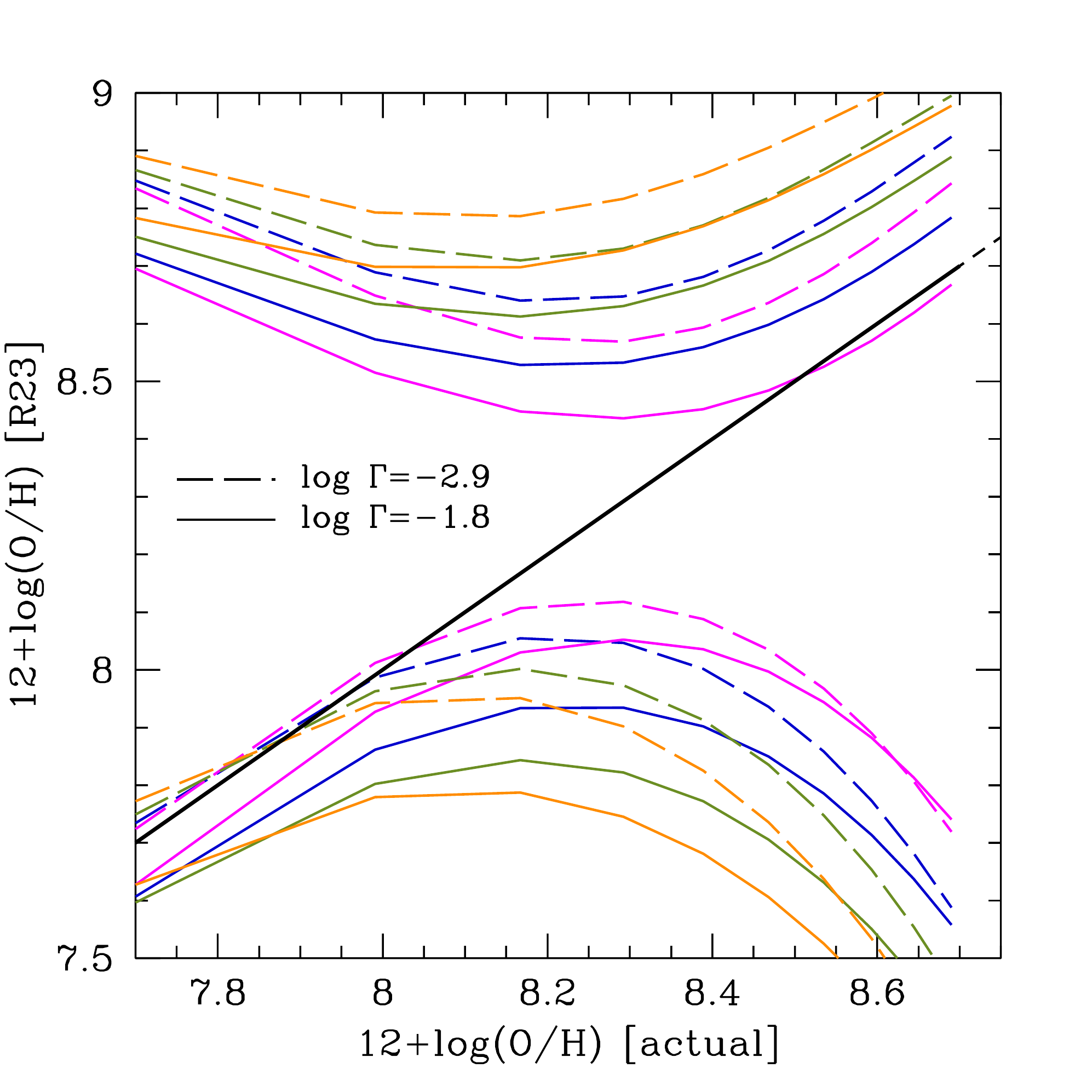}}
\caption{ (Left) Predictions of the photoionization models for the ``R23'' parameter as a function of
gas-phase metallicity, where the shaded regions in each color show the effects of varying $\Gamma$ over the range
indicated. (Right) R23-based metallicities versus actual gas-phase metallicity predicted by the model grids. 
Two curves are plotted for each color-coded set of models: the solid curve corresponds to ${\rm log \Gamma=-1.8}$ and the
long-dashed curve corresponds to ${\rm log \Gamma=-2.9}$. The two distinct sets of curves for a given $T_{\rm eff}$
reflect the well-known
double-valued behavior of R23-based metallicities. 
The lower set of curves uses the lower branch calibration of \cite{mcgaugh91}, while the upper set
of curves uses the upper-branch calibration of \cite{kobulnicky04}-- both calibrations use O32 to correct
for ionization parameter variations in addition to the R23 parameter. The black solid line indicates equality between
the R23-inferred and model values of 12+log(O/H). Note that neither R23 calibration performs well in the $Z \sim 0.2-1.0$ Z$_{\odot}$ range
(i.e., 12+log(O/H) = 8.0-8.7.)}  
\label{fig:R23}
\end{figure*}

For the moment we treat $T_{\rm eff}$ of the ionizing sources independently of the 
metallicity of the ionized gas producing the observed emission lines. The rationale for doing this, explained in more detail
below, is once again that, given the uncertainty in modeling massive star populations as a function of stellar metallicity,
we choose to fix the input spectrum in order to better understand the sensitivity of the strong-line indicators to
{\it gas phase} metallicity.  
Similarly, it seems prudent not to assume that other physical conditions in high-redshift \ion{H}{2} regions are similar
to local ones until it has been shown to be the case; for this reason, our models include no assumptions about dust, depletion of elements
onto dust grains, or non-solar abundance ratios of any elements relative to O.   

The range of electron density $n_{\rm e}$ used in the model grids was chosen based on the approximate
range inferred from observations of the density-sensitive 
${\rm [OII]\lambda3727/[OII]\lambda3729}$ ratio for a sub-sample of 113 KBSS-MOSFIRE galaxies having appropriate J-band spectra\footnote{
While $n_e$ can also be estimated from the ratio ${\rm [SII]\lambda 6718/[SII]\lambda 6732}$, these lines are considerably weaker 
than the [OII] lines for most objects in our sample, limiting the number of objects with ratios determined with sufficient S/N.}
(to be described in more detail in future work). 
For 90 spectra of individual objects with 
$2.06 < z < 2.62$ and significant detections ($>5\sigma$) of both members of the [OII]$\lambda\lambda 3726$, 3729 doublet, we find
$I(3727)/I(3729) = 0.86_{-0.15}^{+0.29}$ (median, with errors corresponding to the 16th and 84th
percentile).  The corresponding electron densities are  
$n_{\rm e}  \simeq 220_{-160}^{+380}$ cm$^{-3}$ for $10000 < T_{\rm e} < 14000$ K,  
with the largest values approaching $n_{\rm e} \simeq 2000$ (see section~\ref{sec:extreme} below)
A stacked spectrum of all 113 J band [OII] spectra also has $I(3726)/I(3729)=0.86 \pm 0.03$ (Figure~\ref{fig:o2ratio}).  

Figure~\ref{fig:bpt_mod} shows model BPT diagrams where the solid curves and shading are 
as in Figure~\ref{fig:bpt} but the KBSS-MOSFIRE data points have been suppressed for clarity. 
Two versions of the model are plotted, representing the approximate range of electron density $n_{\rm e}$ 
among the sample. 
We focus on the results for $n_e=1000$ 
cm$^{-3}$ for the purpose of discussion; the main effect of the lower-density
model grid is to require values of $T_{\rm eff}$ higher by $\sim 5000$ K to reproduce a given value of ${\rm log([OIII]/\Hb)}$.
The left-hand panel of Figure~\ref{fig:bpt_mod} shows that the locus of models 
with $T_{\rm eff} = 50000$ K and $-2.9 \le {\rm log}\Gamma \le -1.8$, with metals $Z/Z_{\odot} = 0.2-1.0$, follows very closely
the global fit to the KBSS-MOSFIRE BPT data presented above (equation~\ref{eqn:bpt_fit}); if the $T_{\rm eff} = 45000$ and $T_{\rm eff} = 55000$ grids
are included, the correspondence with the full distribution of the $z \sim 2.3$ galaxies is remarkably good. 
Similarly, Figure~\ref{fig:met_comp_mod} shows that the same range of model parameters predicts a relationship between
the N2 and O3N2 indices in excellent agreement with the observations (cf. Figure~\ref{fig:met_compare}.)

Unfortunately, in the context of the models that work well to reproduce the observations, neither
N2 nor O3N2 is particularly sensitive to the oxygen abundance in the {\it ionized gas}, which of course is known {\it a priori} for the models. 
In fact, the position of the model locus on the BPT diagram is nearly independent of {\it gas-phase} oxygen abundance 
over the modeled range (0.2-1.0 times solar); the position along the BPT sequence is sensitive primarily to
ionization parameter $\Gamma$, while the maximum value of ${\rm log([OIII]/\Hb)}$ reached in the BPT 
diagram is closely related to $T_{\rm eff}$ of the assumed ionizing radiation field.  
Figure~\ref{fig:bpt_mod}a shows that the ``metallicity sequence'', such as it is, is a very subtle effect,
in which a factor of 5 change in gas-phase metallicity moves the locus primarily vertically, but only by $\simlt \pm0.05$ dex
(and the trend with metallicity is not monotonic). 
Other strong-line methods would be equally problematic; for example, we find
that the same range of model parameters predicts that ${\rm log(([OIII]+[OII])/\Hb)}$, the ratio
upon which the ``R23'' method depends, is also essentially independent of input gas-phase metallicity (Figure~\ref{fig:R23}). 
The implication is that, if the models are reasonable, essentially all galaxies in the KBSS-MOSFIRE sample
are consistent with having anywhere from 0.2-1.0 times solar nebular oxygen abundance, and that the strong-line ratios
are probably not measuring ${\rm 12+log(O/H)}$ {\it of the ionized gas} at high redshifts. The implications for
strong-line metallicity measurements are discussed in section~\ref{sec:abundances_implications} below.

\begin{figure}[htbp]
\centerline{\includegraphics[width=8.5cm]{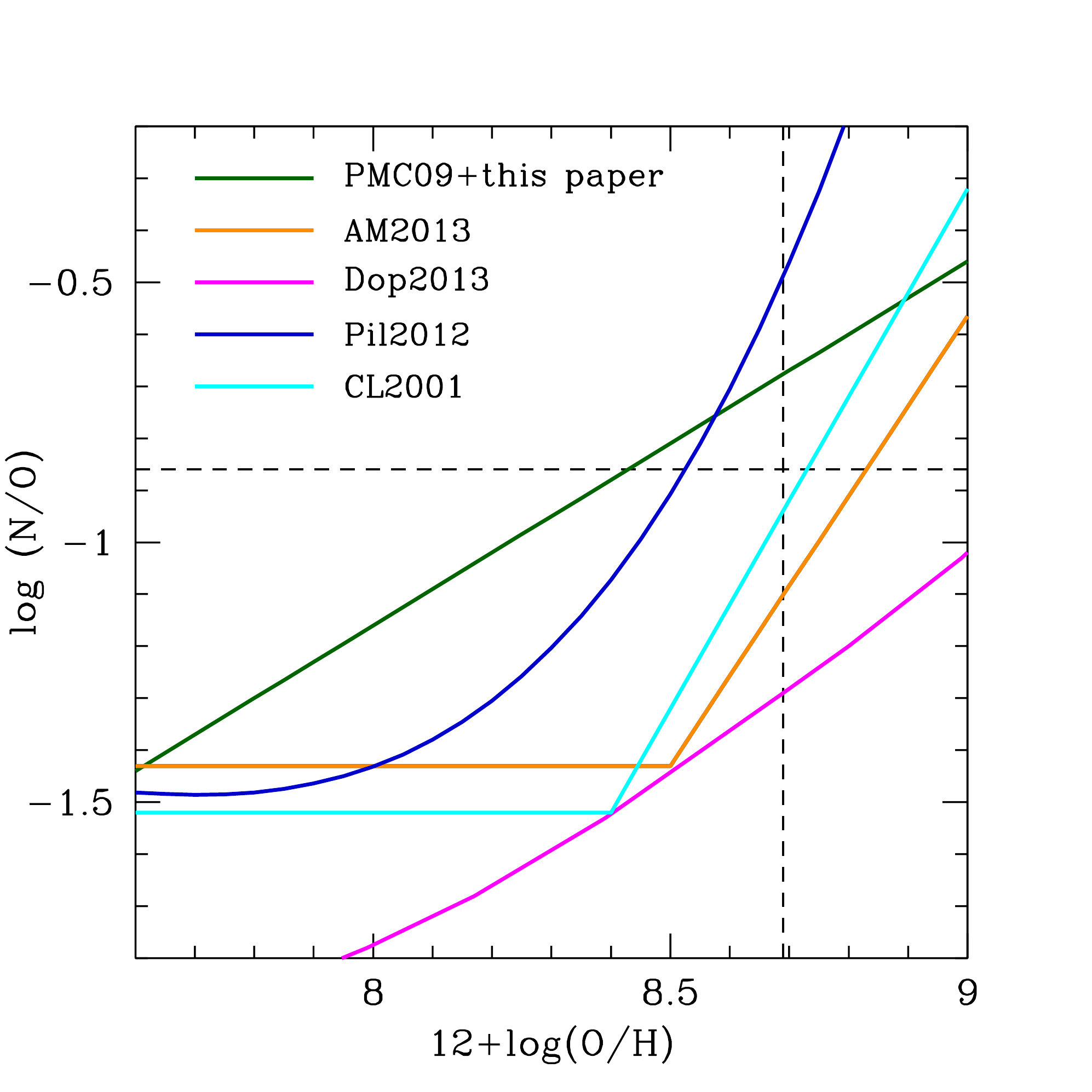}}
\caption{Plot showing examples of N/O versus O/H trends and/or assumptions from the recent literature (PMC09=\citealt{perez-montero+contini09};
AM2013 = \citealt{andrews+martini13}; Dop2013 = \citealt{dopita13}; Pil2012=\citealt{pilyugin12}; CL2001=\citealt{charlot+longhetti2001}. )
The solar values
of 12+log(O/H) and log(N/O) are indicated with vertical and horizontal dashed lines, respectively. At low oxygen abundance, several of the
results show the ``plateau'' near ${\rm log (N/O) \simeq -1.5}$, usually attributed to primary N enrichment. The PMC09 result is based on 
a collection of both extragalactic \ion{H}{2} regions and emission line galaxies having $T_e$-based measurements of 
N and O abundances; Pil2012
is based on a literature sample of extragalactic \ion{H}{2} regions with $T_e$ measurements, AM2013 is based on stacks of SDSS galaxies in bins of ${\rm M_{\ast}}$, also based on the direct method.  CL2001 
is a parametrization of data presented by \cite{henry99}, while Dop2013 is a new fit of data presented by \citet{vanzee98} (note
that \cite{dopita13} assumes a much lower value of the solar N/H than our assumption of ${\rm log(N/H) = 7.83}$). }
\label{fig:no_vs_oh}
\end{figure} 

\begin{figure*}[htbp]
\centerline{\includegraphics[width=9.0cm]{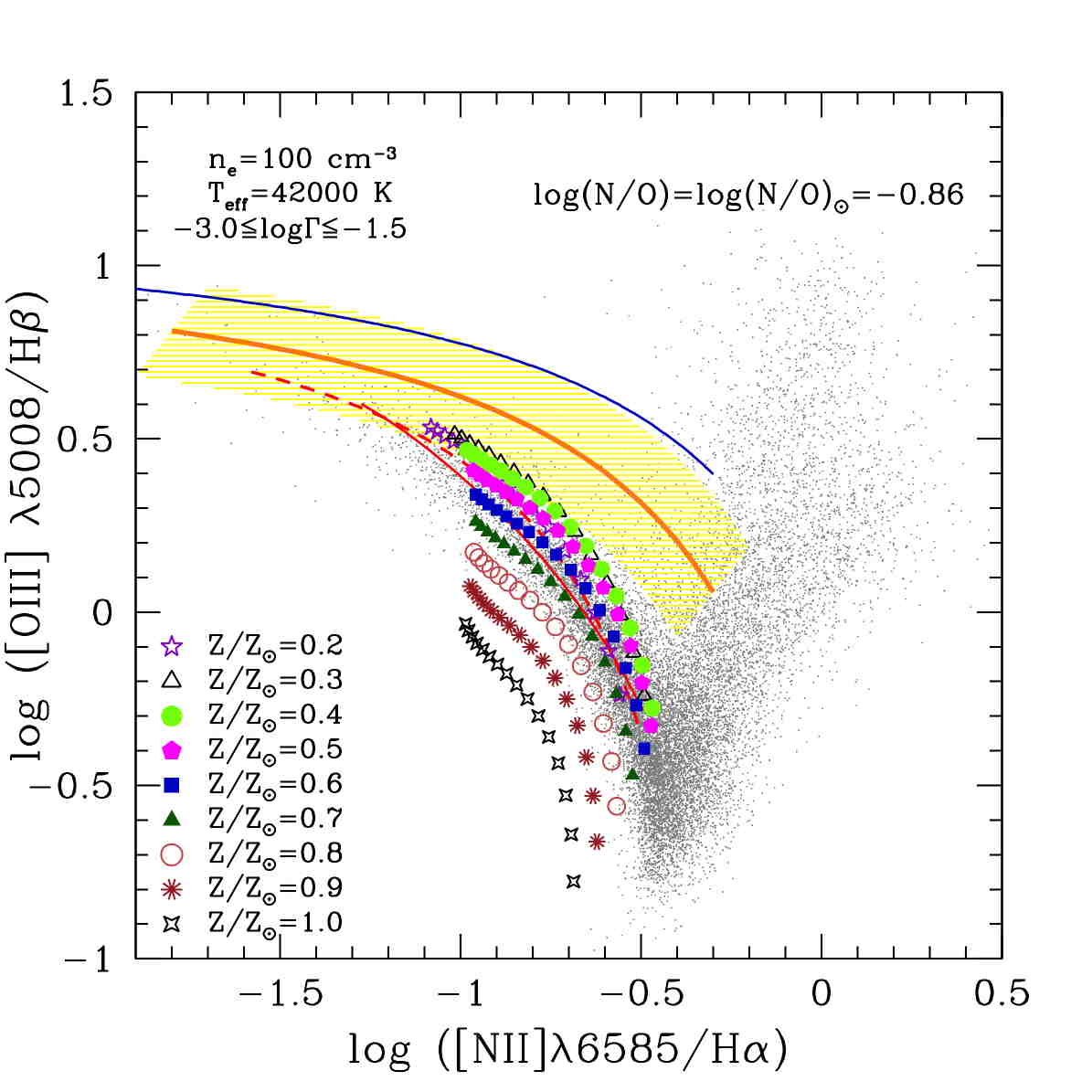}\includegraphics[width=9.0cm]{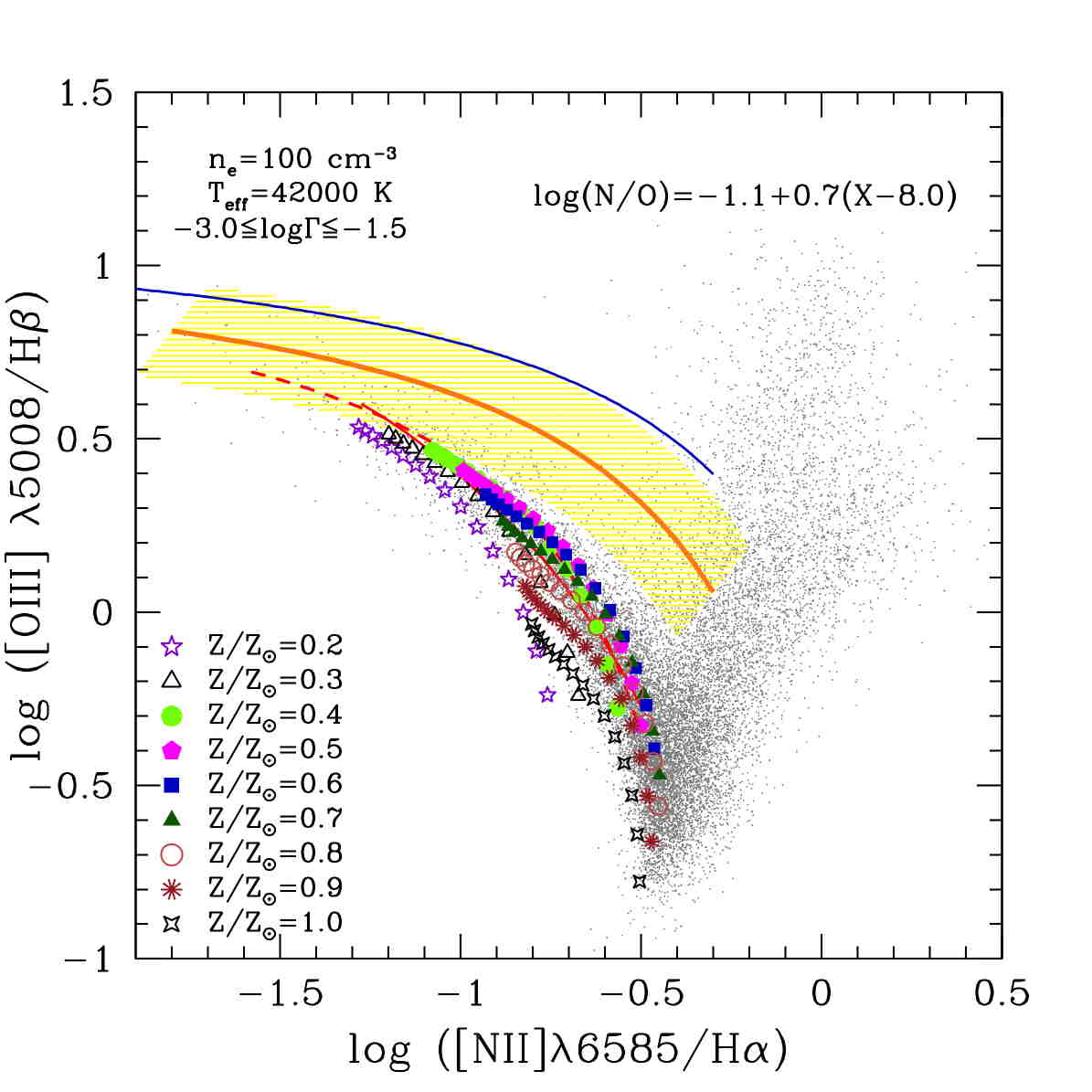}}
\caption{
BPT diagram showing example photoionization model grids capable of  
reproducing the salient features of the star-forming galaxy sequence at low redshift. 
The point colors and symbols depend on assumed O/H relative to the solar values, with
the ionization parameter $\Gamma$ spanning the same range at each metallicity, in steps of ${\rm \Delta log \Gamma=0.1}$.
The assumed radiation field shape is that of a
$T_{\rm eff}=42000$ K blackbody. ({\it Left:}) Models for which the solar ratio of N/O (i.e., independent of O/H) has been assumed
({\it Right:}) Models which have N/O varying with O/H as in equation~\ref{eqn:no_vs_oh}. Note that the assumptions about N/O 
have two effects: one is to shift the model loci onto the low-redshift BPT locus. The other is to introduce a dependence
on O/H of the position along the BPT sequence, at least for $Z/Z_{\odot} \simlt 0.5$.  
Note that in both model sets, no assumptions have been made about depletion of N or O onto grains (both
assume zero depletion). 
 }
\label{fig:bpt_demo}
\end{figure*}

\begin{figure*}
\centerline{\includegraphics[width=9.0cm]{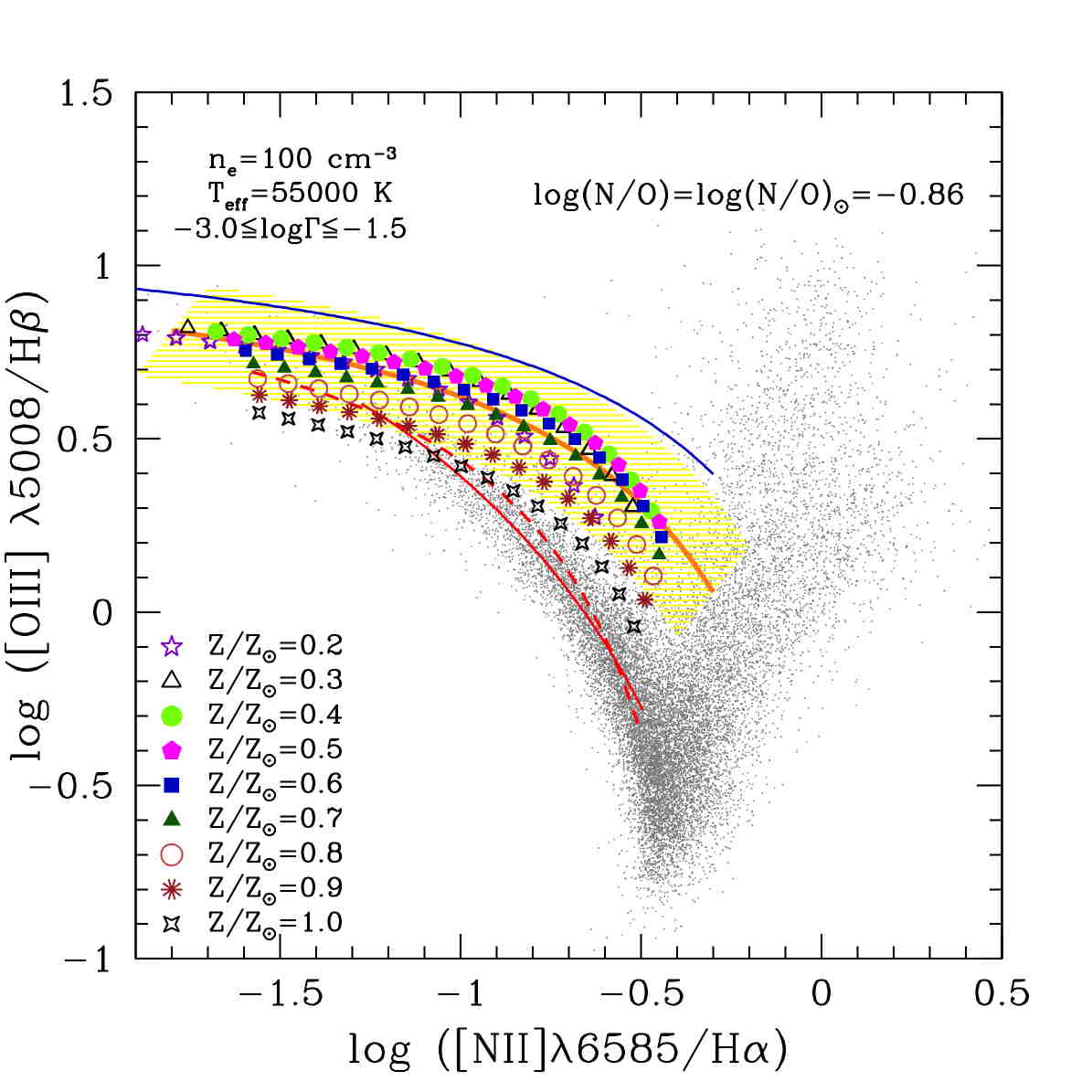}\includegraphics[width=9.0cm]{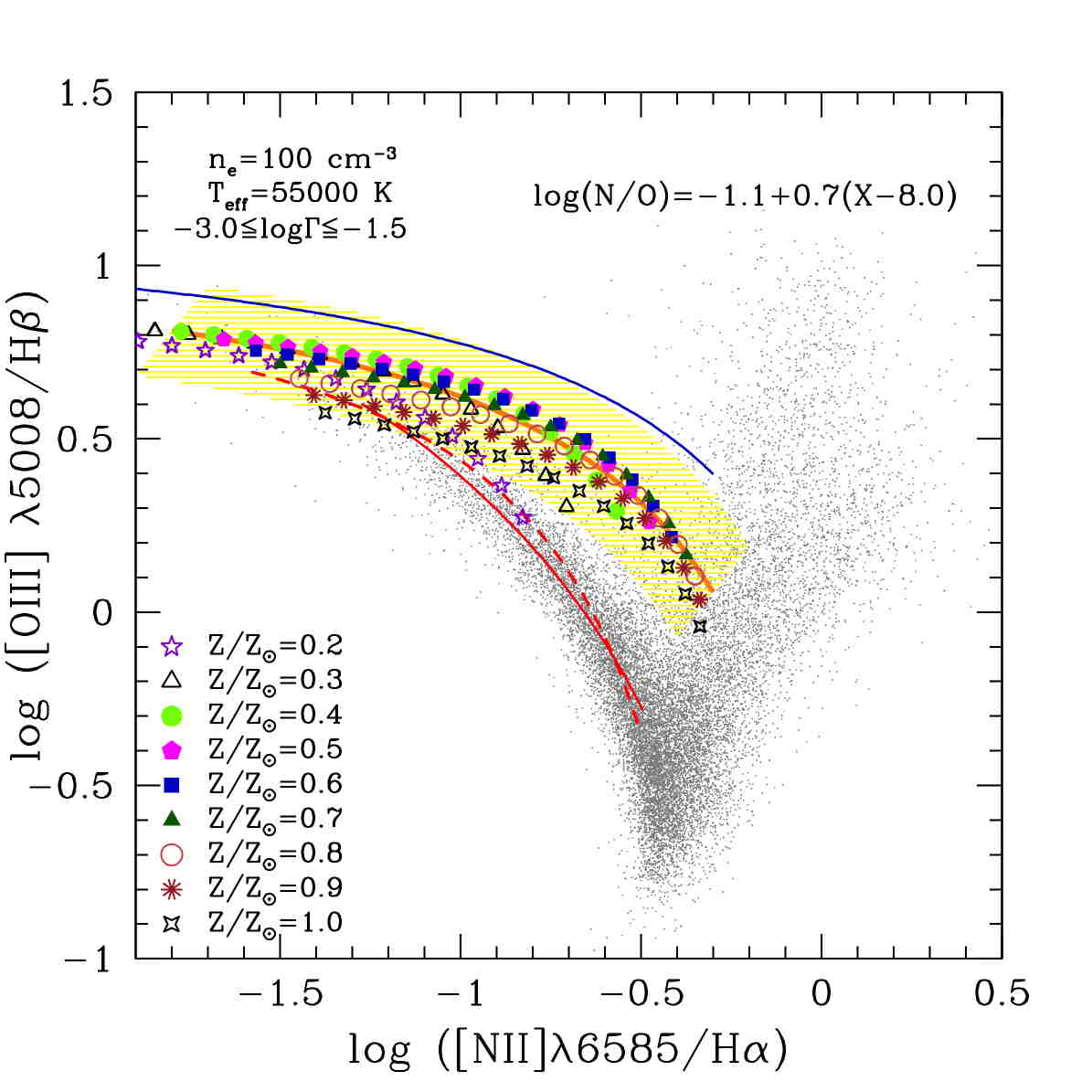}}
\caption{Predictions for the locations on the BPT diagram as in Figure~\ref{fig:bpt_demo}, where the only change is that
$T_{\rm eff}$ of the ionizing sources has been increased from 42000 K to 55000 K. As in Figure~\ref{fig:bpt_demo}, the left-hand
panel is for solar N/O, while the right-hand panel assumes that N/O is dependent on O/H according to equation~\ref{eqn:no_vs_oh}.   }
\label{fig:bpt_demo2}
\end{figure*} 

\subsection{Implications for Strong-Line Metallicity Calibrations}
\label{sec:abundances_implications}

The utility of strong lines for estimating metallicity at high redshift can still be salvaged as long
as $T_{\rm eff}$ and/or $\Gamma$ are monotonically correlated with {\it stellar} metallicity, likely
to closely trace the gas-phase metallicity for such young stars. 
Such correlations are expected at some level, though they
are arguably more model-dependent. It is known
that early O-type Magellanic Cloud stars of a given spectral classification have $T_{\rm eff}$ higher by several
thousand K compared to their Galactic counterparts (e.g., \citealt{massey05}). A systematic shift
from $T_{\rm eff} \simeq 42,000$ K to $T_{\rm eff}= 50,000$ K for the stars dominating the ionizing radiation field 
could produce a vertical shift in the BPT diagram of ${\rm \Delta \left( log([OIII]/\Hb)\right) \simeq 0.3}$ dex
(see, e.g., Figure~\ref{fig:bpt_mod}). In general, stellar metallicity is expected to affect the shape of
the ionizing radiation field \citep{shields+tinsley76} over the critical range $1-4$ Rydberg relevant for ionizing
H, He, O, N, and S and producing the observed nebular lines. Harder ionizing UV spectra 
are expected at lower overall metallicity due to reduced metallic line blanketing in the stellar photospheres; metallicity
also strongly affects the prevalence, composition, and structure of massive star stellar winds (e.g., \citealt{kudritzki+puls2000}),
which in turn have implications for the degree of stellar rotation.  These effects and their possible implications are discussed further
in section~\ref{sec:teff} below.   

If one admits the possibility that the strong line ratios at high redshifts are driven primarily by factors
only indirectly related to gas-phase metallicity, why do they appear to work at low redshift?  
For the empirically-calibrated O3N2 and N2 relations, for example, the oxygen abundances are rather directly
related to galaxy positions along the BPT sequence. Lines of constant O3N2 index are very close to being
perpendicular to the low-redshift BPT star-forming locus. However, as we show in
Figures~\ref{fig:no_vs_oh} and \ref{fig:bpt_demo} and discuss below, a large part of this behavior may be attributable to variations in
N/O, and not O/H. 

In this context, it is of interest to
ask whether the low redshift BPT sequence can be reproduced using simple photoionization models similar to those applied above to the 
high redshift data.  As shown in Figure~\ref{fig:bpt_demo}, the characteristic shape of the low-redshift BPT locus can be reproduced
by assuming ionizing sources with $T_{\rm eff} \simeq 42000$ K, ionization parameter extending to slightly lower values than
those required for the $z \sim 2.3$ locus, and metallicities $Z/Z_{\odot} \simeq 0.5-0.7$, substantially lower than usually
ascribed to the low-excitation branch of the BPT diagram. The typical metallicities required to match the
nearly-vertical portion of the BPT locus depend to a large degree on assumptions built into models-- most commonly, 
the dependence of (N/O) on (O/H). There is ample evidence in local \ion{H}{2} regions for strong systematic variation of N/O
with O/H when both have been determined by the direct $T_{\rm e}$ method; for example, data compiled by \citet{pilyugin12} 
show that for ${\rm 12+log(O/H) \simlt 8.2}$, log(N/O)$\simeq -1.45$ ([N/O]$=-0.6$), but that for higher oxygen abundance, 
${\rm log(N/O) \simeq -1.45 + 1.8[12+log(O/H)-8.2]}$ (see Figure~\ref{fig:no_vs_oh}, where we show a third-order polynomial fit
to the \citealt{pilyugin12} data set). Qualitatively similar results have been obtained by many other
studies (e.g., \citealt{vc+edmunds93,vanzee98,henry99,andrews+martini13}). The generally-accepted interpretation is that, for low oxygen abundances, 
\ion{H}{2} regions are chemically very young, so that only primary N is present
(hence the plateau at ${\rm log(N/O) \simeq -1.5}$), while at higher O/H,  N/H includes contributions from
both primary and secondary N enrichment, with the result that N/O increases more rapidly than O/H.   
\citet{pilyugin12} show that the solar ratio
of N/O (log(N/O)$ \simeq -0.86$) is reached when ${\rm 12+log(O/H) \simeq 8.5}$, with N/O becoming super-solar [${\rm log(N/O)\simeq -0.6}$]
for solar (O/H) (${\rm 12+log(O/H) = 8.69}$). Other results suggest more gradual changes in (N/O) with (O/H) (e.g., \citealt{perez-montero+contini09})
or a significantly higher (O/H) at the transition from primary to primary+secondary N (e.g., \citealt{andrews+martini13})-- suggesting that
the precise behavior depends on the nature of the calibration sample and the details of the methods used to measure the abundances. 
Figure~\ref{fig:no_vs_oh} illustrates the substantial range of N/O versus O/H from the recent literature (by no means exhaustive). 

Clearly, assumptions
concerning the behavior of N/O versus O/H directly affect the predicted locations of models in the BPT plane for a given $\Gamma$ and $T_{\rm eff}$--
for example, lowering N/O by 0.2 dex relative to solar at a given O/H essentially shifts the entire sequence by 0.2 dex in N2, i.e., toward the
left in the BPT diagram\footnote{
In addition to the behavior of (N/O) vs. (O/H), another common assumption in photoionization models 
is that gas-phase N and O are depleted by amounts similar to
those observed in Galactic \ion{H}{2} regions (e.g., \citealt{esteban04}), typically $\simeq 0.07-0.09$ dex for each. In our models,
we have made no attempt to account for O or N outside of the gas phase; assuming the Orion nebula depletions would raise
the inferred total abundance of oxygen by $\sim 20$\% relative to those shown in Figure~\ref{fig:bpt_demo}. }
.  
For the same reason, the dependence of N/O on O/H within samples used for local (empirical) calibration of strong line abundance
indicators is ``built-in'' to any method that makes use of the N2 line ratio-- even when no explicit reference is made to N/O. 
Similarly, the ratio ${\rm log([NII]\lambda 6585/[OII]\lambda3729)}$, often used as an indicator of oxygen abundance, is only weakly dependent
on O/H when N/O is held fixed, and the mapping of strong line ratios to O/H depends almost entirely on what has
been assumed for (N/O) as a function of O/H.  The right-hand panel of Figure~\ref{fig:bpt_demo} shows the effects in the BPT plane of 
the assumption of modest dependence of N/O on O/H, 
\begin{equation}
{\rm log (N/O)} = -1.1 + 0.7(X - 8.0) 
\label{eqn:no_vs_oh}
\end{equation}
where $X = {\rm 12+log(O/H)}$. This relation, which predicts log(N/O)$ = -0.62$ for solar O/H,
is consistent with the local sample of giant extragalactic \ion{H}{2} regions and \ion{H}{2} galaxies 
compiled by \citet{perez-montero+contini09}, as well as with our inferences for the $z \sim 2.3$ KBSS-MOSFIRE sample
as discussed below.

Many commonly-used models have encoded the assumptions into the model grids,
and in some cases they assume a very rapid increase of N/O over the most relevant range of O/H (e.g., \citealt{charlot+longhetti2001,dopita13}.) 
The issue of how N/O affects strong line methods of measuring O/H is discussed in detail by \citet{perez-montero+contini09}; see section
\ref{sec:pp04_calibration}. 
In spite of the well-established (though not necessarily numerically agreed-upon) trends of N/O as a function of O/H 
in nearby \ion{H}{2} regions and emission line galaxies, similar measurements are not yet available for galaxies at high redshift.   

\subsection{Implications for N/O at $z\simeq 2.3$}

An interesting, and potentially important, issue emerging from the KBSS-MOSFIRE data and the photoionization models that
reproduce them is the implication for (N/O) in the \ion{H}{2} regions of the high redshift galaxies:
models that simultaneously produce the observed values of [NII]/\Ha\ and [OIII]/\Hb\ ratios do not obviously
require non-solar (N/O).   
One possibility would be that the models resemble the data by accident, and that, if the correct N/O dependence
on O/H were in the models, a different set of parameters would be required to match the data.  
However, there also exists evidence that (N/O) may behave differently in the high redshift galaxies compared 
to local HII regions with the same range in 12+log(O/H). 
For example, from stacks of SDSS spectra in bins of SFR, \cite{andrews+martini13} found qualitatively
similar (N/O) behavior compared to those of \citet{pilyugin12} for galaxies with ${\rm log(SFR/M_{\odot} yr^{-1}) \le 1}$, with
a ``plateau'' in N/O at low O/H, and a rapid increase in N/O above ${\rm 12+log(O/H) \simeq 8.5}$ (note that the
transition occurs at an oxygen abundance higher by 0.2 dex than \citet{pilyugin12} in spite of the fact that both
samples were based on direct metallicity measurements-- see Figure~\ref{fig:no_vs_oh}). However, for the sub-samples of
galaxies with SFR similar to those of the KBSS-MOSFIRE sample (${\rm log(SFR/M_{\odot} yr^{-1}) \simgt 1}$), N/O does {\it not}
appear to vary with O/H and, moreover, {\it is consistent with solar} (see their Figure 14).  

We have also considered 
the ratio N2S2${\rm \equiv log([NII]\lambda 6585/([SII]\lambda 6718 +\lambda 6732))}$, proposed
by \citet{perez-montero+contini09} as a sensitive measure of (N/O) in \ion{H}{2} regions. N2S2 has the advantage of being insensitive
to extinction and (in the case of KBSS-MOSFIRE) involves lines measured simultaneously in the K-band spectra.  
We find that ${\rm N2S2 \simeq -0.1 \pm 0.1}$, i.e., nearly constant, in both individual spectra for which both [\ion{N}{2}] and [\ion{S}{2}] 
are detected, and in spectral stacks formed from subsets of the KBSS-MOSFIRE $z\sim 2.3$ sample. 
The photoionization models that reproduce the observed N2 and [OIII]/\Hb\  ratios predict  
N2S2 in the observed range if log(N/O)$\simeq -1.0\pm0.1$, independent of gas-phase oxygen abundance for $Z/Z_{\odot}=0.1-1$.  

Thus, at present we do not see evidence for the low values of log(N/O) ($\simeq -1.5$) that 
might be expected for very young systems in which only primary N has enriched the ISM, such as damped Lyman $\alpha$ systems (see e.g., \citealt{pettini08}),
nor of a strong dependence of (N/O) on (O/H) as expected as secondary N production progresses. Rather, most of the galaxies
in our sample appear to have log(N/O) within $\sim 0.2$ dex of log(N/O)$_{\odot}$. 
Of course, we cannot make strong statements about the galaxies with only
upper limits on ${\rm log([NII]/\Ha)}$. 

Figure~\ref{fig:bpt_demo2} illustrates another potentially important set of issues, when viewed together with Figure~\ref{fig:bpt_demo}: 
the higher overall excitation of the high redshift BPT sequence has the effect of ``compressing'' the predicted metallicity
dependence of the models, and it is possible to match both the low redshift and high redshift BPT sequences with the same
model (with the same modest dependence of N/O on O/H) where the only change was to increase $T_{\rm eff}$ from 42000 K to 55000 K. 
Figures~\ref{fig:bpt_demo} and \ref{fig:bpt_demo2} show why the lower portion of the BPT sequence, so    
prominent in the low-redshift galaxy samples where it is extremely sensitive to N/O versus O/H, may be much less apparent 
in high redshift sample.
At both low and high redshifts, it remains the case that the BPT sequence is primarily a sequence in $\Gamma$, 
with the vertical position (i.e., in log([OIII]/\Hb)) 
of the leftward ``bend'' being primarily sensitive to the radiation field shape.  
Comparison of the right-hand panels of Figures~\ref{fig:bpt_demo} and ~\ref{fig:bpt_demo2} shows that  
a galaxy's position in the BPT plane may be significantly less dependent on metallicity (N/O {\it or} O/H) as compared to local galaxies.  It also
suggests that the highest metallicity objects at high redshift might be expected in the region between the two ``branches'' of 
the BPT diagram, where they might ordinarily be classified as AGN or AGN/star-forming composite objects (see section~\ref{sec:agn}).  

\subsection{Is $T_{\rm eff}=50000$ K reasonable?} 

\label{sec:teff}

While more sophisticated modeling is beyond the scope of this paper, we note that 
the upper envelope of the KBSS-MOSFIRE $z = 2-2.6$ sample (Figure~\ref{fig:bpt})   
is well-represented by the models with $T_{\rm eff} \simeq 55000-60000$ K (see Figure~\ref{fig:bpt_mod}), 
which strongly resembles the so-called ``maximum starburst'' model curve of \citet{kewley01}, whose
main distinguishing characteristic was a much ``harder'' stellar ionizing radiation field between 1 and 4 Ryd compared to standard
stellar models. 
The main point is that high ionization parameter {\it and} hard (i.e., high $T_{\rm eff}$) ionizing spectra are both
required to easily match the observations. 

Stellar models capable of producing the inferred harder radiation field have been proposed-- particularly
those including binaries and/or rotation (e.g., \citealt{eldridge09,brott11,levesque12}; see also Figure~\ref{fig:bpass_compare}). 
In fact, the shape of the ionizing spectra of individual massive stars, and the expected net radiation field from massive
star populations, remain very uncertain, both theoretically and
observationally.     
Indeed, there are other areas
where tension exists between observations and stellar evolution models-- for example, the very blue observed
colors of a fraction of star-forming galaxies at $z \simgt 2.7$ which indicate little or no ``break'' shortward
of the rest-frame Lyman limit (e.g., \citealt{iwata09,nestor11,mostardi13}), or the possible short-fall of
H-ionizing photons from known galaxy populations at redshifts relevant for reionization. This issue
is discussed further in section~\ref{sec:discussion} below.   

Increased importance of binarity and rotation (which are expected to be strongly coupled, since mass loss in binary systems naturally 
produces more rapidly-rotating stars-- see, e.g. \citealt{eldridge12}) among massive stars could reasonably explain
a number of qualitative properties observed in the high-redshift \ion{H}{2} regions.  
Aside from producing massive stars that evolve toward {\it hotter} $T_{\rm eff}$ while on the main sequence, rapid rotation also results in  
larger UV luminosities and longer main sequence lifetimes at a given mass and metallicity \citep{brott11}. The effects become much
stronger in stars with metallicities comparable to that of the LMC (assumed in the models to have 12+log(O/H)=8.35, similar to
the mean inferred metallicity of the $z \simeq 2.3$ sample). Models suggest that all rapid rotators with $M \simgt 30 M_{\odot}$ spend a substantial fraction
of their lifetimes with $T_{\rm eff} \sim 50000-60000$ K, much hotter than their slowly-rotating counterparts (see Figure~7 of \citealt{brott11}).
Rapidly-rotating massive stars also produce much more N during their main-sequence evolution, possibly affecting the gas-phase
N/O in the surrounding nebula (see section~\ref{sec:abundances_implications} for further discussion). 
{\it The implication is that it may not be necessary to invoke unusual numbers of Wolf-Rayet stars, extremely young stellar population ages, or extremely
``top-heavy'' initial mass functions (IMFs) to understand the high $T_{\rm eff}$ that appear to be common in high redshift galaxies. }
It thus seems entirely plausible that the BPT sequence observed for $z \sim 2.3$ galaxies could be driven by changes in stellar
evolution that are favored in high redshift star-forming galaxies compared to most galaxies in the local samples. 

Of course, there are other potentially important physical processes that could alter the positions of galaxies
in the BPT plane. For example, 
it has been shown recently that shifts in the BPT diagram can result from \ion{H}{2} regions in 
which radiation pressure dominates over gas pressure, so that the standard assumption of constant electron density breaks down and the
structure of the \ion{H}{2} zone is fundamentally altered (\citealt{verdolini13,yeh13}.) 
Alternatively, shocks almost certainly play some role in modulating the locations of galaxies in the BPT plane (e.g., \citealt{newman13,kewley13}), 
particularly for galaxies with ${\rm log(M_{\ast}/M_{\odot}) \simgt 11}$ where kinematic evidence for AGN activity often accompanies
large values of the N2 ratio (\citealt{forst14,genzel14}; see also section~\ref{sec:agn}).  Objects for which shocks dominate (energetically) 
over star formation in the integrated spectrum are evidently rare at lower stellar masses in the high redshift universe; however, 
it has been shown that, at least in some cases, spatially resolved spectroscopy (particularly with higher spatial resolution observations
assisted by adaptive optics) reveals nuclear regions dominated by shocks and/or AGN excitation in what might otherwise appear
to be a normal star-forming object \citep{wright10}.  
However, the bulk of $z \sim2.3$ BPT locus is most easily explained by photoionization, as described above. 
For the sake of simplicity, since the vast majority of the current sample does not appear to require 
shocks to explain the observations in the context of the BPT 
diagram, we will not consider them further.

\subsection{N2 and O3N2 Calibrations, Revisited}

\label{sec:pp04_calibration}

Thus far we have used the calibrations presented by PP04 for mapping the N2 and O3N2 indices
to oxygen abundances determined from direct $T_{\rm e}$ measurements.  We have seen above that
both of these calibrations are sensitive (through the N2 index) to the behavior of N/O as
a function of O/H, so that differences in this behavior between the local calibration set and
the high-redshift galaxies could produce systematic 
differences in inferred 12+log(O/H). Systematically higher N/O at a given O/H in the high
redshift sample could potentially account for the N2-inferred oxygen abundances being
systematically higher than the corresponding O3N2 values (see Figure~\ref{fig:met_compare}). 

According to the linear versions of the PP04 calibrations,
\begin{equation}
{\rm 12+log(O/H)_{N2} = 8.90+ 0.57\times N2}
\label{eqn:pp04_n2}
\end{equation}
and
\begin{equation}
{\rm 12+log(O/H)_{O3N2} = 8.73-0.32\times O3N2 }
\label{eqn:pp04_o3n2}
\end{equation} 
so the dependence of the inferred metallicity on N2 is shallower
in the case of O3N2. In addition, both PP04 fits intentionally cover a wide range 
of line indices-- considerably wider than the range observed in the current KBSS-MOSFIRE
sample-- in order to calibrate the index over the widest possible metallicity range. 
It may be useful in the case of the $z\sim2.3$ sample to restrict the calibration
data set to the same range of N2 and O3N2 index observed, since it allows 
estimation of the calibration uncertainties most relevant to the high redshift sample 
(see section~\ref{sec:mzr_scatter} below). 

\begin{figure*}[htbp]
\centerline{\includegraphics[width=8.5cm]{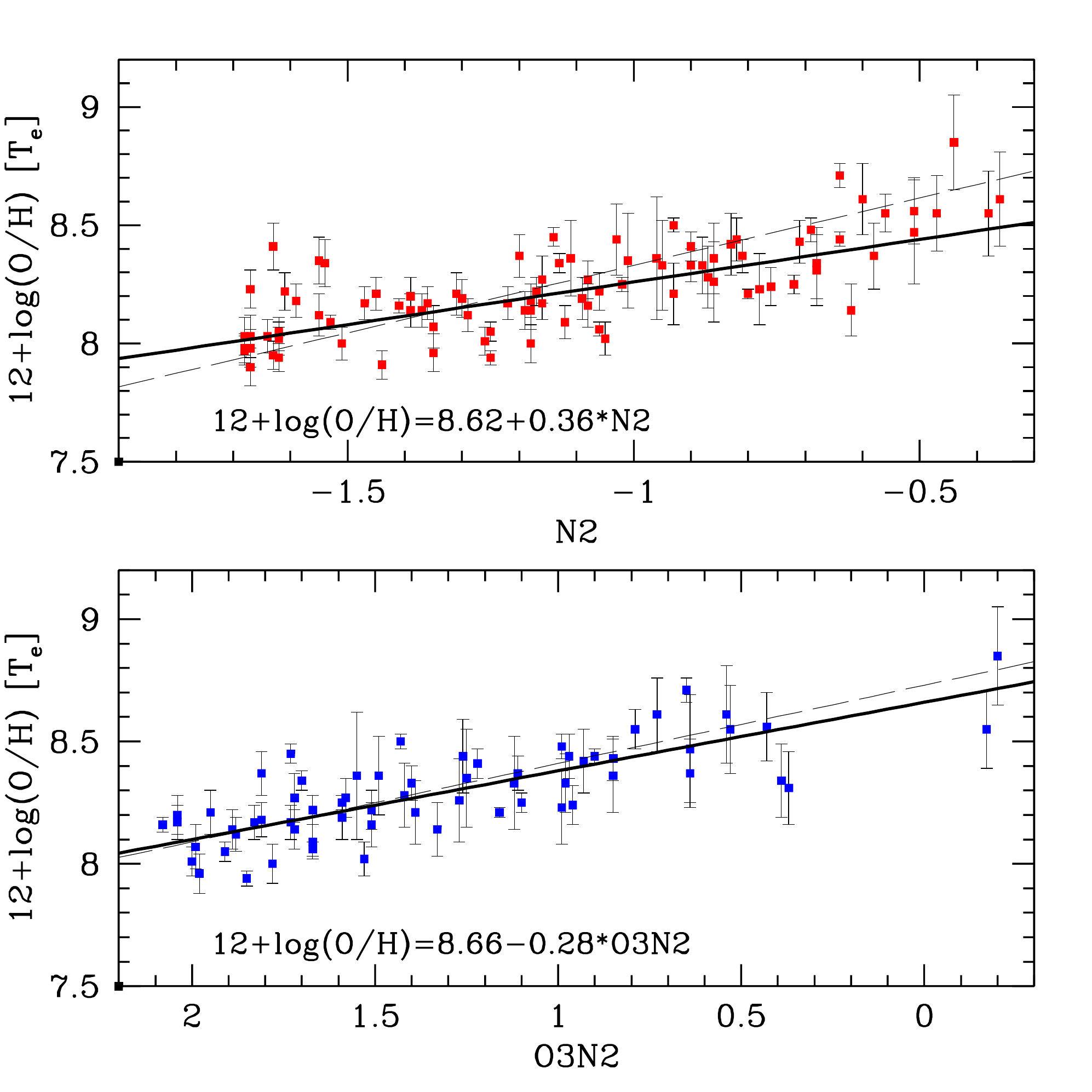}\includegraphics[width=8.5cm]{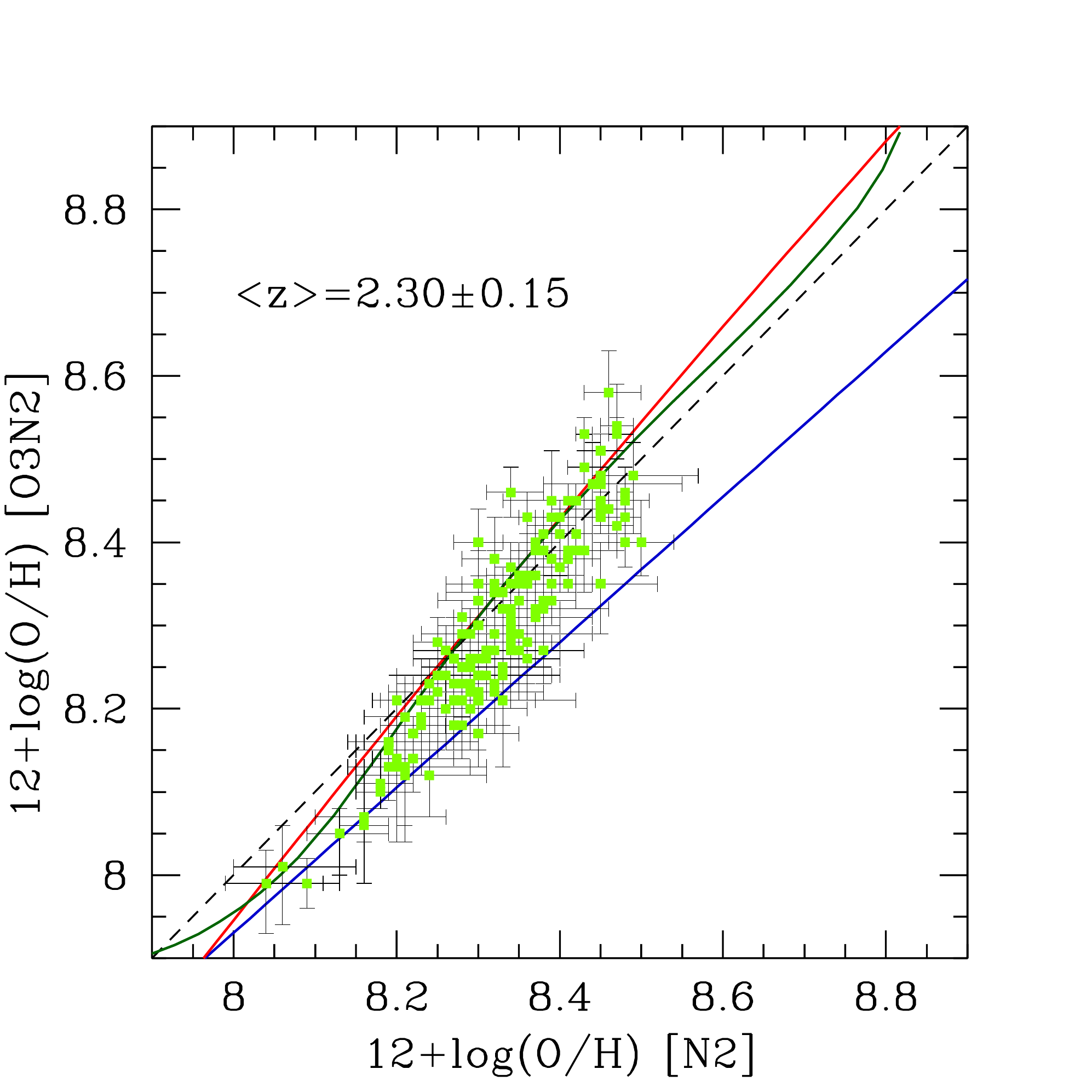}}
\caption{(Left) Linear regressions between $T_{\rm e}$-based metallicity and the N2 line index (${\rm N2 \equiv log([NII]\lambda 6585/\Ha)}$ 
(top) and O3N2${\rm \equiv log([OIII]\lambda 5008/\Hb) - log([NII]\lambda 6585/\Ha)}$ (bottom) for a subset of the data used by PP04, as
described in the text. The modified fits, given as an equation in each panel, 
are shown by the heavy solid lines.  The best linear fits given by PP04 (and used in this paper) 
are shown with lighter, dashed lines. (Right) Same as Figure~\ref{fig:met_compare}, but assuming the modified
calibrations for O3N2 and N2 metallicity measurements given in equations \ref{eqn:N2_new} and \ref{eqn:O3N2_new}.
The blue solid line is best fit linear relationship from Figure~\ref{fig:met_compare}, using the original PP04
calibrations. }
\label{fig:pp04_new}
\end{figure*}

We have repeated the fits to the N2 and O3N2 metallicity calibrations of PP04, using 
{\it the same data set and measurement errors as PP04}, with the following exceptions:
first, we limited the regression to the range of  
N2 and O3N2 line indices observed in the KBSS-MOSFIRE $z\sim 2.3$ sample, and second, we
have included only the data points for which the oxygen abundance was measured using
the direct $T_e$ method, to reduce the effect of systematics on the overall metallicity scales. 
The results of the least-squares fits are as follows (see Figure~\ref{fig:pp04_new}:)

For N2, 
\begin{equation}
{\rm 12+log(O/H)_{N2}}  = {\rm 8.62 + 0.36\times N2}  
\label{eqn:N2_new}
\end{equation}
$$ ({\rm \sigma}  =  {\rm 0.13~dex; ~\sigma_{sc} = 0.10~dex})  $$
where the new fit includes only the PP04 data for which ${\rm -1.7 \le N2 \le -0.3}$ 
(92 measurements). 

For O3N2, 
\begin{equation}
{\rm 12+log(O/H)_{O3N2} = 8.66 - 0.28 \times O3N2}
\label{eqn:O3N2_new}
\end{equation}
$$ ({\rm \sigma = 0.12~dex,~\sigma_{sc} = 0.09~dex})$$
where the fit was restricted to the range ${\rm -0.4 \le O3N2 \le 2.1}$, again including
only direct $T_{\rm e}$-based oxygen abundances (65 measurements). In both relations,
$\sigma$ is the weighted error between the data points and the best fit, and $\sigma_{\rm sc}$ 
is an estimate of the {\it intrinsic} scatter calculated in a manner analogous to that used
above for the $z \sim 2.3$ BPT sequence fits.  The values of $\sigma$ should be compared
to those obtained by PP04, $\sigma =0.18$ dex and $\sigma=0.14$ dex for N2 and O3N2, respectively. 
Both calibrations 
become tighter when considered over the smaller range of line index, with $\sigma_{\rm sc} \simeq 0.10$ dex
and $\sigma_{\rm sc} \simeq 0.09$ dex for N2 and O3N2, respectively. These values should be viewed as the {\it minimum}
uncertainties in the calibration between the N2 and O3N2 line indices and 12+log(O/H) at $z \simeq 0$. 

Using the revised regression formulae in equations
\ref{eqn:N2_new} and \ref{eqn:O3N2_new} (shown in Figure~\ref{fig:pp04_new}) lowers the systematic
offset between the two indicators when applied to the $z\sim 2.3$ sample, primarily
because the coefficient in front of the N2 index in equation~\ref{eqn:N2_new} is substantially reduced 
relative to that in equation~\ref{eqn:pp04_n2}, while the new calibration of [O3N2] (equation
\ref{eqn:O3N2_new}) is nearly identical to the original PP04 solution (equation~\ref{eqn:pp04_o3n2}). 

Although we have used the data set assembled by PP04, a more recent calibration of O3N2-based oxygen
abundances by
\citet{perez-montero+contini09} finds an almost identical linear fit to that of
PP04, ${\rm 12+log(O/H)= 8.74-0.31\times O3N2}$, using a larger sample of $T_e$-based measurements.
In addition, these authors examined how the strong-line calibration of O/H would be affected
systematically by N/O; they find that the overall scatter is reduced substantially if 
a term dependent on N/O is included, 
\begin{equation}
{ \rm 12+log(O/H) = 8.33-0.31\times O3N2-0.35log(N/O) } 
\label{eqn:pmc09}
\end{equation} 
Equation~\ref{eqn:pmc09} becomes identical to that of PP04 if ${\rm log(N/O)=-1.17}$
(${\rm [N/O] = -0.3}$) 
and matches the normalization of equation~\ref{eqn:O3N2_new}  
at the median O3N2 of the calibration data set (corresponding to ${\rm 12+log(O/H)=8.32}$) 
if ${\rm log(N/O)=-1.06}$ (${\rm [N/O] = -0.2}$), close to the values inferred for
the high redshift sample. 

Thus, there is some cause for optimism that, at least in the case of O3N2, the inferred
oxygen abundances are not likely to be strongly biased by differences in N/O between 
the calibration data set compared to that of the high-redshift sample. 

\begin{figure*}[htbp]
\centerline{\includegraphics[width=9cm]{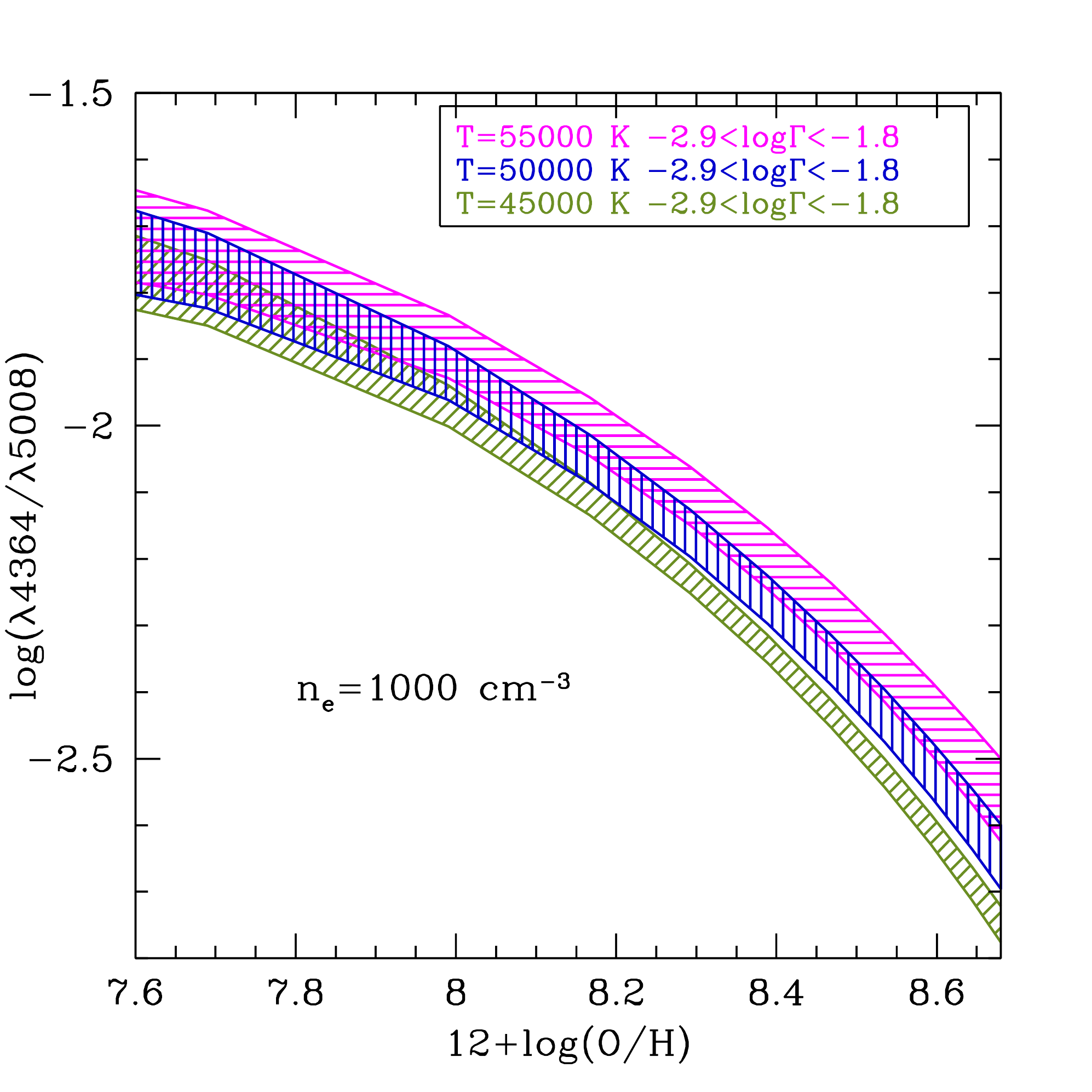}\includegraphics[width=9cm]{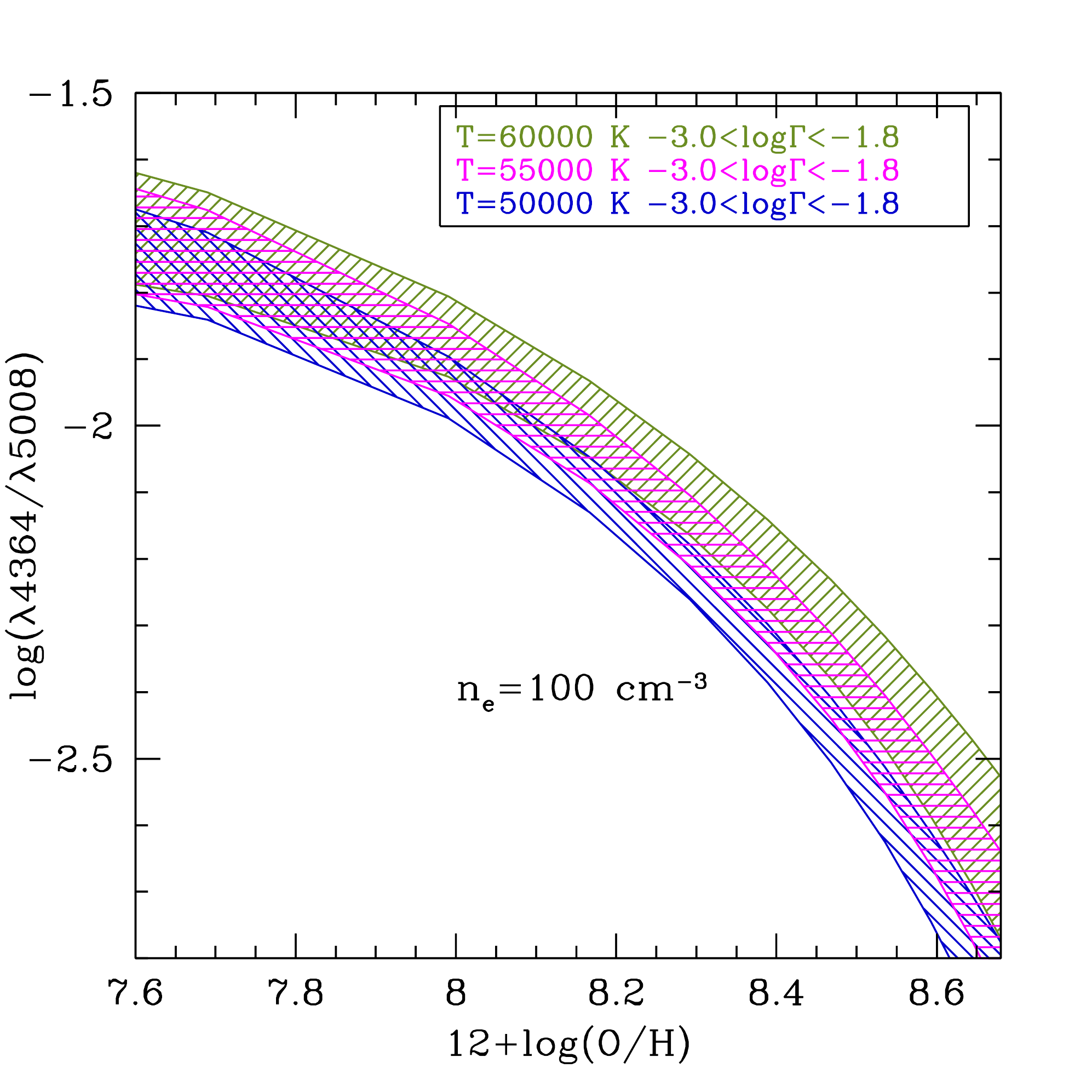}}
\caption{The expected intensity of the ${\rm T_e}$-sensitive auroral line [OIII]$\lambda4364$ relative to [OIII]$\lambda 5008$ 
for the range of photoionization models as in Figures~\ref{fig:bpt_mod} and~\ref{fig:met_comp_mod}.  The dependence
on the radiation field intensity and shape (as well as on $n_{\rm e}$) is modest over this range, so that measurements 
should yield accurate values of 12+log(O/H) suitable for calibrating the strong-line ratios.    
}
\label{fig:o3ratio}
\end{figure*}

\subsection{Direct Metallicity Calibration at $z\sim 2.3$}

At present, there is only a handful of direct metallicity measurements at $z > 1.5$ (\citealt{villar-martin04,yuan+kewley09,erb2010,rigby11,christensen12,
james14,bayliss13}),
some of which are limits only and/or quite uncertain. In any case, as an ensemble they remain 
insufficient to discern any systematic trends. At minimum, a cross-check on strong-line abundance estimates at high 
redshifts will require a substantial sample of galaxies 
covering a range of implied $\Gamma$ for which each galaxy has both  
measurements of the doublet ratio of [OII]$\lambda\lambda 3727$, 3729 and/or [SII]$\lambda\lambda 6718$, 6732 (for estimates
of $n_{\rm e}$) and the ratio
[OIII]$\lambda 4364$/[OIII]$\lambda 5008$ in addition to the strong lines. 
Figure~\ref{fig:o3ratio} shows that, in the context of the models, 
measurement of the weak $\lambda 4364$ feature would (as expected) provide a relatively model-independent
measure of {\it gas phase} oxygen abundance in the high redshift galaxies\footnote{Alternatively, the UV OIII]$\lambda1661,1666$ intercombination feature, 
which
is predicted to be somewhat stronger than $\lambda 4364$ over most of the range in physical conditions covered by the models, can be used
instead, although its use introduces a much stronger dependence on accurate nebular extinction estimates--see section~\ref{sec:analogs} for
examples.}. Figure~\ref{fig:o3ratio} also indicates that uncertainties in the radiation field shape (i.e., $T_{\rm eff}$) may limit
the precision of measuring oxygen abundances to $\sim \pm0.1$ dex; qualitatively, this may be understood as due to the
dependence of equilibrium $T_{\rm e}$ on the mean energy per ionizing photon, at fixed metallicity.   

As for the general detectability of the [OIII]$\lambda 4364$ feature, its predicted strength relative to \Hb\ ranges from 
0.1 to 0.03 for ${\rm 0.05 \simlt Z/Z_{\odot} \simlt 0.5}$;  the median observed \Hb\ flux in the current KBSS-MOSFIRE sample
is ${\rm f(\Hb) \simeq 7.5\times 10^{-18}}$ ergs s$^{-1}$ cm$^{-2}$, with median S/N$\simeq 8.6$.  Thus, typical KBSS-MOSFIRE spectra
are within a factor of a few of the expected flux level $\sim 1\times10^{-18}$ ergs s$^{-1}$ cm$^{-2}$, and the best individual 
spectra, with (S/N)$_{\Hb}> 40$, should allow for detections.  The results of our analysis of the KBSS-MOSFIRE data on this topic will be presented
in future work (see also section~\ref{sec:analogs}.) Importantly, all of the strong lines (of [\ion{O}{2}], [\ion{O}{3}], \Ha, \Hb, [\ion{N}{2}], and [\ion{S}{2}]) 
plus [\ion{O}{3}]$\lambda 4364$ can be observed from
the ground only over the more restrictive range $2.36 \simlt z \simlt 2.57$ (see Figure~\ref{fig:zhist_kbss}); by design, 
a large fraction of the objects in Tables 1-3 (103
of 251 or 41\%)  
falls in this range. Thus, we expect that analysis of the highest-quality spectra, together with stacks formed from those
of typical quality, should allow for the necessary calibration tests using direct $T_{\rm e}$ method measurements of gas-phase oxygen abundances--
at least for $z \simeq 2.36-2.57$.  

\subsection{Physical Interpretation: Summary}

Sections 4.1-4.5 above have attempted to highlight the {\it caveats} associated with interpreting 
the ratios of strong emission lines produced in the \ion{H}{2} regions of high redshift galaxies
in the context of what is known from much more extensively studied ``local'' star-forming galaxies.  

The ``offset'' in the position of the locus of star-forming galaxies in the high redshift sample 
compared to the BPT sequence of local star-forming galaxies appears to have contributions, in rough
order of importance, from: 
\begin{itemize}
 \item harder stellar ionizing radiation field, needed to explain the preponderance
of large observed [OIII]/\Hb\ in the high redshift sample;  
\item higher ionization parameters than inferred for
most low-redshift star-forming galaxies;  
\item shallower dependence of (N/O) on (O/H) than is typically inferred
for galaxies in the local universe, with (N/O) close to the solar value over the full range of inferred (O/H)
(see also \citealt{masters14}, which independently reached a similar conclusion).  
\end{itemize}

The implications of the
BPT shift for measurements of gas-phase abundances from strong emission lines remain uncertain, but the
generally higher level of excitation, and the less-pronounced behavior of (N/O) vs. (O/H), have the combined effect of
{\it reducing} the degree to which the strong line ratios are sensitive to gas-phase (O/H). However, the inferred
higher $T_{\rm eff}$, enhanced (N/O), and higher $\Gamma$ may all be direct consequences of pronounced differences in the evolution
of massive main sequence stars in sub-solar metallicity environments at high redshifts. If so, the differences
will have very broad implications-- perhaps more important than measurement of gas-phase metallicities.  

At present, we suggest that metallicities inferred from strong line ratios should be used with caution until they have
been calibrated directly (i.e., at high redshift) 
using $T_{e}$-based measurements, which has become feasible with the advent of multiplexed near-IR
spectroscopy. Based on the (currently limited) observational constraints together with inferences from photoionization
models, we suggest that the most reliable of the commonly-used strong line indices is O3N2, whose calibration onto the
``$T_{\rm e}$'' abundance scale appears stable with respect to changes in the (low-$z$) samples used
for calibration, and is only moderately sensitive to the behavior of N/O with O/H, unlike N2.  
The commonly-used R23 method is unfortunately of limited use over the actual metallicity range 
most relevant at $z \sim 2.3$, ${\rm 8.0 \simlt 12+log(O/H) \simlt 8.7}$.

\section{AGN versus Stellar Ionization}

\label{sec:agn}

The BPT diagram has been used most often in the literature as a means of separating galaxies 
whose nebular spectra are produced predominantly by HII regions from those for which a significant
contribution of the observed line emission is likely to have been excited by AGN.  
The basic principle in distinguishing the star-forming galaxy sequence from 
that of the so-called ``mixing sequence'' (see, e.g., \citealt{kewley01,kauffmann03,kewley13}) is
that AGN generally have much harder far-UV spectra than stellar populations, 
resulting in a tendency to produce higher [OIII]/\Hb\ relative to [NII]/\Ha. In addition, regions
near the centers of galaxies harboring  AGN tend to be relatively metal-rich, which together with
emission from slow shocks that often accompany such activity, pushes [NII]/\Ha\ toward high values.
Expectations for the behavior of high-redshift AGN in the BPT plane have been explored in
some detail by \citet{kewley13}, who pointed out that AGN in low metallicity hosts (which appear to 
be extremely rare at $z \simeq 0$) could conceivably be found with high [OIII]/\Hb\ but low [NII]/\Ha. If so,
they could fall near to the star-forming sequence in the BPT diagram, possibly leading to ambiguities
in classification of objects falling above the $z \simeq 0$ star-forming sequence.  This is
clearly an important issue to address here, since we have shown that nearly all star-forming galaxies at $z \sim 2.3$
are found in that region. 

\begin{figure*}[htpb]
\centerline{\includegraphics[width=7cm]{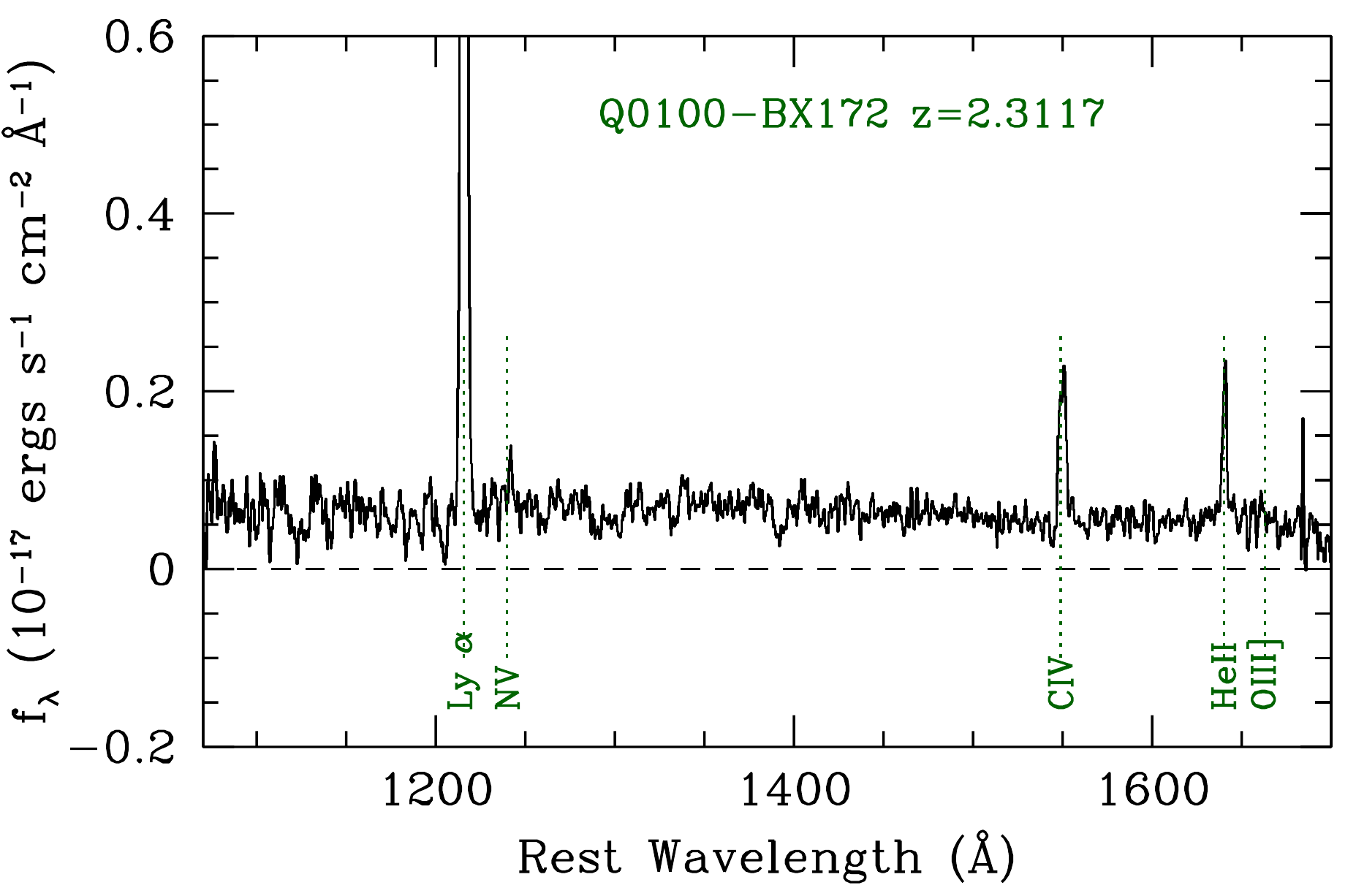}\includegraphics[width=7cm]{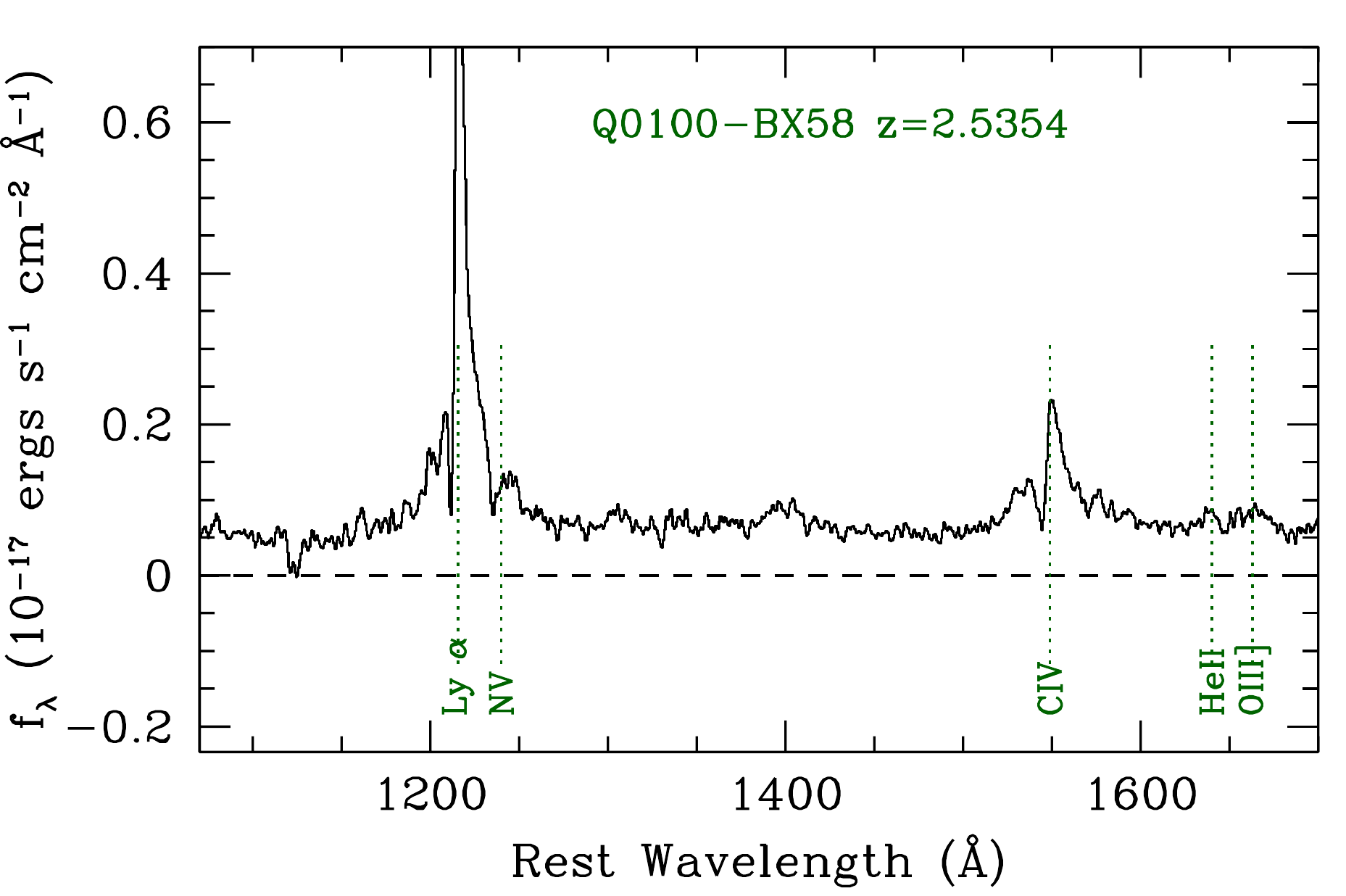}}
\centerline{\includegraphics[width=7cm]{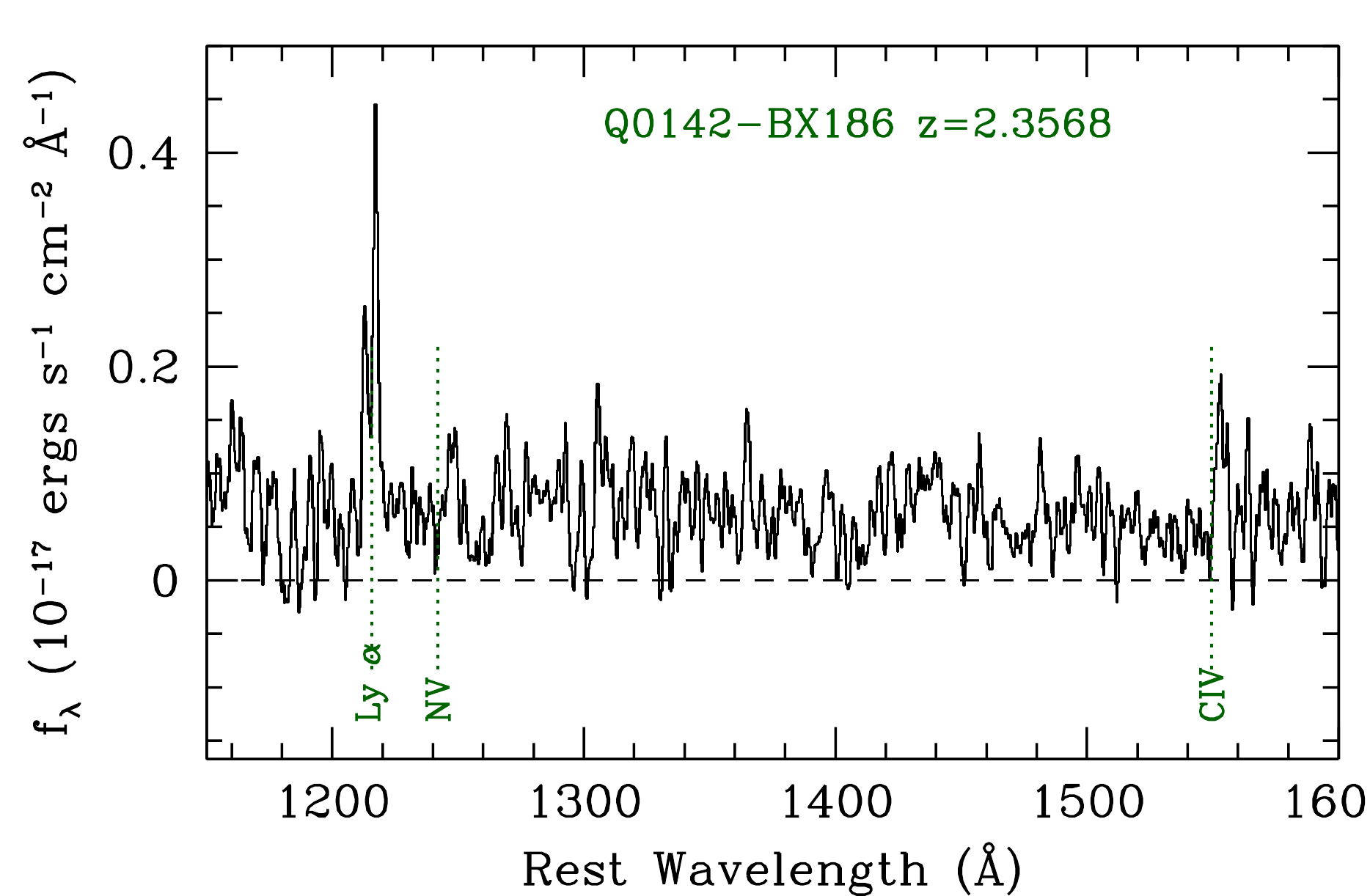}\includegraphics[width=7cm]{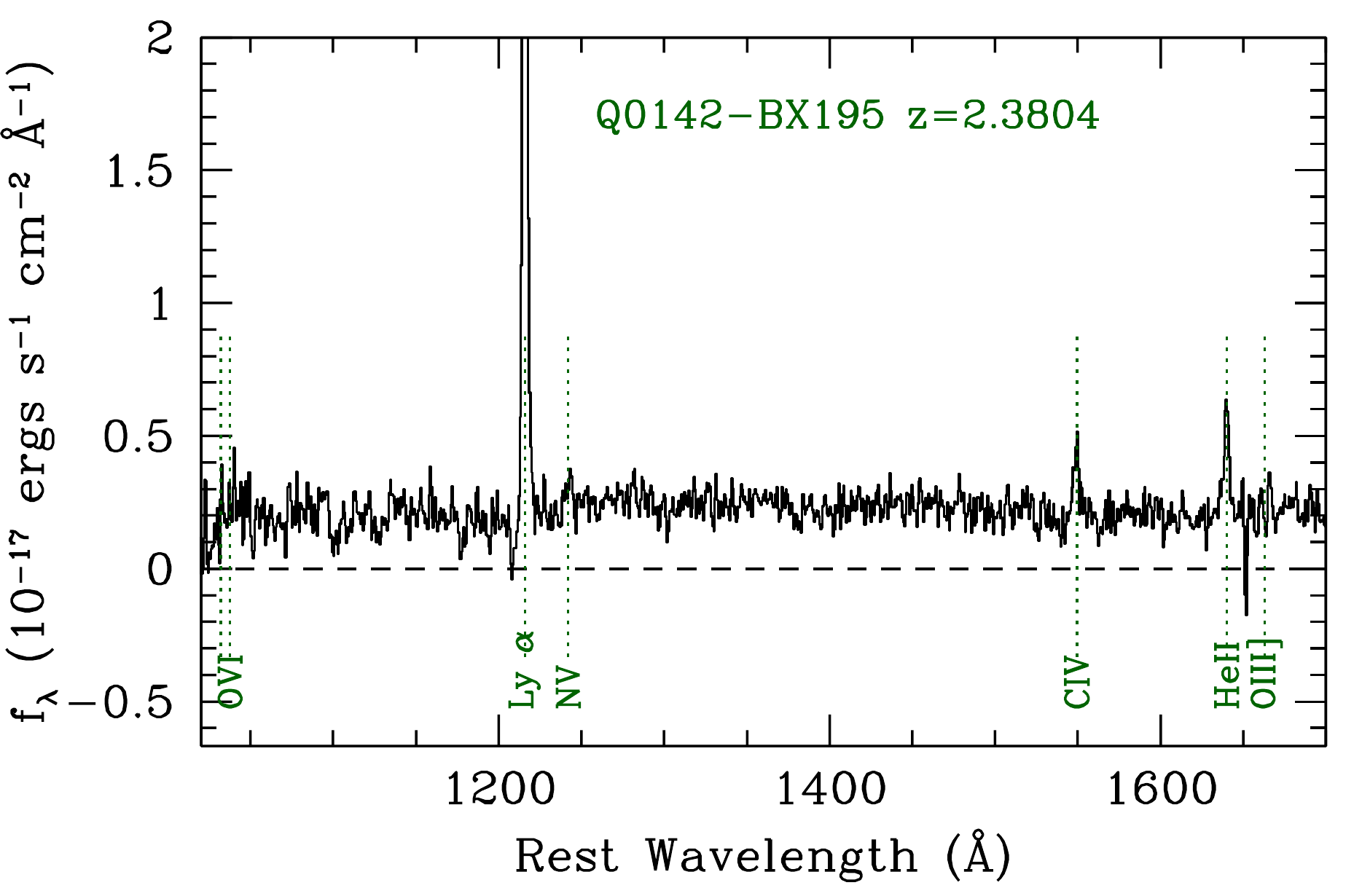}}
\centerline{\includegraphics[width=7cm]{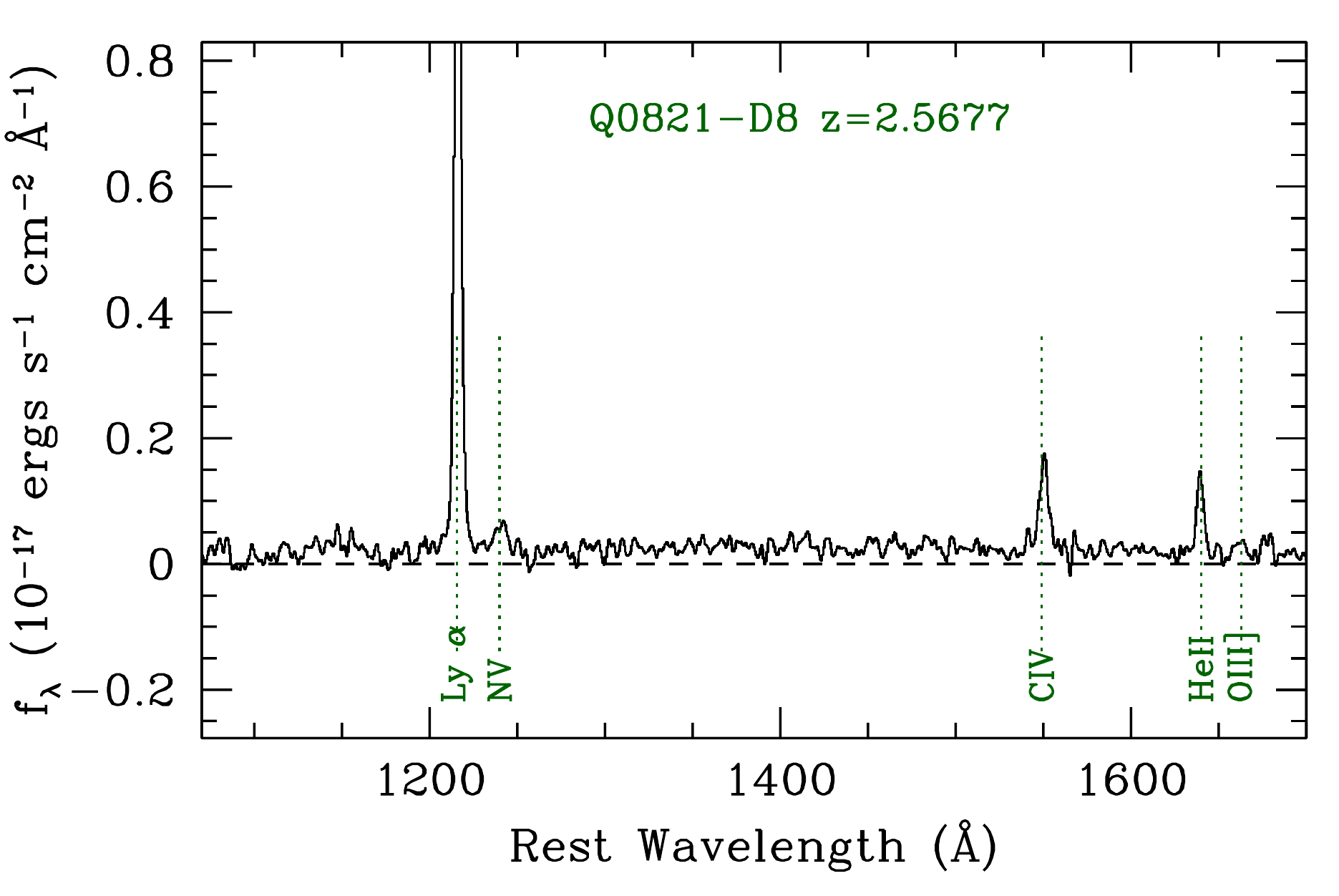}}
\caption{Rest-UV spectra (Keck/LRIS-B) for the 5 objects spectroscopically identified as AGN prior 
to being observed with MOSFIRE;  
their rest-frame optical MOSFIRE spectra
are shown in Figure~\ref{fig:agn_spectra_mos}. Emission lines of \ion{N}{5},
\ion{C}{4}, and \ion{He}{2} clearly indicate the presence of AGN; Q0100-BX58 (top right panel) is a broad-lined AGN, albeit quite
faint (${\cal R} =23.4$).
\label{fig:agn_spectra_uv}.
}
\end{figure*}

\begin{figure*}[htpb]
\centerline{\includegraphics[height=3.7cm]{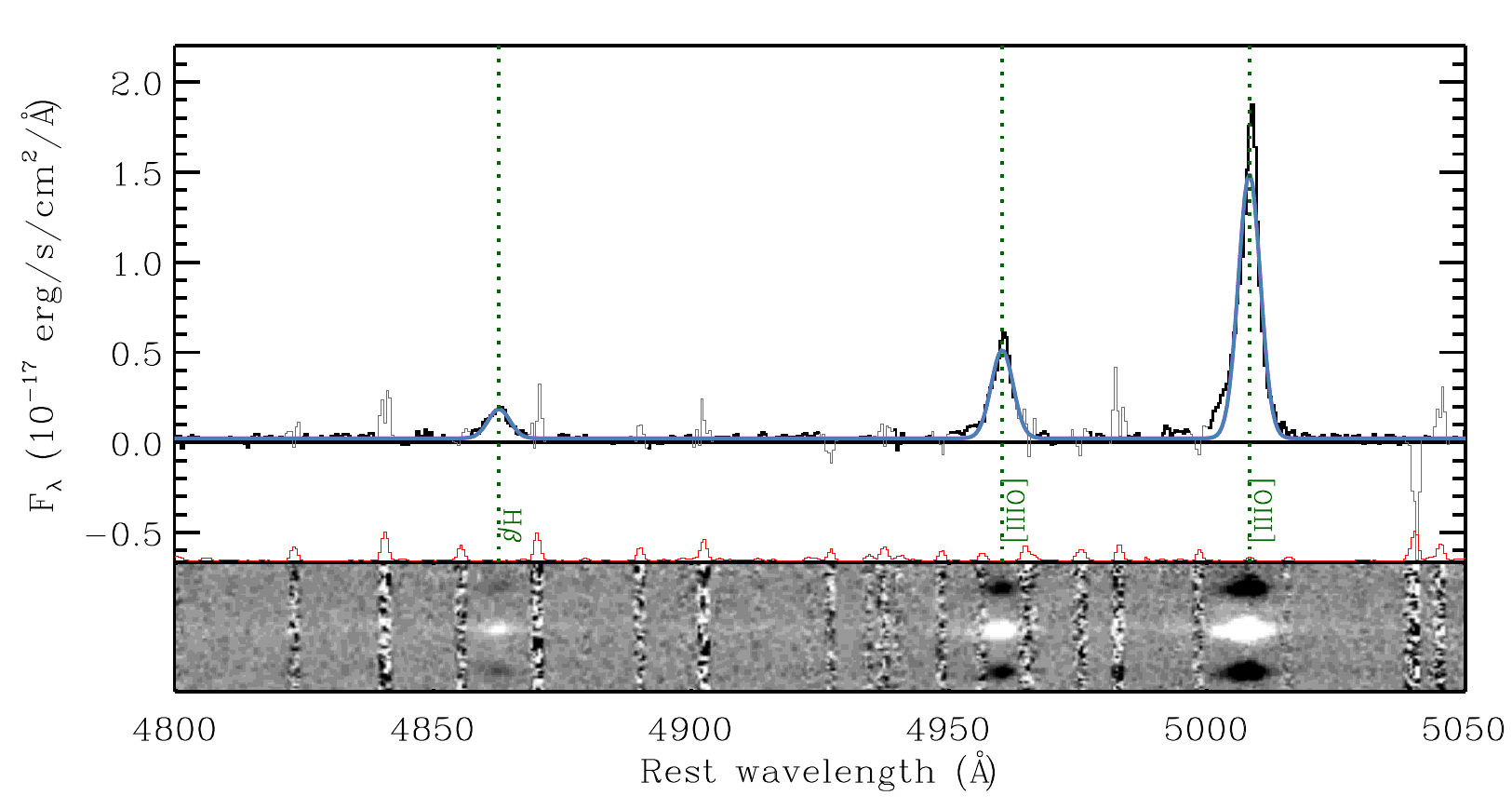}\includegraphics[height=3.7cm]{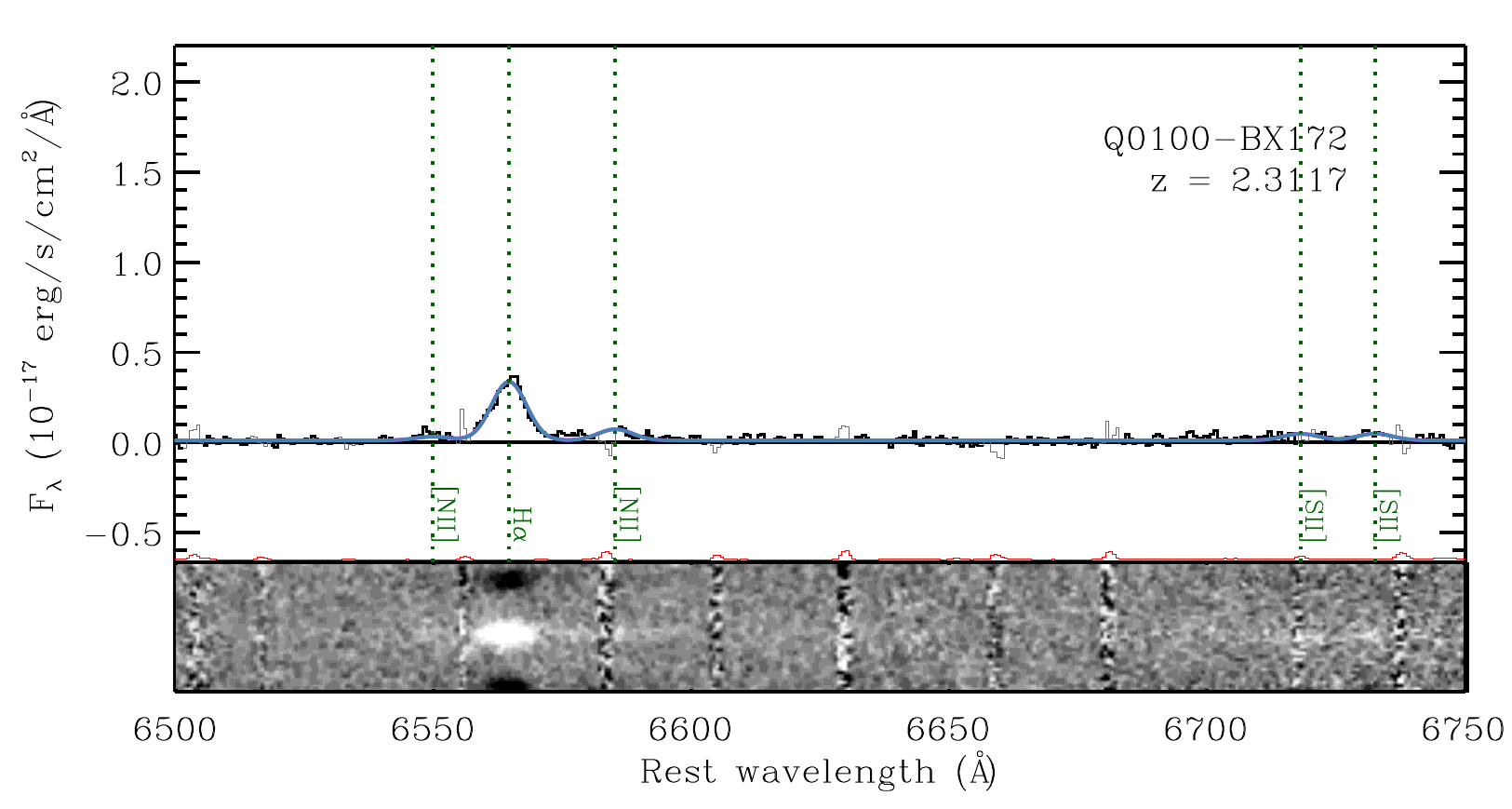}}
\centerline{\includegraphics[height=3.7cm]{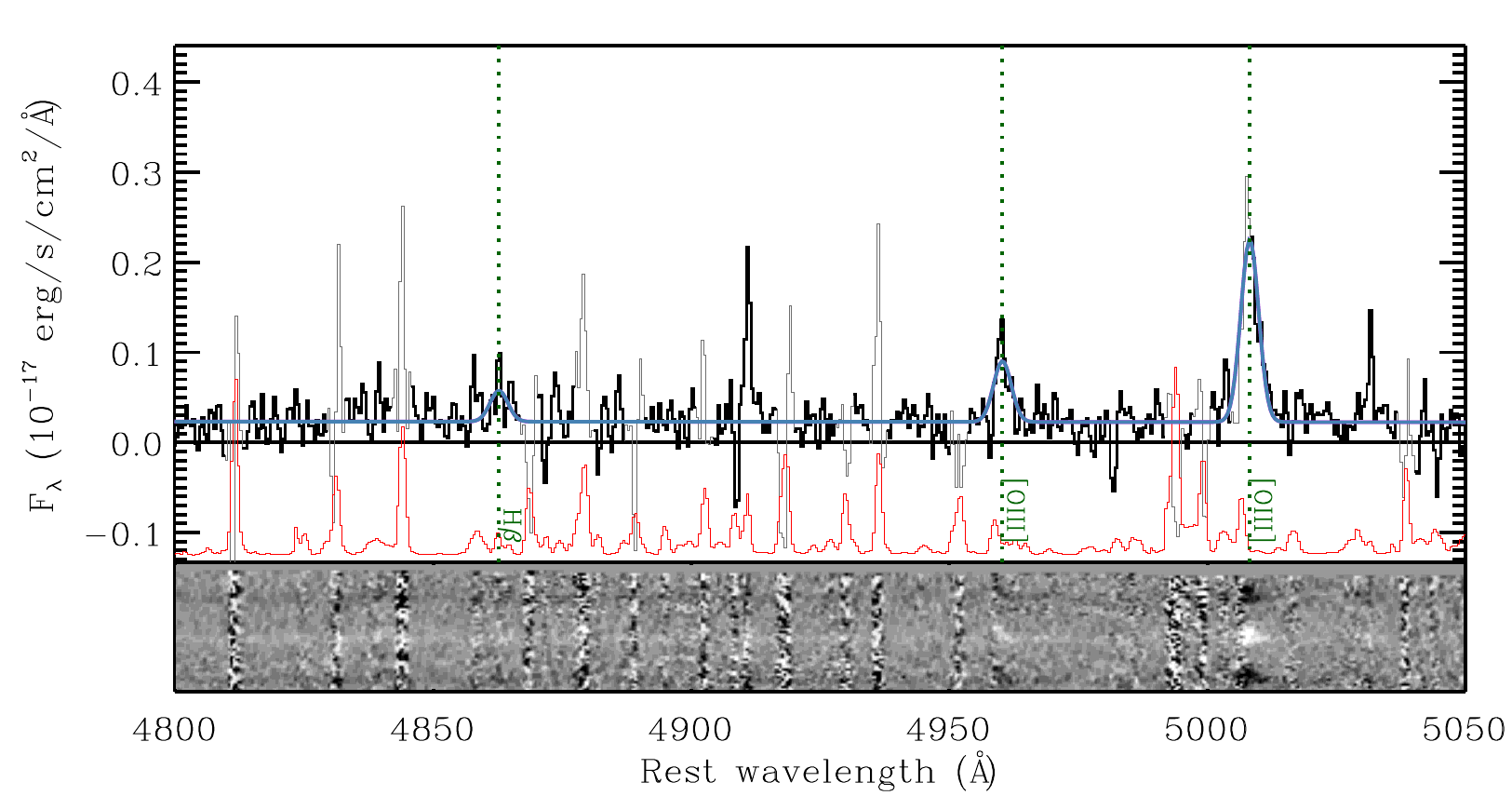}\includegraphics[height=3.7cm]{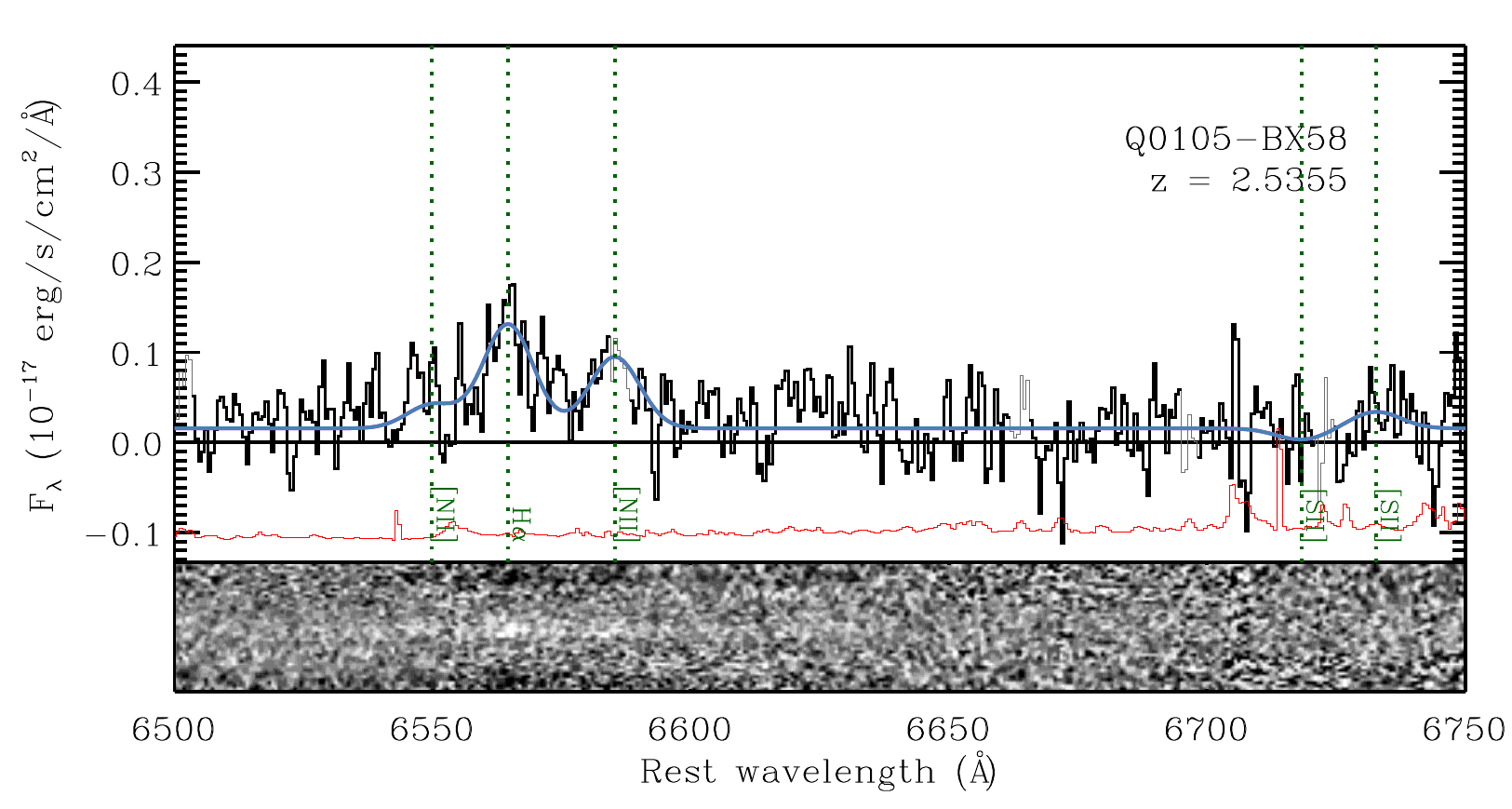}}
\centerline{\includegraphics[height=3.7cm]{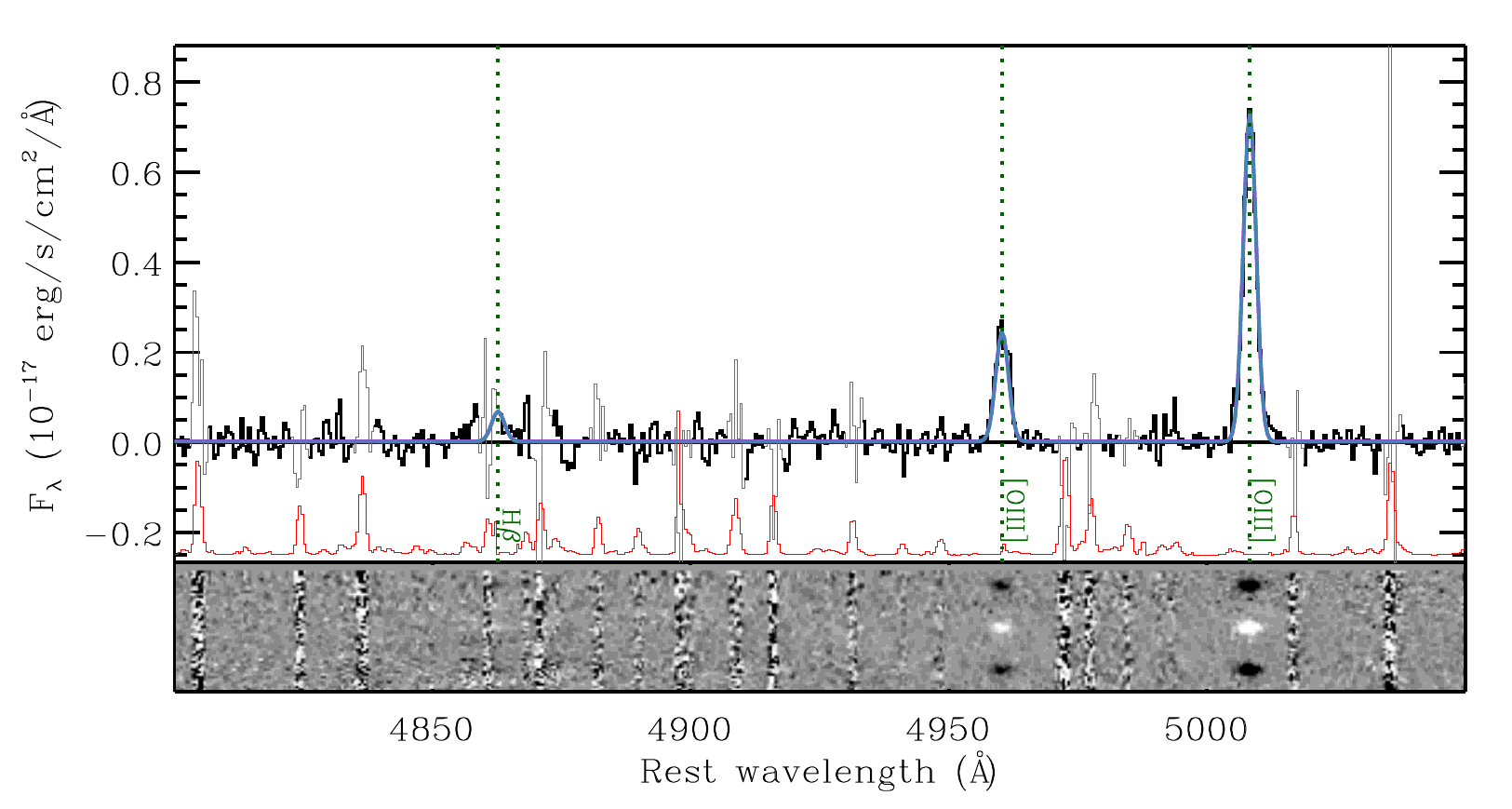}\includegraphics[height=3.7cm]{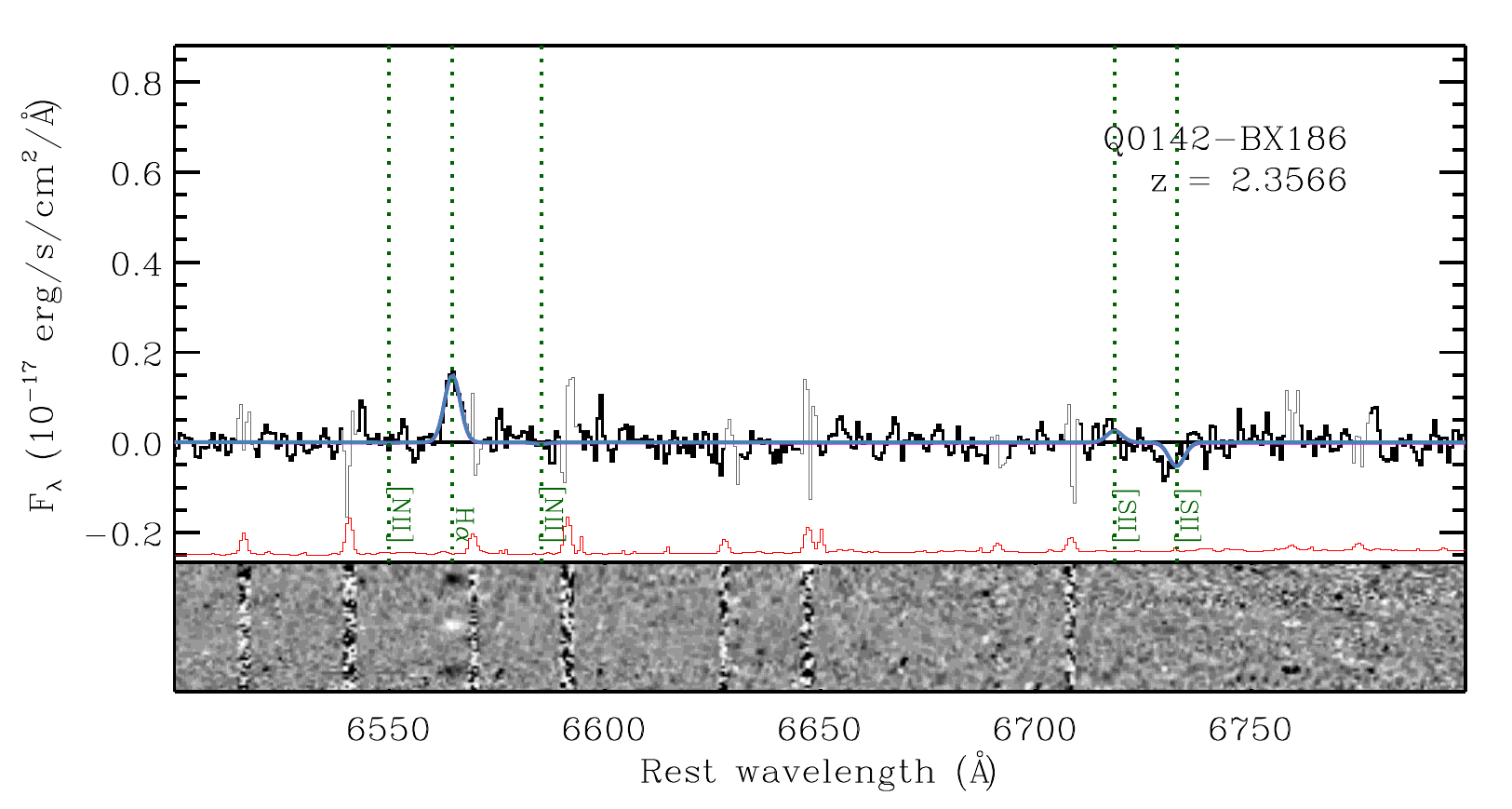}}
\centerline{\includegraphics[height=3.7cm]{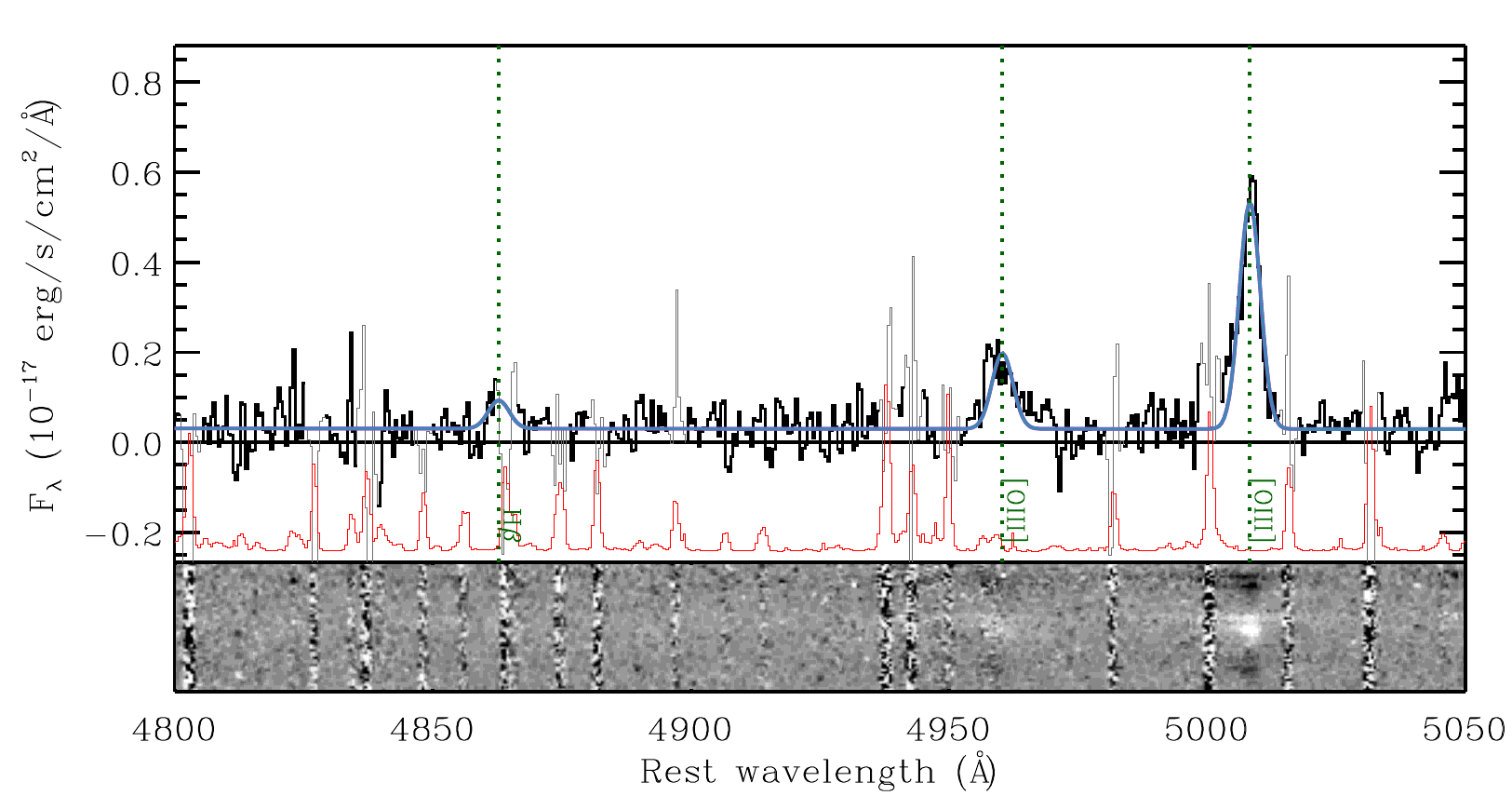}\includegraphics[height=3.7cm]{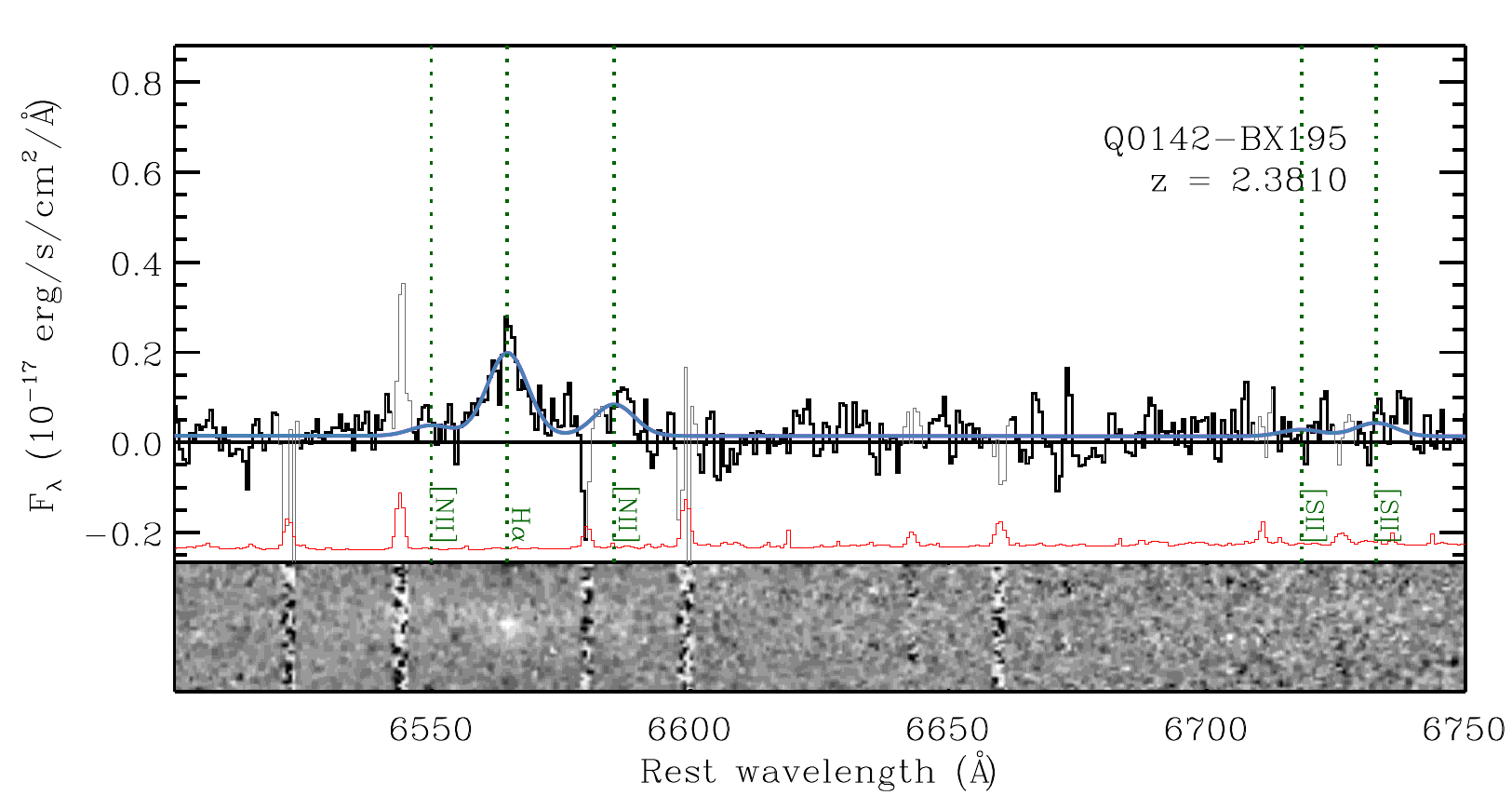}}
\centerline{\includegraphics[height=3.7cm]{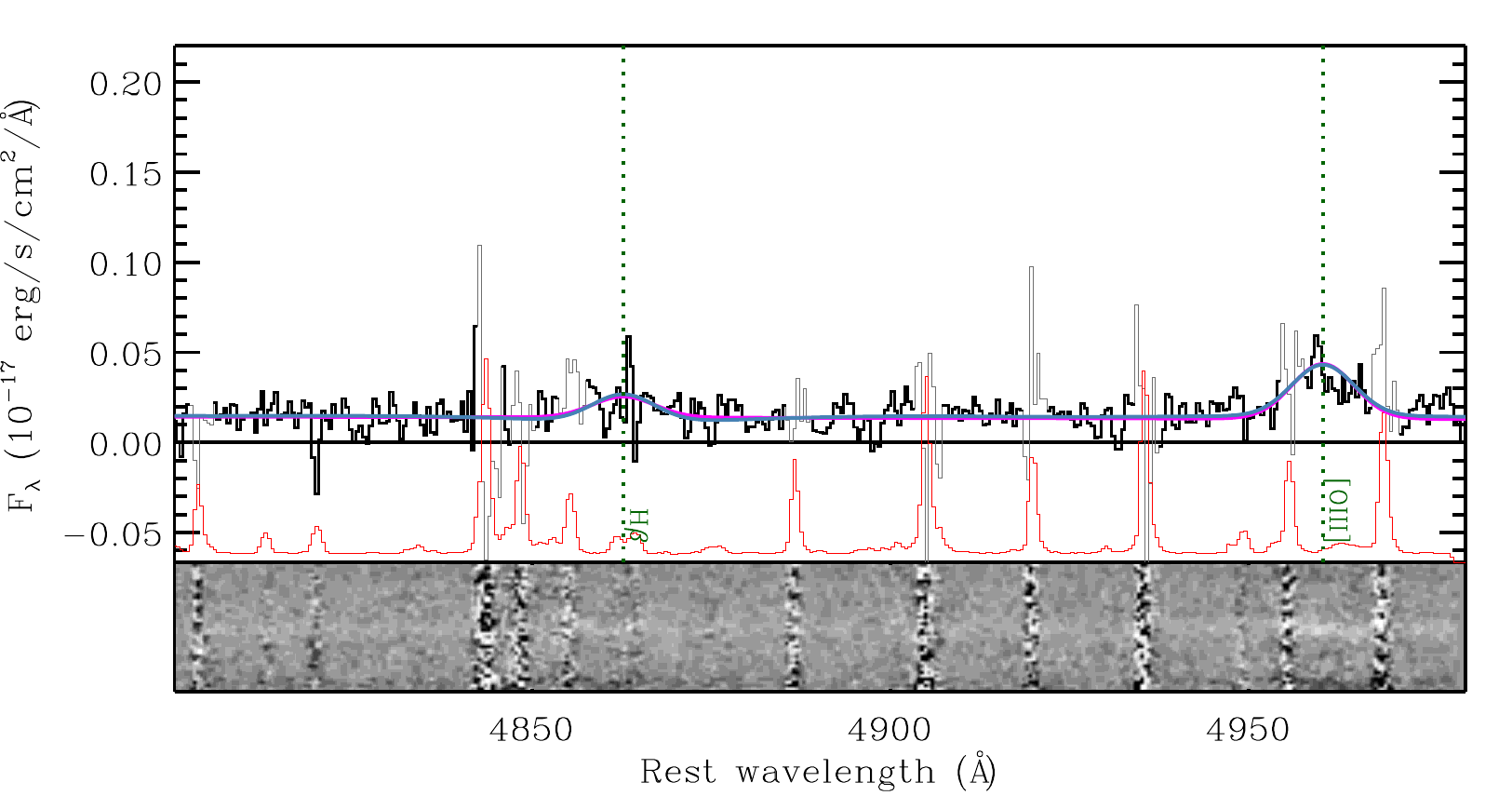}\includegraphics[height=3.7cm]{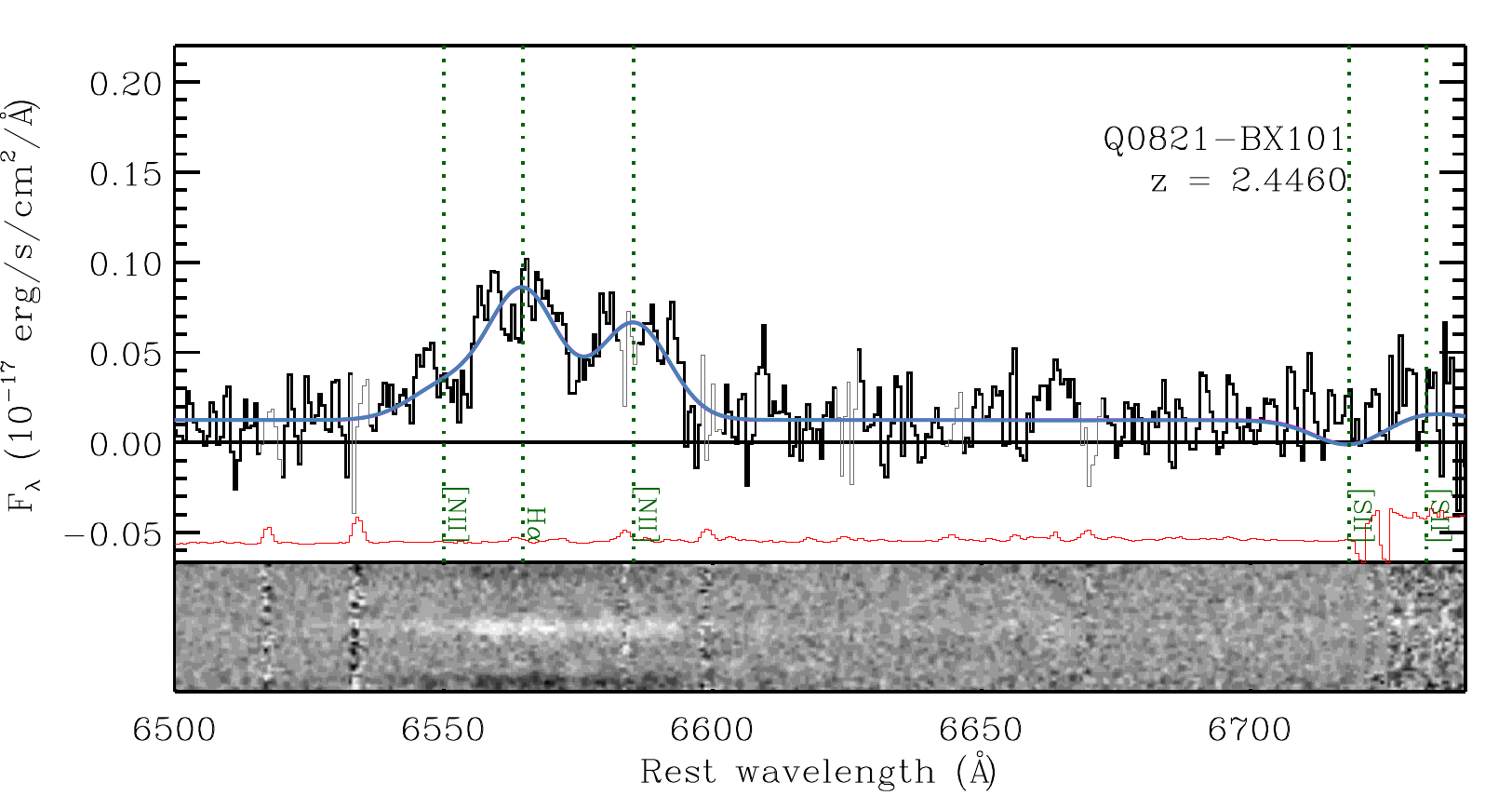}}
\caption{MOSFIRE H-band (left) and K-band (right) spectra of the 8 objects in Tables 1 and 2 identified as AGN (those marked with magenta stars in Figure~\ref{fig:bpt}). 
}
\label{fig:agn_spectra_mos}
\end{figure*}

\begin{figure*}[htpb]
\setcounter{figure}{17} 
\centerline{\includegraphics[height=3.7cm]{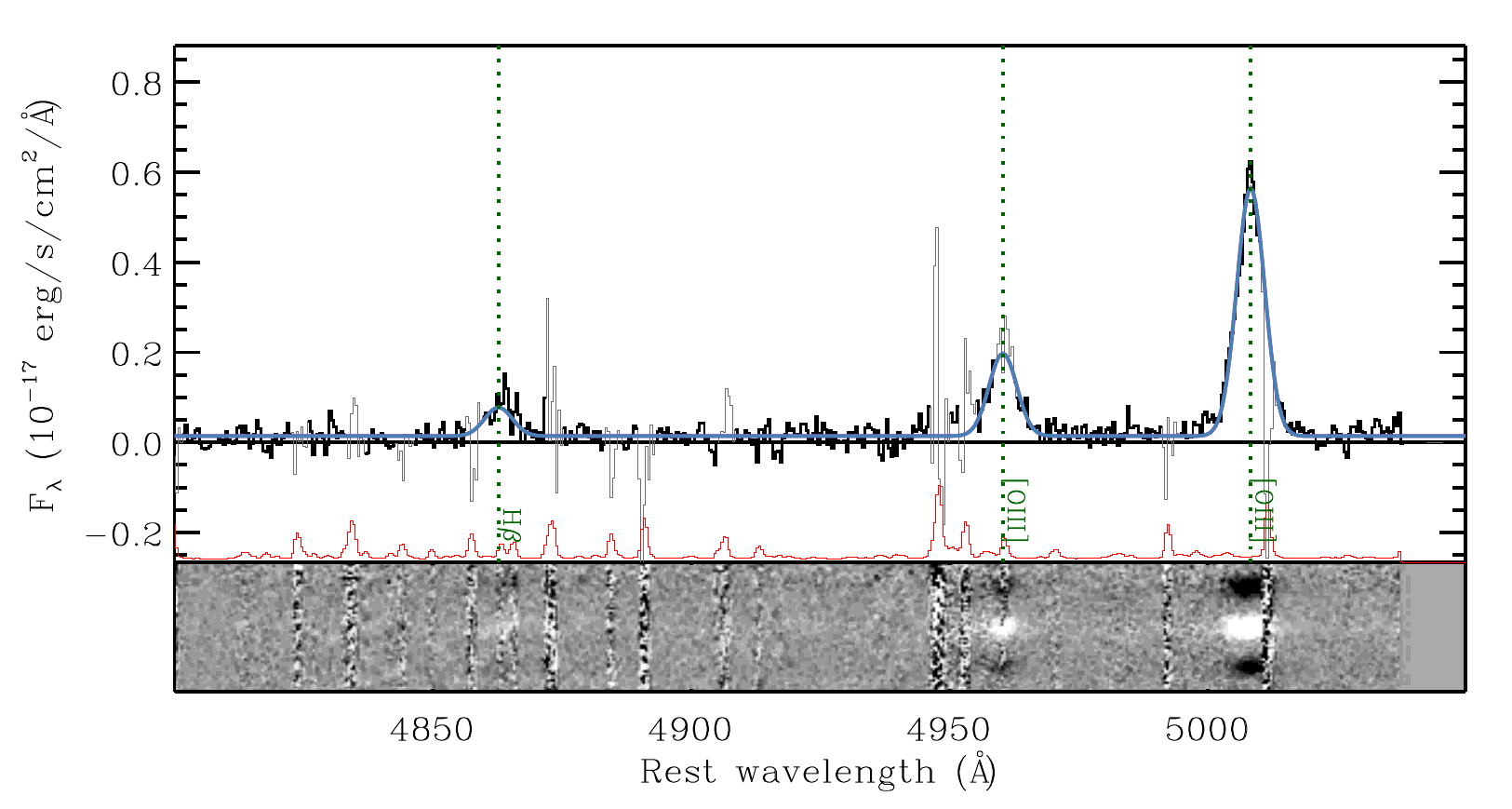}\includegraphics[height=3.7cm]{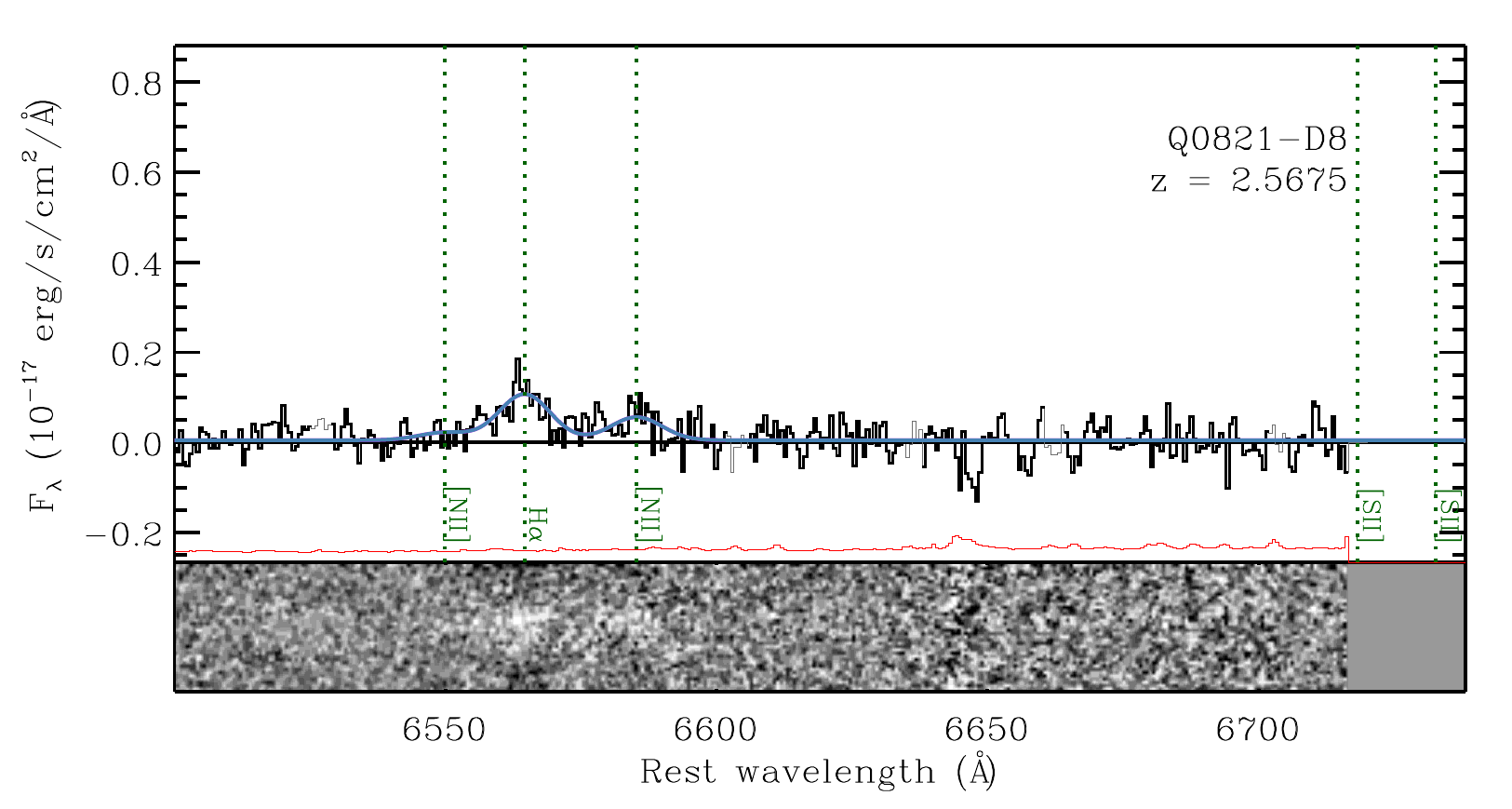}}
\centerline{\includegraphics[height=3.7cm]{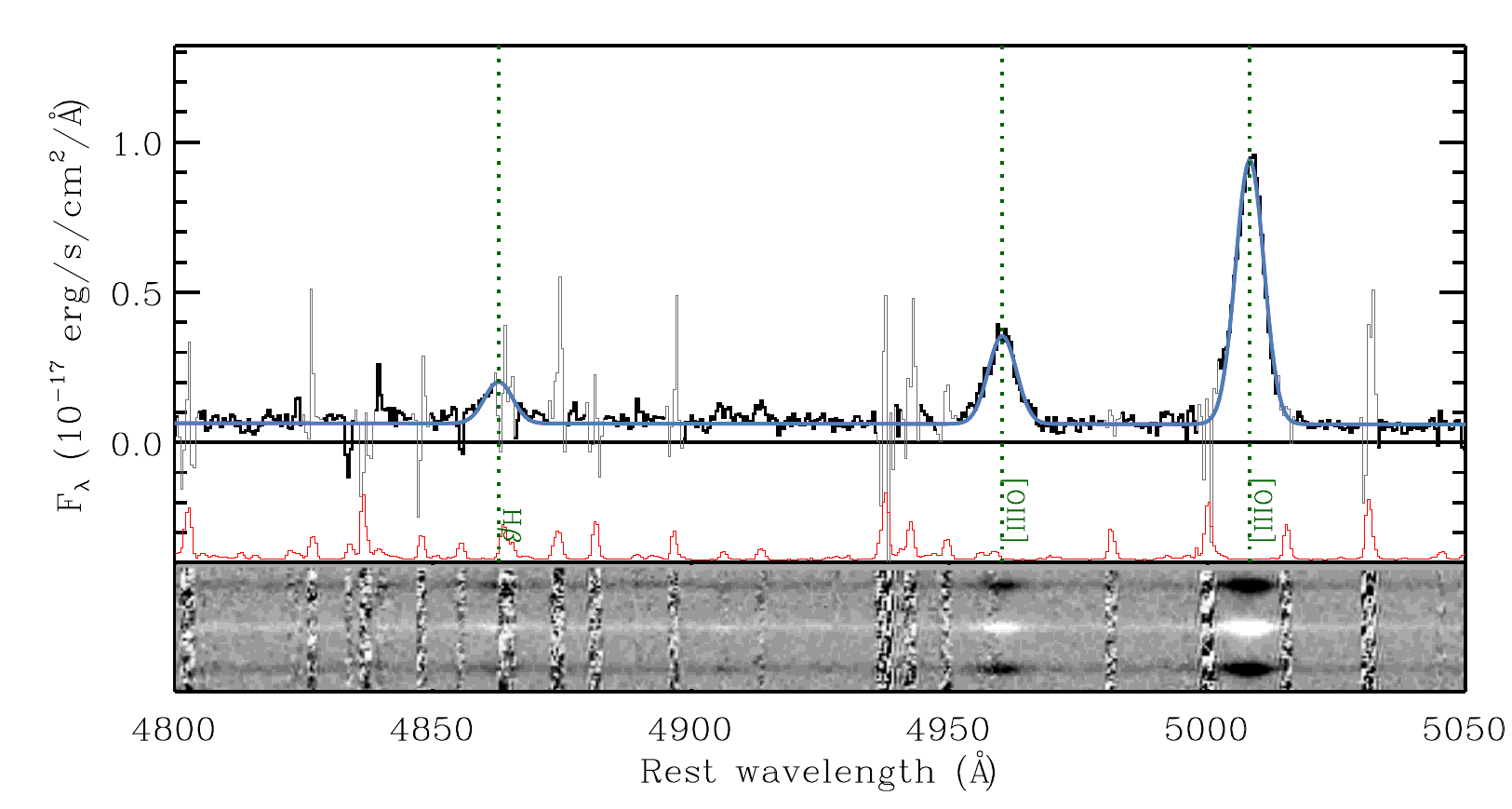}\includegraphics[height=3.7cm]{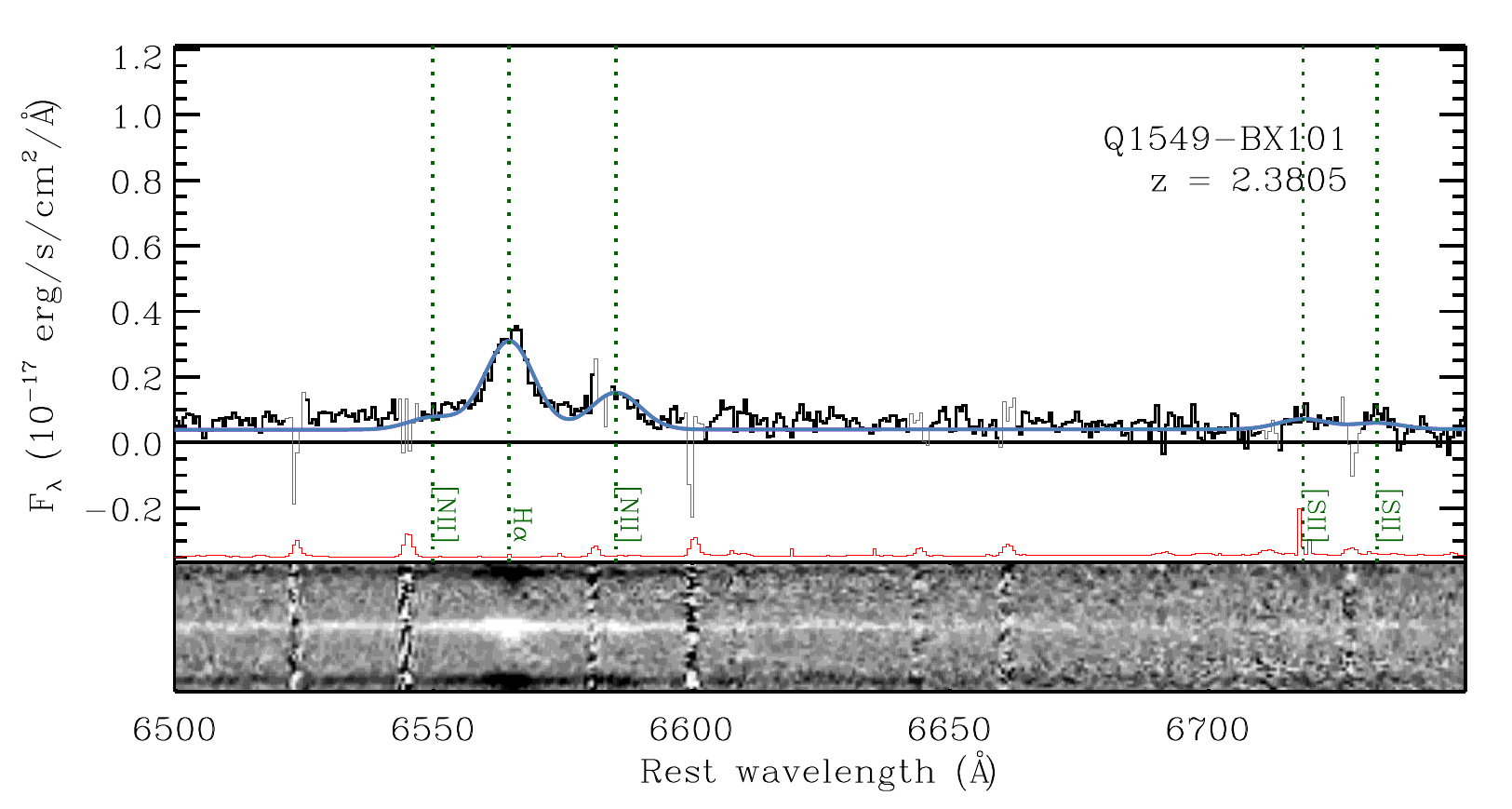}}
\centerline{\includegraphics[height=3.7cm]{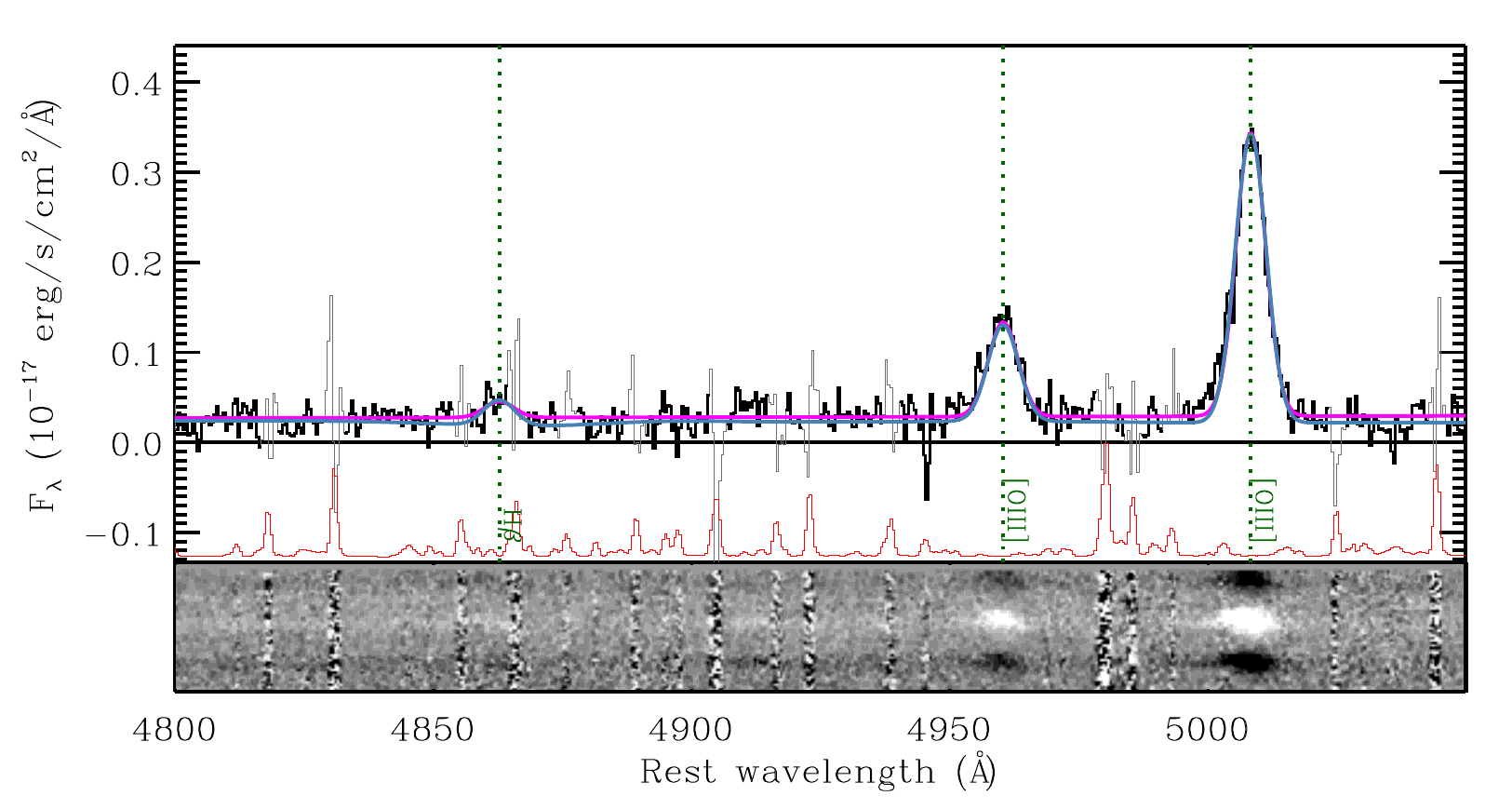}\includegraphics[height=3.7cm]{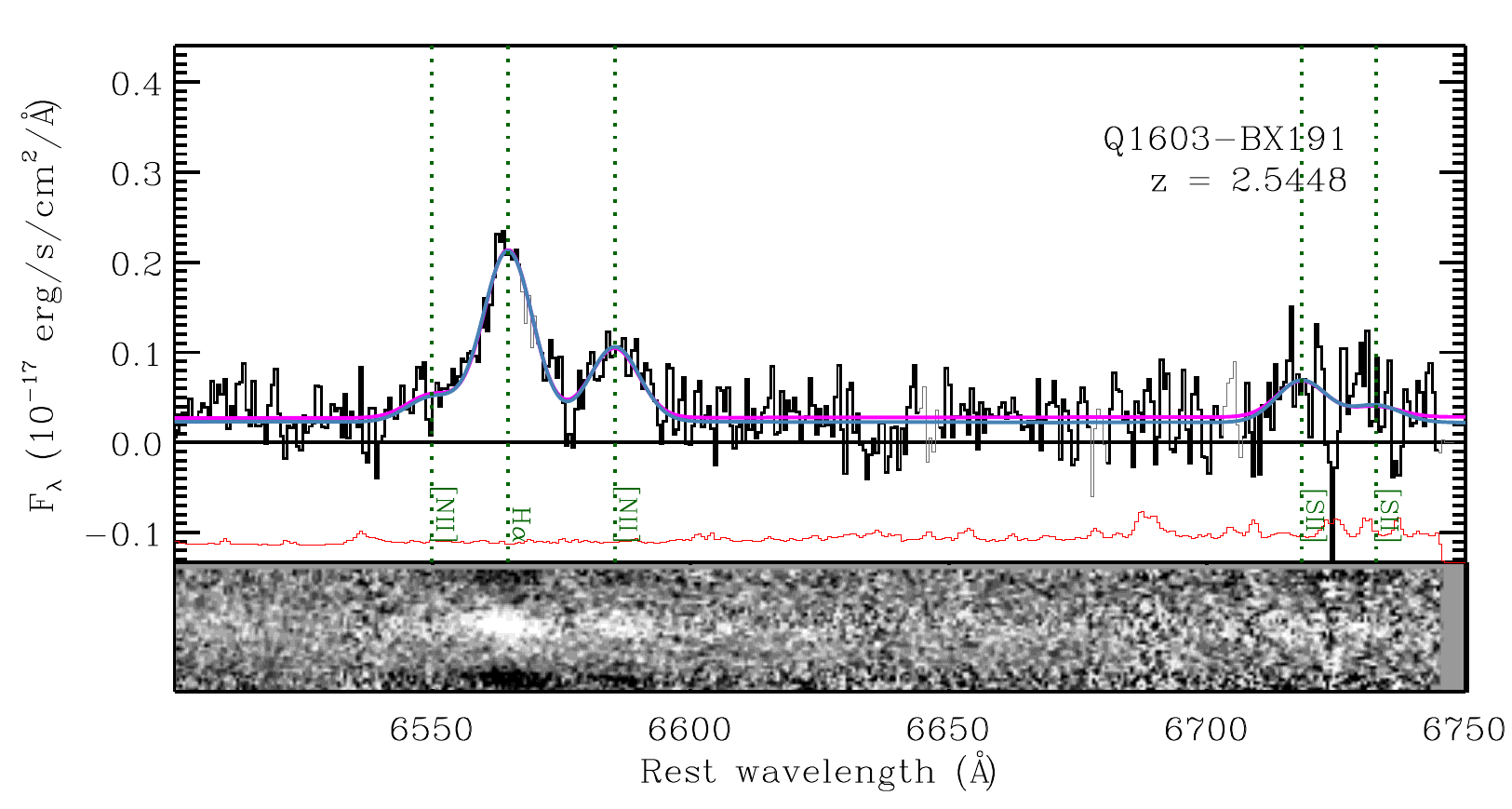}}
\caption{(Continued)}
\label{fig:agn_spectra_mos_cont}
\end{figure*}
Fortunately, a deep survey at high redshifts such as KBSS
provides some advantages for AGN identification over wide-field samples at $z \simeq 0$ 
such as SDSS.
One is that the identification of active galactic nuclei in distant galaxies has been revolutionized in
recent years thanks to pointed, very deep X-ray surveys with CHANDRA and mid-IR photometry with
SPITZER/IRAC, which image, respectively, the X-rays produced in AGN accretion disks and
emission from AGN-heated dust. 
In addition, most of the galaxies in our KBSS-MOSFIRE sample 
have already been observed in the rest-frame far-UV using LRIS; the far-UV provides
access to emission lines of much higher ionization species than are easily observed
in the rest-frame optical (e.g., \ion{C}{4}$\lambda\lambda 1548$, 1550, \ion{N}{5}$\lambda\lambda
1238$,1242.) The presence of nebular emission in such species clearly indicates AGN excitation, since
the relevant ionization potentials are too high to have been produced by hot stars (see, e.g., \citealt{steidel02,hainline11}.)  
By combining what is known from UV spectroscopy with multi-wavelength observations sensitive to 
the presence of AGN, one would normally not need to rely on the BPT diagram as the primary means 
of discrimination. Nevertheless, it is useful to examine the small number of objects identified as 
likely AGN to see where they lie in BPT space.  

Five objects in the current sample were identified as AGN based on existing 
rest-frame UV spectra (Figure~\ref{fig:agn_spectra_uv}.) One of the 5 (Q0105-BX58) is
a faint broad-lined AGN, while the others have relatively narrow rest-UV emission lines 
but were flagged as AGN based on  
strong emission lines of  
\ion{N}{5}, \ion{C}{4}, and \ion{He}{2}$\lambda 1640$. 
Figure~\ref{fig:agn_spectra_mos} shows the MOSFIRE spectra of the 5 objects from Figure~\ref{fig:agn_spectra_uv}, 
along with 3 additional AGN identified as such solely on the basis of their rest-frame optical spectra
(rest-UV spectra have not yet been obtained).  
As can be seen in Figure~\ref{fig:bpt}, all 8 of these objects occupy positions in the BPT plane
that distinguish them from the main locus of $z \sim 2.3$ star-forming galaxies. 
In addition to the
unusual BPT line ratios, the rest-optical emission lines (even for the ``narrow-lined'' AGN) are substantially 
broader than typical among the star-forming
galaxy sample; see Figure~\ref{fig:agn_spectra_mos}. 

%Three additional objects, all among those lacking previous rest-UV spectra, 
%have been classified as AGN: one is Q1603-BX191, the only object in the current KBSS-MOSFIRE 
%sample with a significant measurement of [NII]/\Ha\ but only a lower limit on [OIII]/\Hb\ (i.e., \Hb\ is not
%detected significantly)-- see Figure~\ref{fig:bpt}. It is also distinguished by having very broad emission lines,
%with $\sigma =223$ \kms\ (after correction for the instrumental resolution), placing it in the 98th percentile 
%among the $\sim 300$ objects with $2 \simlt z \simlt 2.6$ observed at \Ha.  
One object from the list in Table~3, Q0821-RK5, is similarly flagged as an AGN because of its 
observed ${\rm log([NII]/\Ha)=+0.02}$ (it has not yet been observed in the H-band, and so does not appear in Figure~\ref{fig:bpt}). 
Such large [\ion{N}{2}]/\Ha\ ratios  are reached only by
galaxies in the ``AGN'' portion of the BPT diagram for local galaxies, and thus are very likely
to harbor AGN; this is supported also by its very red color $({\cal R}-K_s)_{\rm AB}=3.59$, its very broad and diffuse emission lines
($\sigma = 246$ \kms, or 99th percentile of all measured \Ha\ line widths)
and its huge inferred stellar mass (${\rm log(M_{\ast}/M_{\odot}) = 11.79}$, by far the largest in the sample) likely
due to contamination by hot dust emission in the observed near-IR (see \citealt{hainline12}). In this particular case the
SED is ambiguous due to the lack information for observed wavelengths $> 2\mu$m, since Q0821  
is the only KBSS field lacking deep IRAC coverage.   

\begin{figure}
\centerline{\includegraphics[width=9cm]{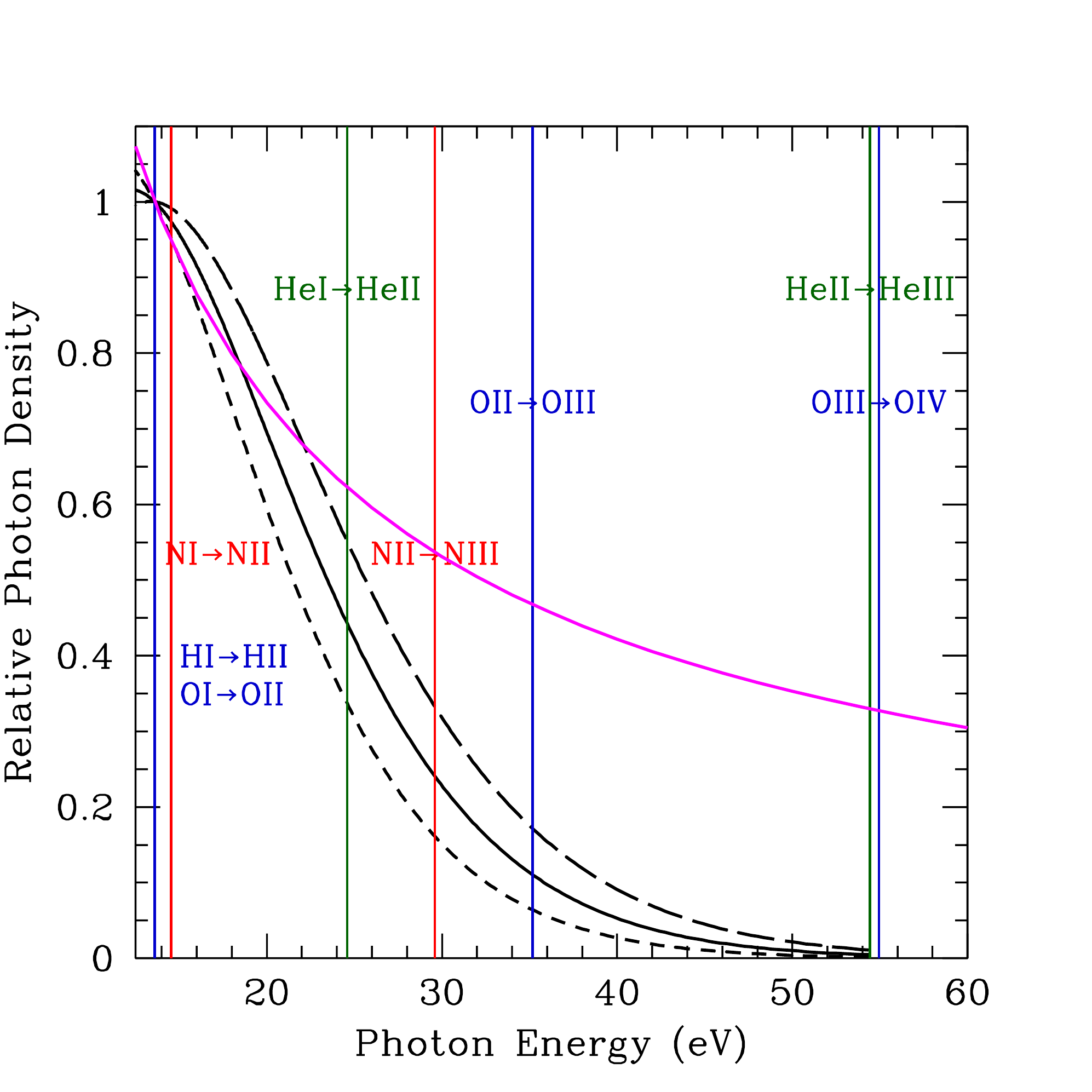}}
\caption{Illustration of the relative photon density of ionizing photons under various
assumptions about the ionizing radiation field.  The vertical lines indicate the ionization
potential for ions most relevant to the observations.  The black curves are for
blackbody spectra with ${\rm T_{eff} = 45000}$, 50000, and 55000 K (short-dashed, solid, and
long-dashed lines, respectively). The magenta curve is for a power law spectrum of the form
${\rm f_{\nu} \propto \nu^{-\alpha}}$, with $\alpha = 0.8$. }
\label{fig:bb_ion}
\end{figure}

Among the KBSS galaxies exhibiting no evidence for AGN (e.g., those falling within or consistent
with the shaded region in Figure~\ref{fig:bpt}), there appears to be a rather sharp upper limit of  
${\rm log([OIII]/\Hb) \simlt 0.9}$. This threshold is exceeded only by two of the 
objects flagged as AGN.   
As discussed below in section~\ref{sec:analogs}, this upper envelope in [\ion{O}{3}]/\Hb\ implies a maximum $T_{\rm eff}$
for sources dominating the ionizing radiation field, $T_{\rm eff,max}\sim 55000-60000$ K based on our modeling.  
AGN can exhibit a wide range of energy distributions over the important 1-4 Ryd range, but
their shape is approximately a power law rather than an exponential (as in the case of stars) over this interval. 
Figure~\ref{fig:bb_ion}
helps to illustrate the difference between the assumed blackbody spectra and power-law AGN spectra. 
Clearly, lines associated with species having ionization potentials above $\simeq 50$ eV would be
more unambiguous signatures of AGN excitation compared to [OII], [OIII], and [NII] transitions available
in the rest-frame optical. On the other hand, 
Figure~\ref{fig:bb_ion} shows that these ions (particularly [OIII]) are extremely sensitive to $T_{\rm eff}$-- e.g.,  
increasing ${\rm T_{eff}}$ from 45000 K to 55000 K approximately triples the 
number density of photons capable of ionizing \ion{O}{2} to \ion{O}{3} relative to those that can ionize 
\ion{H}{1} to \ion{H}{2}. Although the possible range of AGN SEDs is large, and while AGN would tend to produce larger
ratios of [\ion{O}{3}]/\Hb, they would nevertheless be unlikely to produce a {\it consistent} upper envelope at a particular value
([OIII]/\Hb\ $\simeq 0.9$) as observed in the $z \sim 2.3$ sample.  

There is a small number of galaxies (see Figure~\ref{fig:bpt}) whose positions on the BPT plane might
be considered ambiguous in terms of their classification (those lying above the yellow shaded region
but below the objects known to be AGN), but they all have relatively large error bars, so that their
true positions may well lie within the $z \sim 2.3$ star forming sequence.  
Thus, we conclude that AGN excitation plays a significant role in only a small fraction of the 
KBSS-MOSFIRE sample. The 9 identified AGN have been excluded from most analyses in this paper, since their
strong line ratios are unlikely to be related to stellar processes.  

\begin{figure*}[htpb]
\centerline{\includegraphics[width=9cm]{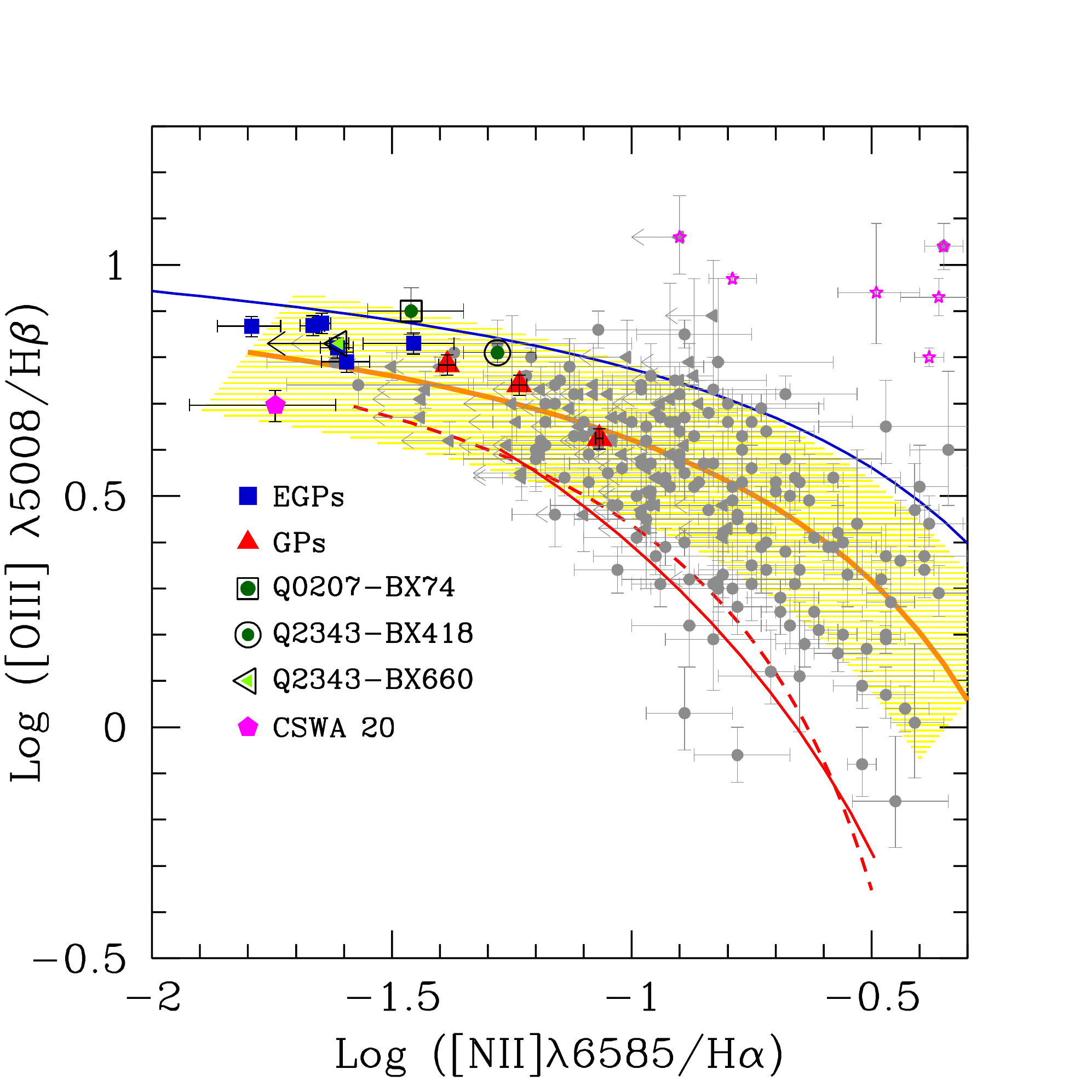}\includegraphics[width=9cm]{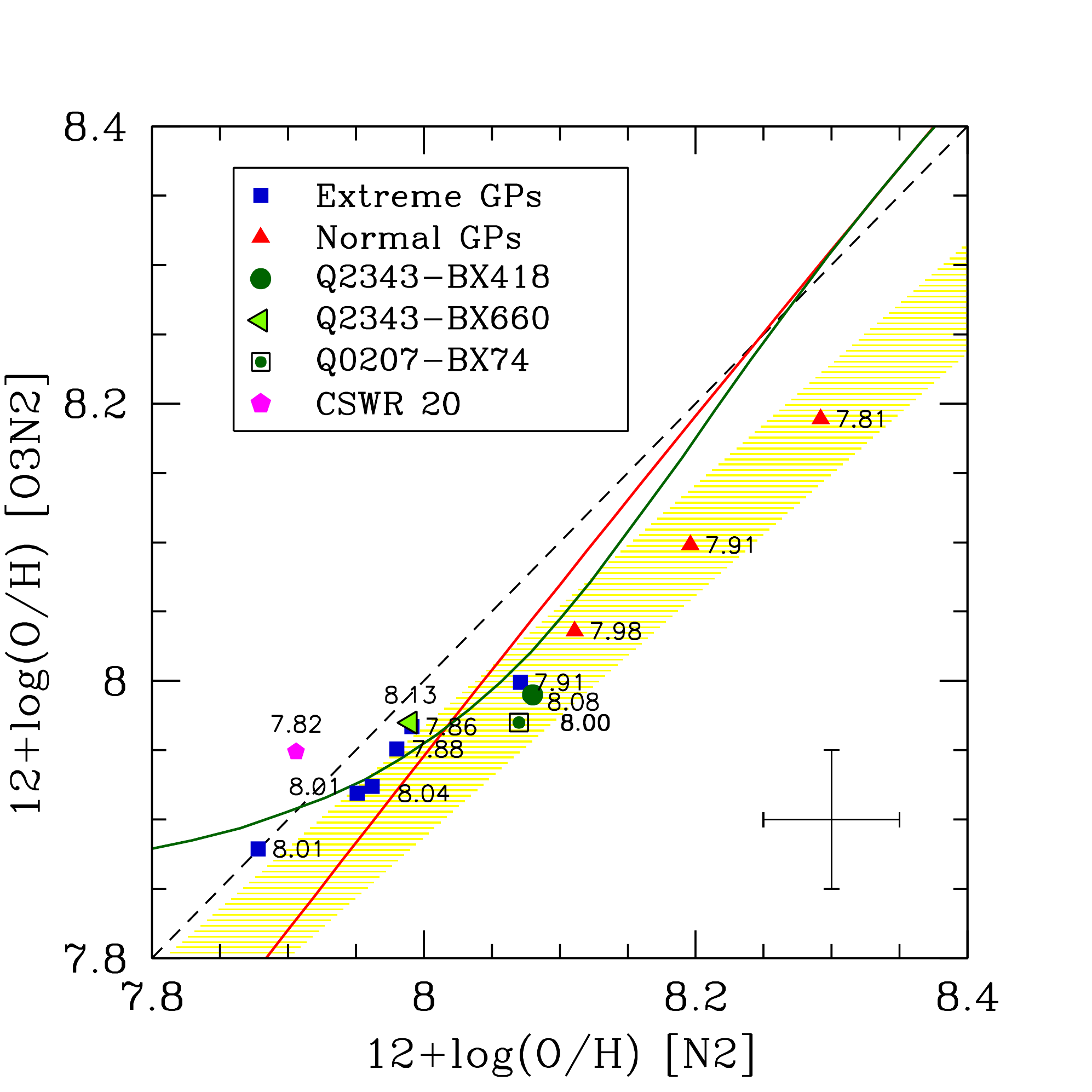}}
\caption{(Left) Plot analogous to Figure~\ref{fig:bpt_mod}, but adding 9 $z \simeq 0.2$ ``green pea'' galaxies
(EGP: dark blue squares; GP: red triangles). (Right) Analogous to Figures~\ref{fig:met_compare} and \ref{fig:met_comp_mod}, showing the
metallicities that would be inferred using the N2 and O3N2 indices for the GPs (blue squares), EGPs (red triangles), the 3 
$z \sim 2$ galaxies with direct metallicity measurements (this work; green symbols), and CSWA 20, a $z=1.4$ lensed galaxy
(\citealt{james14}; magenta pentagon.) For each point, the value of 12+log(O/H) measured from the direct method is indicated
(see text for discussion, and Figure~\ref{fig:met_v_met}. The errorbar in the righthand panel show the typical uncertainties
on the $T_{\rm e}$-based metallicity determinations; the formal uncertainties for the N2- and O3N2-based determinations
are similar or smaller.)
}
\label{fig:bpt_mod_peas}
\end{figure*}

\section{Local Analogs of KBSS-MOSFIRE Galaxies}
\label{sec:analogs}

It is potentially instructive to examine what is known about a relatively local population of galaxies that in many respects resembles
both typical and extreme members of the KBSS-MOSFIRE sample at $z \simeq 2.3$: the so-called 
``green pea'' (GP) galaxies (e.g., \citealt{cardamone09}) are relatively rare $z \simeq 0.2$ objects selected by their distinctive colors
caused by unusually large [\ion{O}{3}] equivalent widths. As shown in Figure~\ref{fig:bpt_mod_peas}, these rapidly star-forming, compact
galaxies occupy much of the same region of the BPT plane as the high redshift objects. 
The sample of 9 GPs in Figure~\ref{fig:bpt_mod_peas} is comprised of 6  
``extreme'' GPs studied by \citet{jaskotandoey13} and 3 ``normal'' GPs with 
very deep follow-up spectroscopy (\citealt{amorin12}). These 9 galaxies have a complete set of strong nebular lines as well
as metallicity measurements based on the direct $T_{\rm e}$ method; they serve as a possible ``preview'' of the efficacy of the
strong line metallicity measurements for similar galaxies at higher redshifts. 

First, we note that our simple photoionization models [more sophisticated, but less general, models were presented by
\citet{jaskotandoey13} and \citet{amorin12} in interpreting their data] presented in section~\ref{sec:bpt_interpretation} above  
are able to reproduce both the position of the GPs in the BPT plane (left panel of Figure~\ref{fig:bpt_mod_peas}) 
and the behavior of the N2 and O3N2 indices 
measured from the GPs' strong line ratios (right panel of Figure~\ref{fig:bpt_mod_peas}).  
The corresponding direct method metallicities (indicated in Figure~\ref{fig:bpt_mod_peas}b beside each point) 
show good agreement when the O3N2 and N2-based numbers are near 12+log(O/H)$\simeq 8.0$ (true for all 6 of
the ``extreme green peas''), but the 3 ``normal'' GP galaxies suggest a possible issue\footnote{We note that
the GP whose direct metallicity is most discrepant with the strong line estimators (GP232539; \citealt{amorin12}) has
$z=0.277$, which places the weak [OIII]$\lambda 4364$ feature at an observed wavelength of $\sim 5572$ \AA, so that
it is possible that its measured intensity has been affected by residuals from the strong 5577 \AA\ night 
sky emission line; there is a positive residual at the position of the (weaker) NaD night sky line in the spectrum.}: their direct metallicities
are comparable to or even lower than those of the ``extreme'' examples, but the strong-line indices
imply higher metallicities-- and the discrepancy may become marginally worse along the sequence
followed by both the $z \sim 2.3$ galaxies and by the ionization parameter sequence in the photoionization
models. Part of the disagreement relative to the local strong-line calibrations, as pointed out by 
\citet{amorin10}, is due to the fact that the GP galaxies appear to have higher (N/O) than typical local
galaxies of the same oxygen abundance. The 3 examples in Figure~\ref{fig:bpt_mod_peas} all have
log(N/O)$\simeq -1.0$, close to the solar ratio. We recall from section~\ref{sec:bpt_interpretation} above that roughly solar (N/O) was also
inferred for most of the $z \simeq 2.3$ galaxies on the basis of the photoionization models. The ``extreme'' GPs, on
the other hand, appear to have (N/O) consistent with that of local metal-poor dwarf galaxies, with
log(N/O)$\sim -1.5$, normally interpreted as systems in which only primary N enrichment has occurred (e.g., \citealt{vanzee98}).   
\begin{figure}[hpbt]
\centerline{\includegraphics[width=9cm]{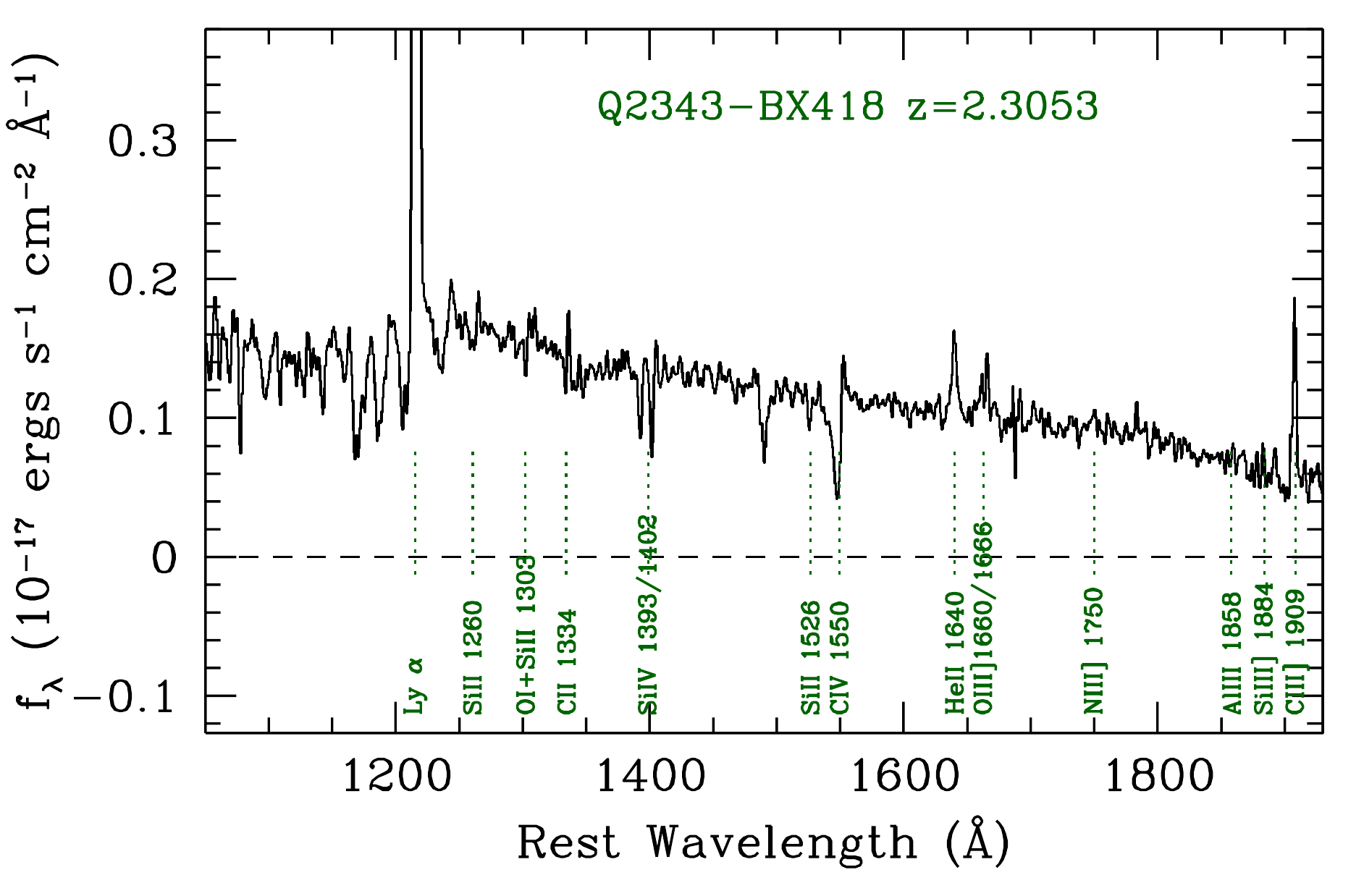}}
\centerline{\includegraphics[width=9cm]{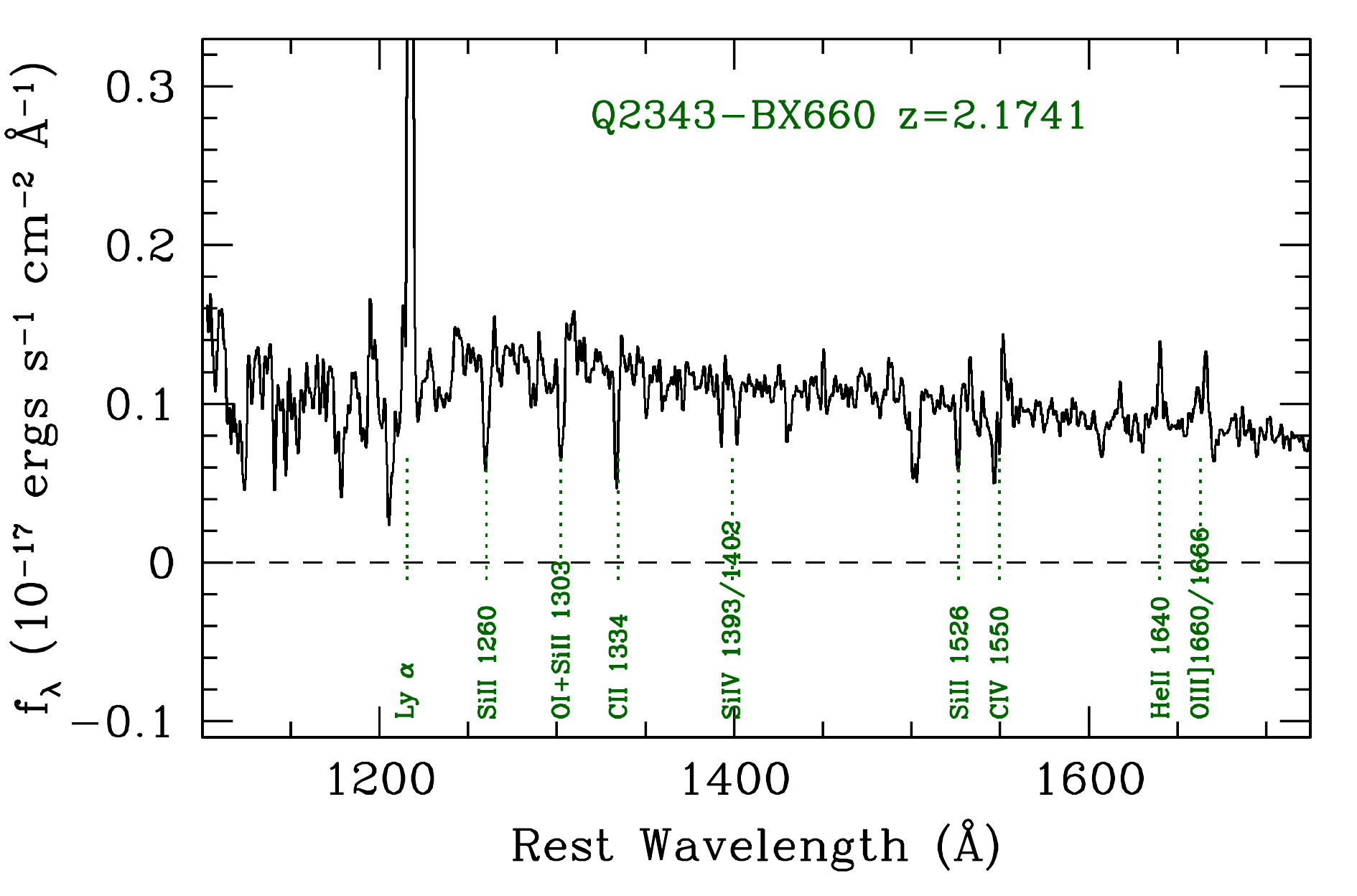}}
\centerline{\includegraphics[width=9cm]{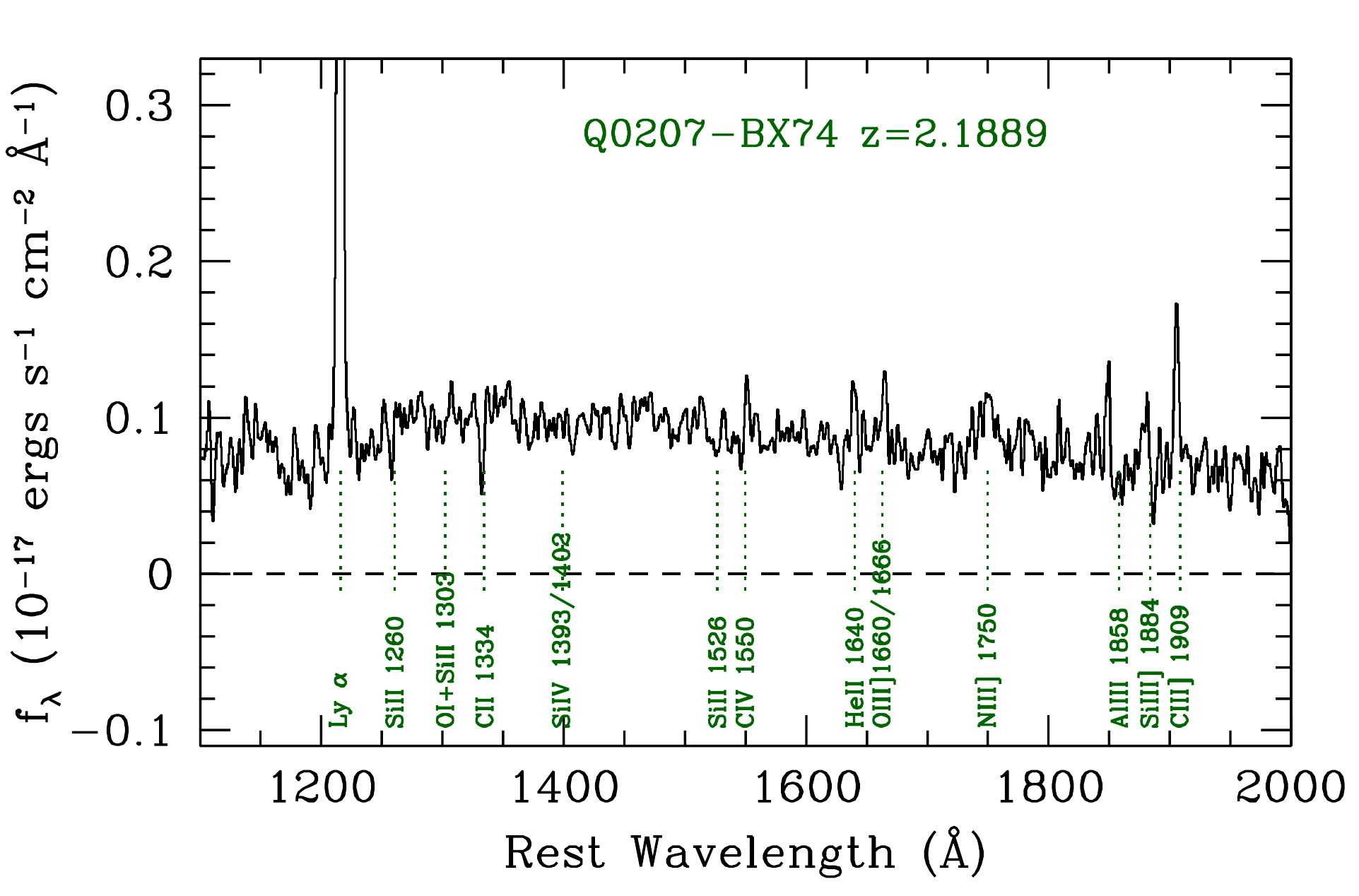}}
\caption{Rest-UV Keck/LRIS-B spectra of (top) Q2343-BX418 (see \citealt{erb2010}), (middle) Q2343-BX660 and (bottom) Q0207-BX74 
Note the presence of unusually strong lines of OIII] $\lambda\lambda 1661$, 1666 and
the [CIII]+CIII] blend near 1908 \AA. }
\label{fig:q2343_bx660_uv}
\end{figure}
\begin{figure*}[hpbt]
\centerline{\includegraphics[width=5.5cm]{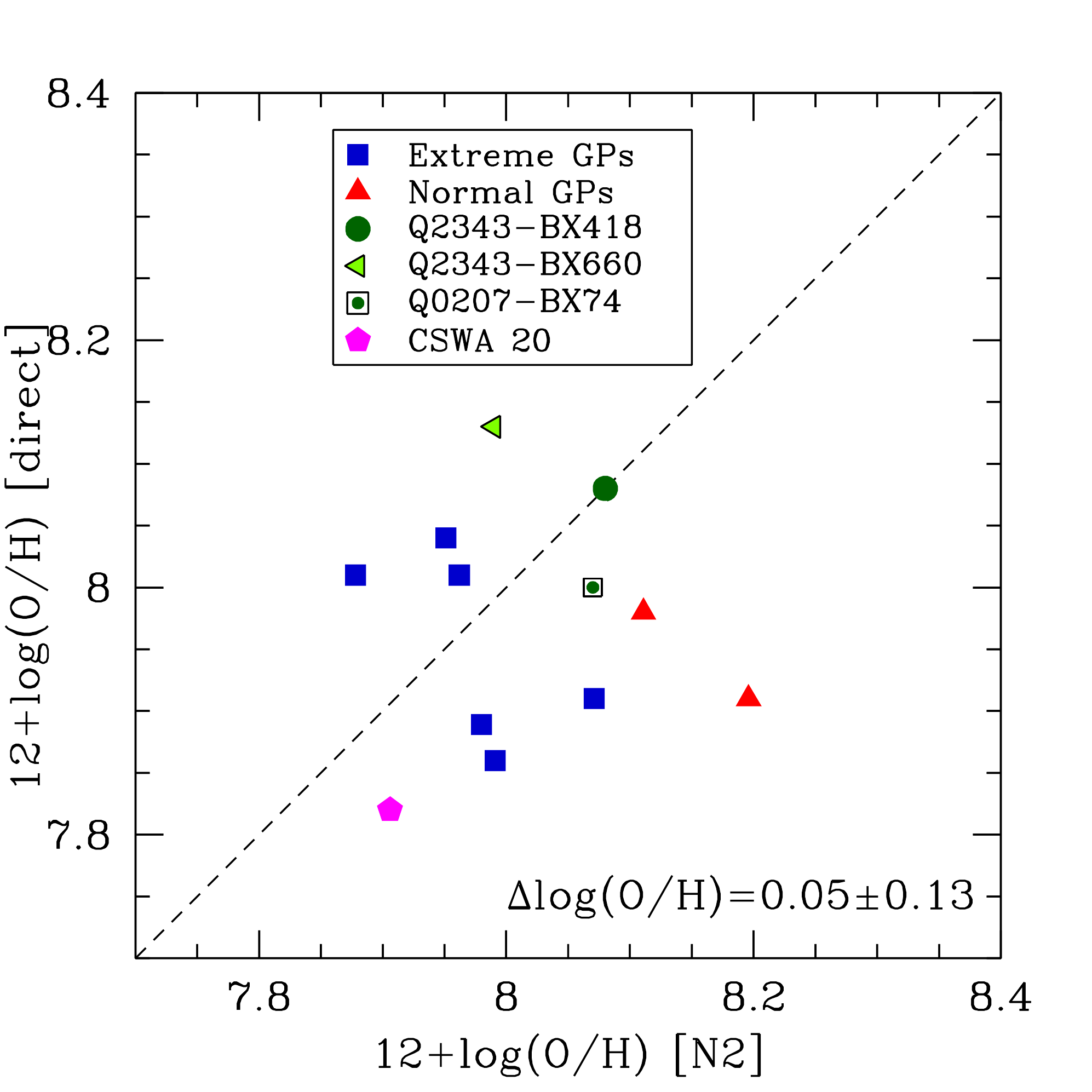}\includegraphics[width=5.5cm]{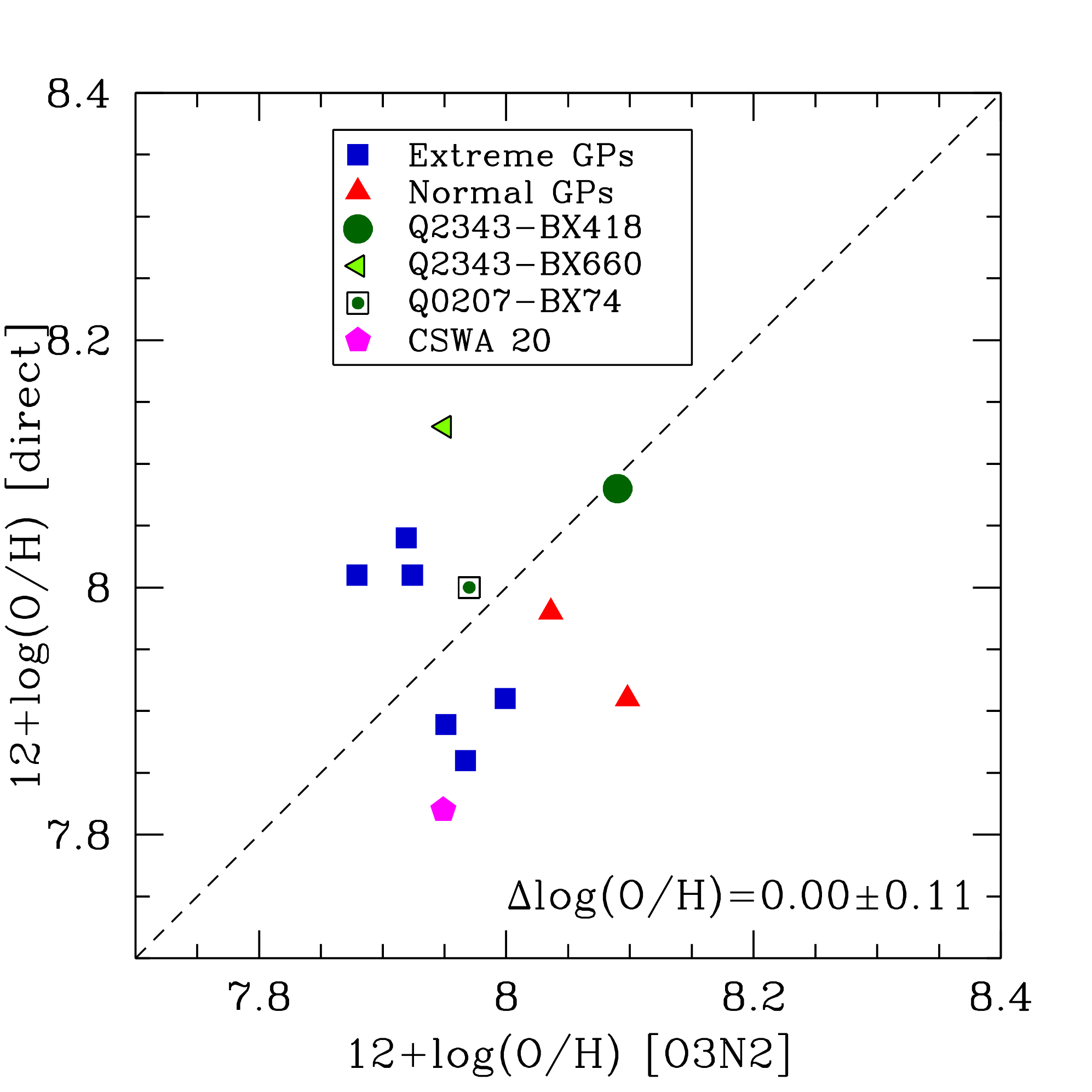}\includegraphics[width=5.5cm]{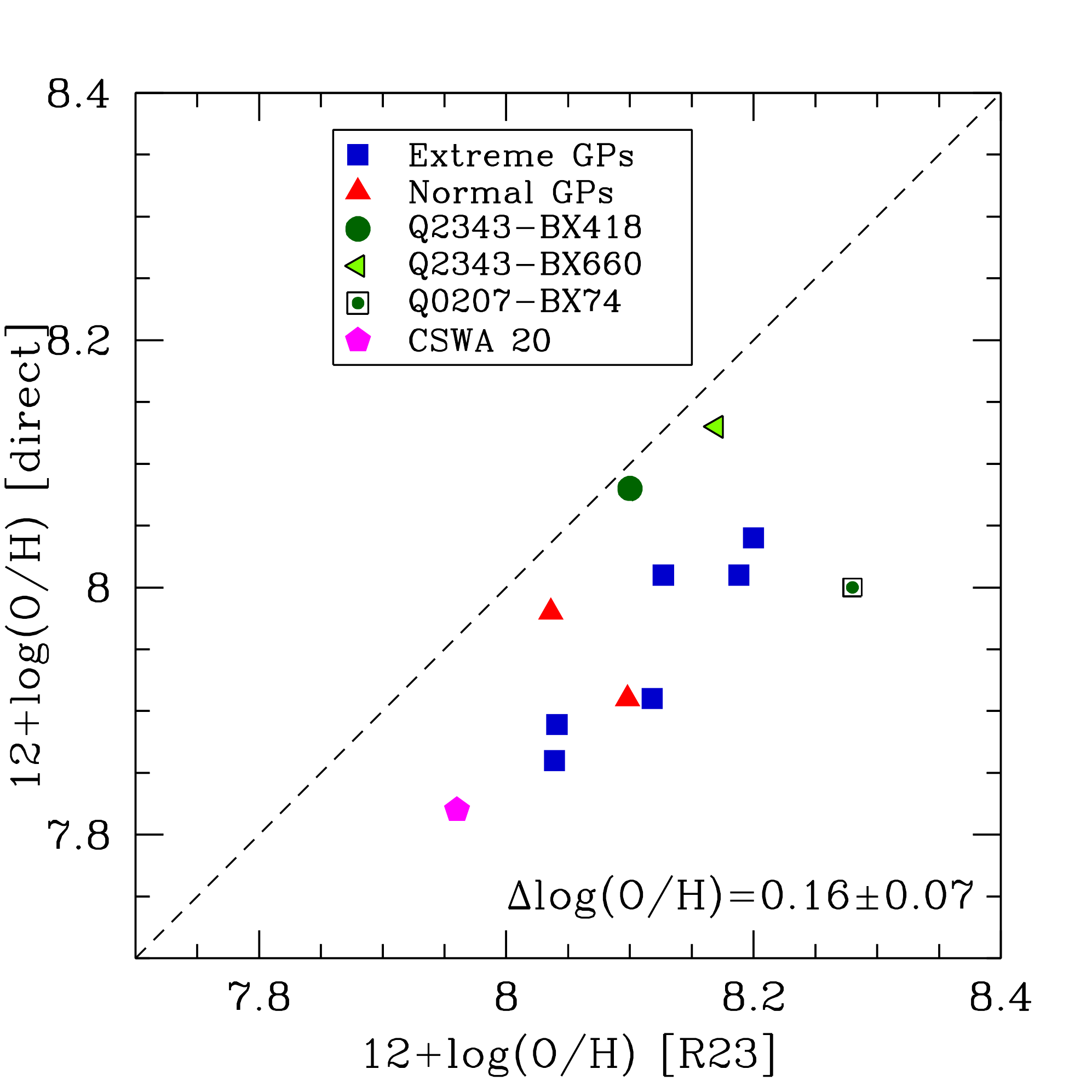}}
\caption{Comparison between direct $T_{\rm e}$ measurements of 12+log(O/H) and values suggested by three different strong-line  
indicators for the same set of galaxies shown in Figure~\ref{fig:bpt_mod_peas}b. Left: N2, Center: O3N2, Right: R23, using
the lower-branch calibration of \citet{mcgaugh91} (see text, and Table~\ref{tab:direct}, for details). 
The mean and rms scatter in the difference between individual measurements of 12+log(O/H) and the direct method measurement is given
in the lower right of each panel. 
}
\label{fig:met_v_met}
\end{figure*}

\subsection{``Extreme'' Galaxies at $z \simeq 2.3$}
\label{sec:extreme}

In fact, the extreme GPs have properties very similar to our most extreme $z \sim 2.3$ galaxies-- most of which have only
upper limits on [\ion{N}{2}/\Ha] (light green triangles in Figure~\ref{fig:bpt_mod_peas}a) as well as the
highest values of [\ion{O}{3}]/\Hb\ in the sample, with ${\rm log([OIII]/\Hb) \simeq 0.9}$.  Thus, it appears that
the extreme galaxies at both low and high redshift share a common ``upper envelope'' in the BPT diagram, previously
discussed in section~\ref{sec:agn}. 

Three galaxies in the current KBSS-MOSFIRE sample (Q2343-BX418, Q2343-BX660, and Q0207-BX74) are found near the EGPs 
in the BPT diagram, and have particularly good MOSFIRE (J, H, and K bands) as well as LRIS-B (rest-frame 
UV) spectra; they are indicated in Figure~\ref{fig:bpt_mod_peas}. Q2343-BX418 ($z=2.3053$) was studied in detail by 
\cite{erb2010} as a prototypical example of a UV-bright galaxy with little or no reddening, strong Lyman $\alpha$ emission, 
and unusually strong rest-UV nebular lines of OIII]$\lambda\lambda1661$,1666 and CIII] $\lambda\lambda 1906$,1909.
Q2343-BX660 ($z=2.1741$) and Q0207-BX74 ($z=2.1889$) have similar rest-UV spectra to that of BX418, as shown in
Figure~\ref{fig:q2343_bx660_uv}, as well as similar rest-optical strong line ratios. 
All 3 galaxies have ${\rm log(M_{\ast}/M_{\odot}) \sim 9.0}$ (very similar to the GP sample introduced above), SFR$\simeq 30-50$
M$_{\odot}$ yr$^{-1}$, and among the lowest inferred oxygen abundances and highest sSFRs in the current KBSS-MOSFIRE sample. 
Both Q2343-BX418 and Q2343-BX660 are known to be compact, both in \Ha\ emission (from Keck/OSIRIS laser guide
star AO IFU observations; \citealt{law09}) and in the rest-optical continuum (from HST/WFC3 F160W observations; \citealt{law2012}). 
Q0207-BX74 appears to be similarly compact, though as yet no AO or HST observations are available.  

\cite{erb2010} used measurements of rest-UV OIII] intercombination lines (at the redshift of BX418, [OIII]$\lambda 4364$ does not
fall within one of the ground-based atmospheric windows) together with rest-optical nebular emission based on Keck/NIRSPEC spectra
to measure $T_{\rm e}$ and thus ``direct'' metallicities, finding 12+log(O/H)$=7.8\pm0.1$, where some of the
uncertainty stems from a non-detection of [OII]$\lambda 3727,3729$ (so that the contribution of O$^{+}$ to O/H could not
be determined).   
We have re-observed Q2343-BX418 with MOSFIRE in J, H, and K bands,  covering, in addition to the BPT line ratios (Table~\ref{tab:n2ha_and_o3n2} 
and Figure~\ref{fig:bpt}), the [OII]$\lambda\lambda 3726,3729$ doublet, detected with S/N$\sim 30$. 
The observations in the 3 near-IR bands were obtained on the same night
and carefully cross-calibrated
to remove any differential slit losses using observations of a calibration star 
placed on one of the slits for all 3 observations. We used
these observations to calculate the electron density $n_{\rm e}$ from the [OII] doublet ratio, the Balmer decrement (\Ha/\Hb), the
ratio 
\begin{equation}
{\rm O32 \equiv[OIII](\lambda4960+\lambda5008)/[OII](\lambda 3727+\lambda 3729)}
\end{equation}
and, using the new [OIII]$\lambda 5008$ measurement together with the rest-UV measurement of the OIII]$\lambda\lambda1661,1666$ 
intercombination feature presented by \citet{erb2010}, the electron
temperature $T_{\rm e}{\rm [OIII]}$, from which (${\rm O^{++}/H^{+}}$) was derived. The ratio ${\rm (O^+/H^+)}$ was inferred
assuming ${\rm T_{\it e}([OII]) \simeq T_{\it e}([OIII])}$ as indicated by photoionization models, yielding a direct measure of
12+log(O/H) assuming that ${\rm (O/H) = (O^{++}/H^+) + (O^{+}/H^{+})}$. 
The results are summarized in Table~\ref{tab:direct}.  
We have assumed zero nebular extinction as in \citet{erb2010}, 
supported by both the Balmer decrement and the SED fitting results, and find 
${\rm 12+log(O/H) = 8.08\pm0.05}$, $\simeq 0.3$ dex higher than that obtained by
\citet{erb2010}. The difference is attributable to a lower derived $T_{\rm e}$ driven by a larger [OIII]$\lambda5008$ flux from the new H-band spectrum,
as well as the detection of the ${\rm [OII]\lambda\lambda 3727}$,3729 doublet, which allowed the contribution to (O/H) from O$^{+}$ to
be included.  
We also note that the measured ${\rm O32\simeq 10}$ is in excellent agreement with the photoionization models that reproduce BX418's position
on the BPT diagram (see section~\ref{sec:bpt_interpretation}). 

We performed similar analyses for Q2343-BX660
and Q0207-BX74. As summarized in Table~\ref{tab:direct}, we infer
high values of $n_{\rm e}=300-1600$ cm$^{-3}$, very high ionization level 
(based on O32), and direct oxygen abundances 12+log(O/H)$\simeq 8.00-8.15$.  Once again the measurement of ${\rm T_{\it e}[OIII]}$ 
is based on
the rest-UV OIII] doublet strength relative to [OIII]$\lambda 5008$, and the oxygen abundance includes the contribution from 
${\rm O^{+}}$; the Keck/LRIS-B spectra used to measure the UV features 
are shown in Figure~\ref{fig:q2343_bx660_uv}. 

The disadvantage of using the UV [\ion{O}{3}] feature instead of [\ion{O}{3}]$\lambda 4364$ to measure $T_{\rm e}$ is that it is
much more sensitive to the nebular extinction correction. Fortunately, two of the 3 galaxies for which the measurements are available
(BX418 and BX660) are consistent with zero nebular extinction based on the observed \Ha/\Hb\ ratio (the ``Balmer decrement''), which
are each consistent with the ``Case B'' expectation of \Ha/\Hb\ = 2.86;  
both are also consistent with zero extinction based on their SED fits. 

For Q0207-BX74, based on the observed Balmer decrement, we obtain 
${\rm E(B-V)_{neb}=0.18}$, while the 
stellar continuum (from SED fitting) has ${\rm E(B-V)_{cont}=0.13}$ assuming the \citet{calzetti00} attenuation relation. 
We corrected the relevant line fluxes in Table~\ref{tab:direct} assuming the former and the \citet{cardelli89} extinction
curve.  
Note that effect of the dust correction to the observed ratio ${\rm OIII](\lambda 1661+\lambda 1666)/[OIII]\lambda 5008}$ 
was to increase it by a factor of 2.04, increasing the inferred $T_{\rm e}$ from $\sim 12140$ K to $\sim 14300$ K and lowering the
inferred oxygen abundance by $\sim 0.23$ dex, from 8.23 to 8.00.  

Figure~\ref{fig:met_v_met} summarizes the comparison of the direct metallicity estimates for the same 13 galaxies 
as in Figure~\ref{fig:bpt_mod_peas}. In addition to the N2 and O3N2-based estimates, we have applied 
the low-metallicity branch of the R23 calibration 
of \citet{mcgaugh91} [as expressed by \citealt{kobulnicky99}] to the measurements of O32 and ${\rm ([OIII]_{tot}+[OII]_{tot})/\Hb}$,
to estimate R23-based metallicities, shown in the rightmost panel of Figure~\ref{fig:met_v_met}. 
Figure~\ref{fig:met_v_met} suggests that, at least for this sample, O3N2 provides a slightly better approximation to the
direct method metallicities, with a relative offset of ${\rm log(O/H)_{O3N2}-log(O/H)_{dir}= 0.00\pm0.11}$ dex,  
while ${\rm log(O/H)_{N2} - log(O/H)_{dir} = 0.04\pm0.14}$ dex and ${\rm log(O/H)_{R23} - log(O/H)_{dir} = 0.16\pm0.08}$ dex. 
We caution that these statistics are based on a very small sample, confined to low metallicities where the various
indicators appear to be in reasonable agreement with one another; however, the finding that the O3N2 index provides the least-biased
estimate of direct-method metallicities for galaxies offset from the BPT excitation sequence 
is consistent with results presented by \cite{liu+shapley08} for $z \simeq 0$ SDSS galaxies; these authors found that
N2 systematically over-estimates 12+log(O/H) compared to the direct method. 
{\it On balance, it seems most likely that the O3N2 index yields more reliable values of 12+log(O/H) than those
derived from the N2 index. }

%The current KBSS-MOSFIRE sample in Tables 1 and 2 have provide measurements
%using both N2 and O3N2, while the 32 galaxies in Table 3 currently lack H-band observations, and thus do not
%have direct measurements of O3N2with N2 measurements is a factor of more
%than two larger than that with O3N2 measurements. In section \ref{sec:mass_met} below, we use the relative
%calibrations between the two metallicity estimates (which has a small scatter-- see Figure~\ref{fig:met_compare}) to
%``convert'' the N2-only measurements onto the O3N2 metallicity scale in the context of discussing the
%relationship between stellar mass, oxygen abundances, and star formation rates at $z \sim 2.3$.

\section{The M$_{\ast}$-Metallicity Relation at $\langle z \rangle = 2.3$}

\label{sec:mass_met}
A correlation between stellar mass M$_{\ast}$ and nebular oxygen abundance has now been well-established 
at $z \simeq 0$ using both strong-line metallicity measurements (e.g., \citealt{tremonti04,kewley+ellison08,maiolino08}) 
as well as direct $T_{\rm e}$-based measures \citep{andrews+martini13}. However, even in the local universe
the quantitative behavior of the mass-metallicity relation (MZR) depends substantially on
the method used to measure 12+log(O/H). 
Apparently similar behavior, with substantial offsets in the sense that galaxies are inferred to have lower ionized gas
metallicities at a given M$_{\ast}$, has been observed for relatively small samples of $z > 2$ galaxies (\citealt{erb+06a,maiolino08,law09,forst09,henry13,newman13,wuyts14,kewley13b}). 

As we have seen above, we do not yet know whether the strong-line determinations of metallicity at high
redshift are directly comparable to those obtained at low redshift, even using the same diagnostic. We showed in the previous section that
there is evidence, albeit limited, 
(see Figure~\ref{fig:met_v_met} and \citealt{liu+shapley08}) that the O3N2 calibration is least
biased with respect to direct-method metallicities for galaxies falling ``above'' the low-redshift BPT ionization sequence. 
We have also shown (see also \citealt{newman13}) that there is a systematic offset of $\Delta \simeq 0.13$ dex between metallicities
inferred from the PP04 N2 and O3N2 indices when they are applied to the same galaxies in the KBSS-MOSFIRE $z \simeq 2.3$ sample, in
spite of the fact that these calibrations were established using direct $T_e$ abundances of the same local galaxy sample. 
There is little doubt that other systematic differences in the MZR plane would be found using other local strong line
calibrations.

In addition to systematics resulting entirely from application of local calibrations, it is also quite probable
that, even at a given redshift, results of different studies may differ in detail either because of the way in which targets
are selected, or by differences in the quality and depth of the resulting spectra (see, e.g., \citealt{juneau14}).  
The degree to which selection and/or observational bias affects global statistics like the MZR will also depend on the extent
to which fundamental galaxy properties (which one is trying to measure) are correlated with a galaxy's ``observability'', as discussed
in section \ref{sec:sample_statistics} above. Measuring evolution of the galaxy population in the MZR plane is potentially
even more problematic, since selection and observation biases may be changing with redshift in a way that could either mask
or exaggerate real differences. 

\begin{figure*}[htpb]
\centerline{\includegraphics[width=8.8cm]{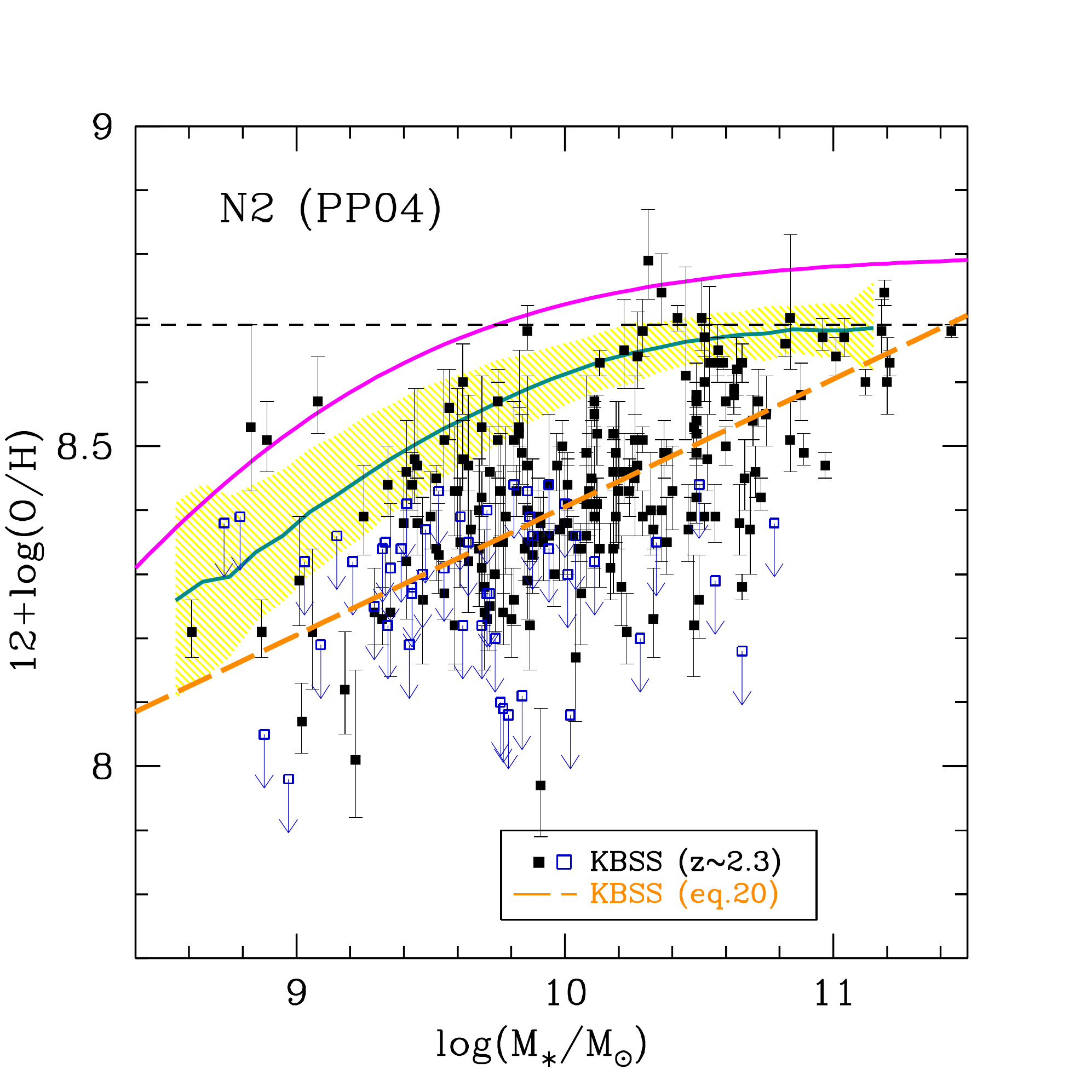}\includegraphics[width=8.8cm]{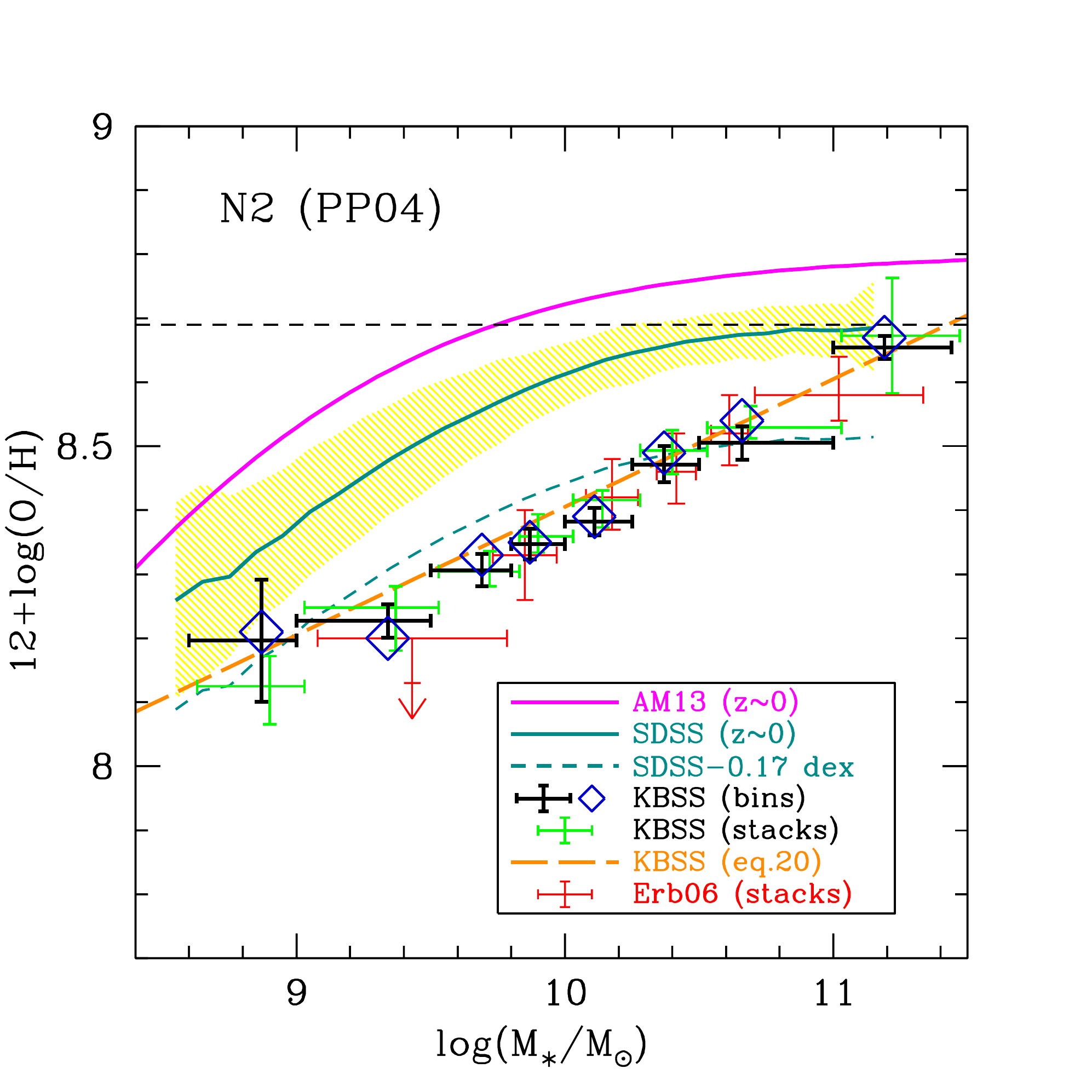}}
\caption{({\it Left}) Observed relation between stellar mass (M$_{\ast}$) and oxygen abundance inferred from the PP04 N2 index calibration, for  
$z \sim 2.3 $ KBSS-MOSFIRE galaxies. The sample includes
192 galaxies with [NII]/\Ha\ measurements (black points) and 50 with $2\sigma$ upper limits on
[NII]/\Ha\ (blue open squares, with downward arrows). 
The long-dashed orange line is the best-fit linear relation between 12+log(O/H)$_{\rm N2}$ and ${\rm log(M_{\ast}/M_{\odot})}$ (equation~\ref{eqn:linear2}; see text for discussion) using the ensemble of individual measurements (i.e, not binned).  
The solid turquoise curve and light shading (representing the approximate scatter) is the best-fit MZR for star-forming galaxies
in SDSS-DR7, assuming the (linear) N2 calibration of PP04. 
The solid magenta curve is the best-fit MZR from \citet{andrews+martini13} 
where metallicities were determined using 
the direct method based on stacked SDSS spectra in bins of stellar mass.
({\it Right}) Same as left panel, but with data points binned by stellar mass (see Table~\ref{tab:mass_met_bins_n2}).  
The black, heavy error bars are the weighted average
metallicities of the individual galaxies in each bin.
In the y-direction, the error bars reflect uncertainty in the bi-weight mean within each bin, with the x-direction error
bars indicating the limits of the ${\rm M_{\ast}}$ bin. The x-location of each point is determined by the median ${\rm log(M_{\ast}/M_{\odot})}$ within
the bin. The blue diamonds are the median inferred metallicity within each stellar mass bin, and the light green error bars are based
on stacked spectra in the same bins (see text for discussion). 
Note that the same linear fit (from equation~\ref{eqn:linear1}) to the KBSS N2 MZR is shown
in both panels, based on the full sample of individual measurements 
as described in the text.  The red error bars show the results of \citet{erb+06a}, based on spectral stacks in bins of stellar mass.
The dashed turquoise curve in the righthand panel shows the local SDSS MZR, shifted to lower inferred oxygen abundance by 0.17 dex. 
\label{fig:mass_met}
}
\end{figure*}

\begin{figure*}[bhtp]
\centerline{\includegraphics[width=8.8cm]{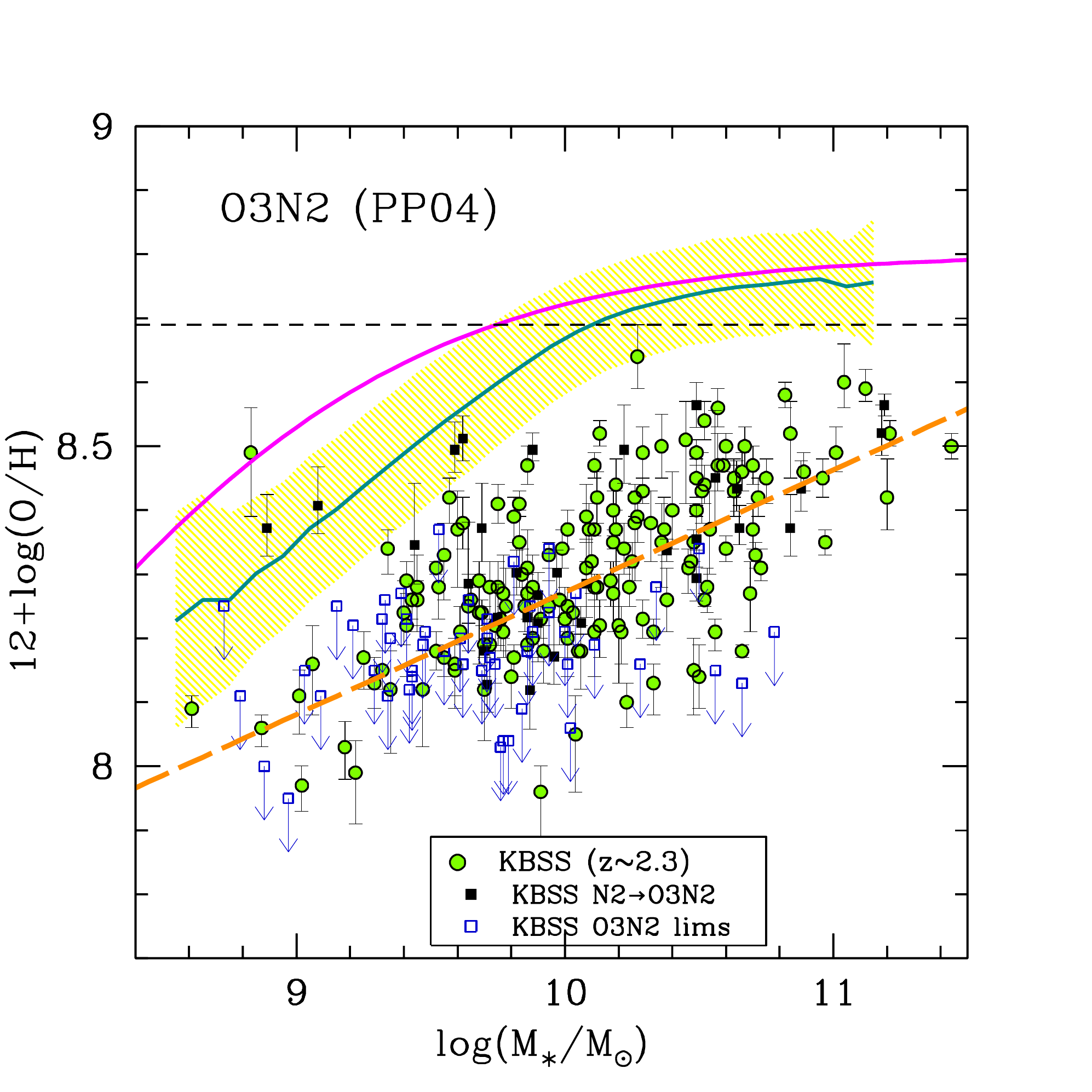}\includegraphics[width=8.8cm]{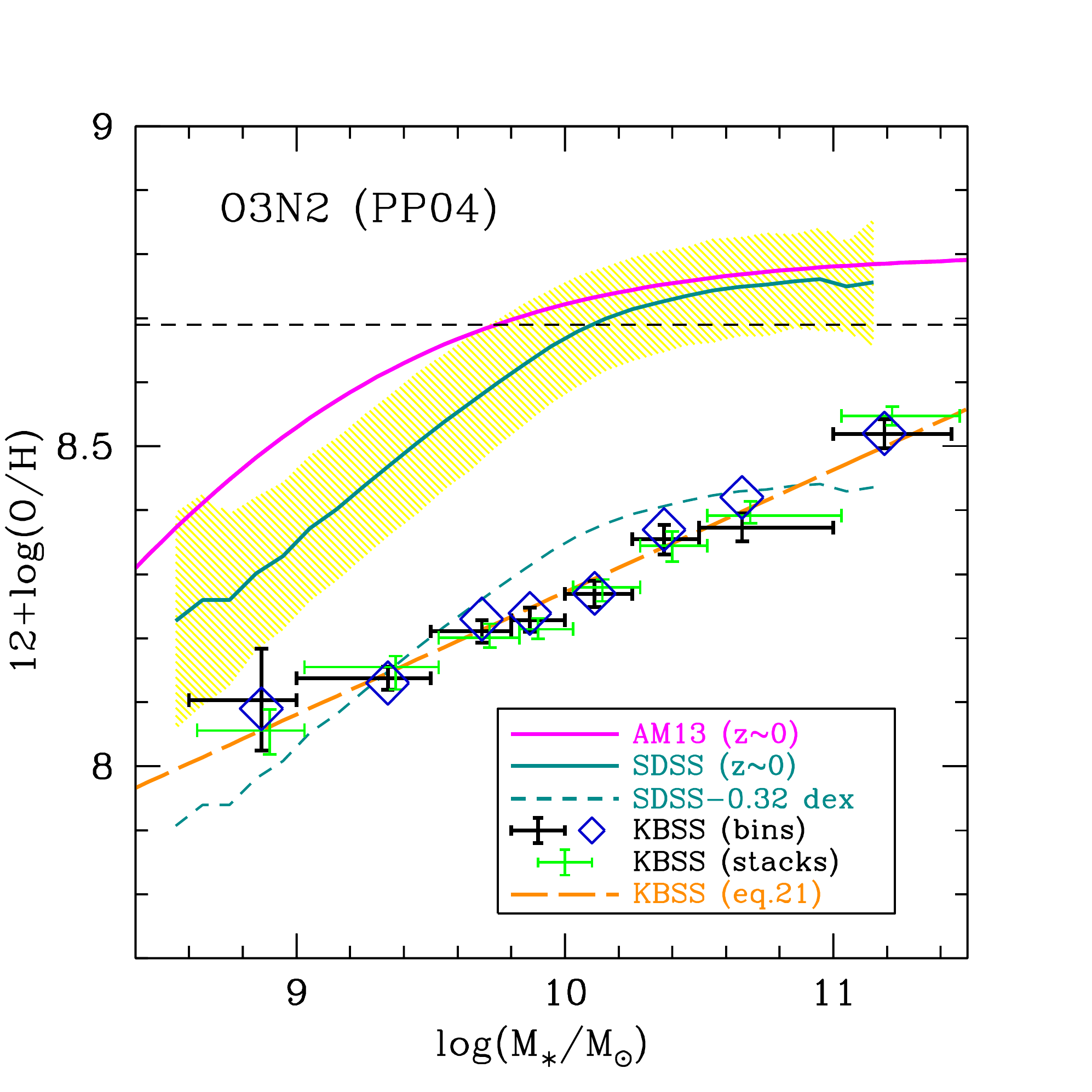}}
\caption{({\it Left}) Same as Figure~\ref{fig:mass_met}, but for metallicities inferred from the PP04 O3N2 calibration. Galaxies with individual 
O3N2 measurements (160 objects; Table 1) are indicated with light green solid dots, while 2$\sigma$ upper limits on the O3N2 metallicity
(51 objects; Table 2) are shown with blue squares. Equation~\ref{eqn:n2_to_o3n2} was used to convert the 31 galaxies with N2 but 
lacking O3N2 measurements (Table 3), represented by solid black squares with error bars). 
The solid turquoise curve and yellow shaded region
represents the $z\simeq 0$ SDSS-DR7 MZR using the same PP04 O3N2 calibration;  
the magenta curve is the \citet{andrews+martini13} $z\sim0$ MZR (as in Figure~\ref{fig:mass_met}).
({\it Right}) As in Figure~\ref{fig:mass_met}, where individual measurements were combined within the same bins of stellar
mass (see Table~\ref{tab:mass_met_bins_o3n2}). In both panels, the long-dashed orange curve is the best linear fit
to the O3N2 MZR from equation~\ref{eqn:linear2}, using all of the individual measurements as described in the text. 
The dashed turquoise curve in the righthand panel is the low-redshift SDSS relation, shifted to lower oxygen abundance
by 0.32 dex. 
 }
\label{fig:mass_met_bins}
\end{figure*}

For the moment, since most of our analysis of the nebular spectra in the KBSS-MOSFIRE sample has been focused
on measuring the BPT line ratios, we have produced MZRs using both the N2 and O3N2 indices and the PP04 calibrations;
the results are shown in Figures~\ref{fig:mass_met} and \ref{fig:mass_met_bins}. 
The same set of 242 galaxies with $\langle z \rangle = 2.30\pm0.16$ (all objects in Tables 1-3 not classified as AGN)
was used for both determinations of the MZR. The N2 data set, shown in Figure~\ref{fig:mass_met}, has 192 N2 detections
(Tables 1 and 3) together with 50 upper limits (Table 2). For O3N2, the sample includes 161 galaxies with detections of
both [OIII]/\Hb and N2 (Table 1), 50 with [OIII]/\Hb\ detections and N2 upper limits (Table 2),
and 31 galaxies for which only N2 has been measured (Table 3). This last sub-sample was
included by using the fact that KBSS-MOSFIRE galaxies with both N2 and O3N2 measurements fall along
a well-defined sequence with small intrinsic scatter
(Figure~\ref{fig:met_compare}); the best-fit linear relation 
\begin{equation}
{\rm 12+log(O/H)_{O3N2}  = 0.87~ [12+log(O/H)_{N2}] + 0.94 }
\label{eqn:n2_to_o3n2}
\end{equation}
was used to convert from N2-based to O3N2-based metallicity scales. The error bars for the converted points include both the uncertainty in N2
and the residual dispersion of the data relative to the fit in equation \ref{eqn:n2_to_o3n2}\footnote{This small subsample
of the O3N2 data set, which represents only 12.8\% of the sample, was included for completeness. Excluding it from the fits
discussed below has no significant effect on the results.}. 
%Figures~\ref{fig:mass_met} and \ref{fig:mass_met_bins} show both versions of the resulting KBSS-MOSFIRE $z \sim 2.3$ MZR, as well as
%previous results from \cite{erb+06a}\footnote{The values of 12+log(O/H)$_{\rm N2}$ from \cite{erb+06a} 
%were converted to 12+log(O/H)$_{\rm O3N2}$ using equation~\ref{eqn:n2_to_o3n2}}. The latter measurements 
%were based on spectral stacks in bins of stellar mass.
%The righthand panels of Figure~\ref{fig:mass_met} and \ref{fig:mass_met_bins} 
%are identical to the corresponding lefthand panels, except that the KBSS-MOSFIRE data are binned
%by M$_{\ast}$ for clarity; the binned data points are collected in Tables~\ref{tab:mass_met_bins_n2} and \ref{tab:mass_met_bins_o3n2} for convenience.  


In both Figures~\ref{fig:mass_met} and~\ref{fig:mass_met_bins} the long-dashed (orange) lines show the best linear 
fits to the ensemble of measurements and limits for the full data set. The fits were obtained using the
Bayesian linear regression method described by \citet{kelly07}, which accounts for measurement errors in both the dependent and 
independent variables, and treats non-detections/limits in a consistent manner. For the purposes of the calculation,
we assumed a characteristic uncertainty in ${\rm log(M_{\ast}/M_{\odot})}$ of $\pm 0.16$ dex\footnote{The results are
insensitive to the exact value adopted; 0.16 dex is the median estimated uncertainty in the stellar mass
estimates for similar galaxy samples (\citealt{shapley05,erb+06b})}. The method returns posterior distributions for each parameter, including 
the variance of the {\it intrinsic} scatter, which is of particular interest since MZR scatter has not been measured previously at high redshift. 
Expressing the MZR as a linear function of the form  
\begin{equation}
{\rm  12+log(O/H)  =  Z_{10} + \gamma ~ [log ({M_{\ast}/M_{\odot})-10]}} 
\label{eqn:linear}
\end{equation}
where ${\rm Z_{10}}$ is the metallicity normalization at ${\rm log(M_{\ast}/M_{\odot}) = 10.0}$ and $\gamma$ is the linear
slope, we find best-fit parameters as follows: 
\begin{equation}
{\rm N2:~~ \quad Z_{10}=8.41\pm0.01 ; \quad \gamma=0.20\pm0.02 ; }
\label{eqn:linear1}
\end{equation}
$$ \quad \sigma_{sc}=0.10\pm0.01$$
and
\begin{equation}
{\rm O3N2: \quad Z_{10}=8.27\pm0.01 ; \quad \gamma=0.19\pm0.02  }
\label{eqn:linear2}
\end{equation}
$$\quad \sigma_{sc}=0.10\pm0.01$$
In both cases, $\sigma_{\rm sc}$ is the best estimate of the {\it intrinsic} scatter in the MZR relative to the fit. 
We will return to a discussion of the low intrinsic scatter in the $z \sim 2.3$ MZRs below (section~\ref{sec:mzr_scatter}. 

Generally, the low-redshift MZRs such as those used for comparison in Figures~\ref{fig:mass_met} and \ref{fig:mass_met_bins} 
reflect a ``flattening'' above a characteristic stellar mass; consequently, the fitting functions used to represent
them include such a characteristic mass as an additional parameter (e.g., \citealt{maiolino08,moustakas11,andrews+martini13,zahid14}),
somewhat akin to ``$L^{\ast}$''  in a luminosity function. However, we found that fitting the more complex functions to the KBSS-MOSFIRE
data at $z\sim 2.3$ could not be justified, since Figures~\ref{fig:mass_met} and \ref{fig:mass_met_bins} clearly show that 
the linear functions in equations~\ref{eqn:linear1} and \ref{eqn:linear2} are good fits, and there is no obvious sign that
either of the MZRs flattens at high ${\rm M_{\ast}}$.   
It is not yet clear how literally one should take apparent differences in shape or normalization of the MZRs
at $z \sim 0$ and $z\sim 2.3$ MZRs, for all of the reasons emphasized above. 
For the same reason, one should probably use caution interpreting similarities or differences between any two galaxy samples 
without a detailed understanding of the systematics of the selection function, the criteria for successful observation, 
and the likely systematic issues inherent in 
mapping strong-line ratios to metallicity.  

It is perhaps encouraging, on the other hand, that the posterior likelihood distributions of both $\gamma$ and $\sigma_{\rm sc}$ 
for the $z\sim 2.3$ MZRs are entirely consistent
with one another; the only significant difference between them (aside from the larger contribution of measurement errors 
for N2 as compared to O3N2) is the aforementioned offset in metallicity normalization of 0.13-0.14 dex.

\subsection{The MZR in Bins of ${\rm M_{\ast}}$}
\label{sec:mstar_bins}

\citet{erb+06a} first showed, based on composite spectra formed from bins of ${\rm M_{\ast}}$ (large open diamonds in
Figure~\ref{fig:mass_met}), that the $z \sim 2.3$ MZR lies substantially below the $z \simeq 0$ relation. The amplitude
of the shift in metallicity depends on the method used to measure it -- \citet{erb+06a} found that the shift
of the $z \sim 2.3$ metallicities (measured using N2) relative to the MZR of \citet{tremonti04} was $-0.56$ dex, 
but decreased to $\simeq -0.3$ dex when the PP04 N2 calibration was applied to the SDSS sample.  
It appears (Figure~\ref{fig:mass_met}) that the KBSS-MOSFIRE N2-based MZR exhibits a slightly shallower 
dependence of the N2 index on M$_{\ast}$ 
than the \citet{erb+06a} sample, at least for low ${\rm M_{\ast}}$. 
We note that, although the galaxies targeted by \citet{erb+06a} came from UV color-selected catalogs defined in
the same way as most of the current sample, 
the KBSS results are nearly independent of 
the \citet{erb+06a} sample, in the sense that all of the nebular line
measurements are based on new observations with MOSFIRE, and only 25 of 251 galaxies ($\simeq 10$\%) 
of the new sample were included in that of \citet{erb+06a}.   

Referring to Figure~\ref{fig:mass_met}, the best fit locus of individual galaxies from
KBSS-MOSFIRE agrees well with the result from the stacked spectra of \citet{erb+06a} for ${\rm log(M_{\ast}/M_{\odot}) \simgt 9.8}$ (i.e.,
in all but the lowest mass bin of the \citealt{erb+06a} data), while
for $\rm log(M_{\ast}/M_{\odot}) \simlt 9.8$, the KBSS data indicate higher values of 
${\rm 12+log(O/H)_{N2}}$ than the upper limit of \citet{erb+06a}. We discuss the significance of and possible reasons for this discrepancy below.  

Most subsequent high-redshift ($z \simgt 1.5$) evaluations of the MZR to date have also relied primarily on stacked spectra
in bins of $M_{\ast}$ as in \citet{erb+06a}
(e.g., \citealt{newman13,henry13,cullen14,troncoso14,wuyts14,sanders14}).   
To facilitate comparison with 
other MZR determinations, we evaluated the KBSS-MOSFIRE data in 8 bins of $M_{\ast}$ covering the full observed range
(see Tables~\ref{tab:mass_met_bins_n2} and \ref{tab:mass_met_bins_o3n2}).  
For the purpose of evaluating stellar mass bins that include objects with metallicity upper limits (in all cases
due to the non-detection of [NII]), we assigned each N2 line index non-detection its nominal 1$\sigma$ upper limit and 
an uncertainty of $\pm 0.3$ dex (i.e., a factor of two). The corresponding metallicity uncertainty was obtained
by propagating 
the assumed line index error to a corresponding error in metallicity. For N2-based metallicities, $\sigma = \pm 0.17$ dex 
on ${\rm 12+log(O/H)_{N2}}$, while for O3N2-based metallicities the N2 line index error contributed 
an uncertainty of $\sim \pm0.10$ dex, which was propagated along with the [OIII]/\Hb\ measurement error to obtain a limiting value.  
We then evaluated the median and
the weighted average metallicity within each mass bin; the results are summarized in Tables~\ref{tab:mass_met_bins_n2} 
and ~\ref{tab:mass_met_bins_o3n2} and plotted in Figures~\ref{fig:mass_met} and \ref{fig:mass_met_bins}.  

Since only $\sim 20$\% of the sample has metallicity limits (50 out of 242 galaxies), the bin values 
are relatively insensitive to the exact metallicity values for the limits. Figures~\ref{fig:mass_met} and \ref{fig:mass_met_bins} 
show both bin mean (black error bars) and bin median (blue diamonds) These values 
are consistent with one another, as well as with the linear fit to the ensemble of individual measurements (equations
\ref{eqn:linear1} and \ref{eqn:linear2} for N2 and O3N2, respectively) described in the previous section. 

We also constructed stacked spectra within the same stellar mass bins 
(to be discussed in
detail elsewhere; Strom et al., in preparation) for the KBSS-MOSFIRE sample. 
Line indices obtained from spectral stacks have distinct advantages, particularly if many of the
individual spectra are not of high enough quality to allow object-by-object line ratio measurements, since
spectra yielding only upper limits on inferred metallicity can be easily included in the
stacks along with those yielding individual detections. However, stacks do require one to choose
how to (or whether to) scale the rest-frame spectra of individual galaxies prior to averaging; 
there are many subtleties to making this choice, and its subsequent effect on the results may depend on the underlying selection
method and the nature of any observational biases. For the present, we made spectral stacks 
for the KBSS-MOSFIRE sample, using the same method employed by \citet{erb+06a}, with
results summarized in Tables~\ref{tab:mass_met_bins_n2} and \ref{tab:mass_met_bins_o3n2} and shown in the righthand panels
of Figure~\ref{fig:mass_met} and \ref{fig:mass_met_bins}. As can be seen in Figures
\ref{fig:mass_met} and \ref{fig:mass_met_bins}, the metallicity values based on stacks are consistent at the $\simlt 1\sigma$ level
with both the median and the average values within each bin; moreover, fits of a linear MZR of the 
form given in equation~\ref{eqn:linear} to the binned data points (whether one chooses the median, mean,
or stacked values) yield values of $Z_{10}$ and $\gamma$ consistent with the fits 
to the full sample ensemble (with no binning).  

Thus, the origin of the apparent difference between the KBSS sample and that of \citet{erb+06a} is probably
not related to binning/stacking, nor
to the details of how one includes spectra with individual N2 upper limits. 
Aside from pure sample variance (the lowest-mass bin in the \citet{erb+06a} sample is based on a spectral stack
of only 15 galaxies, whereas the KBSS sample contains
85 galaxies in the same stellar mass range), some part of the discrepancy 
might be explained by very different spectral resolution and S/N (both are considerably higher 
for the MOSFIRE spectra), when one accounts for the fact that weak emission lines are harder to distinguish from
the continuum level, so that systematic errors in the zero level of the spectra can have a large effect on inferred line strength near the detection
limit.  A related possibility is that there is a real difference in the properties of the galaxy samples at low $M_{\ast}$ that
leads to different line index measures. One obvious possibility is (e.g.) a different average SFR:  
the mean SFR in the lowest-M${\ast}$ bin of the \citet{erb+06a} sample is $\simeq 2.5$ times larger than that of the KBSS galaxies
in the same range of M$_{\ast}$; however, we show in section~\ref{sec:mzr_sfr} below that the inferred metallicities 
at a given stellar mass within the KBSS sample do not obviously depend on SFR. 
 
In any case, the relation between the strong-line metallicity-- using either the N2 or O3N2 indices-- and log(M$_{\ast}$/M$_{\odot}$) is quite shallow
over the range in M$_{\ast}$ spanned by the KBSS-MOSFIRE sample, with best-fit linear slope of $\gamma \simeq 0.20$ that appears to extend
over the full observed range of ${\rm M_{\ast}}$

%In view of
%the $z \sim 0$ MZRs, this is not surprising-- over the same range in stellar mass (${\rm 9 \simlt log(M_{\ast}/M_{\odot}) \simlt 11}$)
%the observed slopes are similarly shallow. For
%example, a linear approximation to the \citet{andrews+martini13} MZR for ${\rm 9 < log(M_{\ast}/M_{\odot})} < 11$) has $\gamma=0.13$,
%while the \citet{maiolino08} O3N2 MZR curve shown in Figure~\ref{fig:mass_met_bins} has $\gamma = 0.15$ over the same
%mass range.          

%Focusing on the O3N2 MZR for definiteness (and because it appears to be closest to the $T_{\rm e}$/direct metallicity
%scale at $z \sim 2.3$ and at $z \sim 0$ as discussed above; see also Figure~\ref{fig:mass_met_bins}),  we find that
%the two functional forms for the $z \sim 0$ MZR shown provide reasonable fits to the observed $z \sim 2.3$ data after
%adjusting the overall metallicity normalization.  The dashed magenta and turquoise curves in Figure~\ref{fig:mass_met_bins} show
%the \citet{andrews+martini13} and
%\citet{maiolino08} MZR fits with $\Delta({\rm 12+log(O/H))}= -0.38$ and $-0.34$ dex, respectively. With these adjustments, 
%either curve provides a fit of similar quality (i.e., $\sigma \simeq 0.12$ dex) to those in equations \ref{eqn:linear2} and \ref{eqn:asympt_o3n2}   
%above. Thus, we find that over the current stellar mass range probed, the MZR at $z\sim2.3$ may be reasonably well-described by 
%the $z \sim 0$ MZR with a $\simeq$0.36 dex reduction in oxygen abundance, {\it independent of stellar mass}.  Qualitatively,
%the same conclusion was reached by \citet{erb+06a}.    

\subsection{Scatter in the $z\sim 2.3$ MZR}
\label{sec:mzr_scatter}

As for the BPT locus discussed in section~\ref{sec:bpt} above, it is interesting to compare
the degree of scatter in inferred metallicity at fixed M$_{\ast}$ at $z \simeq 2.3$ to that observed
at low redshift. We found above that, for both N2- and O3N2-based metallicity determinations, 
the {\it intrinsic} scatter in the MZR was $\sigma_{\rm sc} \simeq 0.10$ dex, compared to $0.08-0.12$ dex for
the SDSS-DR7 sample (Figures~\ref{fig:mass_met} and \ref{fig:mass_met_bins}; the scatter increases toward lower M$_{\ast}$ in the 
SDSS sample).  
Dividing the KBSS galaxy sample in half near the median ${\rm log(M_{\ast}/M_{\odot}) = 10.0}$
and estimating $\sigma_{\rm sc}$ separately for each sub-sample, 
we find no significant difference, 
with ${\rm \sigma_{\rm sc}[log (M_{\ast}/M_{\odot})>10] = 0.11\pm0.01}$ dex and ${\rm \sigma_{sc}
[log(M_{\ast}/M_{\odot})<10] = 0.10\pm0.01}$ dex. 

An obvious point, relevant at both $z \sim 0$ and $z\sim2.3$, is that the scatter in the inferred metallicity at a given stellar mass is {\it smaller} 
than could be reasonably expected even if the ``true'' oxygen abundance were a perfect monotonic function of M$_{\ast}$. 
The scatter in the empirical strong-line metallicity calibration, estimated by PP04 to be $\simeq 0.18$ dex for N2 
and $\simeq 0.14$ dex for O3N2, both {\it exceed} the inferred intrinsic MZR scatter of $\simeq 0.10$ dex.       
We showed in section~\ref{sec:pp04_calibration} that the calibration errors of the N2 and O3N2 methods can be reduced compared
to the numbers given by PP04 by restricting the range of line index included in the linear fit. 
{\it However, even the reduced calibration uncertainties 
would still account for 100\% of the observed scatter in the MZR even if M$_{\ast}$ were perfectly correlated
with oxygen abundance.}   
Taken at face value, this suggests that the relative intensities of the strong emission lines, which we have argued are modulated primarily
by ionization parameter and by the hardness of the UV radiation field, must be {\it more strongly} correlated with
M$_{\ast}$ than is the oxygen abundance.  We will return to the possible implications of this ``M$_{\ast}$-excitation'' relation in section \ref{sec:discussion} below.  

At $z \simeq 0$, the scatter in the correlation between M$_{\ast}$ and metallicity can be reduced significantly by including an
additional parameter that accounts indirectly for the cold gas content of the galaxies, most commonly using the SFR.  A parametrization of
this dependence of the form 
\begin{equation}
{\rm 12+log(O/H) \propto \mu_{\ast} \equiv log~M_{\ast} - \alpha~log(SFR/M_{\odot} yr^{-1})~,}
\label{eqn:fmr}
\end{equation}
where $\alpha$ is a constant that minimizes the scatter in metallicity at a given $\mu_{\ast}$,
was introduced by \cite{mannucci10} as
a convenient ``projection'' of what they called the ``fundamental metallicity relation'' (FMR). According to \citet{mannucci10}, the FMR
is a thin two-dimensional surface in the space defined by M$_{\ast}$, $Z$, and SFR, upon which all star-forming galaxies lie, independent of redshift
for $z \simlt 2.5$.  
To first order, the projection of the FMR parametrized by equation~\ref{eqn:fmr} accounts 
for the clearly observed trend (at $z \simeq 0$) that galaxies with higher SFR have lower gas-phase oxygen abundances at fixed M$_{\ast}$.
In the context of the FMR, high redshift galaxies, known
to have much higher gas fractions and SFRs than most local star-forming galaxies (e.g., \citealt{erb+06b,daddi10,tacconi10,tacconi13}), 
would also be expected to have 
correspondingly lower (O/H) at a given M$_{\ast}$. Thus, the value of $\alpha$ in equation~\ref{eqn:fmr} adjusts the actual M$_{\ast}$ 
to the mass $\mu_{\ast}$ expected for a galaxy with ${\rm log(SFR/ M_{\odot} yr^{-1})=0}$ and the observed metallicity. 
For this form of the projected FMR, \citet{mannucci10} found that $\alpha=0.32$ minimized the scatter in their $z \sim 0$ sample.  
An even stronger dependence on SFR of the M$_{\ast}$-Z relation has been suggested by \citet{andrews+martini13}, who found 
$\alpha = 0.66$ for a local galaxy sample whose oxygen abundances were determined using the ``direct'' method. 

However, the typical $z \sim 2.3$ galaxy in our sample has SFR $\simeq 25$ M$_{\odot}$ yr$^{-1}$ (Figure~\ref{fig:sfr_v_mstar}), 
well beyond the range of SFR well-sampled by
the $z \sim0$ data set used by \citet{mannucci10} and near the high SFR extreme of the $z \simeq 0$ sample
used by \citet{andrews+martini13}. Thus, the assumption that the correlation between SFR and gas-phase metallicity 
extends over the elevated SFR range of the high redshift samples would require a significant (and uncertain) extrapolation.    
We defer a detailed discussion of the relationships among M$_{\ast}$, SFR, and inferred oxygen abundance in the KBSS sample
to future work, for which we plan updates and improvements to the stellar population parameters (benefiting from additional and recently-obtained
ancillary data), extinction estimates, and object-by-object slit loss corrections, as well as increased sample size and dynamic range.  

\begin{figure}[htb]
\centerline{\includegraphics[width=9.0cm]{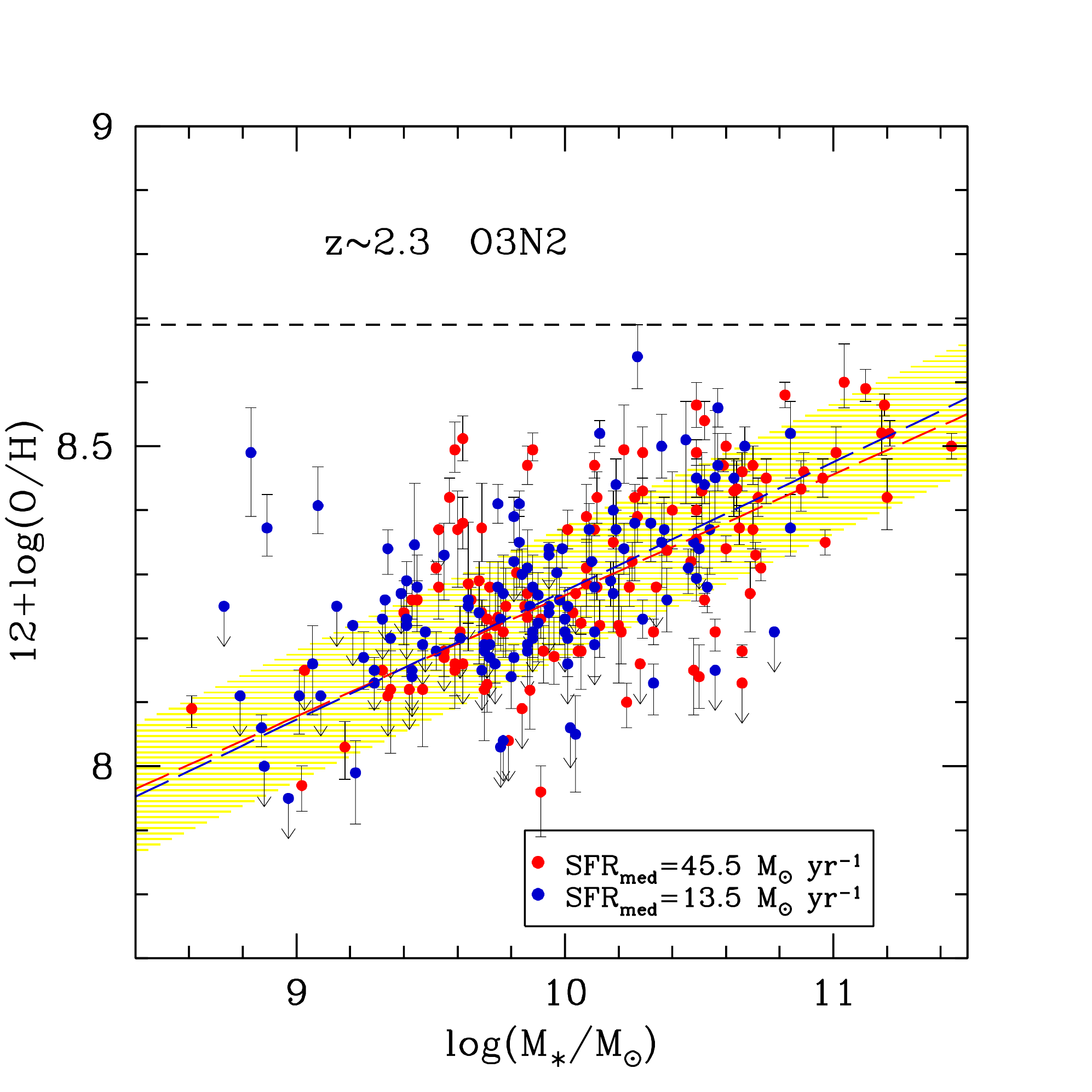}}
\caption{Same as the lefthand panel of Figure~\ref{fig:mass_met_bins}, but here individual points are color-coded
according to whether the galaxy has SFR above or below the sample median, 25.5 ${\rm M_{\odot} yr^{-1}}$; the median SFR of each sub-sample is indicated
in the legend. Each sub-sample includes 121 galaxies, and the best-fit linear MZR (dashed lines) are color-coded in the
same way. The shaded region shows the linear regression for the full sample (with parameters listed in equation
~\ref{eqn:linear2}), with width given by the inferred intrinsic scatter (at a given M$_{\ast}$) relative to the linear fit. The parameter estimates listed
in equations~\ref{eqn:sfr_low} and \ref{eqn:sfr_high} are statistically indistinguishable from one another, and compared to the full sample.
\label{fig:mzr_sfr}
}
\end{figure}

\subsection{The SFR Dependence of the MZR at $z \simeq 2.3$}

\label{sec:mzr_sfr}

Within the current KBSS-MOSFIRE sample, dependence of the MZR relation on SFR must be subtle, if present (see also \citealt{wuyts14,sanders14}).
Figure~\ref{fig:mzr_sfr} shows the O3N2-determined MZR (as in the lefthand panel of Figure~\ref{fig:mass_met_bins}), but where 
the sample has been color-coded according to whether the SFR lies above or below the median SFR of 25.5 ${\rm M_{\odot} yr^{-1}}$. The
two sub-samples, with median SFRs of 13.5 and 45.5 M${\rm_{\odot} yr^{-1}}$, were each fitted independently using the same functional
form (equation~\ref{eqn:linear}) that produced the parameters listed in equation~\ref{eqn:linear2}. 
Note that the overlap in M$_{\ast}$ for the low-SFR and high-SFR subsamples is
substantial, in spite of the well-known overall trend of higher SFR at higher M$_{\ast}$. The fits for the two SFR-based sub-samples
are remarkably similar in normalization, slope, and intrinsic scatter, despite the factor of $\simeq3.4$ difference in median SFR:
\begin{equation}
{\rm SFR_{low}:  Z_{10}=8.27\pm0.01;~~\gamma=0.20\pm0.03; }
\label{eqn:sfr_low}
\end{equation}
$$~~ \sigma_{\rm sc} = 0.11\pm0.01$$
\begin{equation}
{\rm SFR_{high}: Z_{10}=8.27\pm0.01;~~\gamma=0.19\pm0.02;} ~ .
\label{eqn:sfr_high}
\end{equation}
$$~~\sigma_{\rm sc}= 0.10\pm0.01 .$$
The two different best-fit linear relationships are over-plotted in Figure~\ref{fig:mzr_sfr}; clearly they are nearly identical to that
of the full sample (equation~\ref{eqn:linear2} and Figure~\ref{fig:mass_met_bins}, shaded region in Figure~\ref{fig:mzr_sfr}.) and
to one another. 

At first glance this result implies that metallicity and SFR are not strongly linked at $z \simeq 2.3$, at least among galaxies in 
the observed range of ${\rm M_{\ast}}$ and SFR in our current sample. More generally, as will be detailed elsewhere (Strom et al 2014,
in preparation), we have thus far not been able to identify a model in which 
inclusion of SFR as an additional parameter significantly reduces the scatter in the $z\sim2.3$ MZR.

\section{Summary and Discussion}

\label{sec:discussion}

We have presented near-IR spectroscopy for an initial sample of $251$ star-forming galaxies with $2.0 \le z \le 2.6$ observed in the 
15 fields 
of the Keck Baryonic Structure Survey.  All spectra were obtained 
using MOSFIRE, the recently-commissioned near-IR multi-object spectrometer on the Keck 1 10m telescope, during the first
18 months of its operation.  
In addition to the large size of the galaxy sample, the quality of the spectra of individual galaxies is much
higher, and the dynamic range within the sample much larger, than has been possible to achieve previously. 
In this paper, we have explored the quantitative use
of the strong nebular emission lines in the rest-frame optical spectra of high-redshift galaxies, re-examining their
utility for diagnosing the physical conditions in galaxies during the peak of their most active star-forming phase.  
The main conclusions are as follows:

1.  At $z\sim 2.3$, galaxies occupy an almost entirely distinct, but similarly tight, locus in the BPT diagram compared to the vast majority of star-forming
galaxies in the local universe (Figure~\ref{fig:bpt}). The shift in the observed locus can be qualitatively explained if essentially all high-redshift \ion{H}{2} regions are
characterized by both harder ionizing radiation fields and higher ionization parameters  
than apply for all but the most
extreme local galaxies.  

2.  Since all strong-line metallicity indicators and their calibrations are ``tuned'' to reproduce the tight sequence
in the BPT diagram for local galaxies, the shift of the high redshift locus means that the same calibrations among the strong-line
indicators cannot be used at high redshift without introducing systematics in the metallicity scale. Since
ground-based observations are confined to redshift intervals within which particular strong nebular lines fall in the
near-IR atmospheric windows, galaxy samples at different redshifts will necessarily depend on different subsets of the strong lines. 
It is entirely possible, perhaps even likely, that calibration issues could mimic global changes in metallicity or other physical conditions in HII regions with
redshift.  
As an example, we show that metallicities inferred from the N2 and O3N2 indices of $z \sim 2.3$ galaxies differ systematically from
each other, with an offset that averages $\simeq 0.13$ dex, in the sense that N2-inferred
metallicities are higher (Figure~\ref{fig:met_compare}).

3.  Using simple photoionization models (with minimal assumptions about the details of the ionizing sources) we find that
the observed locus of $z\sim2.3$ galaxies in the BPT diagram, as well as the behavior of the N2 and O3N2 indices with respect to one another,
can be reproduced remarkably well if the shape of the net
ionizing radiation field in high-redshift HII regions resembles a blackbody with effective temperature
${\rm T_{eff} = 50000 -60000}$ K and ionization parameter in the
range ${\rm -2.9 \simlt log \Gamma \simlt -1.8}$ (Figures~\ref{fig:bpt_mod} and \ref{fig:met_comp_mod}). 
In the context of the models, most of the variation along the principle axis of the BPT locus is produced by changes
in $\Gamma$, while the overall normalization of [OIII]/\Hb\ is modulated primarily by the effective temperature of the ionizing
radiation field. For the high inferred level of ionization, a galaxy's position in BPT space is nearly independent of the ionized gas metallicity over
the range $0.2 \le Z/Z_{\odot} \le 1.0$--
{\it so that any observed metallicity dependence of the strong-line ratios is more likely caused by correlations 
between the radiation field shape and intensity with the metallicity of the stars themselves.} 
In addition, we find the $z=2.3$ BPT locus is most easily reproduced if the (N/O) ratio in the ionized gas is close to
the solar ratio over the full observed range of (O/H). Such high N/O, as well as high $T_{\rm eff}$, may both be a consequence
of the effects of binaries and rapid rotation on massive main sequence stars. Such effects are predicted to be greatly enhanced at the
sub-solar metallicities that appear to be the rule at high redshift. 

4.  The KBSS-MOSFIRE sample contains a small number of AGN (Figures~\ref{fig:agn_spectra_uv} and \ref{fig:agn_spectra_mos}), most of which had been previously identified 
based on emission lines of high ionization species in their rest-frame UV spectra. The positions of AGN on the BPT diagram (Figure~\ref{fig:bpt})
appear distinct from the vast majority of objects which show no evidence in their rest-UV, rest-optical, or other multi-wavelength
measurements for energetically significant contamination by AGN.  The highest-excitation star-forming galaxies in the KBSS-MOSFIRE sample
exhibit a maximum ${\rm log~([OIII]\lambda 5008/\Hb) \simeq 0.9}$, consistent with the predictions of the photoionization models
with UV ionizing radiation field in the 1-4 Ryd range resembling a blackbody with $T_{\rm eff} = 55000-60000$ K and ${\rm log~\Gamma \simgt -2.0}$ (Figure~\ref{fig:bpt_mod}). 
This upper envelope appears to be the same for the most extreme star-forming galaxies in the local universe, where they are many orders of magnitude rarer.      

5.  We have drawn attention to the similarities between the most extreme galaxies (in terms of their position on the BPT 
diagram) in the $z \sim 0$ and $z \sim 2.3$ samples.  In particular, the so-called ``green pea'' galaxies at $z \simeq 0.2$ 
appear to have strong line ratios placing them directly on the $z\sim2.3$ BPT locus, while the ``extreme green peas''
are coincident with the highest excitation galaxies observed in the $z \simeq 2.3$ sample.  Comparison of the published 
samples of green peas having accurate direct ($T_{\rm e}$) metallicity measurements with the $z \sim 2.3$ galaxies is
also interesting. The strong-line metallicity indices of the GPs follow the same trend as observed among the $z\sim 2.3$ sample
and as predicted by our photoionization models (cf. Figure~\ref{fig:met_compare}, ~\ref{fig:met_comp_mod}, and ~\ref{fig:bpt_mod_peas}).     
The corresponding direct measures of oxygen abundances for the GPs and a small subset of the $z\sim 2.3$ sample suggest that 
among the commonly-applied strong-line calibrations, the least-biased with respect to the direct ($T_{\rm e}$) metallicity measurements is O3N2. 
The differences in N2- and O3N2-based oxygen abundances described above imply that N2 generally over-estimates metallicities at $z \sim 2.3$
(by $\sim 0.13$ dex for $0.2-1.0$ Z$_{\odot}$.). The systematic differences can be reduced considerably (but not entirely eliminated) 
by restricting the low-redshift calibration data sets to the range of line indices observed among the high redshift sample (Figure~\ref{fig:pp04_new}.)   
We propose a simple empirical relation for converting 12+log(O/H)$_{\rm N2}$ to the corresponding O3N2-based value appropriate at $z \sim 2.3$. 

6.  As shown previously using stacked spectra (\citealt{erb+06a}), there is a relationship between ${\rm M}_{\ast}$ and the strong-line indices (N2 or O3N2) 
in place at $z \sim 2.3$ qualitatively similar to those observed at $z \simeq 0$ (Figures~\ref{fig:mass_met} and \ref{fig:mass_met_bins}). 
If one converts the observed line indices into oxygen abundances
using the locally-established calibrations (i.e., under the assumption that the line indices can be used to measure metallicity),
the best-fit $z \sim 2.3$ MZR is somewhat {\it shallower} than some previous studies have suggested, 
${\rm 12+log(O/H) \propto 0.20~[log(M_{\ast}/M_{\odot})-10]}$ using either N2 or O3N2 indices. Both versions of the MZR are consistent with the same linear behavior over the range of ${\rm M_{\ast}}$ observed. 
(Figures~\ref{fig:mass_met} and \ref{fig:mass_met_bins}.)
%However, we show that the $z \simeq 0$ MZRs have slopes consistent with that of the $z \sim 2.3$ sample when evaluated over the same
%range of M$_{\ast}$ and using the same (O3N2) metallicity calibration. 
%Further, we find that the $z \sim 2.3$ MZR is also well-fit by the $z\sim 0$ MZRs with metallicity normalization shifted lower by
%${\rm \Delta log Z=0.34-0.38}$ dex, {\it independent of stellar mass}.     
As for the locus in the BPT diagram, the intrinsic scatter in the MZR 
(i.e., scatter of inferred metallicity at a given ${\rm M_{\ast}}$) is both small and remarkably
similar at $z \sim 2.3$ and $z \sim 0$ when the same metallicity calibration is applied to both: $\sigma_{\rm sc} \simeq 0.10$ dex.     
Over the well-covered range of M$_{\ast}$ observed in the current $z\sim 2.3$ KBSS-MOSFIRE sample (${\rm 9 \simlt log~(M_{\ast}/M_{\odot}) \simlt 11}$), 
there is no obvious M$_{\ast}$ dependence of the MZR scatter. 

7.  We have pointed out that the small values inferred for the {\it intrinsic} scatter in the $z \sim 2.3$ MZR ($\sigma \simeq 0.10$ dex) 
is uncomfortably small compared with the minimum uncertainties inherent in the calibrations of 
the strong-line metallicity methods, even when the latter are re-calibrated only over the range of line indices covered by the $z \sim 2.3$ observations. 
When taken together with the photoionization models showing that the observed line ratios at $z \sim 2.3$ 
are more strongly affected by ionizing radiation field intensity and shape than by ionized gas metallicity, it suggests that the more
fundamental correlation (of which the MZR is a by-product) is between M${\ast}$ and the properties of the massive stars that
determine the ionization/excitation state of the gas in their surroundings.    

8.  We investigated briefly whether there is evidence within our sample for a dependence (at fixed stellar mass) between
inferred oxygen abundance and SFR as observed in the local universe. 
We find nearly identical best-fit MZR relations (normalization, slope, and intrinsic scatter) 
for independent sub-samples (121 galaxies each) median SFRs differing by a factor of $\simeq 3.4$.  
{\it At present, 
over the range spanned by our current $z \sim 2.3$ sample, inferred oxygen abundances appear to be independent of SFR at a given stellar mass. }

%\begin{figure*}[htb]
%\centerline{\includegraphics[width=13cm]{fig26.pdf}}
%\caption{Same as Figure~\ref{fig:bpt}, where points have been color-coded according to their stellar mass. 
%Note the trend of increasing ${\rm M_{\ast}}$ from left to right along the BPT sequence; if one adopts the calibration of strong-line 
%indices at low redshift, this trend would be interpreted as a sequence in metallicity. As explained in the text, it is challenging to
%reconcile the small scatter in the inferred MZR with the calibration uncertainties inherent in the conversion from line index to metallicity. 
%}
%\label{fig:bpt_mstar}
%\end{figure*}

\subsection{Implications and Future Work}
\label{sec:implications}
Among the issues raised, we regard the following as unresolved and particularly interesting to pursue with
future work:

\subsubsection{\it Metallicity measurements at high redshift}
There is currently very limited evidence that {\it any} strong-line abundance estimates at high redshift 
reliably measure gas-phase metallicity, as is universally assumed. However, the prospects for improving the situation are good,
since instruments now exist (like MOSFIRE) capable of obtaining sufficiently sensitive spectra of high redshift
galaxies to measure weak lines such as
[OIII]$\lambda 4364$ whose strengths relative to strong lines provide direct information on physical conditions in the ionized gas.  We have shown that
it should be feasible to obtain such measurements at $z \sim 2.3$ for individual galaxies with metallicities 
as high as $Z \sim 0.5~Z_{\odot}$; it may be possible, using spectral stacks, to extend
the calibrations to higher metallicity (see, e.g., \citealt{andrews+martini13}).  Because of the remaining uncertainty associated with
converting strong-line ratios to oxygen abundance, until direct metallicity cross-checks have been 
completed, we suggest that galaxies should not be ``pre-screened'' for deep follow-up based on their strong-line-implied abundances
(see section~\ref{sec:analogs}). One should also obtain, wherever possible, measurements of the rest-UV
OIII] intercombination lines available from deep ground-based optical spectroscopy (section~\ref{sec:extreme}), which in some cases
may be more sensitive, albeit more dependent on nebular extinction corrections, than measurement of [OIII]$\lambda 4364$ in the  rest-frame optical.  

\subsubsection{\it The dominant ionizing sources in high-redshift \ion{H}{2} regions}
In order to temporarily avoid uncertainties associated with the details of the predicted ionizing spectra
of massive stars in high-redshift galaxies, we have modeled the net radiation field shape using
a single-temperature blackbody, which can be thought of as the effective temperature of whatever
stars are dominating the radiation field for photon energies between 1 and 4 Ryd. It appears that successful population
synthesis models used for future more detailed models of the ionized gas in $z \sim 2.3$ galaxies must be capable of
producing, in steady state, a net luminosity-weighted spectrum resembling a $\simeq 50,000-60,000$ K blackbody in the far-UV.   
This may have implications for the high-mass end of the stellar initial mass function (IMF), as well as for the
details of the models for the most massive stars. Satisfying the constraint that the stars must produce nebulae
with the observed properties may also have implications for the production and transfer of ionizing
photons from young galaxies at high redshift.    

\subsubsection{\it The slope and normalization of the MZR at $z \sim 2.3$}
According to our preferred form of the $z \sim 2.3$ MZR presented in section~\ref{sec:mass_met}, the average metallicity 
of the dominant star-forming galaxy population changes by only $\simlt 0.5$ dex over more than 2.5 orders of magnitude in 
M${\ast}$, ${\rm 8.6 \simlt log(M_{\ast}/M_{\odot}) \sim 11.4}$. 
This shallow dependence of (strong-line-inferred) metallicity on stellar mass (${\rm 12+log(O/H) \propto 0.20~logM_{\ast}}$)  
is comparable to what is observed {\it over the same range in M$_{\ast}$} at $z \simeq 0$. Because the strong nebular lines appear to be relatively
{\it insensitive} to the ionized gas metallicity, one should be cautious in treating inferred oxygen abundances
as direct indications of metallicity in the dominant gas reservoirs of galaxies. Similarly, one should also be cautious 
interpreting changes in strong-line ratios (e.g., as a function of position within galaxies, or scatter among galaxies of similar
stellar mass) as differences in {\it gas-phase} metallicity-- they are perhaps more likely to signal changes in the ionizing sources and their 
distribution, which may have a different origin. 

\subsubsection{\it Fundamental correlations between nebular line ratios and galaxy properties}
As discussed in section~\ref{sec:mzr_scatter}, the tightness of the relationship between inferred oxygen abundance and M$_{\ast}$ is difficult to
understand given the uncertainties in the calibration of the strong line indices onto direct ($T_{\rm e}$) based
oxygen abundance. The observed relationship is more easily understood if a) there is a relatively narrow range of 
radiation field effective temperature across all galaxy masses probed in the current sample and b) there is a monotonic
relationship between effective ionization parameter $\Gamma$ and M$_{\ast}$. 
%Figure~\ref{fig:bpt_mstar} shows the $z\sim2.3$
%BPT diagram, identical to Figure~\ref{fig:bpt} except that galaxies have been color-coded according to their stellar mass. 
Understanding {\it why} ionization level and excitation are so strongly linked to galaxy mass is a key goal 
for future work.   
 
\acknowledgments
This work has been supported in part by the US National Science Foundation through grants
AST-0908805 and AST-1313472 (CCS), as well as by an NSF Graduate Student Research Fellowship (ALS). 
The MOSFIRE instrument was made possible by grants to the W.M. Keck Observatory from the NSF ``Telescope System
Instrumentation Program'' (TSIP) and by a generous donation from Gordon and Betty Moore.  
We thank our colleagues on the MOSFIRE instrument team, particularly Marcia Brown, Khan Bui, John Cromer, 
Jason Fucik, Hector Rodriguez, Bob Weber,
and Jeff Zolkower at Caltech, Ted Aliado, George Brims, John
Canfield, Chris Johnson, Ken Magnone, and Jason Weiss at UCLA, Harland Epps at UCO/Lick Observatory, and Sean Adkins at WMKO. 
Special thanks to all of the WMKO staff who helped make MOSFIRE commissioning successful, especially
Marc Kassis, Allan Honey, Greg Wirth, Shui Kwok, Liz Chock, and Jim Lyke.  
We benefited significantly from an illuminating discussion on the subject of massive stars with Selma de Mink.
Constructive comments from the anonymous referee, which led to significant improvements in the content and presentation
of the results, are gratefully acknowledged.  
Finally, we wish to extend thanks to those of Hawaiian ancestry on whose sacred mountain we are privileged
to be guests.

%\bibliographystyle{apj}
%\bibliography{refs_comb}

%\clearpage

\begin{deluxetable*}{lccccccl}
\setcounter{table}{1}
\tabletypesize{\scriptsize}
\tablewidth{0pc}
\tablecaption{KBSS-MOSFIRE Galaxies with both [NII]/\Ha\ and [OIII]/\Hb\ Measurements}
\tablehead{
\colhead{Name} & \colhead{$z_{\rm neb}$} & \colhead{${\rm log~M_{\ast}}$} & 
%\colhead{${\rm SFR_{\Ha}}$} & 
\colhead{log([NII]/\Ha)} & \colhead{log([OIII]/\Hb)} &
 \colhead{${\rm 12+log(O/H)}$} & \colhead{${\rm 12+log(O/H)}$}   \\
\colhead{} & \colhead{} & \colhead{(${\rm M_{\odot}}$)} & 
\colhead{} & \colhead{} & \colhead{(N2)\tablenotemark{b}} & \colhead{(O3N2)\tablenotemark{c}} & \colhead{Notes}  
}
\startdata
Q0100-BX118  & $ 2.1093 $ & $~9.22 $ & $ -1.57_{-0.15}^{+0.24}$ & $0.74_{-0.05 }^{+0.05}$ & $8.01_{-0.09}^{+0.14}$ & $ 7.99_{-0.08}^{+0.05} $ & 1 \\
Q0100-BX163  & $ 2.2985 $ & $ 10.32 $ & $ -0.88_{-0.12}^{+0.16}$ & $0.22_{-0.09 }^{+0.11}$ & $8.40_{-0.07}^{+0.09}$ & $ 8.38_{-0.06}^{+0.05} $ & 1a \\
Q0100-BX172  & $ 2.3118 $ & $ \ldots $  & $ -0.79_{-0.05}^{+0.05}$ & $0.97_{-0.01 }^{+0.01}$ & $8.45_{-0.03}^{+0.03}$ & $ 8.17_{-0.02}^{+0.02} $ & A1,1 \\
Q0100-BX205  & $ 2.2912 $ & $ ~9.88 $ & $ -1.00_{-0.08}^{+0.10}$ & $0.66_{-0.03 }^{+0.03}$ & $8.33_{-0.05}^{+0.06}$ & $ 8.20_{-0.03}^{+0.03} $ &  \\
Q0100-BX210  & $ 2.2769 $ & $ 10.10 $ & $ -0.79_{-0.12}^{+0.16}$ & $0.49_{-0.05 }^{+0.05}$ & $8.45_{-0.07}^{+0.09}$ & $ 8.32_{-0.05}^{+0.04} $ & 1 \\
Q0100-BX224  & $ 2.1076 $ & $ ~9.40 $ & $ -0.90_{-0.11}^{+0.15}$ & $0.64_{-0.08 }^{+0.09}$ & $8.38_{-0.06}^{+0.08}$ & $ 8.24_{-0.05}^{+0.05} $ & 1a \\
Q0100-BX277  & $ 2.1061 $ & $ 10.18 $ & $ -0.98_{-0.13}^{+0.18}$ & $0.46_{-0.06 }^{+0.07}$ & $8.34_{-0.07}^{+0.10}$ & $ 8.27_{-0.06}^{+0.05} $ & 1 \\
Q0100-BX88  & $ 2.5241 $ & $ ~9.60 $ & $ -0.83_{-0.17}^{+0.27}$ & $0.31_{-0.08 }^{+0.09}$ & $8.43_{-0.09}^{+0.16}$ & $ 8.37_{-0.09}^{+0.06} $ & 1 \\
Q0100-BX90  & $ 2.2850 $ & $ ~9.92 $ & $ -0.96_{-0.10}^{+0.13}$ & $0.76_{-0.06 }^{+0.07}$ & $8.35_{-0.06}^{+0.08}$ & $ 8.18_{-0.05}^{+0.04} $ & 1 \\
Q0100-BX95  & $ 2.2097 $ & $ 10.27 $ & $ -0.75_{-0.07}^{+0.08}$ & $0.31_{-0.08 }^{+0.09}$ & $8.47_{-0.04}^{+0.05}$ & $ 8.39_{-0.04}^{+0.04} $ &  \\
Q0100-MD19  & $ 2.1078 $ & $ 10.27 $ & $ -0.45_{-0.09}^{+0.11}$ & $-0.16_{-0.10 }^{+0.14}$ & $8.64_{-0.05}^{+0.07}$ & $ 8.64_{-0.05}^{+0.05} $ & 1a \\
Q0100-RK17  & $ 2.1076 $ & $ 11.44 $ & $ -0.39_{-0.02}^{+0.02}$ & $0.34_{-0.06 }^{+0.07}$ & $8.68_{-0.01}^{+0.01}$ & $ 8.50_{-0.02}^{+0.02} $ &  \\
Q0100-RK21  & $ 2.0624 $ & $ 10.51 $ & $ -0.34_{-0.08}^{+0.10}$ & $0.60_{-0.12 }^{+0.17}$ & $8.70_{-0.05}^{+0.06}$ & $ 8.43_{-0.05}^{+0.06} $ &  \\
Q0105-BX132  & $ 2.2115 $ & $ 10.75 $ & $ -0.62_{-0.08}^{+0.10}$ & $0.25_{-0.06 }^{+0.06}$ & $8.55_{-0.05}^{+0.06}$ & $ 8.45_{-0.04}^{+0.03} $ & 1 \\
Q0105-BX147  & $ 2.3857 $ & $ ~9.41 $ & $ -0.77_{-0.10}^{+0.13}$ & $0.60_{-0.03 }^{+0.04}$ & $8.46_{-0.06}^{+0.08}$ & $ 8.29_{-0.04}^{+0.03} $ & 1 \\
Q0105-BX186  & $ 2.2003 $ & $ 10.57 $ & $ -0.44_{-0.06}^{+0.07}$ & $0.36_{-0.10 }^{+0.13}$ & $8.65_{-0.04}^{+0.04}$ & $ 8.47_{-0.04}^{+0.05} $ & 1a \\
Q0105-BX57  & $ 2.2589 $ & $ ~9.86 $ & $ -0.89_{-0.05}^{+0.05}$ & $0.55_{-0.06 }^{+0.07}$ & $8.40_{-0.03}^{+0.03}$ & $ 8.27_{-0.03}^{+0.03} $ & 1 \\
Q0105-BX58  & $ 2.5351 $ & $ \ldots $  & $ -0.16_{-0.07}^{+0.08}$ & $0.77_{-0.09 }^{+0.11}$ & $8.81_{-0.04}^{+0.05}$ & $ 8.43_{-0.04}^{+0.04} $ & A1,1 \\
Q0105-BX77  & $ 2.2930 $ & $ 10.01 $ & $ -0.91_{-0.13}^{+0.19}$ & $0.75_{-0.03 }^{+0.03}$ & $8.38_{-0.08}^{+0.11}$ & $ 8.20_{-0.06}^{+0.04} $ & 1a \\
Q0105-BX79  & $ 2.1229 $ & $ 10.52 $ & $ -0.40_{-0.05}^{+0.06}$ & $0.52_{-0.08 }^{+0.09}$ & $8.67_{-0.03}^{+0.03}$ & $ 8.44_{-0.03}^{+0.03} $ & 1a \\
Q0105-MD27  & $ 2.0623 $ & $ 10.36 $ & $ -0.29_{-0.09}^{+0.11}$ & $0.44_{-0.09 }^{+0.12}$ & $8.74_{-0.05}^{+0.06}$ & $ 8.50_{-0.05}^{+0.05} $ & 1a \\
Q0142-BX122  & $ 2.4177 $ & $ ~9.53 $ & $ -0.99_{-0.11}^{+0.15}$ & $0.41_{-0.02 }^{+0.02}$ & $8.33_{-0.06}^{+0.08}$ & $ 8.28_{-0.05}^{+0.04} $ & 1 \\
Q0142-BX169  & $ 2.2824 $ & $ 10.20 $ & $ -0.82_{-0.16}^{+0.24}$ & $0.79_{-0.13 }^{+0.18}$ & $8.43_{-0.09}^{+0.14}$ & $ 8.22_{-0.09}^{+0.08} $ & 1 \\
Q0142-BX188  & $ 2.0602 $ & $ ~9.84 $ & $ -0.72_{-0.08}^{+0.10}$ & $0.64_{-0.08 }^{+0.09}$ & $8.49_{-0.05}^{+0.06}$ & $ 8.30_{-0.04}^{+0.04} $ & 1a \\
Q0142-BX195  & $ 2.3804 $ & $ \ldots $  & $ -0.49_{-0.08}^{+0.09}$ & $0.94_{-0.11 }^{+0.15}$ & $8.62_{-0.04}^{+0.05}$ & $ 8.27_{-0.05}^{+0.06} $ & A1,1 \\
Q0142-BX196  & $ 2.4918 $ & $ ~9.55 $ & $ -0.68_{-0.14}^{+0.20}$ & $0.58_{-0.05 }^{+0.06}$ & $8.51_{-0.08}^{+0.11}$ & $ 8.33_{-0.07}^{+0.05} $ & 1 \\
Q0142-BX214  & $ 2.3865 $ & $ ~9.70 $ & $ -1.09_{-0.11}^{+0.15}$ & $0.59_{-0.02 }^{+0.02}$ & $8.28_{-0.06}^{+0.09}$ & $ 8.19_{-0.05}^{+0.04} $ & 1 \\
Q0142-BX242  & $ 2.2812 $ & $ ~9.68 $ & $ -0.87_{-0.08}^{+0.10}$ & $0.52_{-0.03 }^{+0.03}$ & $8.40_{-0.05}^{+0.06}$ & $ 8.29_{-0.03}^{+0.03} $ & 1 \\
Q0142-BX40  & $ 2.3924 $ & $ 10.84 $ & $ -0.36_{-0.15}^{+0.23}$ & $0.29_{-0.05 }^{+0.06}$ & $8.70_{-0.08}^{+0.13}$ & $ 8.52_{-0.07}^{+0.05} $ & 1a \\
Q0142-BX75  & $ 2.4175 $ & $ ~9.94 $ & $ -0.80_{-0.14}^{+0.20}$ & $0.70_{-0.02 }^{+0.02}$ & $8.44_{-0.08}^{+0.11}$ & $ 8.25_{-0.06}^{+0.04} $ & 1 \\
Q0142-BX81  & $ 2.5026 $ & $ ~9.45 $ & $ -0.90_{-0.10}^{+0.12}$ & $0.57_{-0.02 }^{+0.02}$ & $8.38_{-0.05}^{+0.07}$ & $ 8.26_{-0.04}^{+0.03} $ & 1 \\
Q0142-MD20  & $ 2.5007 $ & $ ~9.57 $ & $ -0.59_{-0.09}^{+0.11}$ & $0.39_{-0.03 }^{+0.03}$ & $8.56_{-0.05}^{+0.06}$ & $ 8.42_{-0.04}^{+0.03} $ & 1a \\
Q0207-BX150  & $ 2.1147 $ & $ 10.37 $&  $ -0.73_{-0.07}^{+0.08}$ & $0.39_{-0.10 }^{+0.13}$ & $8.49_{-0.04}^{+0.05}$ & $ 8.37_{-0.04}^{+0.05} $ & 1a \\
Q0207-BX155  & $ 2.1536 $ & $ ~8.83 $ & $ -0.65_{-0.17}^{+0.28}$ & $0.11_{-0.12 }^{+0.16}$ & $8.53_{-0.10}^{+0.16}$ & $ 8.49_{-0.10}^{+0.07} $ & 1 \\
Q0207-BX285  & $ 2.1504 $ & $ ~9.77 $ & $ -0.96_{-0.14}^{+0.21}$ & $0.48_{-0.09 }^{+0.11}$ & $8.35_{-0.08}^{+0.12}$ & $ 8.27_{-0.07}^{+0.06} $ & 1 \\
Q0207-BX37  & $ 2.0901 $ & $ ~9.81 $ & $ -0.68_{-0.08}^{+0.10}$ & $0.38_{-0.06 }^{+0.07}$ & $8.51_{-0.05}^{+0.06}$ & $ 8.39_{-0.04}^{+0.03} $ & 1 \\
Q0207-BX65  & $ 2.1920 $ & $ 10.19 $ & $ -0.72_{-0.11}^{+0.15}$ & $0.40_{-0.11 }^{+0.15}$ & $8.49_{-0.06}^{+0.08}$ & $ 8.37_{-0.06}^{+0.06} $ &  \\
Q0207-BX67  & $ 2.1954 $ & $ ~9.77 $ & $ -1.16_{-0.09}^{+0.11}$ & $0.46_{-0.07 }^{+0.09}$ & $8.24_{-0.05}^{+0.06}$ & $ 8.21_{-0.04}^{+0.04} $ & 1 \\
Q0207-BX74  & $ 2.1889 $ & $ ~9.02 $ & $ -1.46_{-0.09}^{+0.11}$ & $0.90_{-0.05 }^{+0.05}$ & $8.07_{-0.05}^{+0.06}$ & $ 7.97_{-0.04}^{+0.03} $ & 1 \\
Q0207-BX87  & $ 2.1924 $ & $ 10.04 $ & $ -1.28_{-0.17}^{+0.28}$ & $0.83_{-0.05 }^{+0.05}$ & $8.17_{-0.10}^{+0.16}$ & $ 8.05_{-0.09}^{+0.06} $ & 1 \\
Q0207-MD39  & $ 2.5252 $ & $ ~9.78 $ & $ -0.90_{-0.15}^{+0.24}$ & $0.58_{-0.07 }^{+0.08}$ & $8.39_{-0.09}^{+0.14}$ & $ 8.25_{-0.08}^{+0.06} $ & 1 \\
Q0449-BX128  & $ 2.4604 $ & $ 10.06 $ & $ -1.10_{-0.14}^{+0.20}$ & $0.63_{-0.02 }^{+0.02}$ & $8.27_{-0.08}^{+0.11}$ & $ 8.18_{-0.06}^{+0.04} $ & 1 \\
Q0449-BX40  & $ 2.4008 $ & $ 10.52 $ & $ -0.52_{-0.06}^{+0.06}$ & $0.09_{-0.05 }^{+0.06}$ & $8.60_{-0.03}^{+0.04}$ & $ 8.54_{-0.03}^{+0.03} $ & 1 \\
Q0449-BX68  & $ 2.4972 $ & $ ~9.75 $ & $ -0.68_{-0.11}^{+0.15}$ & $0.72_{-0.03 }^{+0.04}$ & $8.51_{-0.06}^{+0.09}$ & $ 8.28_{-0.05}^{+0.04} $ & 1 \\
Q0449-BX70  & $ 2.4775 $ & $ ~9.86 $ &$ -0.75_{-0.16}^{+0.25}$ & $0.56_{-0.07 }^{+0.08}$ & $8.47_{-0.09}^{+0.14}$ & $ 8.31_{-0.08}^{+0.06} $ & 1 \\
Q0449-BX84  & $ 2.2971 $ & $ ~9.76 $ & $ -0.83_{-0.11}^{+0.16}$ & $0.73_{-0.06 }^{+0.07}$ & $8.43_{-0.07}^{+0.09}$ & $ 8.23_{-0.05}^{+0.04} $ & 1 \\
Q0449-BX92  & $ 2.4021 $ & $ 10.54 $ & $ -0.47_{-0.14}^{+0.21}$ & $0.65_{-0.08 }^{+0.10}$ & $8.63_{-0.08}^{+0.12}$ & $ 8.37_{-0.07}^{+0.06} $ & 1 \\
Q0449-M10  & $ 2.3863 $ & $ 10.72 $ & $ -0.58_{-0.08}^{+0.09}$ & $0.39_{-0.05 }^{+0.05}$ & $8.57_{-0.04}^{+0.05}$ & $ 8.42_{-0.03}^{+0.03} $ & 1 \\
Q0821-BX101  & $ 2.4462 $ & $ 10.87 $ & $ -0.11_{-0.03}^{+0.03}$ & $0.79_{-0.09 }^{+0.11}$ & $8.84_{-0.02}^{+0.02}$ & $ 8.44_{-0.03}^{+0.04} $ & A2 \\
Q0821-BX102  & $ 2.4151 $ & $ ~9.91 $ & $ -1.62_{-0.14}^{+0.21}$ & $0.79_{-0.01 }^{+0.01}$ & $7.97_{-0.08}^{+0.12}$ & $ 7.96_{-0.07}^{+0.04} $ & 1 \\
Q0821-BX207  & $ 2.4133 $ & $ ~9.59 $ & $ -1.20_{-0.11}^{+0.15}$ & $0.58_{-0.01 }^{+0.01}$ & $8.22_{-0.06}^{+0.09}$ & $ 8.16_{-0.05}^{+0.04} $ & 1 \\
Q0821-BX45  & $ 2.1800 $ & $ 10.66 $ & $ -1.08_{-0.03}^{+0.03}$ & $0.64_{-0.02 }^{+0.02}$ & $8.28_{-0.02}^{+0.02}$ & $ 8.18_{-0.01}^{+0.01} $ & 1a \\
Q0821-BX47  & $ 2.4612 $ & $ ~9.25 $ & $ -0.89_{-0.10}^{+0.14}$ & $0.85_{-0.03 }^{+0.03}$ & $8.39_{-0.06}^{+0.08}$ & $ 8.17_{-0.05}^{+0.04} $ & 1 \\
Q0821-BX72  & $ 2.3511 $ & $ 11.20 $ & $ -0.53_{-0.09}^{+0.12}$ & $0.44_{-0.12 }^{+0.16}$ & $8.60_{-0.05}^{+0.07}$ & $ 8.42_{-0.05}^{+0.06} $ &  \\
Q0821-BX77  & $ 2.2942 $ & $ ~9.61 $ & $ -0.97_{-0.12}^{+0.18}$ & $0.65_{-0.03 }^{+0.03}$ & $8.35_{-0.07}^{+0.10}$ & $ 8.21_{-0.06}^{+0.04} $ & 1 \\
Q0821-BX80  & $ 2.4448 $ & $ 10.25 $ & $ -0.77_{-0.07}^{+0.09}$ & $0.53_{-0.04 }^{+0.04}$ & $8.46_{-0.04}^{+0.05}$ & $ 8.32_{-0.03}^{+0.03} $ & 1a \\
Q0821-D10  & $ 2.5178 $ & $ ~9.98 $ & $ -0.93_{-0.15}^{+0.22}$ & $0.53_{-0.03 }^{+0.04}$ & $8.37_{-0.08}^{+0.13}$ & $ 8.26_{-0.07}^{+0.05} $ & 1 \\
Q0821-D8  & $ 2.5675 $ & $ \ldots $  & $ -0.36_{-0.08}^{+0.09}$ & $0.93_{-0.04 }^{+0.04}$ & $8.69_{-0.04}^{+0.05}$ & $ 8.32_{-0.03}^{+0.03} $ & A1,1 \\
Q0821-MD38  & $ 2.0918 $ & $ 10.49 $ & $ -0.56_{-0.03}^{+0.03}$ & $0.20_{-0.06 }^{+0.07}$ & $8.58_{-0.02}^{+0.02}$ & $ 8.49_{-0.02}^{+0.02} $ &  \\
Q0821-RK27  & $ 2.4483 $ & $ 10.08 $ & $ -0.72_{-0.14}^{+0.20}$ & $0.34_{-0.08 }^{+0.10}$ & $8.49_{-0.08}^{+0.12}$ & $ 8.39_{-0.07}^{+0.05} $ &  \\
Q0821-RK29  & $ 2.4681 $ & $ 10.71 $ & $ -0.78_{-0.06}^{+0.07}$ & $0.46_{-0.02 }^{+0.02}$ & $8.46_{-0.04}^{+0.04}$ & $ 8.33_{-0.02}^{+0.02} $ &  
\enddata
%\tablenotetext{a}{Error bars are 1$\sigma$ based on measurement uncertainties only.}
%\tablenotetext{b}{Oxygen abundance assuming the ``N2'' calibration of PP04.}
%\tablenotetext{c}{Oxygen abundance assuming the ``O3N2' calibration of PP04.}
%\tablenotetext{$\ast$}{Object is identified as an AGN on the basis of both its rest-UV and rest-optical spectra.}
\label{tab:n2ha_and_o3n2}
\end{deluxetable*}

\begin{deluxetable*}{lccccccl}
\setcounter{table}{1}
\tabletypesize{\scriptsize}
\tablewidth{0pc}
\tablecaption{KBSS-MOSFIRE Galaxies with both [NII]/\Ha\ and [OIII]/\Hb\ Measurements ({\it cont.})}
\tablehead{
\colhead{Name} & \colhead{$z_{\rm neb}$} & \colhead{${\rm log~M_{\ast}}$} & 
\colhead{log([NII]/\Ha)} & \colhead{log([OIII]/\Hb)} &
 \colhead{${\rm 12+log(O/H)}$} & \colhead{${\rm 12+log(O/H)}$}   \\
\colhead{} & \colhead{} & \colhead{(${\rm M_{\odot}}$)} & 
\colhead{} & \colhead{} & \colhead{(N2)\tablenotemark{b}} & \colhead{(O3N2)\tablenotemark{c}} & \colhead{Notes} 
}
\startdata
Q1009-BX146  & $ 2.2681 $ & $ 10.29 $ & $ -0.69_{-0.04}^{+0.04}$ & $0.25_{-0.05 }^{+0.05}$ & $8.51_{-0.02}^{+0.02}$ & $ 8.43_{-0.02}^{+0.02} $ & 1 \\
Q1009-BX215  & $ 2.5056 $ & $ 10.26 $ & $ -0.69_{-0.07}^{+0.08}$ & $0.28_{-0.03 }^{+0.04}$ & $8.51_{-0.04}^{+0.04}$ & $ 8.42_{-0.03}^{+0.02} $ & 1,6 \\
Q1009-BX218  & $ 2.1090 $ & $ 10.38 $ & $ -0.96_{-0.09}^{+0.12}$ & $0.51_{-0.09 }^{+0.11}$ & $8.35_{-0.05}^{+0.07}$ & $ 8.26_{-0.05}^{+0.05} $ & 1 \\
Q1009-MD36  & $ 2.5048 $ & $ 10.70 $ & $ -0.64_{-0.06}^{+0.08}$ & $0.18_{-0.05 }^{+0.05}$ & $8.54_{-0.04}^{+0.04}$ & $ 8.47_{-0.03}^{+0.03} $ & 1 \\
Q1009-MD39  & $ 2.1425 $ & $ 11.04 $ & $ -0.41_{-0.05}^{+0.06}$ & $0.01_{-0.12 }^{+0.17}$ & $8.67_{-0.03}^{+0.03}$ & $ 8.60_{-0.04}^{+0.06} $ & 1a \\
Q1217-BX102  & $ 2.1936 $ & $ ~9.75 $ & $ -0.57_{-0.07}^{+0.09}$ & $0.42_{-0.05 }^{+0.06}$ & $8.57_{-0.04}^{+0.05}$ & $ 8.41_{-0.03}^{+0.03} $ & 1 \\
Q1217-BX164  & $ 2.3310 $ & $ ~9.72 $ & $ -0.77_{-0.10}^{+0.12}$ & $0.63_{-0.07 }^{+0.09}$ & $8.46_{-0.05}^{+0.07}$ & $ 8.28_{-0.05}^{+0.04} $ & 1 \\
Q1217-BX193  & $ 2.2164 $ & $ ~9.86 $ & $ -1.07_{-0.11}^{+0.15}$ & $0.61_{-0.03 }^{+0.03}$ & $8.29_{-0.06}^{+0.08}$ & $ 8.19_{-0.05}^{+0.04} $ & 1 \\
Q1217-BX95  & $ 2.4244 $ & $ 10.23 $ & $ -1.22_{-0.09}^{+0.11}$ & $0.76_{-0.01 }^{+0.01}$ & $8.21_{-0.05}^{+0.06}$ & $ 8.10_{-0.04}^{+0.03} $ & 1 \\
Q1217-MD13  & $ 2.3826 $ & $ 10.48 $ & $ -1.20_{-0.14}^{+0.20}$ & $0.60_{-0.08 }^{+0.09}$ & $8.22_{-0.08}^{+0.11}$ & $ 8.15_{-0.07}^{+0.05} $ & 1 \\
Q1217-MD15  & $ 2.1272 $ & $ 10.22 $ & $ -0.78_{-0.08}^{+0.10}$ & $0.45_{-0.09 }^{+0.11}$ & $8.46_{-0.05}^{+0.06}$ & $ 8.34_{-0.04}^{+0.04} $ & 1 \\
Q1442-BX108  & $ 2.4280 $ & $ ~9.69 $ & $ -1.04_{-0.07}^{+0.08}$ & $0.48_{-0.01 }^{+0.01}$ & $8.31_{-0.04}^{+0.04}$ & $ 8.24_{-0.02}^{+0.02} $ & 1 \\
Q1442-BX116  & $ 2.0463 $ & $ ~9.74 $ & $ -1.05_{-0.17}^{+0.27}$ & $0.55_{-0.05 }^{+0.05}$ & $8.30_{-0.09}^{+0.16}$ & $ 8.22_{-0.09}^{+0.06} $ & 1 \\
Q1442-BX133  & $ 2.1053 $ & $ ~9.68 $ & $ -0.98_{-0.15}^{+0.24}$ & $0.57_{-0.09 }^{+0.11}$ & $8.34_{-0.09}^{+0.14}$ & $ 8.24_{-0.08}^{+0.06} $ & 1 \\
Q1442-BX160  & $ 2.4418 $ & $ ~9.55 $ & $ -1.10_{-0.07}^{+0.08}$ & $0.66_{-0.01 }^{+0.01}$ & $8.27_{-0.04}^{+0.05}$ & $ 8.17_{-0.03}^{+0.02} $ & 1 \\
Q1442-BX172  & $ 2.4496 $ & $ 10.08 $ & $ -0.95_{-0.13}^{+0.19}$ & $0.37_{-0.04 }^{+0.04}$ & $8.36_{-0.08}^{+0.11}$ & $ 8.31_{-0.06}^{+0.04} $ & 1 \\
Q1442-BX235  & $ 2.4443 $ & $ 10.60 $ & $ -0.70_{-0.05}^{+0.05}$ & $0.53_{-0.02 }^{+0.02}$ & $8.50_{-0.03}^{+0.03}$ & $ 8.34_{-0.02}^{+0.02} $ & 1 \\
Q1442-BX270  & $ 2.3578 $ & $ ~9.52 $ & $ -0.98_{-0.08}^{+0.10}$ & $0.74_{-0.01 }^{+0.01}$ & $8.34_{-0.05}^{+0.06}$ & $ 8.18_{-0.03}^{+0.03} $ &  \\
Q1442-BX277  & $ 2.3125 $ & $ 10.01 $ & $ -0.87_{-0.09}^{+0.11}$ & $0.63_{-0.03 }^{+0.03}$ & $8.41_{-0.05}^{+0.06}$ & $ 8.25_{-0.04}^{+0.03} $ & 1 \\
Q1442-BX350  & $ 2.4422 $ & $ 10.11 $ & $ -0.90_{-0.14}^{+0.20}$ & $0.72_{-0.04 }^{+0.04}$ & $8.39_{-0.08}^{+0.11}$ & $ 8.21_{-0.07}^{+0.05} $ & 1 \\
Q1442-BX351  & $ 2.4518 $ & $ 10.13 $ & $ -0.97_{-0.11}^{+0.14}$ & $0.62_{-0.03 }^{+0.03}$ & $8.34_{-0.06}^{+0.08}$ & $ 8.22_{-0.05}^{+0.04} $ & 1 \\
Q1442-BX69  & $ 2.0888 $ & $ 10.53 $ & $ -0.73_{-0.09}^{+0.12}$ & $0.69_{-0.03 }^{+0.03}$ & $8.48_{-0.05}^{+0.07}$ & $ 8.28_{-0.04}^{+0.03} $ & 1 \\
Q1442-BX69b  & $ 2.1489 $ & $ 10.13 $ & $ -0.47_{-0.04}^{+0.04}$ & $0.19_{-0.03 }^{+0.03}$ & $8.63_{-0.02}^{+0.02}$ & $ 8.52_{-0.02}^{+0.02} $ &  \\
Q1442-C18  & $ 2.3166 $ & $ 10.40 $ & $ -0.83_{-0.10}^{+0.13}$ & $0.19_{-0.11 }^{+0.16}$ & $8.43_{-0.06}^{+0.07}$ & $ 8.40_{-0.06}^{+0.06} $ & 1 \\
Q1442-MD13  & $ 2.4528 $ & $ 10.56 $ & $ -0.90_{-0.07}^{+0.09}$ & $0.72_{-0.01 }^{+0.01}$ & $8.39_{-0.04}^{+0.05}$ & $ 8.21_{-0.03}^{+0.02} $ & 1a \\
Q1442-MD53  & $ 2.2926 $ & $ 10.96 $ & $ -0.41_{-0.04}^{+0.05}$ & $0.47_{-0.08 }^{+0.10}$ & $8.67_{-0.02}^{+0.03}$ & $ 8.45_{-0.03}^{+0.03} $ & 1 \\
Q1442-MD57  & $ 2.4440 $ & $ ~9.86 $ & $ -0.38_{-0.06}^{+0.07}$ & $0.44_{-0.05 }^{+0.06}$ & $8.68_{-0.03}^{+0.04}$ & $ 8.47_{-0.03}^{+0.03} $ &  \\
Q1549-BX101  & $ 2.3806 $ & $ \ldots $ & $ -0.38_{-0.03}^{+0.03}$ & $0.80_{-0.02 }^{+0.02}$ & $8.68_{-0.01}^{+0.02}$ & $ 8.35_{-0.01}^{+0.01} $ & A2 \\
Q1549-BX127  & $ 2.5336 $ & $ ~9.35 $ & $ -1.16_{-0.18}^{+0.30}$ & $0.74_{-0.05 }^{+0.05}$ & $8.24_{-0.10}^{+0.17}$ & $ 8.12_{-0.10}^{+0.06} $ & 1 \\
Q1549-BX180  & $ 2.3870 $ & $ ~9.32 $ & $ -1.18_{-0.08}^{+0.09}$ & $0.61_{-0.01 }^{+0.01}$ & $8.23_{-0.04}^{+0.05}$ & $ 8.15_{-0.03}^{+0.03} $ & 1 \\
Q1549-BX197  & $ 2.4351 $ & $ ~9.62 $ & $ -0.75_{-0.08}^{+0.10}$ & $0.35_{-0.07 }^{+0.08}$ & $8.48_{-0.05}^{+0.06}$ & $ 8.38_{-0.04}^{+0.04} $ & 1 \\
Q1549-BX207  & $ 2.3802 $ & $ ~9.65 $ & $ -0.93_{-0.05}^{+0.06}$ & $0.54_{-0.02 }^{+0.02}$ & $8.37_{-0.03}^{+0.03}$ & $ 8.26_{-0.02}^{+0.02} $ & 1 \\
Q1549-BX221  & $ 2.3407 $ & $ ~9.43 $ & $ -0.80_{-0.07}^{+0.08}$ & $0.66_{-0.06 }^{+0.07}$ & $8.44_{-0.04}^{+0.05}$ & $ 8.26_{-0.03}^{+0.03} $ & 1a \\
Q1549-BX223  & $ 2.3492 $ & $ ~9.52 $ & $ -0.79_{-0.03}^{+0.03}$ & $0.52_{-0.01 }^{+0.01}$ & $8.45_{-0.02}^{+0.02}$ & $ 8.31_{-0.01}^{+0.01} $ & 1 \\
Q1549-BX227  & $ 2.0573 $ & $ 10.69 $ & $ -0.92_{-0.11}^{+0.16}$ & $0.52_{-0.07 }^{+0.09}$ & $8.37_{-0.07}^{+0.09}$ & $ 8.27_{-0.06}^{+0.05} $ & 1 \\
Q1549-BX240  & $ 2.0412 $ & $ 10.63 $ & $ -0.55_{-0.05}^{+0.05}$ & $0.33_{-0.07 }^{+0.08}$ & $8.59_{-0.03}^{+0.03}$ & $ 8.45_{-0.03}^{+0.03} $ & 1a \\
Q1549-BX45  & $ 2.0645 $ & $ ~9.83 $ & $ -0.66_{-0.15}^{+0.22}$ & $0.54_{-0.07 }^{+0.08}$ & $8.52_{-0.08}^{+0.13}$ & $ 8.35_{-0.07}^{+0.05} $ & 1 \\
Q1549-BX51  & $ 2.2895 $ & $ ~9.72 $ & $ -1.14_{-0.13}^{+0.18}$ & $0.54_{-0.04 }^{+0.05}$ & $8.25_{-0.07}^{+0.10}$ & $ 8.19_{-0.06}^{+0.04} $ & 1,6 \\
Q1603-BX101  & $ 2.3202 $ & $ 10.29 $ & $ -0.39_{-0.07}^{+0.09}$ & $0.37_{-0.09 }^{+0.11}$ & $8.68_{-0.04}^{+0.05}$ & $ 8.49_{-0.04}^{+0.04} $ &  \\
Q1603-BX106  & $ 2.2743 $ & $ ~9.34 $ & $ -0.81_{-0.09}^{+0.11}$ & $0.42_{-0.04 }^{+0.05}$ & $8.44_{-0.05}^{+0.06}$ & $ 8.34_{-0.04}^{+0.03} $ &  \\
Q1603-BX173  & $ 2.5490 $ & $ ~9.01 $ & $ -1.07_{-0.13}^{+0.18}$ & $0.86_{-0.04 }^{+0.04}$ & $8.29_{-0.07}^{+0.10}$ & $ 8.11_{-0.06}^{+0.04} $ & 1 \\
Q1603-BX190  & $ 2.0216 $ & $ 10.01 $ & $ -0.81_{-0.06}^{+0.07}$ & $0.33_{-0.05 }^{+0.05}$ & $8.44_{-0.04}^{+0.04}$ & $ 8.37_{-0.03}^{+0.03} $ & 1 \\
Q1603-BX191  & $ 2.5446 $ & $ 10.42 $ & $ -0.35_{-0.04}^{+0.04}$ & $1.04_{-0.05 }^{+0.05}$ & $8.70_{-0.02}^{+0.02}$ & $ 8.28_{-0.02}^{+0.02} $ & A2 \\
Q1603-BX255  & $ 2.4349 $ & $ 10.21 $ & $ -1.09_{-0.11}^{+0.14}$ & $0.53_{-0.03 }^{+0.03}$ & $8.28_{-0.06}^{+0.08}$ & $ 8.21_{-0.05}^{+0.04} $ & 1 \\
Q1603-BX277  & $ 2.5499 $ & $ ~9.91 $ & $ -0.92_{-0.07}^{+0.08}$ & $0.66_{-0.02 }^{+0.02}$ & $8.38_{-0.04}^{+0.04}$ & $ 8.23_{-0.03}^{+0.02} $ & 1 \\
Q1603-BX294  & $ 2.4510 $ & $ 10.03 $ & $ -0.96_{-0.05}^{+0.05}$ & $0.57_{-0.01 }^{+0.01}$ & $8.36_{-0.03}^{+0.03}$ & $ 8.24_{-0.02}^{+0.01} $ & 1 \\
Q1603-BX379  & $ 2.1768 $ & $ 10.11 $ & $ -0.61_{-0.03}^{+0.04}$ & $0.21_{-0.04 }^{+0.04}$ & $8.55_{-0.02}^{+0.02}$ & $ 8.47_{-0.02}^{+0.02} $ & 1 \\
Q1603-MD26  & $ 2.5511 $ & $ ~9.85 $ & $ -0.99_{-0.09}^{+0.12}$ & $0.50_{-0.02 }^{+0.02}$ & $8.34_{-0.05}^{+0.07}$ & $ 8.25_{-0.04}^{+0.03} $ & 1 \\
Q1603-MD42  & $ 2.3806 $ & $ ~9.99 $ & $ -0.70_{-0.06}^{+0.07}$ & $0.51_{-0.05 }^{+0.06}$ & $8.50_{-0.03}^{+0.04}$ & $ 8.34_{-0.03}^{+0.03} $ & 1 \\
Q1603-MD45  & $ 2.3858 $ & $ 10.09 $ & $ -0.82_{-0.12}^{+0.16}$ & $0.30_{-0.05 }^{+0.06}$ & $8.43_{-0.07}^{+0.09}$ & $ 8.37_{-0.05}^{+0.04} $ & 1 \\
Q1603-MD50  & $ 2.4326 $ & $ 10.70 $ & $ -0.63_{-0.05}^{+0.06}$ & $0.49_{-0.08 }^{+0.10}$ & $8.54_{-0.03}^{+0.04}$ & $ 8.37_{-0.03}^{+0.04} $ & 1 \\
Q1603-MD85  & $ 2.4507 $ & $ 10.48 $ & $ -0.65_{-0.07}^{+0.08}$ & $0.53_{-0.04 }^{+0.04}$ & $8.53_{-0.04}^{+0.04}$ & $ 8.35_{-0.03}^{+0.02} $ & 1 \\
Q1623-BX366  & $ 2.4204 $ & $ 10.12 $ & $ -0.66_{-0.08}^{+0.10}$ & $0.31_{-0.06 }^{+0.08}$ & $8.52_{-0.05}^{+0.06}$ & $ 8.42_{-0.04}^{+0.04} $ & 1,4,6 \\
Q1623-BX428  & $ 2.0542 $ & $ 10.45 $ & $ -0.51_{-0.17}^{+0.29}$ & $0.17_{-0.05 }^{+0.06}$ & $8.61_{-0.10}^{+0.17}$ & $ 8.51_{-0.10}^{+0.06} $ & 1,2,4,6 \\
Q1623-BX429  & $ 2.0159 $ & $ ~9.94 $ & $ -0.94_{-0.04}^{+0.04}$ & $0.31_{-0.05 }^{+0.06}$ & $8.36_{-0.02}^{+0.02}$ & $ 8.33_{-0.02}^{+0.02} $ & 1,4,6 \\
Q1623-BX447  & $ 2.1480 $ & $ 10.67 $ & $ -0.78_{-0.09}^{+0.11}$ & $-0.06_{-0.06 }^{+0.06}$ & $8.45_{-0.05}^{+0.06}$ & $ 8.50_{-0.04}^{+0.03} $ & 1,2,4,6,7 \\
Q1623-BX449  & $ 2.4180 $ & $ 10.26 $ & $ -0.79_{-0.17}^{+0.27}$ & $0.30_{-0.08 }^{+0.10}$ & $8.45_{-0.09}^{+0.16}$ & $ 8.38_{-0.09}^{+0.06} $ & 2,4,6 \\
Q1623-BX452  & $ 2.0584 $ & $ 10.57 $ & $ -0.47_{-0.06}^{+0.06}$ & $0.07_{-0.05 }^{+0.06}$ & $8.63_{-0.03}^{+0.04}$ & $ 8.56_{-0.03}^{+0.03} $ & 1,6 \\
Q1623-BX453  & $ 2.1820 $ & $ 10.59 $ & $ -0.48_{-0.01}^{+0.02}$ & $0.32_{-0.02 }^{+0.02}$ & $8.63_{-0.01}^{+0.01}$ & $ 8.47_{-0.01}^{+0.01} $ & 1,4,5,6,10 \\
Q1623-BX472  & $ 2.1141 $ & $ 10.46 $ & $ -0.93_{-0.09}^{+0.12}$ & $0.39_{-0.04 }^{+0.04}$ & $8.37_{-0.05}^{+0.07}$ & $ 8.31_{-0.04}^{+0.03} $ & 1,4,6 
\enddata
\label{tab:n2ha_and_o3n2cont}
\end{deluxetable*}

\begin{deluxetable*}{lccccccl}
\setcounter{table}{1}
\tabletypesize{\scriptsize}
\tablewidth{0pc}
\tablecaption{KBSS-MOSFIRE Galaxies with both [NII]/\Ha\ and [OIII]/\Hb\ Measurements ({\it cont.})}
\tablehead{
\colhead{Name} & \colhead{$z_{\rm neb}$} & \colhead{${\rm log~M_{\ast}}$} & 
\colhead{log([NII]/\Ha)} & \colhead{log([OIII]/\Hb)} &
 \colhead{${\rm 12+log(O/H)}$} & \colhead{${\rm 12+log(O/H)}$}   \\
\colhead{} & \colhead{} & \colhead{(${\rm M_{\odot}}$)} & 
\colhead{} & \colhead{} & \colhead{(N2)\tablenotemark{b}} & \colhead{(O3N2)\tablenotemark{c}} & \colhead{Notes} 
}
\startdata
Q1700-BX490  & $ 2.3958 $ & $ 10.05 $ & $ -0.98_{-0.03}^{+0.03}$ & $0.73_{-0.01 }^{+0.01}$ & $8.34_{-0.02}^{+0.02}$ & $ 8.18_{-0.01}^{+0.01} $ & 1,3,4,5,6 \\
Q1700-BX505  & $ 2.3083 $ & $ 10.66 $ & $ -0.47_{-0.05}^{+0.06}$ & $0.37_{-0.06 }^{+0.07}$ & $8.63_{-0.03}^{+0.03}$ & $ 8.46_{-0.03}^{+0.03} $ & 1,3,4,6 \\
Q1700-BX563  & $ 2.2910 $ & $ 10.33 $ & $ -0.94_{-0.05}^{+0.05}$ & $0.67_{-0.02 }^{+0.02}$ & $8.37_{-0.03}^{+0.03}$ & $ 8.21_{-0.02}^{+0.02} $ & 1,3 \\
Q1700-BX585  & $ 2.3066 $ & $ ~9.06 $ & $ -1.20_{-0.15}^{+0.24}$ & $0.58_{-0.07 }^{+0.08}$ & $8.21_{-0.09}^{+0.13}$ & $ 8.16_{-0.08}^{+0.06} $ & 1,3 \\
Q1700-BX625  & $ 2.0752 $ & $ ~9.80 $ & $ -1.18_{-0.11}^{+0.15}$ & $0.66_{-0.03 }^{+0.04}$ & $8.23_{-0.06}^{+0.08}$ & $ 8.14_{-0.05}^{+0.04} $ & 1,3 \\
Q1700-BX649  & $ 2.2946 $ & $ 10.18 $ & $ -0.78_{-0.05}^{+0.06}$ & $0.26_{-0.06 }^{+0.08}$ & $8.46_{-0.03}^{+0.03}$ & $ 8.40_{-0.03}^{+0.03} $ & 1 \\
Q1700-BX691  & $ 2.1891 $ & $ 10.89 $ & $ -0.71_{-0.04}^{+0.05}$ & $0.12_{-0.07 }^{+0.09}$ & $8.49_{-0.02}^{+0.03}$ & $ 8.46_{-0.03}^{+0.03} $ & 1,2,3,4 \\
Q1700-BX708  & $ 2.3992 $ & $ ~9.70 $ & $ -1.15_{-0.15}^{+0.23}$ & $0.75_{-0.04 }^{+0.04}$ & $8.24_{-0.09}^{+0.13}$ & $ 8.12_{-0.08}^{+0.05} $ & 1,4,6 \\
Q1700-BX710  & $ 2.2946 $ & $ 10.52 $ & $ -0.90_{-0.03}^{+0.03}$ & $0.59_{-0.02 }^{+0.02}$ & $8.39_{-0.02}^{+0.02}$ & $ 8.26_{-0.01}^{+0.01} $ & 1,5 \\
Q1700-BX711  & $ 2.2947 $ & $ ~8.61 $ & $ -1.21_{-0.07}^{+0.08}$ & $0.80_{-0.02 }^{+0.02}$ & $8.21_{-0.04}^{+0.05}$ & $ 8.09_{-0.03}^{+0.02} $ & 1 \\
Q1700-BX713  & $ 2.1381 $ & $ ~9.64 $ & $ -1.03_{-0.14}^{+0.20}$ & $0.48_{-0.05 }^{+0.06}$ & $8.32_{-0.08}^{+0.11}$ & $ 8.25_{-0.07}^{+0.05} $ &  \\
Q1700-BX752  & $ 2.4001 $ & $ 10.60 $ & $ -0.57_{-0.05}^{+0.05}$ & $0.16_{-0.04 }^{+0.05}$ & $8.57_{-0.03}^{+0.03}$ & $ 8.50_{-0.02}^{+0.02} $ & 1a \\
Q1700-BX763  & $ 2.2919 $ & $ 10.11 $ & $ -0.85_{-0.06}^{+0.07}$ & $0.57_{-0.03 }^{+0.03}$ & $8.41_{-0.04}^{+0.04}$ & $ 8.28_{-0.03}^{+0.02} $ & 1,5,6 \\
Q1700-BX879  & $ 2.3065 $ & $ ~9.88 $ & $ -0.71_{-0.06}^{+0.06}$ & $0.06_{-0.17 }^{+0.29}$ & $8.49_{-0.03}^{+0.04}$ & $ 8.48_{-0.06}^{+0.10} $ & 1,3 \\
Q1700-BX913  & $ 2.2905 $ & $ 10.24 $ & $ -0.83_{-0.07}^{+0.09}$ & $0.57_{-0.05 }^{+0.06}$ & $8.43_{-0.04}^{+0.05}$ & $ 8.28_{-0.03}^{+0.03} $ & 1,6 \\
Q1700-BX917  & $ 2.3066 $ & $ 10.73 $ & $ -0.84_{-0.04}^{+0.05}$ & $0.47_{-0.06 }^{+0.07}$ & $8.42_{-0.02}^{+0.03}$ & $ 8.31_{-0.02}^{+0.03} $ & 1,3,4,6 \\
Q1700-BX951  & $ 2.3053 $ & $ 10.49 $ & $ -0.67_{-0.06}^{+0.07}$ & $0.22_{-0.05 }^{+0.06}$ & $8.52_{-0.04}^{+0.04}$ & $ 8.45_{-0.03}^{+0.03} $ & 1 \\
Q1700-BX984  & $ 2.2967 $ & $ 10.19 $ & $ -0.89_{-0.08}^{+0.10}$ & $0.03_{-0.08 }^{+0.10}$ & $8.39_{-0.04}^{+0.05}$ & $ 8.44_{-0.04}^{+0.04} $ & 1,3 \\
Q1700-MD109  & $ 2.2936 $ & $ ~9.88 $ & $ -0.97_{-0.15}^{+0.23}$ & $0.45_{-0.06 }^{+0.07}$ & $8.35_{-0.08}^{+0.13}$ & $ 8.28_{-0.08}^{+0.05} $ & 1,2,3,4,6 \\
Q1700-MD69  & $ 2.2881 $ & $ 11.21 $ & $ -0.47_{-0.03}^{+0.03}$ & $0.20_{-0.04 }^{+0.05}$ & $8.63_{-0.02}^{+0.02}$ & $ 8.52_{-0.02}^{+0.02} $ & 1,3,4 \\
Q1700-MD77  & $ 2.5078 $ & $ ~9.47 $ & $ -1.13_{-0.17}^{+0.28}$ & $0.78_{-0.05 }^{+0.06}$ & $8.26_{-0.10}^{+0.16}$ & $ 8.12_{-0.09}^{+0.06} $ & 1a \\
Q2206-BX140  & $ 2.3517 $ & $ 10.00 $ & $ -0.91_{-0.13}^{+0.19}$ & $0.66_{-0.06 }^{+0.07}$ & $8.38_{-0.08}^{+0.11}$ & $ 8.23_{-0.06}^{+0.05} $ & 1 \\
Q2206-BX145  & $ 2.2349 $ & $ ~9.45 $ & $ -0.75_{-0.14}^{+0.21}$ & $0.66_{-0.05 }^{+0.06}$ & $8.47_{-0.08}^{+0.12}$ & $ 8.28_{-0.07}^{+0.05} $ & 1 \\
Q2206-BX168  & $ 2.1966 $ & $ ~9.70 $ & $ -1.06_{-0.10}^{+0.13}$ & $0.50_{-0.17 }^{+0.27}$ & $8.30_{-0.06}^{+0.08}$ & $ 8.23_{-0.07}^{+0.09} $ & 1 \\
Q2206-BX189  & $ 2.0779 $ & $ 11.01 $ & $ -0.46_{-0.05}^{+0.05}$ & $0.27_{-0.09 }^{+0.12}$ & $8.64_{-0.03}^{+0.03}$ & $ 8.49_{-0.03}^{+0.04} $ & 1 \\
Q2206-BX191  & $ 2.1575 $ & $ 10.50 $ & $ -1.12_{-0.10}^{+0.13}$ & $0.72_{-0.09 }^{+0.11}$ & $8.26_{-0.06}^{+0.07}$ & $ 8.14_{-0.05}^{+0.05} $ & 1a \\
Q2206-BX88  & $ 2.1806 $ & $ 10.29 $ & $ -0.89_{-0.06}^{+0.07}$ & $0.67_{-0.05 }^{+0.05}$ & $8.39_{-0.03}^{+0.04}$ & $ 8.23_{-0.03}^{+0.03} $ & 1 \\
Q2343-BX182  & $ 2.2876 $ & $ ~9.81 $ & $ -1.12_{-0.07}^{+0.09}$ & $0.63_{-0.03 }^{+0.03}$ & $8.26_{-0.04}^{+0.05}$ & $ 8.17_{-0.03}^{+0.02} $ & 1,4,6 \\
Q2343-BX222  & $ 2.2872 $ & $ 10.63 $ & $ -0.56_{-0.05}^{+0.06}$ & $0.40_{-0.09 }^{+0.12}$ & $8.58_{-0.03}^{+0.03}$ & $ 8.43_{-0.04}^{+0.04} $ & 1 \\
Q2343-BX231  & $ 2.4989 $ & $ 10.11 $ & $ -0.58_{-0.02}^{+0.02}$ & $0.54_{-0.03 }^{+0.03}$ & $8.57_{-0.01}^{+0.01}$ & $ 8.37_{-0.01}^{+0.01} $ & 1 \\
Q2343-BX336  & $ 2.5445 $ & $ 10.12 $ & $ -0.86_{-0.05}^{+0.06}$ & $0.53_{-0.01 }^{+0.01}$ & $8.41_{-0.03}^{+0.04}$ & $ 8.28_{-0.02}^{+0.02} $ & 1,4,6 \\
Q2343-BX348  & $ 2.4491 $ & $ 10.49 $ & $ -0.62_{-0.02}^{+0.02}$ & $0.41_{-0.01 }^{+0.01}$ & $8.54_{-0.01}^{+0.01}$ & $ 8.40_{-0.01}^{+0.01} $ & 1 \\
Q2343-BX389  & $ 2.1711 $ & $ 10.97 $ & $ -0.75_{-0.04}^{+0.04}$ & $0.43_{-0.05 }^{+0.06}$ & $8.47_{-0.02}^{+0.02}$ & $ 8.35_{-0.02}^{+0.02} $ & 1,6,7 \\
Q2343-BX418  & $ 2.3054 $ & $ ~8.87 $ & $ -1.28_{-0.07}^{+0.08}$ & $0.81_{-0.02 }^{+0.02}$ & $8.17_{-0.04}^{+0.05}$ & $ 8.06_{-0.03}^{+0.02} $ & 1,4,5,6,8 \\
Q2343-BX442  & $ 2.1752 $ & $ 11.12 $ & $ -0.52_{-0.03}^{+0.03}$ & $-0.08_{-0.07 }^{+0.08}$ & $8.60_{-0.02}^{+0.02}$ & $ 8.59_{-0.02}^{+0.03} $ & 1,4,6,9 \\
Q2343-BX445  & $ 2.5445 $ & $ ~9.69 $ & $ -0.84_{-0.05}^{+0.06}$ & $0.68_{-0.01 }^{+0.01}$ & $8.42_{-0.03}^{+0.04}$ & $ 8.24_{-0.02}^{+0.02} $ & 1 \\
Q2343-BX460  & $ 2.3945 $ & $ ~9.18 $ & $ -1.37_{-0.12}^{+0.17}$ & $0.81_{-0.02 }^{+0.03}$ & $8.12_{-0.07}^{+0.09}$ & $ 8.03_{-0.05}^{+0.04} $ & 1 \\
Q2343-BX473  & $ 2.5437 $ & $ 10.33 $ & $ -1.18_{-0.11}^{+0.14}$ & $0.70_{-0.01 }^{+0.02}$ & $8.23_{-0.06}^{+0.08}$ & $ 8.13_{-0.05}^{+0.03} $ & 1 \\
Q2343-BX480  & $ 2.2316 $ & $ 10.17 $ & $ -1.03_{-0.09}^{+0.12}$ & $0.34_{-0.05 }^{+0.05}$ & $8.31_{-0.05}^{+0.07}$ & $ 8.29_{-0.04}^{+0.03} $ & 1,4,6 \\
Q2343-BX484  & $ 2.1874 $ & $ 10.36 $ & $ -0.88_{-0.14}^{+0.20}$ & $0.32_{-0.10 }^{+0.13}$ & $8.40_{-0.08}^{+0.11}$ & $ 8.35_{-0.07}^{+0.06} $ & 1 \\
Q2343-BX496  & $ 2.3934 $ & $ ~9.29 $ & $ -1.16_{-0.10}^{+0.12}$ & $0.70_{-0.05 }^{+0.06}$ & $8.24_{-0.05}^{+0.07}$ & $ 8.13_{-0.04}^{+0.04} $ & 1a \\
Q2343-BX537  & $ 2.3394 $ & $ ~9.59 $ & $ -1.19_{-0.12}^{+0.18}$ & $0.62_{-0.05 }^{+0.06}$ & $8.22_{-0.07}^{+0.10}$ & $ 8.15_{-0.06}^{+0.04} $ & 1,4,6 \\
Q2343-BX587  & $ 2.2427 $ & $ 10.18 $ & $ -0.67_{-0.02}^{+0.02}$ & $0.50_{-0.04 }^{+0.04}$ & $8.52_{-0.01}^{+0.01}$ & $ 8.35_{-0.01}^{+0.01} $ & 1,4 \\
Q2343-BX601  & $ 2.3768 $ & $ 10.47 $ & $ -0.89_{-0.04}^{+0.05}$ & $0.40_{-0.03 }^{+0.03}$ & $8.39_{-0.02}^{+0.03}$ & $ 8.32_{-0.02}^{+0.02} $ & 1,4,6 \\
Q2343-D29  & $ 2.3866 $ & $ ~9.83 $ & $ -0.65_{-0.05}^{+0.05}$ & $0.34_{-0.03 }^{+0.03}$ & $8.53_{-0.03}^{+0.03}$ & $ 8.41_{-0.02}^{+0.02} $ & 1 \\
Q2343-D35  & $ 2.3986 $ & $ 10.82 $ & $ -0.43_{-0.04}^{+0.05}$ & $0.04_{-0.05 }^{+0.05}$ & $8.66_{-0.02}^{+0.03}$ & $ 8.58_{-0.02}^{+0.02} $ & 1a \\
Q2343-MD86  & $ 2.3976 $ & $ ~9.41 $ & $ -1.02_{-0.16}^{+0.25}$ & $0.56_{-0.03 }^{+0.03}$ & $8.32_{-0.09}^{+0.14}$ & $ 8.22_{-0.08}^{+0.05} $ & 1 
\enddata
\tablenotetext{a}{Error bars are 1$\sigma$ based on measurement uncertainties only.}
\tablenotetext{b}{Oxygen abundance assuming the ``N2'' calibration of PP04.}
\tablenotetext{c}{Oxygen abundance assuming the ``O3N2' calibration of PP04.}
\tablenotetext{A1}{Object identified as an AGN on the basis of both rest-UV (LRIS-B) and rest-optical (MOSFIRE) spectra.}
\tablenotetext{A2}{Object identified as an AGN on the basis of near-IR (MOSFIRE) spectra.}
\tablenotetext{1}{Objects having optical (rest-UV) spectra obtained using Keck/LRIS-B; galaxies whose LRIS-B spectra yielded 
spectroscopic redshifts are marked ``1'', while ``1a'' denotes objects that were attempted spectroscopically in the rest-UV 
without yielding a secure redshift. }
\tablenotetext{}{References to other spectroscopic/photometric measurements: (2) \citet{erb03} (3) \citet{shapley05} (4) \citet{erb+06b} 
(5) \citet{law09} (6) \citet{steidel2010} (7) \citet{forst09} (8) \citet{erb2010} (9) \citet{law2012} (10) \citet{sep+04}} 
\label{tab:n2ha_and_o3n2cont}
\end{deluxetable*}

\begin{deluxetable*}{lccccccl}
\setcounter{table}{2}
\tabletypesize{\scriptsize}
\tablewidth{0pc}
\tablecaption{KBSS-MOSFIRE Galaxies with [OIII]/\Hb\ Measurements and [NII]/\Ha\ Limits\tablenotemark{a}}
\tablehead{
\colhead{Name} & \colhead{$z_{\rm neb}$} & \colhead{${\rm log~M_{\ast}}$} & 
\colhead{log([NII]/\Ha)} & \colhead{log([OIII]/\Hb)} &
 \colhead{${\rm 12+log(O/H)}$} & \colhead{${\rm 12+log(O/H)}$}   \\
\colhead{} & \colhead{} & \colhead{(${\rm M_{\odot}}$)} & 
\colhead{} & \colhead{} & \colhead{(N2)\tablenotemark{b}} & \colhead{(O3N2)\tablenotemark{c}} & \colhead{Notes}
}
\startdata
Q0100-BX167  & $ 2.2894 $ & $ ~9.39 $ & $ < -0.98$ & $0.47_{-0.06 }^{+0.06}$ & $ <8.34 $ & $ <8.27$ &  \\
Q0100-BX185  & $ 2.3659 $ & $ ~9.53 $ & $ < -0.82$ & $0.32_{-0.12 }^{+0.17}$ & $ <8.43 $ & $ <8.37$ &  \\
Q0100-BX187  & $ 2.2660 $ & $ ~9.03 $ & $ < -1.01$ & $0.80_{-0.07 }^{+0.08}$ & $ <8.32 $ & $ <8.15$ & 1 \\
Q0100-BX57  & $ 2.2706 $ & $ ~9.41 $ & $ < -0.86$ & $0.70_{-0.04 }^{+0.05}$ & $ <8.41 $ & $ <8.23$ & 1 \\
Q0105-BX163  & $ 2.2912 $ & $ 10.01 $ & $ < -1.05$ & $0.72_{-0.03 }^{+0.03}$ & $ <8.30 $ & $ <8.16$ & 1a \\
Q0105-BX49  & $ 2.1144 $ & $ ~9.33 $ & $ < -0.97$ & $0.51_{-0.07 }^{+0.08}$ & $ <8.35 $ & $ <8.26$ & 1 \\
Q0105-BX89  & $ 2.2278 $ & $ ~9.84 $ & $ < -1.38$ & $0.62_{-0.03 }^{+0.04}$ & $ <8.11 $ & $ <8.09$ & 1 \\
Q0105-MD12  & $ 2.5053 $ & $ ~9.55 $ & $ < -1.04$ & $0.67_{-0.02 }^{+0.02}$ & $ <8.31 $ & $ <8.18$ & 1a \\
Q0142-BX138  & $ 2.4177 $ & $ ~9.48 $ & $ < -0.94$ & $0.70_{-0.03 }^{+0.03}$ & $ <8.37 $ & $ <8.21$ & 1 \\
Q0142-BX182  & $ 2.3555 $ & $ 10.78 $ & $ < -0.92$ & $0.71_{-0.16 }^{+0.25}$ & $ <8.38 $ & $ <8.21$ & 1 \\
Q0142-BX186  & $ 2.3568 $ & $ ~8.79 $ & $ < -0.90$ & $1.06_{-0.08 }^{+0.09}$ & $ <8.39 $ & $ <8.11$ & A1,1 \\
Q0142-BX212  & $ 2.3781 $ & $ ~9.81 $ & $ < -0.81$ & $0.48_{-0.03 }^{+0.04}$ & $ <8.44 $ & $ <8.32$ & 1a \\
Q0207-BX119  & $ 2.0588 $ & $ 10.28 $ & $ < -1.23$ & $0.54_{-0.02 }^{+0.02}$ & $ <8.20 $ & $ <8.16$ & 1 \\
Q0207-BX144  & $ 2.1682 $ & $ ~8.88 $ & $ < -1.50$ & $0.78_{-0.03 }^{+0.03}$ & $ <8.05 $ & $ <8.00$ & 1 \\
Q0207-BX211  & $ 2.5468 $ & $ ~9.71 $ & $ < -1.10$ & $0.46_{-0.08 }^{+0.10}$ & $ <8.27 $ & $ <8.23$ &  \\
Q0207-BX243  & $ 2.0385 $ & $ ~9.61 $ & $ < -0.90$ & $0.75_{-0.04 }^{+0.04}$ & $ <8.39 $ & $ <8.20$ & 1 \\
Q0449-BX138  & $ 2.3934 $ & $ ~9.86 $ & $ < -0.83$ & $0.89_{-0.10 }^{+0.12}$ & $ <8.43 $ & $ <8.18$ & 1 \\
Q0821-BX221  & $ 2.3958 $ & $ ~9.76 $ & $ < -1.40$ & $0.78_{-0.02 }^{+0.02}$ & $ <8.10 $ & $ <8.03$ & 1 \\
Q0821-BX52  & $ 2.1767 $ & $ 10.56 $ & $ < -1.08$ & $0.74_{-0.08 }^{+0.09}$ & $ <8.29 $ & $ <8.15$ &  \\
Q0821-BX61  & $ 2.3526 $ & $ ~9.87 $ & $ < -0.89$ & $0.61_{-0.10 }^{+0.13}$ & $ <8.39 $ & $ <8.25$ &  \\
Q0821-BX92  & $ 2.4163 $ & $ ~9.21 $ & $ < -1.02$ & $0.59_{-0.04 }^{+0.04}$ & $ <8.32 $ & $ <8.22$ & 1 \\
Q0821-MD5  & $ 2.5367 $ & $ ~9.88 $ & $ < -0.95$ & $0.68_{-0.03 }^{+0.03}$ & $ <8.36 $ & $ <8.21$ & 1 \\
Q1009-BX155  & $ 2.1448 $ & $ ~9.74 $ & $ < -1.23$ & $0.55_{-0.06 }^{+0.06}$ & $ <8.20 $ & $ <8.16$ & 1 \\
Q1009-BX177  & $ 2.0949 $ & $ ~9.09 $ & $ < -1.25$ & $0.70_{-0.13 }^{+0.19}$ & $ <8.19 $ & $ <8.11$ & 1 \\
Q1217-BX220  & $ 2.3225 $ & $ ~9.79 $ & $ < -1.44$ & $0.71_{-0.03 }^{+0.03}$ & $ <8.08 $ & $ <8.04$ & 1 \\
Q1442-BX138  & $ 2.4336 $ & $ ~9.62 $ & $ < -1.19$ & $0.59_{-0.04 }^{+0.04}$ & $ <8.22 $ & $ <8.16$ & 1 \\
Q1442-BX199  & $ 2.2938 $ & $ ~9.29 $ & $ < -1.13$ & $0.69_{-0.03 }^{+0.03}$ & $ <8.25 $ & $ <8.15$ & 1 \\
Q1442-BX290  & $ 2.4318 $ & $ ~9.69 $ & $ < -1.19$ & $0.61_{-0.04 }^{+0.05}$ & $ <8.22 $ & $ <8.15$ & 1 \\
Q1442-BX295  & $ 2.4514 $ & $ ~9.35 $ & $ < -1.04$ & $0.62_{-0.03 }^{+0.03}$ & $ <8.31 $ & $ <8.20$ &  \\
Q1442-BX305  & $ 2.5165 $ & $ ~9.71 $ & $ < -0.88$ & $0.79_{-0.04 }^{+0.04}$ & $ <8.40 $ & $ <8.20$ & 1 \\
Q1442-BX346  & $ 2.4473 $ & $ ~9.43 $ & $ < -1.08$ & $0.72_{-0.07 }^{+0.08}$ & $ <8.28 $ & $ <8.15$ & 1 \\
Q1549-BX102  & $ 2.1934 $ & $ ~9.43 $ & $ < -1.11$ & $0.72_{-0.02 }^{+0.02}$ & $ <8.27 $ & $ <8.14$ & 1 \\
Q1549-BX121  & $ 2.4983 $ & $ ~8.73 $ & $ < -0.92$ & $0.59_{-0.08 }^{+0.10}$ & $ <8.38 $ & $ <8.25$ & 1 \\
Q1549-BX170  & $ 2.3836 $ & $ ~9.77 $ & $ < -1.43$ & $0.73_{-0.04 }^{+0.04}$ & $ <8.09 $ & $ <8.04$ & 1 \\
Q1549-MD18  & $ 2.5116 $ & $ 10.00 $ & $ < -0.87$ & $0.76_{-0.14 }^{+0.21}$ & $ <8.41 $ & $ <8.21$ & 1 \\
Q1603-BX389  & $ 2.4266 $ & $ 10.66 $ & $ < -1.26$ & $0.61_{-0.02 }^{+0.02}$ & $ <8.18 $ & $ <8.13$ & 1 \\
Q1603-BX55  & $ 2.3706 $ & $ ~9.42 $ & $ < -1.24$ & $0.66_{-0.03 }^{+0.03}$ & $ <8.19 $ & $ <8.12$ & 1 \\
Q1603-MD16  & $ 2.5475 $ & $ 10.34 $ & $ < -0.97$ & $0.43_{-0.13 }^{+0.20}$ & $ <8.35 $ & $ <8.28$ & 1 \\
Q1623-BX431  & $ 2.1127 $ & $ ~9.15 $ & $ < -0.95$ & $0.54_{-0.04 }^{+0.04}$ & $ <8.36 $ & $ <8.25$ & 1 \\
Q1623-BX432  & $ 2.1824 $ & $ 10.02 $ & $ < -1.44$ & $0.67_{-0.02 }^{+0.03}$ & $ <8.08 $ & $ <8.06$ & 1,2,4,6 \\
Q1623-BX469  & $ 2.5499 $ & $ ~9.32 $ & $ < -0.98$ & $0.58_{-0.05 }^{+0.05}$ & $ <8.34 $ & $ <8.23$ & 1 \\
Q1623-MD127  & $ 2.4592 $ & $ ~9.94 $ & $ < -0.80$ & $0.43_{-0.04 }^{+0.05}$ & $ <8.44 $ & $ <8.34$ & 1 \\
Q1700-BX609  & $ 2.5697 $ & $ ~9.64 $ & $ < -0.96$ & $0.50_{-0.07 }^{+0.09}$ & $ <8.35 $ & $ <8.26$ & 1,3 \\
Q1700-BX717  & $ 2.4358 $ & $ ~9.47 $ & $ < -1.05$ & $0.62_{-0.10 }^{+0.13}$ & $ <8.30 $ & $ <8.19$ & 1,2,3,4,6 \\
Q1700-BX772  & $ 2.3416 $ & $ ~9.72 $ & $ < -1.11$ & $0.65_{-0.08 }^{+0.10}$ & $ <8.27 $ & $ <8.17$ & 1,3 \\
Q2206-BM64  & $ 2.1942 $ & $ ~9.34 $ & $ < -1.19$ & $0.73_{-0.04 }^{+0.05}$ & $ <8.22 $ & $ <8.11$ & 1 \\
Q2206-BX128  & $ 2.3484 $ & $ 10.04 $ & $ < -0.95$ & $0.48_{-0.08 }^{+0.10}$ & $ <8.36 $ & $ <8.27$ & 1a \\
Q2343-BX236  & $ 2.4341 $ & $ 10.50 $ & $ < -0.81$ & $0.41_{-0.15 }^{+0.23}$ & $ <8.44 $ & $ <8.34$ & 1,4,6 \\
Q2343-BX660  & $ 2.1742 $ & $ ~8.97 $ & $ < -1.61$ & $0.81_{-0.02 }^{+0.02}$ & $ <7.99 $ & $ <7.96$ & 1,4,5,6 \\
Q2343-D34  & $ 2.4538 $ & $ 10.11 $ & $ < -1.02$ & $0.67_{-0.04 }^{+0.04}$ & $ <8.32 $ & $ <8.19$ & 1 \\
Q2343-MD37  & $ 2.4246 $ & $ ~9.94 $ & $ < -0.97$ & $0.56_{-0.03 }^{+0.04}$ & $ <8.34 $ & $ <8.24$ & 1 
\enddata
\tablenotetext{a}{Upper limits on log([NII]/\Ha) are 2$\sigma$; error bars are otherwise 1$\sigma$ based on measurement
errors only.}
\tablenotetext{b}{2$\sigma$ upper limit on oxygen abundance assuming the ``N2'' calibration of PP04.}
\tablenotetext{c}{2$\sigma$ upper limit on oxygen abundance assuming the ``O3N2' calibration of PP04.}
\tablenotetext{A1}{Object identified as an AGN on the basis of both rest-UV (LRIS-B) and rest-optical (MOSFIRE) spectra.}
\tablenotetext{A2}{Object identified as an AGN on the basis of near-IR (MOSFIRE) spectra.}
\tablenotetext{1}{Objects with optical (rest-UV) spectra obtained using Keck/LRIS-B; galaxies whose LRIS-B spectra yielded 
spectroscopic redshifts are marked ``1'', while ``1a'' denotes objects that were observed in the rest-UV 
but did not yield a secure spectroscopic redshift. }
\tablenotetext{}{References to other spectroscopic/photometric measurements: (2) \citet{erb03} (3) \citet{shapley05} (4) \citet{erb+06b} 
(5) \citet{law09} (6) \citet{steidel2010} (7) \citet{forst09} (8) \citet{erb2010} (9) \citet{law2012} (10) \citet{sep+04}} 
\label{tab:n2ha_lim_o3hb_det}
\end{deluxetable*}

\begin{deluxetable*}{lccccl}
\setcounter{table}{3}
\tabletypesize{\scriptsize}
\tablewidth{0pc}
\tablecaption{KBSS-MOSFIRE Galaxies with [NII]/\Ha\ and Stellar Mass Measurements}
\tablehead{
\colhead{Name} & \colhead{$z_{\rm neb}$} & \colhead{log~M$_{\ast}$} & 
\colhead{log([NII]/\Ha)} & \colhead{${\rm 12+log(O/H)}$\tablenotemark{b}} \\
\colhead{} & \colhead{} & \colhead{${\rm M_{\odot}}$} & 
\colhead{} & \colhead{(N2 PP04)} & 
\colhead{Notes}
}
\startdata
Q0100-BX53  & $ 2.0009 $ & $ ~9.87 $ & $ -1.19_{-0.12 }^{+0.17}$ & $8.22_{-0.07}^{+0.10} $ & 1a \\
Q0100-MD37  & $ 2.3899 $ & $ 11.18 $ & $ -0.39_{-0.07 }^{+0.08}$ & $8.68_{-0.04}^{+0.05} $ &  \\
Q0105-BX52  & $ 1.9717 $ & $ 10.64 $ & $ -0.50_{-0.07 }^{+0.09}$ & $8.62_{-0.04}^{+0.05} $ & 1 \\
Q0105-BX93  & $ 2.0301 $ & $ 10.06 $ & $ -0.98_{-0.04 }^{+0.05}$ & $8.34_{-0.02}^{+0.03} $ & 1 \\
Q0105-BX95  & $ 2.0304 $ & $ 10.38 $ & $ -0.72_{-0.03 }^{+0.04}$ & $8.49_{-0.02}^{+0.02} $ & 1,5 \\
Q0142-BX116  & $ 2.1131 $ & $ 10.31 $ & $ -0.19_{-0.10 }^{+0.13}$ & $8.79_{-0.06}^{+0.08} $ & 1 \\
Q0142-BX119  & $ 2.2237 $ & $ 10.84 $ & $ -0.68_{-0.09 }^{+0.11}$ & $8.51_{-0.05}^{+0.06} $ & 1 \\
Q0142-BX120  & $ 2.2241 $ & $ ~9.96 $ & $ -1.05_{-0.10 }^{+0.12}$ & $8.30_{-0.05}^{+0.07} $ &  \\
Q0142-BX241  & $ 2.2843 $ & $ ~9.82 $ & $ -0.82_{-0.07 }^{+0.08}$ & $8.43_{-0.04}^{+0.04} $ & 1 \\
Q0142-BX248  & $ 2.4980 $ & $ ~9.64 $ & $ -0.76_{-0.13 }^{+0.19}$ & $8.47_{-0.07}^{+0.11} $ & 1 \\
Q0142-BX61  & $ 2.0702 $ & $ ~9.62 $ & $ -0.53_{-0.08 }^{+0.10}$ & $8.60_{-0.05}^{+0.06} $ & 1 \\
Q0821-RK5  & $ 2.1831 $ & $ 11.80 $ & $ +0.02_{-0.04 }^{+0.05}$ & $8.91_{-0.03}^{+0.03} $ & A2 \\
Q1009-BX93  & $ 2.0450 $ & $ ~9.44 $ & $ -0.74_{-0.13 }^{+0.19}$ & $8.48_{-0.07}^{+0.11} $ &  \\
Q1442-BX159  & $ 1.9957 $ & $ ~9.97 $ & $ -0.76_{-0.07 }^{+0.08}$ & $8.47_{-0.04}^{+0.05} $ & 1 \\
Q1442-BX317  & $ 2.0273 $ & $ ~9.71 $ & $ -1.18_{-0.09 }^{+0.11}$ & $8.23_{-0.05}^{+0.07} $ & 1a \\
Q1549-BX42  & $ 2.2194 $ & $ ~9.08 $ & $ -0.57_{-0.09 }^{+0.12}$ & $8.57_{-0.05}^{+0.07} $ & 1 \\
Q1549-BX94  & $ 2.0074 $ & $ 10.49 $ & $ -0.72_{-0.08 }^{+0.10}$ & $8.49_{-0.05}^{+0.06} $ & 1 \\
Q1603-BX127  & $ 2.5521 $ & $ ~9.59 $ & $ -0.83_{-0.13 }^{+0.18}$ & $8.43_{-0.07}^{+0.10} $ & 1 \\
Q1623-BX410  & $ 2.5395 $ & $ 10.49 $ & $ -0.57_{-0.09 }^{+0.11}$ & $8.57_{-0.05}^{+0.06} $ & 1 \\
Q1700-BX475  & $ 2.4027 $ & $ ~9.86 $ & $ -0.96_{-0.11 }^{+0.14}$ & $8.35_{-0.06}^{+0.08} $ & 1a \\
Q1700-BX535  & $ 2.6366 $ & $ ~9.69 $ & $ -0.65_{-0.11 }^{+0.15}$ & $8.53_{-0.06}^{+0.08} $ & 1a,3 \\
Q1700-BX684  & $ 2.2922 $ & $ ~8.89 $ & $ -0.69_{-0.09 }^{+0.11}$ & $8.51_{-0.05}^{+0.06} $ &  \\
Q1700-BX801  & $ 2.0380 $ & $ 10.22 $ & $ -0.44_{-0.10 }^{+0.13}$ & $8.65_{-0.06}^{+0.08} $ & 1 \\
Q1700-BX810  & $ 2.2923 $ & $ 10.08 $ & $ -0.87_{-0.10 }^{+0.13}$ & $8.41_{-0.06}^{+0.08} $ & 1,3 \\
Q1700-BX909  & $ 2.2934 $ & $ 10.65 $ & $ -0.92_{-0.08 }^{+0.10}$ & $8.38_{-0.05}^{+0.06} $ & 1,6 \\
Q1700-BX939  & $ 2.2971 $ & $ ~9.75 $ & $ -0.97_{-0.09 }^{+0.11}$ & $8.35_{-0.05}^{+0.06} $ & 1,3 \\
Q2206-BX102  & $ 2.2099 $ & $ 11.19 $ & $ -0.29_{-0.03 }^{+0.03}$ & $8.74_{-0.02}^{+0.02} $ & 1,6 \\
Q2206-BX166  & $ 1.9742 $ & $ ~9.90 $ & $ -0.95_{-0.11 }^{+0.16}$ & $8.36_{-0.07}^{+0.09} $ & 1 \\
Q2206-BX169  & $ 2.0960 $ & $ 10.88 $ & $ -0.56_{-0.07 }^{+0.09}$ & $8.58_{-0.04}^{+0.05} $ &  \\
Q2206-BX68  & $ 2.0971 $ & $ 10.49 $ & $ -0.84_{-0.07 }^{+0.09}$ & $8.42_{-0.04}^{+0.05} $ &  \\
Q2343-BX390  & $ 2.2311 $ & $ ~9.90 $ & $ -0.89_{-0.06 }^{+0.07}$ & $8.39_{-0.04}^{+0.04} $ & 1,4,6 \\
Q2343-D25  & $ 2.1865 $ & $ ~9.50 $ & $ -0.89_{-0.09 }^{+0.12}$ & $8.39_{-0.05}^{+0.07} $ & 1  
\enddata
\tablenotetext{a}{Error bars are 1$\sigma$ based on measurement uncertainties only.}
\tablenotetext{b}{Oxygen abundance assuming the ``N2'' calibration of PP04.}
\tablenotetext{A1}{Object identified as an AGN on the basis of both rest-UV (LRIS-B) and rest-optical (MOSFIRE) spectra.}
\tablenotetext{A2}{Object identified as an AGN on the basis of near-IR (MOSFIRE) spectra.}
\tablenotetext{1}{Objects with optical (rest-UV) spectra obtained using Keck/LRIS-B; galaxies whose LRIS-B spectra yielded 
spectroscopic redshifts are marked ``1'', while ``1a'' denotes objects that were observed in the rest-UV 
but did not yield a secure spectroscopic redshift. }
\tablenotetext{}{References to other spectroscopic/photometric measurements: (2) \citet{erb03} (3) \citet{shapley05} (4) \citet{erb+06b} 
(5) \citet{law09} (6) \citet{steidel2010} (7) \citet{forst09} (8) \citet{erb2010} (9) \citet{law2012} (10) \citet{sep+04}} 
\label{tab:n2ha}
\end{deluxetable*}

\begin{deluxetable*}{lcccccccc}
\setcounter{table}{4}
\tabletypesize{\scriptsize}
\tablewidth{0pc}
\tablecaption{Properties of ``Extreme'' KBSS-MOSFIRE Galaxies}
\tablehead{
\colhead{Object} & 
\colhead{[OII] Ratio\tablenotemark{a}} & 
\colhead{$n_{\rm e}$\tablenotemark{b}} & 
\colhead{[OIII] Ratio\tablenotemark{c}} &
\colhead{${\rm H\alpha/H\beta}$} &
\colhead{${\rm T_e (K)}$} &
\colhead{${\rm O32}$\tablenotemark{d}} &
\colhead{${\rm [OIII]_{tot}/\Hb}$ \tablenotemark{e} } &
\colhead{${\rm 12+log(O/H)}$\tablenotemark{f}} 
}
\startdata
Q0207-BX74\tablenotemark{g}  & $1.68\pm 0.11$ & $1575_{-200}^{+300}$ & $0.037\pm0.005$ & $3.46\pm0.25$ & $14300\pm 400$ & $~8.23\pm 0.41$ & $10.06\pm0.45$ & $8.00\pm0.05$ \\ 
Q2343-BX418 & $1.13\pm0.05$ & $580_{-70}^{+80}$ & $0.023\pm0.003$ & $2.81\pm0.20$ & $12830\pm500$ & $~9.66\pm0.38$ & $~8.67\pm0.42$ & $8.08\pm0.05$ \\ 
Q2343-BX660 & $0.93\pm0.04$ & $300_{-40}^{+40}$ & $0.022\pm0.004$ & $2.77\pm0.20$ & $12650\pm500$ & $10.98\pm0.50$ & $~9.58\pm0.45$ & $8.13\pm0.06$  \\
\enddata
\tablenotetext{a}{Intensity ratio ${\rm [OII]\lambda 3726/[OII]\lambda3729}$.}
\tablenotetext{b}{Electron density in cm$^{-3}$ determined from the intensity ratio of the [OII] doublet.}
\tablenotetext{c}{Measured intensity ratio ${\rm OIII](\lambda1661+\lambda1666)/[OIII]\lambda 5008}$. }
\tablenotetext{d}{Ratio ${\rm [OIII](\lambda 4960+\lambda5008)/[OII](\lambda 3726+\lambda3729})$. }
\tablenotetext{e}{Ratio of [OIII]($\lambda 4960+\lambda5008$)/\Hb. }
\tablenotetext{f}{Inferred oxygen abundance from the direct ${\rm T_e}$ method.} 
\tablenotetext{g}{Line intensity ratios (other than \Ha/\Hb) corrected for nebular extinction assuming ${\rm E(B-V)_{neb}=0.18}$ 
and the \citet{cardelli89} attenuation relation.}
\label{tab:direct}
\end{deluxetable*}

\begin{deluxetable*}{cccccc}
\setcounter{table}{5}
\tabletypesize{\scriptsize}
\tablewidth{0pc}
\tablecaption{$z \sim 2.3$ Binned Mass-Metallicity Relation (N2)}
\tablehead{
\colhead{Bin Range} & \colhead{Median} & \colhead{${\rm N_{gal}}$} & \colhead{${\rm 12+log(O/H)_{N2}}$\tablenotemark{a}} 
& \colhead{${\rm 12+log(O/H)_{N2}}$\tablenotemark{b}} & \colhead{${\rm 12+log(O/H)_{N2}}$\tablenotemark{c}}   \\
\colhead{${\rm log(M_{\ast}/M_{\odot})}$} & \colhead{${\rm log(M_{\ast}/M_{\odot})}$} & \colhead{} & \colhead{(Mean)} & \colhead{(Median)} & \colhead{(Stack)} 
}
\startdata
      $~8.60-~9.00$ & $~8.87$  &    ~8    &  $8.20\pm   0.10$ &  8.21 & $8.13_{-0.06}^{+0.05}$ \\ 
      $~9.00-~9.50$ & $~9.34$  &    35    &  $8.23\pm   0.03$ &  8.20 & $8.25_{-0.07}^{+0.03}$  \\
      $~9.50-~9.80$ & $~9.69$  &    48    &  $8.31\pm   0.03$ &  8.33 & $8.30_{-0.02}^{+0.03}$  \\
      $~9.80-10.00$ & $~9.87$  &    34    &  $8.35\pm   0.02$ &  8.35 & $8.36_{-0.03}^{+0.04}$  \\
      $10.00-10.25$ & $10.11$  &    39    &  $8.38\pm   0.02$ &  8.39 & $8.42_{-0.04}^{+0.02}$  \\
      $10.25-10.50$ & $10.37$  &    32    &  $8.47\pm   0.03$ &  8.49 & $8.49_{-0.04}^{+0.03}$  \\
      $10.50-11.00$ & $10.66$  &    39    &  $8.51\pm   0.03$ &  8.54 & $8.53_{-0.02}^{+0.03}$  \\
      $11.00-11.60$ & $11.19$  &    ~8    &  $8.65\pm   0.02$ &  8.67 & $8.67_{-0.09}^{+0.09}$   
\enddata
\tablenotetext{a}{Bi-weight mean of individual measurements in bins of ${\rm M_{\ast}}$, determined using the PP04 N2 calibration;
y-axis error bars are uncertainties in the weighted mean within each bin, while x-axis error bars reflect the range
of ${\rm M_{\ast}}$ within each bin.} 
\tablenotetext{b}{Median inferred N2-based oxygen abundance in bin.} 
\tablenotetext{c}{Oxygen abundance inferred using PP04 N2, from spectral stacks within each bin of ${\rm M_{\ast}}$; error bars 
reflect both formal measurement uncertainties and sample variance within each mass bin. }
\label{tab:mass_met_bins_n2}
\end{deluxetable*}

\begin{deluxetable*}{cccccc}
\setcounter{table}{6}
\tabletypesize{\scriptsize}
\tablewidth{0pc}
\tablecaption{$z \sim 2.3$ Binned Mass-Metallicity Relation (O3N2)}
\tablehead{
\colhead{Bin Range} & \colhead{Median} & \colhead{${\rm N_{gal}}$} & \colhead{${\rm 12+log(O/H)_{O3N2}}$\tablenotemark{a}} 
& \colhead{${\rm 12+log(O/H)_{O3N2}}$\tablenotemark{b}} & \colhead{${\rm 12+log(O/H)_{O3N2}}$\tablenotemark{c}}   \\
\colhead{${\rm log(M_{\ast}/M_{\odot})}$} & \colhead{${\rm log(M_{\ast}/M_{\odot})}$} & \colhead{} & \colhead{(Mean)} & \colhead{(Median)} & \colhead{(Stack)} 
}
\startdata
      $~8.60-~9.00$ & $~8.87$  &    ~8    &  $8.10\pm   0.08$ &  8.09 & $8.06 \pm 0.03$  \\ 
      $~9.00-~9.50$ & $~9.34$  &    35    &  $8.14\pm   0.02$ &  8.13 & $8.16 \pm 0.02$  \\
      $~9.50-~9.80$ & $~9.69$  &    48    &  $8.21\pm   0.02$ &  8.23 & $8.20 \pm 0.01$  \\
      $~9.80-10.00$ & $~9.87$  &    34    &  $8.23\pm   0.02$ &  8.24 & $8.21 \pm 0.01$  \\
      $10.00-10.25$ & $10.11$  &    39    &  $8.27\pm   0.02$ &  8.27 & $8.28 \pm 0.01$  \\
      $10.25-10.50$ & $10.37$  &    32    &  $8.35\pm   0.02$ &  8.37 & $8.35 \pm 0.01$  \\
      $10.50-11.00$ & $10.66$  &    39    &  $8.37\pm   0.02$ &  8.42 & $8.39 \pm 0.01$  \\
      $11.00-11.60$ & $11.19$  &    ~8    &  $8.52\pm   0.02$ &  8.52 & $8.55 \pm 0.03$ 
\enddata
\tablenotetext{a}{Bi-weight mean of individual measurements in bins of ${\rm M_{\ast}}$, determined using the PP04 O3N2 calibration; 
y-axis error bars are uncertainties in the weighted mean within each bin, with x-axis error bars reflecting the range of ${\rm M_{\ast}}$ within
each bin.}
\tablenotetext{b}{Median inferred O3N2-based oxygen abundance in bin.} 
\tablenotetext{c}{Oxygen abundance inferred using PP04 O3N2, from spectral stacks within each bin of ${\rm M_{\ast}}$; y-axis error bars 
reflect formal measurement uncertainties only. }
\label{tab:mass_met_bins_o3n2}
\end{deluxetable*}

\end{document}